\documentclass{elsart}
\usepackage{epsf}
\usepackage{epsfig}
\usepackage{amssymb}\usepackage{latexsym}\usepackage{amsfonts}
\usepackage{longtable}
\usepackage{Abbreviations}
\begin{document}
\setcounter{tocdepth}{4}
\setcounter{secnumdepth}{4}
\renewcommand{\theequation}{\arabic{section}.\arabic{equation}}

\begin{frontmatter}
\title{New Developments in the Casimir Effect}
\author{M. Bordag\thanksref{eb}}, \author{U. Mohideen\thanksref{emo}},
\author{V.M. Mostepanenko\thanksref{emost}} 
\address[eb]{Institute for Theoretical Physics, Leipzig University,\\
Augustusplatz 10/11, 04109, Leipzig, Germany} 
\address[emo]{Department of Physics, University of California, \\
Riverside, California 92521, USA} 
\address[emost]{Department of Physics, Federal University of Paraiba, \\
Caixa Postal 5008, CEP 58059--970, Jo\~{a}o Pessoa, Para\'{\i}ba, Brazil\\
(on leave from A.~Friedmann Laboratory for Theoretical Physics, \\
St.Petersburg, Russia)}
\begin{abstract}
We provide a review of both new experimental and theoretical developments
in the Casimir effect. The Casimir effect results from the alteration by
the boundaries of the zero-point electromagnetic energy. Unique to the
Casimir force is its strong dependence on shape, switching from
attractive to repulsive as function of the size, geometry and topology
of the boundary. Thus the Casimir force is a direct manifestation of
the boundary dependence of quantum vacuum. We discuss in depth the 
general structure of the infinities in the field theory which are removed 
by a combination of zeta-functional regularization and heat kernel expansion.
Different representations for the regularized vacuum energy are given.
The Casimir energies and forces in a number of configurations of
interest to applications are calculated. We stress the development 
of the Casimir force for real media including effects of nonzero
temperature, finite conductivity of the boundary metal and surface
roughness. Also the combined effect of these important factors is 
investigated in detail on the basis of condensed matter physics and quantum
field theory at nonzero temperature. The experiments on measuring
the Casimir force are also reviewed, starting first with the older
measurements
and finishing with a detailed presentation of modern precision experiments.
The latter are accurately compared with the theoretical results for real
media. At the end of the review we provide the most recent constraints on
the corrections
to Newtonian gravitational law and other hypothetical long-range
interactions at submillimeter range obtained from the Casimir
force measurements.
\end{abstract}

\end{frontmatter}  
\begin{flushleft}
Corresponding author: Prof. U.~Mohideen, Department of Physics, 
University of California, 
Riverside, California 92521, USA\\
E-mail: Umar.Mohideen@ucr.edu\\
Fax: 1(909)787--4529
\end{flushleft}

\setcounter{page}{2}
\noindent
{\normalsize
\begin{tabular}{llllr}
\multicolumn{5}{l}{{\bf Contents}}\\[1mm]
1 & \multicolumn{3}{l}{Introduction} & \pageref{sec1}\\
&1.1 &  \multicolumn{2}{l}
{Zero-point oscillations and their manifestation}& 
\pageref{sec1.1}\\                        
&1.2 &  \multicolumn{2}{l}
{The Casimir effect as a macroscopic quantum effect} &\pageref{sec1.2} \\
&1.3 &  \multicolumn{2}{l}
{The role of the Casimir effect in different fields of physics} & 
\pageref{sec1.3}\\
&1.4 &  \multicolumn{2}{l}
{What has been accomplished during the last years?} & \pageref{sec1.4}\\
&1.5 &   \multicolumn{2}{l}
{The structure of the review} &\pageref{sec1.5}\\
2 & \multicolumn{3}{l} {The Casimir effect in simple models}&
\pageref{sec2}   \\
&2.1 &  \multicolumn{2}{l}
{Quantized scalar field on an interval} & \pageref{sec2.1}\\
&2.2 &  \multicolumn{2}{l}
{Parallel conducting planes} &\pageref{sec2.2}\\
&2.3 &  \multicolumn{2}{l}
{One- and two-dimensional spaces with nontrivial topologies} &
\pageref{sec2.3}\\
&2.4 &  \multicolumn{2}{l}
{Moving boundaries in a two-dimensional space-time} &\pageref{sec2.4}\\
3&  \multicolumn{3}{l}
 {Regularization and renormalization of the vacuum energy} & 
\pageref{sec3}\\
&3.1 &  \multicolumn{2}{l}
{Representation of the regularized vacuum energy} & \pageref{sec3.1}\\
&&3.1.1 & Background depending on one Cartesian coordinate &\pageref{sec3.1.1}
 \\
&&3.1.2 & Spherically symmetric background & \pageref{sec3.1.2}
 \\
&3.2 & \multicolumn{2}{l}
{The heat  kernel expansion} & \pageref{sec3.2}\\
&3.3 & \multicolumn{2}{l}
{The divergent part of the ground state energy} & \pageref{sec3.3}\\
&3.4 & \multicolumn{2}{l}
{Renormalization and normalization conditions} & \pageref{sec3.4}\\
&3.5 & \multicolumn{2}{l}
{The photon propagator with boundary conditions} & \pageref{sec3.5}\\
&&3.5.1 & Quantization in the presence of boundary conditions & 
\pageref{sec3.5.1}\\
&&3.5.2 & The photon propagator & \pageref{sec3.5.2}\\
&&3.5.3 & The photon propagator in plane parallel geometry &
\pageref{sec3.5.3}\\
4 &  \multicolumn{3}{l}{Casimir effect in various configurations} 
& \pageref{sec4}\\
&4.1 & \multicolumn{2}{l}
{Flat boundaries} & \pageref{sec4.1}\\
&&4.1.1&Two semispaces and stratified media  & \pageref{sec4.1.1}\\
&&4.1.2&Rectangular cavities: attractive or repulsive force?  
& \pageref{sec4.1.2}\\
&4.2 & \multicolumn{2}{l}
{Spherical and cylindrical boundaries} & \pageref{sec4.2}\\
&&4.2.1 &Boundary conditions on a sphere & \pageref{sec4.2.1}\\
&&4.2.2 & Analytical continuation of the regularized ground state energy
& \pageref{sec4.2.2}
\end{tabular}
\begin{tabular}{llllr}
&&4.2.3 & Results on the Casimir effect on a sphere& \pageref{sec4.2.3}\\
&&4.2.4 & The Casimir effect for a cylinder& \pageref{sec4.2.4}\\
&4.3 & \multicolumn{2}{l}
{Sphere (lens) above a disk: additive methods and proximity
forces} & \pageref{sec4.3}\\
&4.4 & \multicolumn{2}{l}
{Dynamical Casimir effect} & \pageref{sec4.4}\\
&4.5 & \multicolumn{2}{l}
{Radiative corrections to the Casimir effect} & \pageref{sec4.5}\\
&4.6 & \multicolumn{2}{l}
{Spaces with non-Euclidean topology} & \pageref{sec4.6}\\
&&4.6.1 & Cosmological models& \pageref{sec4.6.1}\\
&&4.6.2 & Vacuum interaction between cosmic strings& \pageref{sec4.6.2}\\
&&4.6.3 & Kaluza-Klein compactification of extra dimensions 
& \pageref{sec4.6.3}\\
5 &  \multicolumn{3}{l}{Casimir effect for real media} 
& \pageref{sec5}\\
&5.1 & \multicolumn{2}{l}
{The Casimir effect at nonzero temperature} & \pageref{sec5.1}\\
&&5.1.1 & Two semispaces& \pageref{sec5.1.1}\\
&&5.1.2 & A sphere (lens) above a disk& \pageref{sec5.1.2}\\
&&5.1.3 & The asymptotics of the Casimir force at high and low temperature
& \pageref{sec5.1.3}\\
&5.2 & \multicolumn{2}{l}
{Finite conductivity corrections} & \pageref{sec5.2}\\
&&5.2.1 & Plasma model approach for two semispaces  & \pageref{sec5.2.1}\\
&&5.2.2 &Plasma model approach for a sphere (lens) above a disk   
& \pageref{sec5.2.2}\\
&&5.2.3 & Computational results using the optical tabulated data  
& \pageref{sec5.2.3}\\
&5.3 & \multicolumn{2}{l}
{Roughness corrections} & \pageref{sec5.3}\\
&&5.3.1 & Expansion in powers of relative distortion amplitude:& \\
&&& two semispaces &\pageref{sec5.3.1}\\
&&5.3.2 & Casimir force between nonparallel plates and plates covered &\\
&&& by large scale roughness& \pageref{sec5.3.2}\\
&&5.3.3 & Casimir force between plates covered by short scale roughness
& \pageref{sec5.3.3}\\
&&5.3.4 &Expansion in powers of relative distortion amplitude: &\\
&&& a spherical lens above a plate & \pageref{sec5.3.4}\\
&&5.3.5 & Corrections to the Casimir force between a plate and a lens &\\
&&& due to different kinds of roughness& \pageref{sec5.3.5}\\
&&5.3.6 & Stochastic roughness& \pageref{sec5.3.6}
\end{tabular}
\begin{tabular}{llllr}
&5.4 & \multicolumn{2}{l}
{Combined effect of different corrections} & \pageref{sec5.4}\\
&&5.4.1 & Roughness and conductivity& \pageref{sec5.4.1}\\
&&5.4.2 & Conductivity and temperature: two semispaces& \pageref{sec5.4.2}\\
&&5.4.3 &Conductivity and temperature: lens (sphere) above a disk 
& \pageref{sec5.4.3}\\
&&5.4.4 & Combined effect of roughness, conductivity and temperature
& \pageref{sec5.4.4}\\ 
6 &  \multicolumn{3}{l}{Measurements of the Casimir force} 
& \pageref{sec6}\\
&6.1 & \multicolumn{2}{l}
{General requirements for the Casimir force measurements} & \pageref{sec6.1}\\
&6.2 & \multicolumn{2}{l}
{Primary achievements of the older measurements} & \pageref{sec6.2}\\
&&6.2.1 &  Experiments with parallel plates by M.J.~Sparnaay
& \pageref{sec6.2.1}\\
&&6.2.2 &  Experiments by Derjaguin et al.& \pageref{sec6.2.2}\\
&&6.2.3 &  Experiments by D.~Tabor, R.~Winter and J.~Israelachvili using&\\
&&& mica cylinders& \pageref{sec6.2.3}\\
&&6.2.4 & Experiments of P.~van Blokland and J.~Overbeek & \pageref{sec6.2.4}\\
&&6.2.5 &  Dynamical force mesurement techniques& \pageref{sec6.2.5}\\
&6.3 & \multicolumn{2}{l}
{Experiment by S.K.~Lamoreaux} & \pageref{sec6.3}\\
&6.4 & \multicolumn{2}{l}
{Experiments with the Atomic Force Microscope by Mohideen et al.} 
& \pageref{sec6.4}\\
&&6.4.1 & First AFM experiment with aluminium surfaces& \pageref{sec6.4.1}\\
&&6.4.2 & Improved precision measurement with aluminium surfaces &\\
&&&using the
AFM& \pageref{sec6.4.2}\\
&&6.4.3 & Precision measurement with gold surfaces using the AFM
& \pageref{sec6.4.3}\\
&6.5 & \multicolumn{2}{l}
{Demonstration of the nontrivial boundary properties of} & \\
&&\multicolumn{2}{l}{the Casimir force} & \pageref{sec6.5}\\
&&6.5.1 & Measurement of the Casimir force due to the corrugated plate
& \pageref{sec6.5.1}\\
&&6.5.2 & Possible explanation of the nontrivial boundary dependence of&\\
&&& the Casimir force& \pageref{sec6.5.2}\\
&6.6 & \multicolumn{2}{l}
{The outlook for the measurements of the Casimir force} & \pageref{sec6.6}
\end{tabular}
\begin{tabular}{llllr}
7 &  \multicolumn{3}{l}{Constraints for non-Newtonian gravity from the
Casimir effect} & \pageref{sec7}\\
&7.1 & \multicolumn{2}{l}
{Constraints from the experiments with dielectric test bodies} 
& \pageref{sec7.1}\\
&7.2 & \multicolumn{2}{l}
{Constraints from S.K.~Lamoreaux experiment} & \pageref{sec7.2}\\
&7.3 & \multicolumn{2}{l}
{Constraints from the Casimir force measurements by means} & \\
& & \multicolumn{2}{l}
{of atomic force microscope} & \pageref{sec7.3}\\
8 &  \multicolumn{3}{l}{Conclusions and discussion} & \pageref{sec8}\\
\multicolumn{4}{l}{Acknowledgements} & \pageref{ackn}\\
\multicolumn{4}{l}{Appendix A. Application of the Casimir force
in nanotechnology} & \pageref{appA}\\
 A1 &\multicolumn{3}{l}
{Casimir force and nanomechanical devices} & \pageref{A1}\\
 A2 &\multicolumn{3}{l}
{Casimir force in nanoscale device fabrication} & \pageref{A2}\\
\multicolumn{4}{l}{References} & \pageref{bibl}
\end{tabular}
}

\section{Introduction}\label{sec1}
More than 50 years have passed since H.B.G. Casimir published his famous paper
\cite{1} where he found a simple yet profound explanation for the retarded van
der Waals interaction (which was described by him along with D. Polder \cite{2}
only a short time before) as a manifestation of the zero-point energy of a
quantized field. For a long time, the paper remained relatively unknown. But 
starting 
from the 70ies this effect quickly received increasing attention and in the
last few years has become highly popular. New high precision experiments on
the demonstration of the Casimir force have been performed and more are under 
way. In theoretical developments considerable progress had been made in
the investigation of the structure of divergencies in general, non-flat
background and in the calculation of the effect for more complicated
geometries and \bc including those due to the real structures of the 
boundaries. In the Introduction we discuss the fundamental problems
connected with the concept of physical vacuum, the role of the Casimir
effect in different domains of physics and the scope of the review.

\subsection{Zero-point oscillations and their manifestation}
\label{sec1.1}
The Casimir effect in its simplest form is the interaction of a pair of
neutral, parallel conducting planes due to the disturbance of the vacuum of
the \elm field. It is a pure quantum effect -- there
is no force between the plates (assumed to be neutral) in classical 
electrodynamics. In the ideal
situation, at zero temperature for instance, there are no real photons in
between the plates. So it is only the vacuum, i.e., the ground state of
quantum electrodynamics (QED) which causes the plates to attract each other.
It is  remarkable  that a macroscopic quantum effect appears in this way.

In fact the roots of this effect date back to the introduction by Planck in
1911 of the half quanta \cite{3}. In the language of quantum
mechanics one has to consider a harmonic oscillator with energy levels
$\E_{n}=\hbar \om (n+\frac12)$, where $n=0,1,\dots$, and $\hbar$ is the Planck
constant. It is the energy
\beq
\label{gse}\E_{0}={\hbar \om \over 2} 
\eeq
of the ground state ($n=0$) which matters here. From the point of view of the
canonical quantization procedure this is connected with the arbitrariness of
the operator ordering in defining the Hamiltonian operator $\hat{H}$ by
substituting in the classical Hamiltonian $H(p,q)$ the dynamical
variables by the corresponding operators, $\hat{H}=H(\hat{p},\hat{q})$. It
must be underlined that the \gse $\E_{0}$ cannot be observed by measurements
within the quantum system, i.e. in transitions between different quantum
states, or for instance in scattering experiments. However, the frequency $\om$
of the oscillator may depend on parameters external to the quantum system. It
was as early as 1919 that this had been noticed in the explanation
for the vapor pressure of certain isotopes, the different masses provide the
necessary change in the external parameter (for a historical account see
\cite{4}).

In \qft one is faced with the problem of \uv divergencies  which come into play
when one tries to assign a \gse to each mode of the field. One has to consider
then
\be\label{gseqft}\E_{0}={\hbar\over 2 }\sum_{J}\om_J \,,
\ee
where the index $J$ labels the quantum numbers of the field modes. For
instance, for the \elm field in Minkowski space the modes are labeled by a
three vector ${\bf k}$ in addition to the two polarizations.  The sum
\Ref{gseqft} is clearly infinite. It was Casimir who was the first 
to extract  the finite force acting between the two parallel neutral plates
\be\label{FCas}F(a)=
-{\pi^{2}\over 240} \ {\hbar c\over a^{4}}S
\ee
from the infinite zero-point energy of the quantized electromagnetic field
confined in between the plates.  Here $a$ is the separation between the
plates, $S>>a^{2}$ is their area and $c$ is the speed of light.  

To do this Casimir had subtracted away from the infinite vacuum energy of Eq.
\Ref{gseqft} in the presence of plates, the infinite vacuum energy of 
quantized \elmf
in free Minkowski space. Both infinite quantities were regularized and after
subtraction, the regularization was removed leaving the finite result.

Note that in standard textbooks on \qft the dropping of the infinite
vacuum energy of free Minkowski space is motivated by the fact that energy is
generally defined up to an additive constant. Thus it is suggested that
all physical energies be measured starting from the top of this infinite 
vacuum energy 
in free space. 
In this manner effectively the infinite energy of free space is 
set to zero. Mathematically it
is achieved by the so called normal ordering procedure. This operation 
when applied
to the operators for physical observables puts all creation operators to the
left of annihilation operators as if they commute \cite{5,6,7}.
It would be incorrect, however, to apply the normal ordering procedure in this
simplest form when there are external fields or boundary conditions, e.g., on
the parallel metallic plates placed in vacuum.  In that case there is an
infinite set of different vacuum states and corresponding annihilation and 
creation
operators for different separations between plates.  These states turn into
one another under adiabatic changes of separation.  Thus,
it is incorrect to pre-assign zero energy values to several states between
which transitions are possible. Because of this, the finite difference between
the infinite vacuum energy densities in the presence of plates and in free
space is observable and gives rise to the Casimir force.

It is important to discuss briefly the relation of the Casimir effect
to other effects of \qft connected with the existence of zero-point
oscillations. It is well known that there is an effect of vacuum
polarization by external fields. The characteristic property of this
effect is some nonzero vacuum energy depending on the field strength.
Boundaries can be considered as a concentrated external field. In this
case the vacuum energy in restricted quantization volumes is
analogous to the vacuum polarization by an external field. We can then say that
material boundaries polarize the vacuum of a quantized field, and the force
acting on the boundary is a result of this polarization.

The other vacuum quantum effect is the creation of particles from
vacuum by external fields. In this effect energy is transfered from
the external field to the virtual particles (vacuum oscillations)
transforming them into the real ones. There is no such effect in
the case of static boundaries. However, if the boundary conditions
depend on time there is particle creation, in addition to a force (this is the
so called non-stationary or dynamic Casimir effect).
 
A related topic to be mentioned is \qft with \bcp The most common part of that
is \qft at finite temperature in the Matsubara formulation (we
discuss this subject here only in application to the Casimir force).  
The effects to be considered in this context 
can be divided
into pure vacuum effects like the Casimir effect and those where excitations of
the quantum fields are present, i.e., real particles in addition to virtual
ones.  An example is an atom whose spontaneous emission is changed in a
cavity.  Another example is the so called apparatus correction to the electron
g-factor. Here, the virtual photons responsible for the anomaly
$a_{e}=(g-2)/2$ of the magnetic moment are affected by the boundaries. In this
case by means of the electron a real particle is involved and the quantity to
be considered is the expectation value of the energy in a one electron state.
The same holds for cavity shifts of the hydrogen levels. This topic, together
with a number of related ones, is called ``cavity QED''. In the methods used, a
photon propagator obeying boundary conditions, this is very closely related to
the quantum field theoretic treatment of the Casimir effect. However, the
difference is ``merely'' that expectation values are considered in the vacuum
state in one case and in one (or more) particle states in the other.

\subsection{The Casimir effect as a macroscopic quantum effect}
\label{sec1.2}
The historical path taken by Casimir in his dealings with vacuum fluctuations
is quite different from the approaches discussed in the previous subsection
(see for example \cite{8}).  In investigating long-range van der Waals forces
in colloids together with his collaborator D. Polder he took the retardation
in the \elm interaction of dipoles into account and arrived at the so called
Casimir-Polder forces between polarizable molecules \cite{2}. This was later
extended by E.M. Lifshitz \cite{9} to forces between dielectric macroscopic
bodies usually characterized by a dielectric constant $\ep_{0}$
\be\label{Lifshitz}F(a)=-{\pi^{2}\over 240}{\hbar c\over
  a^{4}}{(\ep_{0}-1)^{2}\over (\ep_{0}+1)^{2}}\varphi(\ep_{0}) \ S\,,
\ee
where $\varphi(\ep_{0})$ is a tabulated function.  In this microscopic
description, the ideal conductor is obtained in the limit
$\ep_{0}\to\infty$, the same Casimir force \Ref{FCas} emerges
just as in the zero-point energy
approach. The point is that in the limit of ideal conductors only the surface
layer of atoms thought of as a continuum interacts with the \elm field.
Clearly, in this idealized case, \bc provide an equivalent description with
the known consequences on the vacuum of the \elm field.
These alternative descriptions also work for deviations from the
ideal conductor limit.
For example, the vacuum interaction of two bodies with finite conductivity can
be described approximately by impedance \bc with finite penetration depth in
one case and by the microscopic model on the other.  For two
dielectric bodies of arbitrary shape the summation of the Casimir-Polder
interatomic potentials was shown to be approximately equal to the exact
results if special normalizations accounting for the non-additivity 
effects are
performed \cite{10}.  Only recently has an important theoretical advance
occured in our understanding of this equivalence in the example 
of a spherical body
(instead of two separate bodies). Here the equivalence of the 
Casimir-Polder
summation and vacuum energy has been shown, at least in the dilute gas
approximation \cite{11}.

The microscopic approach to the theory of both van der Waals and Casimir
forces can be formulated in a unified way.  It is well known that the van der
Waals interaction appears between neutral atoms of condensed bodies separated
by distances which are much larger than the atomic dimensions. It can be
obtained non-relativistically in second order perturbation theory from
the dipole--dipole interaction energy \cite{12}. Because the expectation
values of the dipole moment operators are zero, the van der Waals interaction
is due to their dispersions, i.e. to quantum fluctuations. Thus, it is
conventional to speak about fluctuating electromagnetic field both inside the
condensed bodies and also in the gap of a small width between them.  Using the
terminology of Quantum Field Theory, for closely spaced macroscopic
bodies the virtual photon emitted by an atom of one body reaches an atom of
the second body during its lifetime. The correlated oscillations of the
instantaneously induced dipole moments of those atoms give rise to the
non-retarded van der Waals force \cite{13,13a}.

Let us now increase the distance between the two macroscopic bodies to be so
large that the virtual photon emitted by an atom of one body cannot reach the
second body during its lifetime. In this case the usual van der Waals force is
absent.  Nevertheless, the correlation of the quantized electromagnetic field
in the vacuum state is not equal to zero at the two points where the atoms
belonging to the different bodies are situated. Hence nonzero correlated
oscillations of the induced atomic dipole moments arise once more, resulting
in the Casimir force. In this theoretical approach the latter is also referred
to as the retarded van der Waals force \cite{7}. In the case of perfect metal
the presence of the bounding condensed bodies can be reduced to boundary
conditions at the sides of the gap. In the general case it is necessary to
calculate the interaction energy in terms of the frequency dependent
dielectric permittivity (and, generally, also the magnetic permeability) of
the media. For the case of two semispaces with a gap between them this was
first realised in \cite{9} where the general expressions for both the van der
Waals and Casimir force were obtained.  Needless to say that this theoretical
approach is applicable only for the electromagnetic Casimir effect caused by
some material boundaries having atomic structure. The case of quantization
volumes with non-trivial topology which also lead to the \bc \cite{14,14a,15}
is not covered by it.

An important feature of the Casimir effect is that even though it is quantum
in nature, it predicts a force between macroscopic bodies. For two
plane-parallel metallic plates of area $S=1\,{\rm cm}^{2}$ separated by a large
distance (on the atomic scale) of $a=1\,\mu$m the value of the attractive
force given by Eq. \Ref{FCas} is $F(a)\approx 1.3\times
10^{-7}\,$N.  This force while small, is now within the range of modern 
laboratory force measurement techniques. Unique
to the Casimir force is its strong dependence on shape, switching from
attractive to repulsive as a function of the geometry and topology of a
quantization manifold \cite{16,17}. This makes the Casimir effect a likely
candidate for applications in nanotechnologies and nanoelectromechanical
devices. The attraction between
neutral metallic plates in a vacuum was first observed experimentally in
\cite{18}. This and other recent experimental developments in the 
measurement of the
Casimir force is discussed in Sec.6.

There exist only a few other macroscopic manifestations of quantum
phenomena. Among them there are the famous ones such as Superconductivity,
Superfluidity, and the Quantum Hall Effect. In the above macroscopic
quantum effects the coherent behaviour of large number of quantum
particles plays an important role. In line with the foregoing the Casimir
force arises due to coherent oscillations of the dipole moments of
a great number of atoms belonging to the different boundary bodies. For this
reason the Casimir effect can be considered also as a macroscopic quantum 
effect. The clearest implication of the above is that the greater attention
traditionally given to the macroscopic quantum effects will also be eventually 
received by the Casimir effect.

\subsection{The role of the Casimir effect in different fields of physics}
\label{sec1.3}

The Casimir effect is an interdisciplinary subject. It plays an important
role in a variety of fields of physics such as Quantum Field Theory,
Condensed Matter Physics, Atomic and Molecular Physics, Gravitation and
Cosmology, and in Mathematical Physics \cite{18a}.

In Quantum Field Theory, the Casimir effect finds three main applications.
In the bag model of hadrons in Quantum Chromodynamics the Casimir energy
of quark and gluon fields makes essential contributions to the total nucleon 
energy. In Kaluza-Klein field theories Casimir effect offers one of the
most effective mechanisms for spontaneous compactification of extra
spatial dimensions. Moreover, measurements of the Casimir force provide
opportunities to obtain more strong constraints for the parameters of
long-range interactions and light elementary particles predicted by
the unified gauge theories, supersymmetry, supergravity, and string theory.

In Condensed Matter Physics, the Casimir effect leads to attractive and
repulsive forces between the closely spaced material boundaries which depend
on the configuration geometry, on temperature, and on the electrical and
mechanical properties of the boundary surface. It is responsible for some
properties of thin films and should be taken into account in investigations of
surface tension and latent heat. The Casimir effect plays an important role in
both bulk and surface critical phenomena.

In Gravitation, Astrophysics and Cosmology, the Casimir effect arises in 
space-times with non-trivial topology. The vacuum polarization resulting 
from the Casimir effect can drive the inflation process. In the theory of 
structure formation of the Universe due to topological defects, the Casimir 
vacuum polarization near the cosmic strings may play an important role.

In Atomic Physics, the long-range Casimir interaction leads to corrections
to the energy levels of Rydberg states. A number of the Casimir-type effects
arise in cavity Quantum Electrodynamics when the radiative processes and
associated energy shifts are modified by the presence of the cavity walls.

In Mathematical Physics, the investigation of the Casimir effect has 
stimulated the development of powerful regularization and renormalization 
techniques based on the use of zeta functions and heat kernel expansion.

The majority of these applications will be discussed below and the references
to the most important papers will be also given.

\subsection{What has been accomplished during the last years?}\label{sec1.4}

This review is devoted to new developments in the Casimir effect.
In spite of the extensive studies on the subject performed in the more
than 50 years there is only a  small number of review publications.
The first two large reviews \cite{19,20} were published more than ten 
years ago. There is a single monograph \cite{21} specially devoted to
the different aspects of the Casimir effect (the first, Russian, edition
of it was published in 1990). The other monograph \cite{22} is mostly
concerned with the condensed matter aspects of the subject. Several chapters
of the monograph \cite{7} are also devoted to the Casimir effect.

There are at least three very important new developments in the Casimir effect
which have made their appearance after the publication of the above mentioned
reviews. The first falls within the domain of Quantum Field Theory. It has
been known that in the case of flat boundaries the vacuum energy turns into
infinity at large momenta in the same way as in free Minkowski space. 
Thus it is apparently enough 
to subtract the contribution of Minkowski space in order to obtain
the final physical result for the Casimir energy. For arbitrary compact
domains bounded by closed surfaces (for example, the interior of a sphere)
this is, however, not the case.  
Except for the highest infinity (which is proportional
to the fourth power of a cut-off momentum) there exist lower order infinities.
The investigation of the general structure of these infinities for an
arbitrary domain was a  theoretical problem which had been  solved 
by the combination of
zeta-functional regularization \cite{Dowker1976bs,Hawking} and
heat kernel expansion \cite{DeWitt}. However,
 these results had been obtained
mostly in the context of curved space-time.
The explicit application to the Casimir effect was done later, first in
\cite{Blau:1988kv,Elizalde:1990dd}, 
where as an example the known divergencies for the Casimir
effect for a massive field with \bc on a sphere had been related to the
corresponding {\hkksp} 
In the nineties essential progress had been made in the
understanding and application of zeta functions \cite{sIV62-7,ten} as well as
in the calculation of the Casimir effect for massive fields for nonplane
boundaries, e.g., in \cite{Bordag:1997ma}.

The second important development in the Casimir effect during the last years
is concerned with Condensed Matter Physics. It has long been known that there
are large corrections to the ideal field-theoretical expressions for the
Casimir force due to several factors which are necessarily present
in any experimental situation. The most important factors of this sort are
those due to the finiteness of the conductivity of the boundary metal, 
surface roughness, and nonzero temperature. In the papers 
\cite{26,27,28,29,30,31}
the Casimir force including these factors was investigated in detail.  In
doing so not only the influence of each individual factor was examined, but
also their combined action was determined. This gave the possibility to 
increase
the degree of agreement between theory and experiment.

Probably, the third development is the most striking. It consists in new, more
precise measurements of the Casimir force between metallic surfaces.  In
\cite{32} a torsion pendulum was used to measure the force between Cu plus Au
coated quartz optical flat, and a spherical lens. In
\cite{33,34,35,36} an atomic force microscope was first
applied to measure the Casimir force between Al plus Au/Pd and Au coated
sapphire disk and polystyrene sphere.  Considerable progress has been made
towards the improving the accuracy of the Casimir force measurements. The
results of these measurements have allowed the stringent calculation of 
constraints
on hypothetical forces such as ones predicted by supersymmetry, supergravity,
and string theory \cite{37,37a,38,39}.  Other important
results in the Casimir effect obtained during the last few years are also 
discussed
below (see the review paper \cite{40}, and resource letter \cite{Lam-res}).

\subsection{The structure of the review}\label{sec1.5}

In the present review both the theoretical and  experimental developments
mentioned above are considered in detail. In Section 2 the simplified
overview of the subject is provided. The main theoretical concepts
used in the theory of the Casimir effect are illustrated here by
simple examples, where no technical difficulties arise and all calculations
can be performed in a closed form. Thus, the concepts of regularization
and renormalization are demonstrated for the case of a scalar field on an 
interval and for the simplest spaces with nontrivial topology.  
The famous Casimir formula \Ref{FCas} for the force between perfectly 
conducting parallel plates is derived by two methods. The additional 
effects arising for 
moving boundaries are considered in two-dimensional space-time.
The presentation is designed to be equally accessible to field theorists,
specialists in condensed matter and experimentalists.

Section 3 contains the general field-theoretical analyses of regularization
and renormalization procedures for the quantized field in an arbitrary
quantization domain with boundaries. Here the divergent parts of the vacuum
state energy and effective action are found by a combination of heat kernel
expansion and zeta-functional regularization. Different representations for the
regularized vacuum energy are obtained. The correspondence between the massive
and massless cases is discussed in detail.

In Section 4 the Casimir energies and forces in a number of different 
configurations are calculated, among which are stratified media, rectangular
cavities, wedge, sphere, cylinder, sphere (lens) above a disk and others.
Different kinds of boundary conditions are considered and possible 
applications to the bag model of hadrons, Kaluza-Klein field theories,
and cosmology are discussed. Radiative corrections to the
Casimir effect are also presented. Both exact and approximate calculation
methods are used in Sec.4. Some of the obtained results (especially the ones
for the stratified media and a sphere above a disk) are of principal
importance for the following sections devoted to aspects of condensed matter
physics and of the experiments.

Section 5 is devoted to the consideration of the Casimir force for the real
media. Here the Casimir force with account of nonzero temperature, finite
conductivity of the boundary metal and surface roughness is investigated.
The finite conductivity corrections are found both analytically in the
framework of the plasma model of metals and numerically using the optical tabulated data
for the complex refractive index. Surface roughness is taken into account
by means of perturbation theory in powers of small parameter which is
relates the effective roughness amplitude to the distance between the
boundary surfaces. Special attention is paid to the combined effect
of roughness and conductivity corrections, conductivity and temperature
corrections, and also of all three factors acting together. It is shown that
there are serious difficulties when applying the well known general 
expression for the temperature Casimir force \cite{9} in the case of
real metals. A line of attack on this problem is advanced.

In Section 6 the experiments on measuring the Casimir force are first
reviewed. The presentation begins with the discussion of experimental problems
connected with the measuring of small forces and small separations. Different
background effects are also considered in detail. The historical experiments
on measuring the Casimir force between metals and dielectrics are presented
starting from \cite{18}. The major part of Sec.6 is devoted to the
presentation of the results of modern experiments
\cite{32,33,34,35,36} and their comparison with the theoretical
results for the Casimir force between real media represented in Sec.5. The
prospects for further improving the accuracy of Casimir force measurements are
outlined.

In Section 7 the reader finds new interesting applications of the Casimir 
effect for obtaining constraints on the parameters of hypothetical long-range
interactions including corrections to Newtonian gravitational law 
and light elementary particles predicted by the modern
theories of fundamental interactions. Both the constraints following from
the historical and modern experiments on measuring the Casimir force are
presented. They are the best ones in comparison to all the other laboratory 
experiments in a wide interaction range. With further improvements in 
the Casimir
force measurements the obtained constraints can further be strengthened.

The presentation is organized in such a way that the specialists in different
fields of physics and also students could restrict their reading to some
selected sections. For example, those who are interested in condensed matter
aspects of the Casimir effect could read only the Secs.1, 2, 4.1.1, 4.3 and 5.
Those who are also interested in experimental aspects of the problem may add
to this Secs.6 and 7. Except for the purely theoretical Sec. \ref{sec3} we are
preserving in all formulas the fundamental constants $\hbar$ and $c$, as
experimentalists usually do, which helps physical understanding.

\setcounter{equation}{0}

\section{The Casimir effect in simple models}
\label{sec2}
In this section we present the elementary calculation of the Casimir energies
and forces for several simple models.  These models are mainly
low-dimensional ones. Also the classical example of two perfectly conducting
planes is considered. Such important concepts as regularization and
renormalization are illustrated here in an intuitive manner readily accessible
to all physicists, including non-specialists in Quantum Field Theory.
Introduction into the dynamical Casimir effect is given at the end of the
section.

\subsection{Quantized scalar field on an interval} \label{sec2.1}
We start with a real scalar field $\varphi(t,x)$ defined on an interval
$0<x<a$ and obeying \bc
\be\label{2.1}\varphi(t,0)= \varphi(t,a)=0 \,. 
\ee
This is the typical case where the Casimir effect arises. The simplicity of the
situation (one dimensional space and one component field) gives the
possibility to discuss the problems connected with the calculation of the
Casimir force in the most transparent manner. In Sec. \ref{sec2.2} a more 
realistic case
of the quantized \elm field between perfectly conducting planes will be
considered.

The scalar field equation is as usual \cite{5,6} 
\be\label{2.2}\frac{1}{c^{2}}{\pa^{2} \varphi(t,x)\over \pa t^{2}}-{\pa^{2}
  \varphi(t,x)\over \pa x^{2}}+{m^{2}c^{2}\over
  \hbar^{2}}\varphi(t,x)=0\,, 
\ee
where $m$ is the mass of the field. The indefinite scalar product associated
with Eqs. \Ref{2.1}, \Ref{2.2} is
\be\label{2.3}(f,g)=i\int_{0}^{a}d x
\left(f^{*}\pa_{x_{0}}g-\pa_{x_{0}}f^{*}g\right)\,,   
\ee
where $f,g$ are two solutions of \Ref{2.2}, $x_{0}\equiv ct$. We remind the 
reader that the
scalar field in two-dimensional space-time is dimensionless. 
It is easy to check that the positive- and  negative-frequency solutions of
\Ref{2.2} obeying \bc \Ref{2.1} are as follows   
\bea\label{2.4} \varphi_{n}^{(\pm)}(t,x)=\left({c\over a\om_{n}}\right)
^{1/2}e^{\pm i\om _{n}t} \sin k_{n}x \,, \\
\om_{n}=\left({m^{2}c^{4}\over \hbar^{2}}+c^{2}k_{n}^{2}\right)^{1/2}\,, \qquad
k_{n}={\pi n\over a}, \quad n=1,2,\dots \ . \nn
\eea
They are orthonormalized in accordance with the scalar product \Ref{2.3}
\be\label{2.5}\left(\varphi_{n}^{(\pm)},\varphi_{n'}^{(\pm)}\right)=
\mp\delta_{nn'}\,, ~~ \left(\varphi_{n}^{(\pm)},
  \varphi_{n'}^{(\mp)}\right)=0 \,.
\quad  
\ee
We consider here a free field. Soliton-type solutions for the
self-interacting field between the boundary points in two-dimensional
space-time with different boundary conditions are considered in
\cite{IJMP}.

Now the standard quantization of the field is performed by means of the
expansion 
\be\label{2.6} \varphi(t,x)=\sum_{n}\left[ \varphi_{n}^{(-)}(t,x)a_{n}+
  \varphi_{n}^{(+)}(t,x)a_{n}^{+}\right] \,, 
\ee
where the quantities $a_{n}$, $a_{n}^{+}$ are the annihilation and creation
operators obeying the commutation relations
\be\label{2.7}\left[a_{n},a_{n'}^{+}\right]=\delta_{n,n'}, \quad
\Big[a_{n},a_{n'}\Big]= \left[a^{+}_{n},a_{n'}^{+}\right]=0 \,.
\ee

The vacuum state in the presence of \bc is defined by
\be\label{2.8}a_{n}\mid 0 \rangle =0 \,.
\ee

We are interested in investigating the energy of this state in
comparison with the vacuum energy of the scalar field defined on an
unbounded axis $-\infty <x < \infty$. The operator of the energy
density is given by the $00$-component of the energy-momentum tensor
of the scalar field in the two-dimensional space-time
\be\label{2.9}T_{00}(x)={\hbar c\over 2}\left\{ {1\over
    c^{2}}\left[\pa_{t}\varphi(x)\right]^{2}+
\left[\pa_{x}\varphi(x)\right]^{2}\right\}    \,.  
\ee

Substituting Eq. \Ref{2.6} into Eq. \Ref{2.9} with account of \Ref{2.4},
\Ref{2.7}, and \Ref{2.8} one easily obtains
\be\label{2.10} \langle 0 \mid T_{00}(x)\mid 0 \rangle ={\hbar\over 2 a}
\sum_{n=1}^{\infty}\om_{n}- {m^{2}c^{4}\over 2a\hbar}\sum_{n=1}^{\infty}{\cos
  2k_{n}x\over \om_{n}}\,.
\ee
The total vacuum energy of the interval $(0,a)$ is obtained by the integration
of \Ref{2.10}

\be\label{2.11}\E_{0}(a)=\int_{0}^{a} \langle 0 \mid T_{00}(x)\mid 0 \rangle
dx = \frac\hbar{2}\sum_{n=1}^{\infty}\om_{n}\,.
\ee
The second, oscillating term in the right-hand side of \Ref{2.10} does not
contribute to the result. 

The expression \Ref{2.11} for the vacuum state energy of the quantized field
between boundaries is the standard starting point in the theory of the
Casimir effect. Evidently the quantity $\E_{0}(a)$ is infinite. It can be
assigned a meaning by the use of some regularization procedure. There are many
such regularization procedures discussed below. Here we use one of 
the simplest ones, i.e.,
we introduce an exponentially damping function $\exp(-\delta \om_{n})$ 
after the summation sign. 
In the limit $\delta\to 0$ the regularization is
removed. For simplicity let us consider the regularized vacuum energy of the
interval for a massless field ($m=0$). In this case
\be\label{2.12} \E_{0}(a,\delta)\equiv  \frac\hbar{2}\sum_{n=1}^{\infty}{c\pi
  n\over a}\exp\left(-{\delta c \pi n\over a}\right) = {\pi \hbar c\over 8a}
\sinh^{-2}{\delta c \pi\over 2a} \,.
\ee
In the limit of small $\delta$ one obtains
\be\label{2.13} \E_{0}(a,\delta)={\hbar a\over 2\pi
  c\delta^{2}}+\E(a)+O(\delta^{2}), \ \ \E(a)=-{\pi\hbar c\over 24 a}\,,  \ee
i.e., the vacuum energy is represented as a sum of a singular term and a
finite contribution.

Let us compare the result \Ref{2.13} with the corresponding result for the
unbounded axis. Here instead of \Ref{2.4} we have the positive and negative
frequency solutions in the form of traveling waves
\bea\label{2.14}&&\varphi_{k}^{(\pm)}(t,x)=\left({c\over 4\pi \om}\right)
^{1/2}e^{\pm i(\om t-k x)} \,,  \
\om=\left({m^{2}c^{4}\over \hbar^{2}}+c^{2}k^{2}\right)^{1/2}, \nn\\&& \ -\infty<k<\infty \,.
\eea
The sum in the field operator \Ref{2.6} is interpreted now as an integral with
the measure $dk/2\pi$, and the commutation relations contain delta functions
$\delta(k-k')$ instead of the Kronecker symbols. Let us call the vacuum state
defined by 
\be\label{2.15}a_{k}\mid 0_{{\rm M}} \rangle =0
\ee
the Minkowski vacuum to underline  the fact that it is defined in free space
without any boundary conditions.

Repeating exactly the same simple calculation which was performed for the
interval we obtain the divergent expression for the vacuum energy density in
Minkowski vacuum 
\be\label{2.16}  \langle 0_{M} \mid T_{00}(x)\mid 0_{M} \rangle ={\hbar\over
  2\pi}\int_{0}^{\infty}\om dk\,,
\ee
and for the total vacuum energy on the axis
\be\label{2.17}\E_{0M}(-\infty,\infty)={\hbar\over 2\pi}\int_{0}^{\infty}\om
dk L\,, 
\ee
where $L\to\infty$ is the normalization length.

Let us separate the interval $(0,a)$ of the whole axis whose energy should be
compared with \Ref{2.11}
\be\label{2.18}\E_{0M}(a)={\E_{0M}(-\infty,\infty)\over L}a = {\hbar a \over
  2\pi}\int_{0}^{\infty}\om dk
\,.\ee
To calculate \Ref{2.18} we use the same regularization as above, i.e., we
introduce the exponentially damping function  under the integral. For
simplicity consider once more the massless case 
\be\label{2.19} \E_{0M}(a)={c\hbar a\over 2\pi}
\int_{0}^{\infty}k e^{-\delta c k}dk={\hbar a \over 2\pi c \delta^{2}} \,.
\ee

The obtained result coincides with the first term in the right-hand
side of \Ref{2.13}. Consequently, the renormalized vacuum energy of the
interval $(0,a)$ in the presence of \bc can be defined as
\be\label{2.20}\Ern (a)\equiv\lim_{\delta\to
  0}\left[\E_{0}(a,\delta)-\E_{0M}(a,\delta)\right]=\E(a)=
-{\pi\hbar c\over 24 a}\,. 
\ee
It is quite clear that in this simplest case the renormalization corresponds to
removing a quantity equal to the vacuum energy of the unbounded space in the
given interval. The general structure of the divergencies of the vacuum energy
will be considered in Sec.~3.3. The renormalized energy $\E(a)$
monotonically decreases as the boundary points approach each other. This 
points to the
presence of an attractive force between the conducting planes,
\be\label{2.21}F(a)=-{\pa\E(a)\over\pa a}=-{\pi\hbar c\over 24a^{2} \,.} 
\ee

In the massive case $m\ne 0$
 the above calculations lead to the result%
\be\label{2.22}\E(a,m)=-{mc^{2}\over 4}-{\hbar c\over 4\pi a }
\int_{2\mu}^{\infty}{\sqrt{y^{2}-4\mu^{2}}\over \exp(y)-1}dy  
\ee
with $\mu\equiv mca/\hbar$. Here the first, constant contribution is
associated with the total energy of the wall (boundary point). It does not
influence the force. 

For $m=0$, Eq. \Ref{2.22} gives the same result as \Ref{2.20}. It is possible
to find the asymptotic behaviours of \Ref{2.22} in the case of small and large
$\mu$. Thus, for $\mu<<1$ we have
\be\label{2.23} \E(a,m)\approx -{mc^{2}\over 4}-{\pi\hbar c\over 24 a} +{\hbar
c\over 23\pi a}\mu^{2}\ln \mu \,,
\ee
and for $\mu>>1$  
\be\label{2.24}  \E(a,m)\approx -{mc^{2}\over 2}-{\sqrt{\mu}\hbar c\over
  4\sqrt{\pi}a}e^{-2\mu} \,.
\ee
Note, that the exponentially small value of the distance dependent
term in the Casimir energy for $\mu>>1$ is typical.  The same small
value is also obtained for parallel planes in three-dimensional space
and for fields of higher spin.  It is, however, an artefact of plane
boundaries. If some curvature is present, either in the boundary or in
space-time, the behaviour is, generally speaking, in powers of the
corresponding geometrical quantity, for example the radius of a
sphere.  There are only accidental exceptions to this rule, e.g., the
case of a three sphere (see Sec. \ref{sec4.6.2}). Therefore it is
primarily for plane boundaries and flat space that the Casimir effect
for massless fields is larger and more important than that for massive
fields.

Evidently, the physical results like those given by Eq. \Ref{2.20} 
or \Ref{2.22}
should not depend on the chosen regularization procedure. We reserve the
detailed discussion of this point for Sec. \ref{sec3}. It is not difficult,
however, to make sure that the results \Ref{2.20}, \Ref{2.22} actually do not
depend on the specific form of the damping function. Let us, instead of the
exponential function used above, use some function $f(\om,\delta)$ with the
following properties be given. The function $f(\om,\delta)$ monotonically
decreases with increasing $\om$ or $\delta$ and satisfies the conditions
$f(\om,\delta)\le 1$, $f(\om,0)= 1$, $f(\om,\delta)\to 0$ for all $\delta\ne
0$ when $\om\to\infty$.

The non-dependence of the obtained results on the form of  $f(\om,\delta)$ can
be most easily demonstrated by the use of the Abel-Plana formula \cite{41}
\be\label{2.25}\sum\limits_{n=0}^{\infty}F(n)- \int\limits_{0}^{\infty}\d
t~F(t)= \frac{1}{2}F(0)+ i \int\limits_{0}^{\infty}{\d t\over
   e^{2\pi t}-1} \ {\left[F(it)-F(-it)\right]} \,,
\ee
where $F(z)$ is an analytic function in the right half-plane. 

One can substitute $F(n)$ by $\om_{n}$ multiplied by the damping function
$f(\om_{n},\delta)$. Then the left-hand side of Eq. \Ref{2.25} can be
interpreted as the difference in the regularized energies in the presence of
boundaries and in free space from the Eq. \Ref{2.25} defining the
renormalization procedure.  The independence of the integral in the right-hand
side of \Ref{2.25} on the form of $f(\om_{n},\delta)$ follows from the
exponentially fast convergence which permits taking the limit $\delta\to 0$
under the integral (note that the Abel-Plana formula was first applied for the
calculation of the Casimir force in \cite{15}).

In the case of the Casimir effect for both scalar and spinor fields a modification of
\Ref{2.25} is useful for the summation over the half-integer numbers
\be\label{2.26} \sum\limits_{n=0}^{\infty}F(n+\frac12)- \int\limits_{0}^{\infty}\d
t~F(t)= -i \int\limits_{0}^{\infty}{\d t\over
   e^{2\pi t}+1} \ {\left[F(it)-F(-it)\right]} \,.
\ee
Other generalizations of the Abel-Plana formula can be found in \cite{21}.

The Abel-Plana formulas are used in Secs.~2.2, 2.3, 4.1.2, and
4.6.1 to calculate the Casimir energies in different configurations.

\subsection{Parallel conducting planes} \label{sec2.2}
As it was already mentioned in the Introduction, in its simplest case the
Casimir effect is the reaction of the vacuum of the quantized \elm field to
changes in external conditions like conducting surfaces. The simplest case is
that of two parallel perfectly conducting planes with a distance $a$ between
them at zero temperature. They provide conducting \bc to the \elm field. These
\bc can be viewed as an idealization of the interaction of the metal surfaces
with the \elm field. In general, this interaction is much more complicated and
is modified by the finite conductivity of the metal (or alternatively the skin
depth of the electromagnetic field into the metal) and the surface roughness.
But the idealized conducting \bc are a good starting point for the
understanding as they provide a complete problem and one that can be easily
modified for the case of realistic metals. So it is possible to treat real
metals with their finite conductivity and surface roughness as small
perturbations (see Sec. \ref{sec5}). Here we focus on understanding the simple
case of ideal metal boundaries.

It is well known in classical electrodynamics that both polarizations of
the photon field have to satisfy \bc 
\be\label{embc}{\mbox{${\bf{E}}_{t}$}}_{\bigl|_{S}}=
{\mbox{${\bf{H}}_{n}$}}_{\bigl|_{S}}=0 \ee
on the surface $S$ of perfect conductors. Here $\bf{n}$ is the outward normal
to the surface. The index $t$ denotes the tangential component which is
parallel to the surface $S$. The conditions \Ref{embc} imply that the \elmf
can exist outside the ideal conductor only.

To proceed we imagine the \elm field as a infinite set of harmonic oscillators
with frequencies $\omega_{J}=c\sqrt{{\bf k}^{2}}$. 
Here the index of the photon momentum in free space (i.e., without
boundaries) is
$J={\bf k}=(k_{1},k_{2},k_{3})$ 
where all $k_i$ are continuous.
In the presence of boundaries
$J=(k_{1},k_{2},{\pi n\over a})=({\bf k}_{\perp},{\pi n\over a})$,
where ${\bf k}_{\perp}$ is a two-dimensional vector, $n$ is integer. 
In the latter case the frequency
results in
\be\label{freq}
 \omega_{J}=\om_{{\bf k_{\perp}},n}=c \sqrt{k_{1}^{2}+k_{2}^{2}+\left({\pi
      n\over a}\right)^{2}}  \,.
\ee
This has to be inserted into the half sum over  frequencies to get the vacuum
energy of the \elmf between the plates
\be\label{gse1}
\E_{0}(a)=\frac{\hbar}{2} \int{dk_{1}dk_{2}\over
  (2\pi)^{2}}\sum_{n=-\infty}^{\infty}\omega_{{\bf k_{\perp}},n} S\,,
\ee
where $S\to\infty$ is the area of plates.  In contrast to Eq. \Ref{2.11} the
sum is over negative integers also, so as to account for the two photon
polarizations.

The expression obtained is \uv divergent for large momenta. Therefore we have
to introduce some regularization like discussed 
in the preceding section. This procedure
is well known in \qft and consists in changing the initial expression in a way
that it becomes finite.  This change depends on the so called regularization
parameter and it is assumed that it can be removed formally by taking 
the limit 
value of this 
parameter to some appropriate value.  Here, we perform the regularization by
introducing a damping function of the frequency which was used in the original
paper by Casimir (see also Sec. \ref{sec2.1}) and the modern zeta functional
regularization. We obtain correspondingly
\be\label{Eregde}
\E_{0}{(a,\delta)}= \frac{\hbar}{2} \int{dk_{1}dk_{2}\over
  (2\pi)^{2}}\sum_{n=-\infty}^{\infty}\omega_{{\bf k_{\perp}},n}  
e^{-\delta\omega_{{\bf k_{\perp}},n}}S
\ee
and
\be\label{Eregz}
\E_{0}(a,s)=\frac{\hbar}{2} \sum_{n=-\infty}^{\infty}\int{dk_{1}dk_{2}\over
  (2\pi)^{2}} \ \omega_{{\bf k_{\perp}},n}^{1-2s} \ S\,.
\ee
These expressions are finite for $\delta>0$ \resp for $\Re s >\frac32$ and the
limits of removing the regularization are $\delta\to 0$ and $s\to 0$
correspondingly.\footnote{In fact one has to exclude the mode with $n=0$ but
  this does not change the physical result, see below Sec.~\ref{sec3}.} 
The first
regularization has an intuitive physical meaning. As any real body
becomes transparent for high frequencies their contribution should be
suppressed in some way which is provided by the exponential function. The
regularization parameter $\delta$ can be viewed as somewhat proportional to
the inverse plasma frequency.  In contrast, the zeta-functional regularization
does not provide such an explanation.  Its advantage is more mathematical. The
\gse in zeta-functional regularization $\E_{0}(a,s)$ is the zeta function of an
elliptic differential operator which is well known in spectral
geometry. It is a meromorph function of $s$ with simple poles on the real axis
for $\Re s \le \frac32$. To remove this regularization one has to construct
the analytic continuation to $s=0$. In $s=0$ it may or may not have a pole
(see examples below in Sec. \ref{sec3}).  These properties give the zeta
functional regularization quite important technical advantages and allow to
simplify the calculations considerably.  Together with this it must however be
stressed that all regularizations must be equivalent as in the end they
must deliver one and the same physical result.

Let us first consider the regularization 
done by introducing a damping function. The
regularized vacuum energy of the \elm field in free 
Min\-ko\-w\-s\-ki space-time is
given by
\be\label{2.32}\E_{0M}(-\infty,\infty)={\hbar\over (2\pi)^{3}}\int d^{3}k
\om_{\bf k}e^{-\delta \om_{\bf k}}LS\,,
\ee
where $L\to\infty$ is the length along the z-axis which is perpendicular to
the plates, $\om_{\bf k}=c|{\bf k}|=c\sqrt{k_{1}^{2}+k_{2}^{2}+k_{3}^{2}}$,
${\bf k}=(k_{1},k_{2},k_{3})$. 

The renormalized vacuum energy is obtained by the subtraction from
\Ref{Eregde} of the Minkowski space contribution in the volume between the
plates. After that the regularization can be removed. It is given by (compare
\Ref{2.20}) 
\bea\label{42a}\Ern(a)&=&\lim_{\delta\to 0}{\hbar\over
  2}\int{dk_{1}dk_{2}\over(2\pi)^{2}}\left(\sum_{n=-\infty}^{\infty}\om_{{\bf
      k}_{\perp},n}e^{-\delta\om_{{\bf k}_{\perp},n}}-
2a\int{\d k_{3}\over 2\pi}\om_{\bf
  k}e^{-\delta\om_{{\bf k}}}\right)S\nn \\
&=&{c\hbar\pi\over a}\lim_{\delta\to
  0}\int{dk_{1}dk_{2}\over(2\pi)^{2}}
\left(
\sum_{n=0}^{\infty}\sqrt{{k_{\perp}^{2}a^{2}\over
  \pi^{2}}+n^{2}} \ e^{\delta \om_{{\bf k}_{\perp},n}}   \nn \right.\\
&&\left. -\int_{0}^{\infty}dt \sqrt{{k_{\perp}^{2}a^{2}\over
  \pi^{2}}+t^{2}} \ e^{-\delta\om_{{\bf k}}}-{k_{\perp}a\over 2\pi}\right)S\,,
\eea
where $k_{\perp}^{2}\equiv k_{1}^{2}+k_{2}^{2}$, $t\equiv ak_{3}/\pi$. 

To calculate \Ref{42a} we apply the Abel-Plana formula \Ref{2.25} and obtain
\be\label{2.35}\Ern(a)=-{c\hbar\pi^{2}\over
  a^{3}}\int_{0}^{\infty}ydy
\int_{y}^{\infty}{\sqrt{t^{2}-y^{2}}\over e^{2\pi t}-1}dtS\,,
\ee
where $y=k_{\perp}a/\pi$ is the dimensionless radial coordinate in the
$(k_{1},k_{2})$-plane. Note that we could put $\delta=0$ under the sign of the
integrals in \Ref{2.35} due to their convergence. Also the signs when rounding
the branch points $t_{1,2}=\pm iA$ of the function $F(t)=\sqrt{A^{2}+t^{2}}$
by means of
\be\label{2.36}F(it)-F(-it)=2i\sqrt{t^{2}-A^{2}} ~ (t\ge A)
\ee
were taken into account.

To calculate \Ref{2.35}
 finally we change the order of integration and obtain%
\bea\label{2.37}\Ern(a)&=&-{c\hbar\pi^{2}\over
  a^{3}}\int_{y}^{\infty}{dt\over e^{2\pi t}-1} 
\int_{0}^{t}y{\sqrt{t^{2}-y^{2}}}dyS \nn \\
&=&-{c\hbar\pi^{2}\over
  3a^{3}}{1\over (2\pi)^{4}}\int_{0}^{\infty}{dx \ x^{3}\over e^{x}-1} S= 
-{c\hbar\pi^{2}\over
  720a^{3}}S \,.
\eea
The force \Ref{FCas} acting between the plates is obtained as derivative with
respect to their distance
\be\label{2.38}F(a)=-{\pa \Ern(a)\over \pa a}=-{\pi^{2}\over 240}{\hbar c\over
  a^{4}}S\,.
\ee
Now we demonstrate the calculation of the \gse in zeta functional
regularization starting from Eq. \Ref{Eregz}. Using polar coordinates
$(k_{\perp},\varphi_{k})$ in the plane $(k_{1},k_{2})$ and performing the
substitution $k_{\perp}=y{\pi n\over a}$ we obtain
\be \label{Espar} \E_{0}(a,s)={\hbar c\over 2\pi} \int_{0}^{\infty}dy \ y
\left(y^{2}+1\right)^{\frac12-s}\sum_{n=1}^{\infty}\left({\pi n\over
    a}\right)^{3-2s} S\,.  \ee
Note that we put $s=0$ in the powers of some constants, e.g., $c$.
The integration can be performed easily. The sum reduces to the well known
Riemann zeta function ($t=2s-3$)
\be\label{Riezeta}
\zeta_{R}(t)=\sum_{n=1}^{\infty}{1\over n^{t}} \,,
\ee
which is defined for $\Re t>1$, i.e., $\Re s>\frac32$, by this sum. 

We need, however, the value of $\zeta_{R}(-3)$ in the limit of removing the
regularization $s\to 0$. If we use the definition of $\zeta_{R}(t)$ according
to the right-hand side of \Ref{Riezeta} this value evidently diverges. It is
known that there exists a meromorph function with a simple pole in $t=-1$
($s=1$) which can be obtained by analytic continuation of the right-hand side
of Eq. \Ref{Riezeta} to the whole complex plane.  Such analytic continuation
is unique and well defined, for instance, at a point $t=-3$, although needless
to say that its values for $\Re \ t\le 1$ are not represented by the right-hand
side of Eq.  \Ref{Riezeta}.
It can be shown that the use of the value
$\zeta_{R}(-3)=1/120$ instead of the infinite when $s\to 0$ value \Ref{Espar}
is equivalent to the renormalization of the vacuum energy under consideration
(the reasons for the validity of this statement are presented in Sec.
\ref{sec3}). In this simplest case the value of the analytically continued
zeta function can be obtained from the reflection relation
\be\label{Riezeta,a}\Gamma\left(\frac{z}2\right)
\pi^{-z/2}\zeta_R(z)=
\Gamma\left({1-z\over 2}\right)\pi^{z-1\over 2}\zeta_R(1-z) \,,
\ee
where $\Gamma(z)$ is gamma function, taken at $z=4$. 
Substituting
$\zeta_{R}(-3)=1/120$ into \Ref{Espar} and putting $s=0$ one obtains once more
the renormalized physical energy of the vacuum \Ref{2.37} and attractive
force acting between plates \Ref{2.38}. 

Here a remark must be added concerning the renormalization. The result which
we obtained from the regularization by means of  analytical continuation of
zeta function into the complex $s$-plane is finite. This is particular to the 
case under consideration. In
more complicated configurations the result will in general be infinite in the
limit of removing the regularization so that some additional renormalization is
needed. The situation for two plane plates considered above is referred
to sometimes as renormalization by zeta functional regularization. 
It should be noted that it works in
some special cases only. 

\subsection{One- and two-dimensional spaces with nontrivial topologies}
\label{sec2.3}
\begin{figure}[!htbp]\unitlength=1cm
\begin{picture}(5,4.5) 
\put(0,0){\epsfxsize=10cm \epsffile{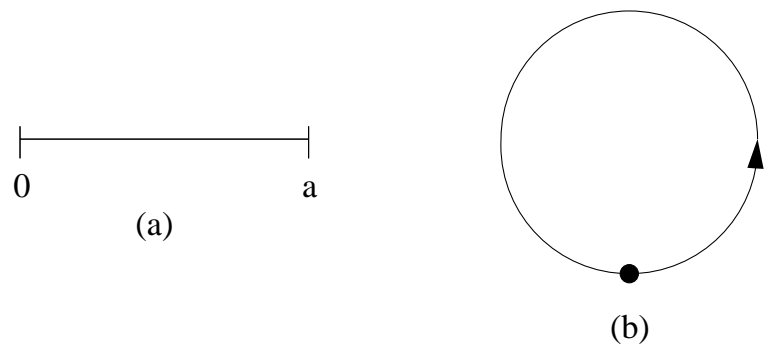}}
\end{picture}
\caption{Illustration of two flat  manifolds with Euclidean topology (a) and
  topology of a circle (b).}
\label{fig2}
\end{figure}

As noted in the Introduction, when the space is topologically
nontrivial the \bc are imposed on the quantized field similar to the case
of material boundaries. As a consequence, a nonzero vacuum energy appears,
though there are no boundaries and therefore no force can act upon them. 
Let us return to
the interval $0\le x\le a$ and impose \bc
\be\label{t0}\varphi(t,0)=\varphi(t,a) , \quad
\pa_{x}\varphi(t,0)=\pa_{x}\varphi(t,a) \,,
\ee
which describe the identification of the boundary points $x=0$ and $x=a$. As a
result we get the scalar field on a flat manifold with topology of a circle
$S^{1}$ (see Fig. \ref{fig2}).

Comparing with \Ref{2.1} now solutions are possible with $\varphi \ne 0$ at
the points $x=0,a$. The orthonormal set of solutions to \Ref{2.2}, \Ref{t0}
can be represented in the following form
\bea\label{t1}
&&\varphi^{(\pm)}_{n}(t,x)=\left(c\over 2 a
  \om_{n}\right)^{1/2}\exp \left[\pm i(\om_{n}t-k_{n}x)\right]\,,  \\
&& \om_{n}=\left({m^{2}c^{4}\over \hbar^{2}}+c^{2}k_{n}^{2}\right)^{1/2}\,,
  \qquad k_{n}={2\pi n\over a}, \quad n=0,\pm1,\pm2,\dots \ . \nn
\eea
Substituting \Ref{t1} into Eqs. \Ref{2.6} and \Ref{2.9} we obtain the vacuum
energy density of a scalar field on $S^{1}$
\be\label{t2} \langle 0\mid T_{00}(x)\mid 0 \rangle  ={\hbar\over
  2a}\sum_{n=-\infty}^{\infty}\om_{n} \,. 
\ee
Here, as distinct from Eq. \Ref{2.10}, no oscillating contribution is
contained. 

The total vacuum energy is
\bea\label{t3} \E_{0}(a,m)&=&\int_{0}^{a}\langle 0\mid T_{00}(x)\mid 0 \rangle
\ dx
={\hbar\over 2}\sum_{n=-\infty}^{\infty}\om_{n}\\
&=&\hbar \sum_{n=0}^{\infty}\om_{n}-{mc^{2}\over 2} \,. \nn
\eea
The renormalization of this infinite quantity is performed by subtracting the
contribution of the Minkowski space in accordance with \Ref{2.20}. The
simplest way to perform the calculation of the renormalized vacuum energy
without introducing an explicit renomalization function  is the use of the
Abel-Plana formula \Ref{2.25}. Substituting \Ref{t1}, \Ref{t3} and \Ref{2.18}
into \Ref{2.20} one obtains
\bea\label{t4} \Ern(a,m)&=&\hbar \left[\sum_{n=0}^{\infty}\om_{n}-{a\over
    2\pi}\int_{0}^{\infty}\om(k)dk\right]-{mc^{2}\over 2} \\
&=&{2\pi\hbar c\over
    a}\left[\sum_{n=0}^{\infty}\sqrt{A^{2}+n^{2}}-\int_{0}^{\infty}\sqrt{A^{2}
+t^{2}}dt\right]-{mc^{2}\over
    2} \,, \nn
\eea
with $A\equiv amc/(2\pi\hbar)$ and the substitution  $t=ak/(2\pi)$ was made.

Now we put $F(t)=\sqrt{A^{2}+t^{2}}$ into Eq. \Ref{2.25} and take account of
Eq. \Ref{2.36}.
Substituting \Ref{2.25} into \Ref{t4}, we finally obtain 
\be\label{t6}\Ern (a,m)=-{4\pi\hbar c\over a
  }\int_{A}^{\infty}{\sqrt{t^{2}-A^{2}\over e^{2\pi t}-1}}dt = -{\hbar
  c\over \pi a}
\int_{\mu}^{\infty}{\sqrt{\xi^{2}-\mu^{2}}\over e^{\xi}-1}\d\xi \,,
\ee
where $\xi=2\pi t$, $\mu\equiv mca/\hbar=2\pi A$. Here the constant term
describing the energy of a wall like  in Eq. \Ref{2.22} is absent as there are
no walls in space with non-Euclidean topology. 

In the massless case we have $\mu=0$, and the result corresponding to
\Ref{2.20} for the interval reads \cite{17}
\be\label{t7}\Ern (a)=-{\hbar c\over \pi a} \int_{0}^{\infty}{\xi\over
  \exp(\xi)-1}d\xi =-{\pi\hbar c\over 6a} \,.
\ee
For $\mu >> 1$ it follows from \Ref{t6}
\be\label{t8}\Ern (a,m)
\approx -{\sqrt{\mu}\hbar c\over\sqrt{2\pi}a }e^{-\mu} \,,
\ee
i.e., the vacuum energy of the massive field is exponentialy small which also 
happens in the case of flat spaces. 

In the one-dimensional case $S^{1}$ is a single topologically nontrivial
manifold. In two-dimensional spaces, i.e., (2+1)-dimensional space-times, there
exist both flat and curved manifolds with non-Euclidean topology. Below we
discuss one example of each. 

A plane with the topology of a cylinder $S^{1}\times R^{1}$
is a flat manifold. This topology implies that points with Cartesian
coordinates $(x+na,y)$, where $n=0,\pm1,\pm2,\dots$ are identified. For the
scalar field $\varphi$ defined on that manifold the following boundary
conditions hold
\be\label{t9}\varphi(t,0,y)= \varphi(t,a,y), \quad \pa_{x}\varphi(t,0,y)=
\pa_{x}\varphi(t,a,y) \,.
\ee

Bearing in mind future applications to curved space-times for different
dimensionality  we remind the reader of the scalar wave equation in 
($N+1$)-dimensional
Riemannian space-time
\be\label{t10}\left(\nabla_{\kappa}\nabla^{\kappa} +\xi R +{m^{2}c^{2}\over
  \hbar^{2}}\right)\varphi(x)=0 \,,
\ee
where $\nabla_{\kappa}$ is the covariant derivative, and $R$ is the scalar
curvature of the space-time, $\xi =(N-1)/4N$, $x=(x_{0},x_{1},\dots ,x_{N})$.
This is the so called ``equation with conformal coupling''. For zero mass it is
invariant under conformal transformations (see \cite{42,43}). The equation
with minimal coupling is obtained from \Ref{t10} with $\xi=0$ for all $N$.

The metric energy-momentum tensor is obtained by varying the Lagrangian
corresponding to \Ref{t10} with respect to the metric tensor $g^{ik}$. Its
diagonal components are \cite{42,43}
\bea\label{t11} T_{ii}&=&\hbar c \left\{ (1-2\xi)\pa_{i}\varphi \pa_{i}\varphi
  +(2\xi-\frac12)g_{ii}\pa_{k}\varphi\pa^{k}\varphi -\xi(\varphi
  \nabla_{i}\nabla_{i}\varphi+ \nabla_{i}\nabla_{i}\varphi \   \varphi
  ) \right.  \nn \\ 
&& \left. +\left[ (\frac12-2\xi){m^{2}c^{2}\over\hbar^{2}}g_{ii}-\xi G_{ii}-
2\xi^{2}Rg_{ii}\right]\varphi^{2}\right\}\,,
\eea
where $G_{ik}=R_{ik}-\frac12Rg_{ik}$ is the Einstein tensor, and $R_{ik}$ is
the Ricci tensor. 

Now let $N=2$, and the curvature be zero as in the case of $S^{1}\times
R^{1}$. It is not difficult to find the orthonormalized solutions to
Eq. \Ref{t10} with the boundary conditions \Ref{t9}. All the preceding
procedure described above for the case of $S^{1}$ can be repeated thereafter
with the result \cite{44}
\be\label{t12}\Ern (a,m)=-{\hbar c\over 4 \pi
  a^{2}}\int_{\mu}^{\infty}{\xi^{2}-\mu^{2}\over \exp(\xi)-1}d\xi \ L \,, 
\ee
where $L\to\infty$ is the normalization length along the $y$-axis as in
Eq. \Ref{2.17}. Note that this result is valid for an arbitrary value of
$\xi$ and not for $\xi=\frac18$ only. 

In the massless case the integral in \Ref{t12} is easily calculated with the
result
\be\label{t13} \Ern (a)=-{\hbar c \zeta_{R}(3)\over 2\pi a^{2}}L \,,
\ee
where $\zeta_{R}(z)$ is the Riemann zeta function with 
$\zeta_{R}(3)\approx 1.202$. The
calculation of the vacuum energy of the scalar field in other topologically
non-trivial two-dimensional  flat manifolds (a 2-torus, a Klein
bottle, a M\"obius strip of infinite width) can be found in \cite{21}. 

We now consider the Casimir effect for a scalar field on a
two-dimensional sphere $S^{2}$ with radius $a$. This is a curved manifold with
scalar curvature $R=2a^{-2}$. In spherical coordinates the space-time metric
reads
\be\label{t14}ds^{2}=c^{2}dt^{2}-a^{2}(d\theta^{2}+\sin^{2}\theta d\varphi^{2})
\,.  
\ee
The scalar field equation \Ref{t10} with $\xi=\frac18$, $R=2a^{-2}$ after the
transformations takes the form
\be\label{t15} {a^{2}\over c^{2}}\pa^{2}_{t}\varphi(x)-\Delta^{(2)}\varphi(x)
+\left({m^{2}c^{2}a^{2}\over\hbar^{2}}+\frac14\right) \varphi(x)=0\,,
\ee
where $\Delta^{(2)}$ is the angular part of the Laplace operator. 

The orthonormal set of solutions to Eq. \Ref{t10} obeying periodic boundary
conditions in both $\theta$ and $\varphi$ can be represented as
\bea\label{t16}
 &&\varphi^{(+)}_{lM}(t,\theta,\varphi)={\sqrt{c}\over
  a\sqrt{2\om_{l}}}\exp(i\om_{l}t)Y_{lM}(\theta, \varphi), \quad
\varphi^{(-)}_{lM}(t,\theta,\varphi)=
\left(\varphi^{(+)}_{lM}(t,\theta,\varphi)\right)^{*}
\,,\nn ~~~ \\
&& \om_{l}=\left[{m^{2}c^{4} \over\hbar^{2}}+{c^{2}\over
    a^{2}}(l+\frac12)^{2}\right]^{1/2}, \quad l=0,1,2,\dots, ~~
  M=0,\pm1,\dots,\pm l,~~~
\eea
where $Y_{lM}(\theta,\varphi)$ are the spherical harmonics. 

Substituting the field operator in the form \Ref{2.6} with eigenfunctions
\Ref{t16} into Eq. \Ref{t11} we find the still non-renormalized vacuum energy
density 
\be\label{t17}  \langle 0\mid T_{00}(x)\mid 0 \rangle =
{\hbar\over 4\pi a^{2}}\sum_{l=0}^{\infty}\left(l+\frac12\right)\om_{l}\,,
\ee
and the total vacuum energy of the sphere $S^{2}$ to be
\be\label{t18}
\E_{0}(a,m)=\hbar\sum_{l=0}^{\infty}\left(l+\frac12\right)\om_{l}\,. 
\ee
The renormalization is performed according to \Ref{2.20}, i.e., by introducing
a regularization by means of a damping function and subtracting the
contribution of the vacuum energy of the 3-dimensional Minkowski
space-time. The final result which does not depend on the specific form of the
damping function is most easily obtained by use of the Abel-Plana formula
\Ref{2.26} for the summation over half-integers resulting in \cite{45}
\be\label{t19}\Ern
(a,m)=2mc^{2}\left({mca\over\hbar}\right)^{2}\int_{0}^{1}{\xi\sqrt{1-\xi^{2}}\over
  \exp(2\pi mca\xi/\hbar)+1} d\xi\,. 
\ee
It is significant that here the Casimir energy is positive in contrast to the
case of a flat manifold considered above. 

When $\mu=mca/\hbar <<1$ it follows from \Ref{t19} that
\be\label{t20}\Ern (a,m)\approx\frac13 mc^{2}\mu^{2}\,. 
\ee

By this means, for a massless scalar field the Casimir energy on $S^{2}$ is
equal to zero. 

In the opposite limiting case $\mu >>1$ we have 
\be\label{t21} \Ern (a,m)\approx {mc^{2}\over 24}\left[ 1-{7\over 40
    \mu^{2}}\right] \,. 
\ee
As seen from \Ref{t21} the Casimir energy on the surface of $S^{2}$ diminishes
as a power of $\mu^{-1}$, whereas in the examples considered above it was
exponentially small for $\mu>>1$. This is because the manifold $S^{2}$ has a
nonzero curvature (see Sec. \ref{sec3}). For a sphere of infinite radius the
total Casimir energy of a scalar field on $S^{2}$  takes the value $\Ern
(m)=mc^{2}/24$ (see the discussion of additional normalization
condition for $m\to\infty$ in Sec.3.4).



\subsection{Moving boundaries in a two-dimensional space-time}\label{sec2.4}
In the preceding sections we considered the case when the boundaries and \bc
are static. The corresponding vacuum energies and forces were also static. 
If the geometrical
configuration, and, respectively, \bc depend on time the so-called
dynamic Casimir effect arises.  The most evident manifestation of
dynamic behavior is the dependence of the force on time. Let us return to a
massless scalar field on an interval $(0,a)$ considered in Sec.2.1.
Now let the right boundary depends on time: $a=a(t)$. It is obvious that in the
first approximation the force also depends on time according to the same law
as in \Ref{2.21}
\be\label{2.68}F(a(t))=-{\pi\hbar c\over 24 a^{2}(t)}\,.
\ee
This result is valid under the condition that the boundary velocity $a'(t)$
is small compared with the velocity of light. The proof of this statement and
the calculation of the velocity dependent corrections to the force can be
found in Sec. \ref{sec4.4} where more realistic dimensionalities are 
considered.

The other, and more interesting manifestation of the dynamic behavior is the
creation of particles from vacuum by a moving boundary (this
effect was first discussed in \cite{46,47})
The effect of
creation of particles from vacuum by non-stationary electric and gravitational
fields is well known (see, e.g., \cite{42,43}). As noted above, \bc are
idealizations of concentrated external fields. It is not surprising, then,
that moving boundaries act in the same way as a non-stationary external
field. We outline the main ideas of the effect of particle creation by moving
boundaries with the same example of a massless scalar field defined on an
interval $(0,a)$ with $a=a(t)$ depending on time when $t>0$. 

Instead of \Ref{2.1} the \bc now read
\be\label{2.69}\varphi(t,0)=\varphi(t,a(t))=0\,.
\ee

For $t<0$ the orthonormalized set of solutions to Eq. \Ref{2.2} with $m=0$ is
given by Eq. \Ref{2.4} in which one should substitute $a$ by $a_{0}\equiv
a(0)$. The field operator which is understood now as in-field (i.e.,
field defined for $t<0$ when the boundary point is at rest) is given by the
Eq. \Ref{2.6}. For any moment the set of solutions to Eq. \Ref{2.2},
$\chi_{n}^{(\pm)}(t,x)$ should satisfy the \bc \Ref{2.69} and the initial
condition
\be\label{2.70}\chi_{n}^{(\pm)}(t\le
0,x)=\varphi^{(\pm)}_{n}(t,x)=\frac1{\sqrt{\pi n}}e^{\pm i\om_{n}t}\sin{\pi n
  x\over a_{0}}\,. 
\ee
The field operator at any moment  is given by 
\be\label{2.71}\varphi(t,x)=\sum_{n}\left[ \chi^{(-)}_{n}(t,x)a_{n}+
  \chi^{(+)}_{n}(t,x)a_{n}^{+}\right] \,.
\ee
The functions $\chi_{n}^{(\pm)}(t,x)$ which are unknown for $t>0$ can be found
in the form of a series (see, e.g., \cite{48,49})
\be\label{2.72} \chi^{(-)}_{n}(t,x)=\frac1{\sqrt{\pi
  n}}\sum_{k}Q_{nk}(t)\sqrt{{a_{0}\over a(t)}}\sin{\pi k x\over a(t)}
\ee
with the initial conditions
\be\label{2.73}Q_{nk}(0)=\delta_{nk}, ~~Q'_{nk}(0)=-i\om_{n}\delta_{nk} \,.
\ee
Here $Q_{nk}(t)$ are the coefficients to be determined, $n,k=1,2,3,\dots$. The
positive-frequency functions $\chi^{(+)}_{n}$ are obtained as the complex
conjugate of \Ref{2.72}.

It is obvious that both \bc \Ref{2.69} and the initial conditions \Ref{2.70}
are satisfied automatically in \Ref{2.72}. Substituting Eq. \Ref{2.72} into
the field equation \Ref{2.2} (with $m=0$) we arrive after conversion to
an infinite coupled system of differential equations with respect to the
functions $Q_{nk}(t)$\cite{49} 
\bea\label{2.74}&&Q_{nk}''(t)+\om_{k}^{2}(t)Q_{nk}(t)  \nn \\
&&= \sum_{j}\left
  [ 2\nu(t)h_{kj}Q_{nj}'(t)+\nu'(t)h_{kj}Q_{nj}(t)+
  \nu^{2}(t)\sum_{l}h_{jk}h_{jl}Q_{nl}(t)\right] \,. 
\eea
Here the following notations are introduced
\bea\label{2.75}  \om_{k}(t)&=&{c\pi k\over a(t)}, ~~~~ 
         \nu(t)={a'(t)\over a(t)} \,, \nn \\
  h_{kj}&=&-h_{jk}=(-1)^{k-j}{2kj\over j^{2}-k^{2}}, ~~ j\ne k\,.
\eea

Let after some time $T$ the right boundary of the interval return to its
initial position $a_{0}$. For $t>T$ the right-hand side of Eq. \Ref{2.74} is
equal to zero and the solution with initial conditions \Ref{2.73} can be
represented as a linear combination of exponents with different frequency
signs 
\be\label{2.76}Q_{nk}(t)=\alpha_{nk}e^{-i\om_{k}t}+\beta_{nk}e^{i\om_{k}t}\,.
\ee

This is a familiar situation which is well-known in the theory of particle
creation from vacuum by a non-stationary external field. Substituting
Eqs. \Ref{2.72} and \Ref{2.76} into the field operator \Ref{2.71} we
represent it once more as the expansion in terms of the functions
$\varphi^{(\pm)}_{k}$ from \Ref{2.4} where $m=0$, $a=a_{0}$ but with the new
creation and annihilation operators 
\be\label{2.77}b_{k}=\sum_{n}\left(\sqrt{k\over
    n}\alpha_{nk}a_{n}+\sqrt{k\over n}\beta^{*}_{nk}a_{n}^{+}\right) 
\ee
and the Hermitian conjugate for $b^{+}_{k}$. 

Eq. \Ref{2.77} is the Bogoliubov transformation connecting in- creation and
annihilation operators $a_{n}^{+}, a_{n}$ with the out-ones $b^{+}_{n},
b_{n}$. Its coefficients satisfy the equality
\be\label{2.78}\sum_{k}k\left(|\al_{nk}|^{2}-|\beta_{nk}|^{2}\right)=n\,,
\ee
which is the unitarity condition (see \cite{42,43} for details). 

Different vacuum states are defined for $T<0$ and for $t>T$ due to
non-stationarity of the \bc
\be\label{2.79}a_{k}|0_{\rm in}>=0
~ \mbox{for} ~ t<0; ~~~~~ b_{k}|0_{\rm out}>=0 ~
\mbox{for} ~ t>T\,.
\ee
The number of particles created in the $k$th mode is equal to the
vacuum-vacuum matrix element in the in-vacuum of the out-operator for the
number of particles. It is calculated with the help of Eqs. \Ref{2.77} and
\Ref{2.79} 
\be\label{2.80}n_{k}=<0_{\rm in}|b^{+}_{k}b_{k}|0_{\rm
    in}>=k\sum_{n=1}^{\infty}\frac1n |\beta_{nk}|^{2}\,.
\ee
The total number of particles created by the moving boundary during time $T$ is
\be\label{2.81}N=\sum_{k=1}^{\infty}n_{k}=
\sum_{k=1}^{\infty}k\sum_{n=1}^{\infty}\frac1n
|\beta_{nk}|^{2}\,.  \ee
It should be mentioned that in the original papers \cite{46,47} another method
was used to calculate the number of created particles. There property
that in two dimensional space-time the classical problem with
non-stationary \bc can be reduced to a static one by means of conformal
transformations was exploited. This method, however, does not work in 
four-dimensional
space-time. 

To calculate the quantities \Ref{2.80} and \Ref{2.81} it is necessary to
solve the system \Ref{2.74} which is a rather complicated problem. It is
possible, however, to obtain a much more simple system in the case when the
boundary undergoes small harmonic oscillations under the condition of a
parametric resonance. Let us consider, following \cite{50}, the motion of the
boundary according to the law 
\be\label{2.82}a(t)=a_{0}\left[1+\ep\sin(2\om_{1}t)\right]\,,
\ee
where $\om_{1}=c\pi/a_0$, and the non-dimensional amplitude of the oscillations
is $\ep<<1$ (in realistic situations $\ep \sim 10^{-7}$). In the framework
of the theory of the parametrically excited systems \cite{51} the
coefficients $\al_{nm}$, $\beta_{nm}$ can be considered as slowly varying
functions of time. Substituting \Ref{2.76} into \Ref{2.74} we neglect by all
the terms of the order $\ep^{2}$ and perform the averaging over fast
oscillations with the frequencies of the order $\om_{k}$. As a result the
simplified system \Ref{2.74} takes the form \cite{50}
\be\label{2.83}\begin{array}{rclrcl}
{d\al_{n1}\over d\tau}&=&-\beta_{n1}+3\al_{n3}, &
{d\al_{nk}\over d\tau}&=&(k+2)\al_{n,k+2}-(k-2)\al_{n,k-2}, ~ k\ge 2 \,, \\
{d\beta_{n1}\over d\tau}&=&-\al_{n1}+3\beta_{n3}, &
{d\beta_{nk}\over d\tau}&=&(k+2)\beta_{n,k+2}-(k-2)\beta_{n,k-2}, ~ 
k\ge 2 \,, \end{array}
\ee
where the initial conditions are
\be\label{2.84}\al_{nk}(0)=\delta_{nk}, ~~~\beta_{nk}(0)=0\,.
\ee
Here we introduce the so-called ``slow time'' 
\be\label{2.85}\tau=\frac12\ep\om_{1}t\,.
\ee
Note that even modes are not coupled to the odd modes in \Ref{2.83}. Due to
the initial conditions \Ref{2.84}
\be\label{2.86}\beta_{n,2k}(t)=\beta_{2l,k}(t)=0\,,
\ee
which is to say that the particles are created in odd modes only.

The solution of the differential system \Ref{2.83} and \Ref{2.84} and of the
integral equation which is equivalent to it can be found in \cite{50}. 
Here we present only the final result for the particle creation rate. With
the proviso that $\tau<<1$ the number of created particles and the creation
rate in the lowest mode with $k=1$ are \cite{50}
\be\label{2.87}n_{1}(t)\approx \frac14(\ep\om_{1}t)^{2}, ~~ {d n_{1}(t)\over
  dt}\approx \frac12\ep^{2}\om_{1}^{2}t \,.
\ee
In the opposite limiting case $\tau>>1$ the results are
\be\label{2.88}n_{1}(t)\approx \frac4{\pi^{2}}(\ep\om_{1}t)+\frac2{\pi^{2}}\ln
  4-\frac12, ~~ {d n_{1}(t)\over
  dt}\approx \frac4{\pi^{2}}\ep\om_{1}\,.
\ee
The total number of created particles in all modes is $N(t)\approx n_{1}(t)$ if
$\tau<<1$, i.e., the lowest mode alone determines the result. However, if
$\tau>>1$ we have $N(t)\sim\tau^{2}>>n_{1}(t)$. 

For the energy of the particles created in the lowest mode one obtains
evidently the result $\om_{1}n_{1}(t)$. The total energy of particles created
in all modes is
\be\label{2.89}E(t)=\om_{1}\sum_{k}kn_{k}(t)=\frac14\om_{1}\sinh^{2}(2\tau)\,.
\ee
As is seen from this result the total energy increases faster than the total
number of photons, i.e., the pumping of energy takes place into the
high-frequency modes at the expense of the low-frequency ones.

In Sec. \ref{sec4.4} where the three dimensional configurations will be
considered we discuss the possibility of experimental observation of the
photons created by the moving mirrors. Additional
factors such as imperfectness of the boundary mirrors, back reaction of the
radiated photons upon the mirror etc. will be discussed. The influence of 
the detector placed into
the cavity will be also touched upon.

\setcounter{equation}{0}
\section{Regularization and renormalization of the vacuum energy}\label{sec3}

This section is devoted to the theoretical foundation of the Casimir
effect. It contains the general regularization and renormalization
procedures formulated in the frames of Quantum Field Theory under the
influence of boundary conditions. The divergent part of the vacuum
state energy is found in an arbitrary quantization domain.
Different representations for the regularized vacuum energy are obtained.
The photon propagator in the presence of boundary conditions is presented.
The mathematical methods set forth in this section give the possibility
to calculate the Casimir energies and forces for variety of
configurations and quantized fields of different spin.
Remind that in this section the units are used in which $\hbar=c=1$.
\subsection{Representation of the regularized \ve}\label{sec3.1}

The basic quantities appearing in \qft in connection with the Casimir effect
are the \vev of the energy operator in the ground state (vacuum) of the
quantum field under consideration and the corresponding effective action.  In
zeta functional regularization the \gse can be written as half sum over the
one-particle energies $\epsilon_{J}$ labeled by 
some general index $J$
\be\label{Ereg1}
\E_{0}(s)=\pm \frac12 \mu^{2s}\sum_{J}{\epsilon_{J}}^{1-2s} \,.
\ee
The one-particle energies $\epsilon_{J}$ are by means of 
$\epsilon_{J}^2=\Lambda_{J}$
connected with the eigenvalues of the corresponding one particle Hamiltonian
\be\label{H1pa}
H\Phi_{J}({\bf x})= \Lambda_{J}\Phi_{J}({\bf x}) \,.
\ee
In case of a first order (in the derivatives) theory, e.g., for a spinor
field, they have to be taken with positive sign, 
$\epsilon_{j}\to |\epsilon_{j}|$. We
assume the spectrum to be discrete for the moment. Within the given
regularization we are free to introduce an arbitrary constant $\mu$ with the
dimension of a mass which assures the energy $\E_{0}(s)$ to have the correct
dimension. The different signs in Eq.  \Ref{Ereg1} correspond to bosonic
and fermionic fields respectively.

The mathematical background of the regularization used in \Ref{Ereg1} is the
zeta function
\be\label{zetaH}\zeta_{H}(s)=\sum_{J}{\Lambda_{J}}^{-s}
\ee
associated with the operator $H$. Then the \gse \Ref{Ereg1} reads
\be\label{Ezeta}\E_{0}(s)=\pm{\mu^{2s}\over 2}\zeta_{H}(s-\frac12)\,.
\ee
This zeta function is one of the functions defined on an operators spectrum
and is used in field theory as well as in geometry. It is a well investigated
object with a clearly defined meaning, especially for hyperbolic
pseudo-differential operators.  It is known to be a meromorphic function of
$s$ and it has simple (sometimes double \cite{Gilkeya}) poles. This
function coincides for $\Re s>d/p$ (where $d$ is the dimension of the manifold
and $p$ the order of the differential operator, we consider operators with
$p=2$ only) with the sum in the \rhs of Eq.  \Ref{zetaH}.

In a broader context, in \qft instead of the \gse one considers the effective
action $\Gamma$ which is defined by means of
\be\label{effaction}\Gamma=-i\ln Z\,,
\ee
where $Z$ is the vacuum-to-vacuum transition amplitude. It can be represented
by the functional integral
\be\label{FI}Z=\int D\Phi\  e^{iS(\Phi)}
\ee
taken over fields satisfying corresponding \bcp In the context of \gse one
usually restricts oneself to an action quadratic in the fields, i.e., to a free
theory.\footnote{Note that free is meant in this context as a free field
  theory whereby the background potential may be taken into account exactly.}
A typical action reads
\be\label{tyact}S(\Phi)=-\frac12\int dx  \ \Phi(x) (\Box +V(x)+m^{2})\Phi(x)\,,
\ee
where $V(x)$ is a background potential. So the integral \Ref{FI} is Gaussian
and can be carried out delivering
\be\label{Zallg}Z=(\det(\Box +V(x)+m^{2}))^{-\frac12}=e^{-\frac12\Tr \ln (\Box
  +V(x)+m^{2})}\,. 
\ee
The box operator may be the usual wave operator or may be more complicated,
for example $\Box \to D_{\mu}D^{\mu}$ with the covariant derivative
$D_{\mu}=\pa_{\mu}+ieA_{\mu}(x)$ in case of the complex Klein-Gordon field in
the background of an electromagnetic potential. The functional integral 
can also be
over Grassmann fields too describing quantized fermion fields. In that case we
have the usual changes in the sign so that the effective action may be
written as
\be\label{Gallg}\Gamma=\pm\frac{i}2 \Tr\ln (\Box +V(x)+m^{2})\,.
\ee

We need to give this expression a more precise meaning. Again, we assume the
spectrum of the one particle Hamiltonian \Ref{H1pa} to be discrete. Moreover,
we assume the background to be static. In that case we can immediately switch
to the Euclidean formulation and separate the time dependence by means of a
Fourier transform to a momentum $p_{0}$. Then the effective action \Ref{Gallg}
becomes diagonal. Finally, we introduce the zeta functional regularization and
the effective action can be written as
\be\label{Greg}\Gamma=
\mp {\pa\over\pa s} \mu^{2s} \int_{-\infty}^{\infty}{\d p_{0}\over
  2\pi}\sum_{J}\left(p_{0}^{2}+\Lambda_{J}\right)^{-s}\,, ~~ (s\to 0)\,.  \ee
The integral  over $p_{0}$ may be carried out and by means of Eq. \Ref{zetaH}
the effective action can be expressed in terms of the zeta function of the
operator $H$
\be\label{Gzeta}\Gamma=\mp{\pa\over\pa
  s}{\mu^{2s}\over\sqrt{4\pi}}{\Gamma(s-\frac12)\over\Gamma(s)} \
  \zeta_{H}(s-\frac12) \,.
\ee
In this way, a connection between the \gse and the effective potential is
established. In general, the physical consequences resulting from both
quantities are expected to be the same. This will be seen after the discussion
of the renormalization and the corresponding normalization conditions.

The quantities introduced so far have a precise mathematical meaning. However,
they have a restricted region of applicability which is due to the assumption
of a discrete spectrum which is here equivalent to a finite quantization
volume. In order to go beyond that we are faced with the infinite Minkowski
space contribution in problems with a flat background manifold. A typical
example is the Casimir effect for the exterior of a ball. More precisely, let
$L$ be the size of the quantization volume. Then the \gse (and the zeta
function as well) for $L\to\infty$ contains a contribution which does not
depend on the background and, by means of translational invariance, it will be
proportional to the infinite volume of the Minkowski space. This contribution
is independent of the background and thus of the boundary conditions. 
Because of this, it does not
carry any
information of interest. In the following we drop it without changing
the notation of the corresponding quantities like $\E_{0}$.  Next, there is a
contribution which does not depend on $L$ but depends on the background and
which we are interested in. Finally there are contributions vanishing for
$L\to\infty$.

To put this procedure into a mathematical framework it is useful to transform
the sum, say in Eq. \Ref{Ereg1}, into an integral.  A second reason for this
is the need to construct the analytic continuation of the zeta function to the
left of $\Re s>s_{0}$ which is necessary 
for removing the regularization. There is
no general procedure to do this and we are left with some special assumptions
which are, however, still quite general and 
allow consideration of a wide class of problems.
So we assume that the variables separate in the problems considered.
The simplest example in this recpect is a background depending on one
Cartesian coordinate only, 
e.g. parallel mirrors. The next example is
a spherically symmetric background, say \bc on a sphere or a potential $V(r)$.
As known, separation of variables 
is connected with some symmetry. So the same can
be done for a problem with cylindrical symmetry, for a generalized cone, for
monopoles and quite a large number of other problems not considered
here. We restrict ourselfs   to the two most simple, but typical
cases.
\subsubsection{Background depending on one Cartesian coordinate}
\label{sec3.1.1}
\begin{figure}[!htbp]\unitlength=1cm
\begin{picture}(5,4.5)
\put(0,0){\epsfxsize=12cm 
\epsffile{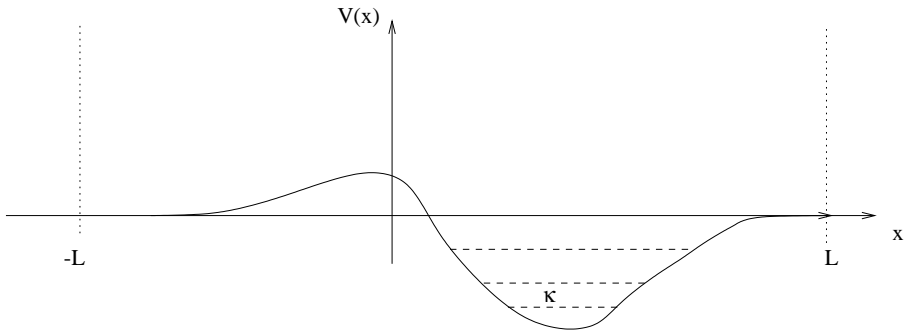}}
\end{picture}
\caption{A typical one dimensional potential, the bound state levels are shown
  by broken lines.}
\label{fig2lost}
\end{figure}

Here we assume the potential to depend on the coordinate $x_{3}$ only. In that
case, by means of translational invariance, the vacuum energy is proportional
to the volume of the directions $x_{\perp}$ perpendicular to $x_{3}$ and we
have in fact to consider the corresponding energy density.  With the obvious
ansatz $\Phi_{J}({\bf x})=\exp (i{\bf k_{\perp}x_{\perp}})\Phi_{n}(x_{3})$ the
equation for the one particle energy becomes one dimensional
\be\label{evp1}
H \Phi_{n}(x_{3})=\Lambda_{n}\Phi_{n}(x_{3})
\ee
with the operator 
\be\label{hevp1} H=k_{\perp}^{2}-{d^{2}\over d x_{3}^{2}}+V(x_{3})+m^{2} \,.
\ee
It is meaningful to define the pure Schr\"odinger operator 
\be\label{levp1}\P=-{d^{2}\over  d x^{2}}+V(x)
\ee
with the eigenvalue problem which looks like a one dimensional Schr\"odinger
equation 
\be\label{evp2}\P\varphi_{n}(x)=\lambda_{n}\varphi_{n}(x)
\ee
 so that Eq. \Ref{Ereg1} for the \ve
takes the form
\be\label{Ereg2}
\E_{0}(s)=\pm\frac{\mu^{2s}}2 \sum_{n}\int{dk_{1}dk_{2}\over
 (2\pi)^{2}}\left(\lambda_{n}+k_{\perp}^{2}+m^{2}\right)^{1/2-s} \,.
\ee
Here we assume that the potential $V(x)$ for $x\to\pm\infty$ tends to zero
(see Fig.~\ref{fig2lost}).
Otherwise, if it tends to a nonzero constant this constant can be absorbed
into a redefinition of the mass $m$. The case when it tends to different
constants at both infinities can be treated in a similar way and is not
considered here.

Formula \Ref{Ereg2} can be rewritten by integrating out $k_{1}$ and $k_{2}$.
By means of the substitution $k_{1,2}\to \sqrt{\lambda_{n}+m^{2}}k_{1,2}$ the
expression factorizes. The $k$-integration reads simply
\[\int{dk_{1}dk_{2}\over
  (2\pi)^{2}}\left(k_{\perp}^{2}+1\right)^{1/2-s}={1\over 4\pi} \
{1\over s-3/2}\,. 
\]
The integral is defined for $\Re s >3/2$. The analytic continuation to the
whole $s$-plane is given by the right-hand side. In more general cases, for
instance when the number of dimensions of the directions perpendicular to
$x_{3}$ is odd, a combination of gamma functions results.  After this
transformation the \gse takes the form
\be\label{Ereg3}
\E_{0}(s)=\pm{\mu^{2s}\over 8\pi}{1\over
  s-3/2}\sum_{n}\left(\lambda_{n}+m^{2}\right)^{3/2-s}\,.  
\ee

Now, with the discrete eigenvalues $\la_{n}$, Eq. \Ref{Ereg3} is a well
defined expression and we could start to construct its analytic continuation
to $s =0$.  An example for such a problem is the Casimir effect between two
planes with Dirichlet boundary conditions at $x=\pm L$ and no potential, i.e.,
with $V(x)=0$ in Eq. \Ref{levp1}, which was considered in Sec.  \ref{sec2.2}.
There, the remaining sum delivered the Riemann zeta function with its well
known analytic continuation.  But this is not the general case and in order to
separate the translational invariant part (in $x_{3}$-direction) we proceed as
follows.

We consider the one dimensional scattering problem on the whole axis ($x\in
(-\infty,\infty)$) associated with the operator
$\P$ \Ref{evp2}   
\be\label{Scpr}
\P \varphi(x)=k^{2}\varphi(x) \,. 
\ee
It is well investigated, please refer to
the textbook \cite{chadanSabatier} for
example. We note the following properties. Eq. \Ref{Scpr} has two linear
independent solutions which can be chosen as to have the asymptotics
\be\label{Scpr2}
\begin{array}{lcrrcl}
\varphi_{1}(x)  &  \raisebox{-0.9ex}{${\sim\atop x\to -\infty}$} &
e^{ikx}+s_{12}e^{-ikx}, & \qquad \varphi_{1}(x) & 
\raisebox{-0.9ex}{${\sim\atop x\to
    \infty}$} & s_{11}e^{ikx} \, , \\[14pt]
\varphi_{2}(x)  &  \raisebox{-0.9ex}{${\sim\atop x\to -\infty}$} &
s_{22}e^{-ikx}, & \varphi_{2}(x)& \raisebox{-0.9ex}{${\sim\atop x\to
    \infty}$}  & s_{21}  e^{ikx}+e^{-ikx} \,.
\end{array}
\ee
The matrix ${\bf s}=(s_{ij})$ composed from the coefficients $s_{ij}$ in \Ref
{Scpr2} is unitary. From this for real $k$ the relation 
\be\label{unitary1}
s_{11}^{2}(k)-s_{21}^{2}(k)={s_{11}(k)\over s_{11}(-k)}=e^{2i\delta(k)} 
\ee
follows, where $\delta(k)$ is the scattering phase.  The first solution,
$\varphi_{1}(x)$, describes a wave incident from the left which is scattered
by the potential, $t(k)=s_{11}(k)$ is the transmission coefficient,
$r(k)=s_{12}(k)$ is the reflection coefficient and the second power of their
modules $R=\mid r(k)\mid^{2}$ and $T=\mid t(k)\mid^{2}$ are connected by
$R+T=1$. The second solution, $\varphi_{2}(x)$, has the same meaning for a
wave incident from the right.  The function $s_{11}(k)$ is a meromorphic
function. Its poles on the upper half plane (if any) are located on the
imaginary axis and correspond to bound states in the potential $V(x)$ with
binding energy $\kappa=ik$.

\begin{figure}[!htbp]\unitlength=1cm
\begin{picture}(5,5.5)
\put(0,-0.4){\epsfxsize=12cm 
\epsffile{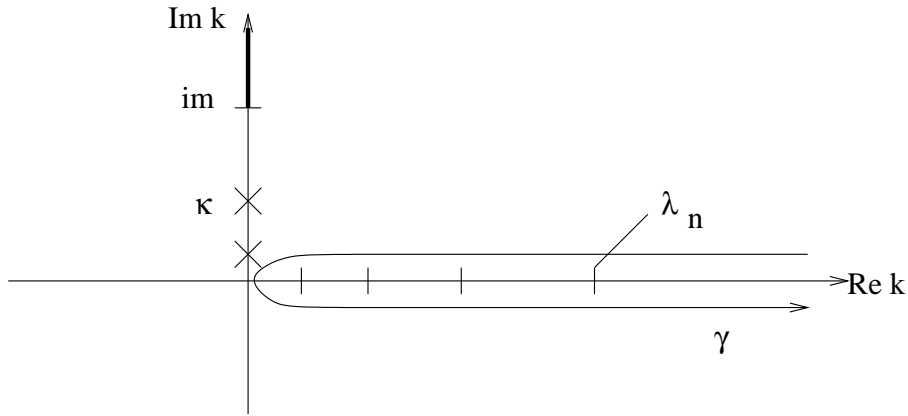}}
\end{picture}
\caption{The complex $k$-plane}
\label{fig1d}
\end{figure}

Now we consider the following two linear combinations of the solutions
\be\label{} 
\varphi_{\pm}(x)=\varphi_{1}(x)\pm\varphi_{2}(x) \,.
\ee
Imposing the \bc  $\varphi_{\pm}(L)=0$ delivers the discrete eigenvalues
$k^2\to\la_{n}$ as solution to these equations. With other words, the functions
$\varphi_{\pm}(L)$, considered as functions of $k$ have zeros in
$k^2=\la_{n}$. By means of this we rewrite the sum in \Ref{Ereg3} 
by an integral
\bea\label{Ereg4}
 \E_{0}(s)&=&{1\over 8\pi}{1\over s-3/2}\left\{
\sum_{i}\left(-\kappa_{i}^{2}+m^{2}\right)^{3/2-s} \right.\\&& + \left.
\int_{\gamma}{dk\over 2 i
 \pi}\left(k^{2}+m^{2}\right)^{3/2-s}  {\pa\over\pa k}\ln
 \left(\varphi_{+}(L)\varphi_{-}(L)\right) \right\}  \nn
\eea
where the first sum is over the bound states $\kappa_{i}$ of the potential
$V(x)$ and the contour $\gamma$ encloses the continuous spectrum of $\P$,
i.e., all eigenvalues $\la_{n}$ on the real axis (see Fig. \ref{fig1d}). We
are interested in the limit of large $L$.  Let $\epsilon$ 
be the imaginary part of
the integration variable $k$ in \Ref{Ereg4}. We note 
$\epsilon>0$ respectively
$\epsilon<0$ on the upper 
respectively lower half of the path $\gamma$. Than we
have using the asymptotic expansion \Ref{Scpr2}.
\[ 
\ln \left(\varphi_{+}(L)\varphi_{-}(L)\right) = \left\{ 
\begin{array}{lr}
-2ikL+2\epsilon L -i\pi +O(e^{-\epsilon L}), &\ (\epsilon>0), \\[7pt]
2ikL-2\epsilon L +
\ln({s_{11}^{2}(k)}-{s_{21}^{2}(k)})+O(e^{\epsilon L}),& (\epsilon<0)
\end{array}  \right.
\label{quatch1}
\]
correspondingly.
The contributions proportional to $L$ constitute the so called ``Minkowski
space contribution'' and are thrown away. Then we obtain,
 from the part depending on the background potential,  in the limit
$L\to\infty$
\bea\label{Ereg5}  \E_{0}(s)&=&{1\over 8\pi}{1\over
  s-3/2} \left\{
\sum_{i}\left(-\kappa_{i}^{2}+m^{2}\right)^{3/2-s} \right.\\ && \left.
 +\int_{0}^{\infty}{dk\over 2 i
 \pi}\left(k^{2}+m^{2}\right)^{3/2-s}  {\pa\over\pa k}\ln
 \left({s_{11}^{2}(k)}-{s_{21}^{2}(k)}\right) \right\}\,. \nn
\eea
Now there are two ways to proceed. By means of relation
\Ref{unitary1} we obtain
\bea\label{EregRe}  \E_{0}(s)&=&{1\over 8\pi}{1\over
  s-3/2} \left\{
\sum_{i}\left(-\kappa_{i}^{2}+m^{2}\right)^{3/2-s} \right.\\&&\left.
 +\int_{0}^{\infty}{dk\over 
 \pi}\left(k^{2}+m^{2}\right)^{3/2-s}  {\pa\over\pa k}
  \delta(k)\right\}\,, \nn 
\eea
where the integration is over the real $k$-axis. Another representation can be
obtained using  \Ref{unitary1} in the form
$\ln\left({s_{11}^{2}(k)}-{s_{21}^{2}(k)}\right)=\ln s_{11}(k)-\ln s_{11}(-k)$ 
and turning the
integration contour towards the positive imaginary axis in the contribution
from $\ln s_{11}(k)$ and to the negative axis in $\ln s_{11}(-k)$. Taking into
account the cut resulting from the factor $(k^{2}+m^{2})^{3/2-s}$ we arrive at
the representation
\be\label{EregIm}  \E_{0}(s)=-{1\over 8\pi}{1\over
  s-3/2} {\cos \pi s\over\pi}\int_{m}^{\infty}dk \
  (k^{2}-m^{2})^{3/2-s}{\pa\over\pa k} \ln s_{11}(ik) \,.
\ee
Here the integration is over the imaginary axis. The explicit contribution
from the bound states is canceled from the extra terms arising when moving the
integration contour to the imaginary axis.  

The two representations, \Ref{EregRe} and \Ref{EregIm}, are connected by the
known dispersion relation \cite{disprel}
\be\label{disprel}
\ln s_{11}(k)={-1\over 2\pi i}\int_{-\infty}^{\infty}dq \ {\ln (1-\mid
  s_{12}(q)\mid^{2}) \over k-q+i\epsilon} +\sum_{i}\ln{k+i\kappa_{i}\over
  k-i\kappa_{i}} 
\ee
representing the analytic properties of the scattering matrix. 
 
Now we return to a problem whose spectrum is discrete from the very beginning.
Here we do not need to separate a translationally invariant part, but in order
to perform the analytic continuation in $s$ it is useful to rewrite the sum as
an integral. Again, as example we consider the interval $x\in [0,a]$ with
Dirichlet boundary conditions, i.e., the Casimir effect between planes. We
define a function $s(k)$ such that the solutions of the equation $s(k)=0$ are
the eigenvalues $k^2\to\la_{n}$. Further we assume that $s(ik)$ and $s(-ik)$
differ by a factor which is independent of $k$. In the example we can choose
$s(k)=\sin ka$. Then by deforming the integration contour as above we arrive
just at formula \Ref{EregIm} with $s$ instead of $s_{11}$. In this way the
representation \Ref{EregIm} for the \gse is valid in both cases, for discrete
and continuous spectra. 

We illustrate this by another simple example. Consider a potential given by a
delta function, $V(x)=\al \delta(x)$. The transmission coefficient is known
from quantum mechanics textbooks. It reads $t(k)\equiv
s_{11}(k)=\left(1-{\al\over 2i k}\right)^{-1}$. For $\al<0$ it has a pole on
the positive imaginary axis corresponding to the single bound state in the
attractive delta potential. The problem of an equation \Ref{Scpr} with the
delta potential present in the operator $\P$ \Ref{levp1} can be reformulated
as problem with no potential but a matching condition
\be\label{mcond} \varphi(x-0)=\varphi(x+0), \ \
\varphi'(x+0)-\varphi'(x-0)=\al\varphi(0)\,,
\ee
stating that the function is continuous and its derivative has a jump in
$x=0$.  In the formal limit $\al\to+\infty$ we obtain Dirichlet \bc at $x=0$.
In this sense the delta potential can be viewed as a 'semi-transparent'
boundary condition. We note the corresponding formula if two
delta potentials, located at $x=\pm a/2$, are present:
\be\label{twodelta}t(k)=\left( \left(1-{\al\over
      2ik}\right)-\left({\al\over 2ik}\right) \ e^{i k a}\right)^{-1} \,.
\ee
This problem had been considered in \cite{Jaekel1991a,Bordag:1992cm}.
%
%

\subsubsection{Spherically symmetric background}\label{sec3.1.2}
Here we assume the background potential $V(r)$ to depend on the radial variable
alone with the boundary conditions being given on a sphere. For example
Dirichlet \bc read $\Phi({\bf x})=0$ for $r=R$. With the ansatz
\be\label{varsep}\Phi({\bf x})=\frac1r Y_{lm}(\theta,\varphi)
\varphi_{nl}(r)\, ,  \ee
where
$Y_{lm}(\theta,\varphi)$ are the spherical harmonics, the equation for the
one particle energies takes the form similar to Eq. \Ref{evp1}
\be\label{eve} H\varphi_{nl}(r)=\om_{nl}\varphi_{nl}(r)
\ee
with   $H=\P+m^{2}$ and the operator $\P$ reads
\be\label{sevp1} \P=-{\pa^{2}\over \pa r^{2}}+{l(l+1)\over r^{2}}+V(r)
\ee
defining the eigenvalue problem 
\be\label{sevp2} \P \varphi_{nl}(r)=\la_{nl}\varphi_{nl}(r)\,.
\ee
The \ve takes the form now
\be\label{sevp3}\E_{0}(s)=\frac12\sum_{l=0}^{\infty} \left(2l+1\right) 
\sum_{n}\left(\la_{nl}+m^{2}\right)^{1/2-s}\,,
\ee
where the factor $\left(2l+1\right)$ accounts for the multiplicity of the
eigenvalues $\la_{nl}$. Again, we assume the potential $V(r)$ to vanish for
$r\to\infty$ otherwise we had to redefine the mass $m$. 

In the case of a continuous spectrum we are faced with the same problem of
separating the translational invariant contribution as in the preceding
subsection. Here we consider as the ``large box'' a sphere of radius $R$ for
$R\to\infty$. For this task we consider the scattering problem on the half
axis $r\in[0,\infty)$ associated with the operator $\P$, Eq.  \Ref{sevp1},
\be\label{Scprs1} \P \varphi_{nl}(r)= k^{2} \varphi_{nl}(r)\,.
\ee
We need to know several facts from the three dimensional scattering
theory related to the standard partial wave analysis. They can 
be found in
familiar textbooks, see \cite{Taylor} for example, and
can be formulated as follows. Let 
$\varphi_{nl}^{\rm  reg}(r)$ 
be the so called ``regular scattering solution''. It is
defined as that solution to Eq. \Ref{Scprs1} which for $r\to 0$ becomes
proportional to the   solution of the free equation, i.e., of the equation
with $V=0$: 
\be\label{reg0} \varphi_{nl}^{\rm  reg}(r)\raisebox{-0.9ex}{$ \sim \atop r\to
  0$} \  \hat{j}_{l}(kr)\,,
\ee
where $\hat{j}_{l}(z)=\sqrt{\pi z\over 2}J_{l+\frac12}(z)$ is the
Riccati-Bessel function. This solution is known to have for $r\to\infty$ the
asymptotic behavior
\be\label{reginf} \varphi_{nl}^{\rm reg}(r) \raisebox{-0.9ex}{$ \sim \atop
  r\to \infty$} \ \frac{i}2 \left(f_{l}(k) \hat{h}^{-}_{l}(kr)-f_{l}^{*}(k)
  \hat{h}^{+}_{l}(kr)\right) \,,
\ee
where $\hat{h}^{\mp}_{l}(z)=\pm i \sqrt{{\pi z\over 2}}H^{1,2}_{l+1/2}(z)$
 are the Riccati-Hankel functions and
the coefficients $f_{l}(k)$ and  $f_{l}^{*}(k)$ are the   ``Jost
function'' and its complex conjugate respectively. We note the property
\be\label{unitary2} f_{l}(-k)=f_{l}^{*}(k)
\ee
for real $k$ and the relation to the scattering phases $\delta_{l}(k)$
\be\label{unitary3} {f_{l}(k)\over f_{l}(-k)}=e^{-2i\delta_{l}(k)}\,.
\ee

Next, we impose Dirichlet \bc on the regular solution $\varphi_{nl}^{\rm
  reg}(R)=0$. Considered as function of $k$ this equation has the discrete
eigenvalues $\la_{nl}$ as solution, $k^{2}\to\la_{nl}$ and we rewrite the
sum in \Ref{sevp3} as   contour integral
\bea\label{ci1} \E_{0}(s)&=&\frac12\sum_{l=0}^{\infty} \left(2l+1\right) 
\left\{
  \sum_{i}\left(-\kappa_{il}^{2}+m^{2}\right)^{1/2-s}  \nn \right.\\
&&\left.  + \int_{\gamma}{d
    k\over2 \pi i}\left(k^{2}+m^{2}\right)^{1/2-s} {\pa\over\pa k}\ln
  \varphi_{nl}^{\rm reg}(R) \right\} \,.  
\eea
Here the sum over $i$ accounts for the bound states which may be present in
the potential $V(r)$. Their binding energy is $\kappa_{il}$ which in general
depends on the orbital quantum number $l$. The integration path $\gamma$ is
the same as in the preceding section and encloses the positive real axis, see
Fig. \ref{fig1d}.

Now, in order to perform the limit $R\to\infty$, we use the asymptotic
expression \Ref{reginf}. We rewrite it in the form
\bea\label{}  && \ln \left( f_{l}(k) \hat{h}^{-}_{l}(kR)-f_{l}^{*}(k)
  \hat{h}^{+}_{l}(kR)\right)  \\ 
&& =
\left\{
\begin{array}{l}
\ln f_{l}(k)+\ln
\hat{h}^{-}_{l}(kR)+
\ln\left(1-{f_{l}^{*}(k)\hat{h}^{+}_{l}(kR)\over f_{l}(k)
\hat{h}^{-}_{l}(kR)}\right),\\[7pt]
\ln f_{l}^{*}(k)+\ln\hat{h}^{+}_{l}(kR)+
\ln\left(-1+
{f_{l}(k)\hat{h}^{-}_{l}(kR)\over f_{l}^{*}(k)\hat{h}^{+}_{l}(kR)}\right) \,.
\end{array}  \right.   \nn
\eea
Keeping in mind the behavior of the Hankel functions for large arguments,
$\hat{h}^{\pm}_{l}(z)\sim \exp(\pm i (z-l\pi/2))$, we use the first
representation in the \rhs on the upper half of the integration path $\gamma$,
i.e., for $\Im k>0$ and the second one on the lower one. So in both cases we
keep the first contribution in the limit of $R\to\infty$. The second one
doesn't depend on the background and represents the Minkowski space
contribution and must be dropped. The third contribution vanishes at
$R\to\infty$. So we arrive at the representation
\bea\label{Eregs}  \E_0(s)&=&\frac12\sum_{l=0}^{\infty} \left(2l+1\right) 
\left\{
  \sum_{i}\left(-\kappa_{il}^{2}+m^{2}\right)^{1/2-s} \nn \right.\\
&&\left. - \int_{0}^{\infty}{d
    k\over2 \pi i}\left(k^{2}+m^{2}\right)^{1/2-s} {\pa\over\pa k}\ln
  {f_{l}(k)\over f_{l}^{*}(k)} \right\} \,.
\eea
To proceed we have two choices.  By means of Eq. \Ref{unitary3} we obtain the
representation which is analogous to 
\Ref{EregRe} with the integration over the
real axis
\bea\label{EregRes}  \E_{0}(s)&=&\frac12\sum_{l=0}^{\infty} \left(2l
+1\right) \left\{
  \sum_{i}\left(-\kappa_{il}^{2}+m^{2}\right)^{1/2-s}\right. \nn \\
&&  \left. + \int_{0}^{\infty}{d
    k\over\pi }\left(k^{2}+
m^{2}\right)^{1/2-s} {\pa\over\pa k}\delta_{l}(k) \right\} \,.
\eea
By turning the integration contour $\gamma$ in \Ref{Eregs} towards the
imaginary axis we obtain with \Ref{unitary2}
\be\label{EregIms} \E_{0}(s)=- {\cos \pi s\over 2\pi}\sum_{l=0}^{\infty}
\left(2l+1\right) \int_{m}^{\infty}{d k }\left(k^{2}-m^{2}\right)^{1/2-s}
{\pa\over\pa k}\ln f_{l}(ik) \,.  \ee
Again, the contributions arising from moving the contour across the poles from
the zeros of the Jost function on the imaginary axis cancel the explicite
contributions from the bound states. In fact, representation \Ref{EregIms} has
to be handled with care. So for instance, while representations \Ref{Eregs}
and \Ref{EregRes} are valid for $\Re s>3/2$, in \Ref{EregIms} $s$ cannot be
made too large otherwise the integration diverges for $k\to m$. This can be
avoided by encircling the $k=m$ at some small, but finite distance, for
example. The advantage of representation \Ref{EregIms}, as well as
\Ref{EregIm}, is that the integrand is not oscillating for large $k$ and that
its behavior for large argument can be found quite easily. 

To complete these representations of the regularized \gse it remains to note
that both representations, Eqs. \Ref{EregRes} and \Ref{EregIms}, are connected
by the standard dispersion relation which reads in this case 
\be\label{disprels} \ln f_{l}(ik)= \sum_{i}\left(1-{\kappa_{il}^{2}\over
    k^{2}}\right)-\frac{2}{\pi}\int_{0}^{\infty}{d q \ q\over
    q^{2}+k^{2}}\delta_{l}(k) \,.
\ee

In case the spectrum is discrete from the beginning, we do not need to
separate a translational invariant part and can use the sum in Eq. \Ref{sevp3}
directly. However, in order to construct the analytic continuation in $s$ it
is useful anyway to transform it into an integral. Consider as example the
Casimir effect inside a conducting sphere. One of the modes of the \elm field
obeys Dirichlet \bc at $r=R$. The radial wave functions are Bessel functions
like \Ref{reg0} and the solutions of the \bc $j_{l}(kR)=0$ are the roots of
the Bessel functions: $\la_{nl}=j_{n,l+1/2}/R$. Although this sum is
explicite so far, its analytic continuation is not easy to construct. It is
advantageous to obtain an integral representation which can be continued 
much more easily. 
The procedure is essentially the same as in the preceeding subsection.
For definiteness we demonstrate it here on the given example. So consider the
\gse \Ref{sevp3}. The sum can be rewritten as integral
\be\label{exam1} \E_0(s)=\frac12\sum_{l=0}^{\infty} \left(2l+1\right)
\int_{\gamma}{d k\over2 \pi i}\left(k^{2}+m^{2}\right)^{\frac12-s} 
{\pa\over\pa  k}\ln \left(k^{-(l+\frac12)} J_{l+\frac12}(kR)\right) \,.  \ee
The zeros of the function in the logarithm are just the roots of the Bessel
function and the contour $\gamma$ must include all of them. Now we note the
following technical point. We are free to modify the function within the
logarithm by any function which is analytic inside the contour $\gamma$
without changing the integral. Also we note that a constant there does not
contribute due to the derivative $\pa/\pa k$. We used this freedom to
introduce the factor $k^{-(l+1/2)}$.  The reason is that we want to turn the
contour towards the imaginary axis. Then it crosses the point $k=0$. Without
the factor we introduced the integrand by means of $\pa/\pa k \ln
\left(J_{l+1/2}(kR)\right)\sim {l+\frac12\over k}$ had a pole there delivering
an extra contribution. With the factor introduced this is a regular point and
we can move the contour towards the imaginary axis.  With
$j_{\nu}(iz)=\exp(i\nu\pi/2)I_{\nu}(z)$ where $I_{\nu}(z)$ is the modified
Bessel function we obtain just the same expression \Ref{EregIms} as in the
case of a continuous spectrum with the substitution
\be\label{subst}f_{l}(ik)\to
k^{-(l+1/2)}I_{l+1/2}(k)
\ee
for the Jost function. 

We want to conclude this section by another example. First of all let us
consider the exterior of a sphere with Dirichlet boundary conditions. We can
use the formulas given above for the related scattering problem. We need the
regular solution of the scattering problem \Ref{Scprs1}. Because of $r\ge R$
in this case we don't need the condition \Ref{reg0}. The operator $\P$ is
given by Eq. \Ref{sevp1} with $V(r)=0$ and the solution coincides with a free
one. There are two independent solutions and we have to choose the linear
combination which fits the asymptotic expression \Ref{reginf}. The solution
in this case is just given by the \rhs of Eq. \Ref{reginf}. Now we impose the
\bc and arrive at the equation $\left(f_{l}(k)
  \hat{h}^{-}_{l}(kR)-f_{l}^{*}(k) \hat{h}^{+}_{l}(kR)\right)=0$ from which
we determine the Jost function as
\be\label{kugela}f_{l}(k)=i(kR)^{-(l+1/2)}\hat{h}^{+}_{l}(kR) \,,
\ee
where the factor in front of the Riccati-Bessel function is chosen in a way
that $f_{l}(k)$ is regular for $k\to 0$.

Frequently, for the modified spherical Riccati-Bessel functions the notations 
\be\label{s_e}\hat{j}_{l}(iz)=i^{l}s_{l}(z) \ \mbox{and} \
\hat{h}_{l}^{+}(iz)=(-i)^{l}e_{l}(z) \ee
are used. So we can represent the Jost functions \Ref{subst} and \Ref{kugela} 
as
\be\label{}f_{l}(ik)=\left\{\begin{array}{c}(kR)^{-(l+1/2)}s_{l}(kR), \\
(kR)^{l+1/2}e_{l}(kR),\end{array}\right.
\ee
for the problem inside and outside the sphere with radius $R$
respectively. Note that
the factors $\left(kR\right)^{\pm(l+1/2)}$ compensate each other when
considering the two problems together. In that case we have to add  the
corresponding vacuum energies and, consequently the logarithms, so, that we
obtain the same expression given by Eq. \Ref{EregIms} with
$f_{l}(ik)=s_{l}(kR)e_{l}(kR)$. 

\subsection{The \hke}\label{sec3.2}
Here we consider the heat kernel expansion as the most suited tool to
investigate the divergence structure of the vacuum energy.  We start
from a general operator $\P$. Specific examples are those given by
Eqs. \Ref{levp1} or \Ref{sevp1} on a manifold {\Mf}. However, the
formulas given below are valid in general for elliptic operators which
may be pseudo differential operators also.  
Likewise, an extension to first
order operators, i.e. the Dirac operator are possible, where
some special considerations must be taken into account. Roughly
speaking one has to take some quadratic combination of the Dirac
operator $D$ such as $D^{\dagger}D$.  
Readers interested in more details should
refer to the paper \cite{DeWittdyn}. In this and the following two
sections we write the formulas for a massive scalar field in order to
avoid unnecessary technical details like internal indices. 
We would like
to promote the understanding of the underlying essential ideas
which are the same for all theories.

In case this manifold has a boundary, \dMf, one has to impose some \bc in
order to have a symmetric operator $\P$. We consider here Dirichlet  and
Neumann \bcp  
Let $\la_{J}$
and  $\varphi_{J}(x)$ be the corresponding eigenvalues and
eigenfunctions respectively. 
The local heat kernel is defined then by
\be\label{lhk}K(x,y|t)=\sum_{J}\varphi_{J}(x)\varphi_{J}^{*}(y)
e^{-t\la_{J}} \,, \ee
It is defined for operators with a discrete and/or continuous spectrum.  The
global heat kernel can be obtained formally as trace over the local one
\be \label{ghk} K(t)\equiv Tr K(x,y|t) =\sum_{J}e^{-t\la_{J}}\,,
\ee
where the trace means an integral over the manifold, $\Tr K(x,y|t)=\int dx \tr
K(x,x|t)$ where $\tr$ is over the internal indices, e.g., corresponding to
some symmetry group. 

The heat kernel obeys the Helmholtz equation
\be\label{hkeq} \left({\pa\over\pa t}+\P\right) K(x,y|t)=0
\ee
with the initial condition
\be \label{indat} K(x,y|t=0)=\delta(x-y)
\ee
and describes the diffusion of heat from a pointlike source. It can be
defined for a very wide class of operators and on quite general manifolds. The
appropriate language is that of differential geometry and geometric objects on
a Riemannian manifold equipped with a connection and some endomorphism (scalar
background field).  However, we do not use this in full generality and restrict
the discussion 
to the necessary minimum for the understanding of the divergencies of
the Casimir energy.

The zeta function $\zeta_{H}(s)$, Eq. \Ref{zetaH}, can be expressed 
as an integral
over the heat kernel. By representing the power of the eigenvalues in
\Ref{zetaH} by
\be\label{euga}{A}^{-s}={1\over \Gamma(s)}\int_{0}^{\infty} dt \ t^{s-1} \
e^{-tA} \,
\ee
($\Re s>0$, $\Re A >0$),
which is in fact an integral representation of the Euler gamma function and
changing the order of summation and integration (this is possible due to the
convergence of both for $\Re s >s_{0}$) we obtain with \Ref{ghk}
\be\label{zausK} \zeta_{H}(s)=
{1\over \Gamma(s)}\int_{0}^{\infty} dt\  t^{s-1} K(t)\,.
\ee

Now, from an inspection of Eq. \Ref{zausK} it is clear that the integral is
well behaved for $t\to\infty$ and possible singularities
of the zeta function result only from the lower $t\to 0$  behavior of the
integrand. Therefore we need information on $K(t)$ for $t\to 0$. 
This is given by
the heat kernel expansion \cite{gilkey} which we note in the form
\be\label{hke1}K(t)=
{1\over\left({4\pi t}\right)^{\frac{d}2}}\sum_{n=0,1/2,1,\dots}a_{n}t^{n} \,,
\ee
where $d$ is the dimension of the manifold \Mf.  Eq. \Ref{hke1} is an
asymptotic expansion for $t\to 0$ and it is in general not a converging
series. The coefficients $a_{n}$ are called \hkksp

There is of course an analogous expansion for the local heat kernel.  Let us
first consider the heat kernel of a free problem for a scalar field, i.e.,
without potential or \bcp The operator is $\P=\Delta+m^{2}$. In that case Eq.
\Ref{hkeq} has the solution
\be\label{hk0} K^{(0)}(x,y|t)={1\over (4\pi
  t)^{d/2}}\exp\left(-{(x-y)^{2}\over 4t}-t  m^{2}\right) \,,
\ee
where $m$ is the mass of the field.  This solution can be easily verified by
inserting into Eq. \Ref{hkeq}. The initial condition \Ref{indat} is 
also satisfied
because for $t\to 0$, the \rhs of Eq.  \Ref{hk0} is a representation of
the delta function.
 
The idea for the behavior of the heat kernel for $t\to 0$ comes from the
observation that the leading \uv divergence in the presence of a background is
the same as in  free space. For that reason the ansatz for the expansion
\be\label{hkel}K(x,y|t)=K^{(0)}(x,y|t) \, \sum_{n\ge 0}a_{n}(x,y)t^{n}
\ee
is meaningful. The coefficients $a_{n}(x,y)$ are called 'local heat kernel
coefficients'. 

The connection between the local and the global coefficients is very simple on
a manifold without boundary
\be\label{glo}a_{n}=\Tr \ a_{n}(x,y)=\int_{M}dx \ \tr \ a_{n}(x,x)\,.
\ee
In that case, from inserting the expansion \Ref{hke1} into Eq. \Ref{hkeq}, the
recurrence relations
\bea\label{recurr}(x-y)_{i}{\pa\over\pa x^{i}} a_{0}(x,y)&=&0\,,\\
((x-y)_{i} {\pa\over\pa x^{i}}+n+1)a_{n+1}(x,y)&=&(\Delta-V(x))a_{n}(x,y) ~~~
(n=1,2,\dots) \nn
\eea
follow, see for example \cite{DeWittdyn}. They allow the determination of
these coefficients and their derivatives very easily.  
We note here as illustration
the first coefficients for a flat manifold with a background potential
$V(x)$ in the so called coincidence limit, i.e., for $y=x$:
\bea\label{reccoeffs}
a_{0}(x,x)&=&1\,, \nn\\
a_{1}(x,x)&=&-V(x)  \,,  \\
a_{2}(x,x)&=&-\frac16 \Delta V(x)+\frac12 V^{2}(x)   \,. \nn
\eea

However, these recurrence formulas do not work on manifolds with boundary or
with singular background potentials. The simplest example is a background
potential given by a delta function. While the coefficient $a_{1}$ is well
defined, whereas in $a_{2}$ the delta function appears squared and the
expression in \Ref{reccoeffs} becomes meaningless. Also, it is well known that
on a manifold with boundary there are in addition coefficients with half
integer numbers and the corresponding powers of $t$ in the expansion
\Ref{hke1}. There is a general framework to determine these coefficients based
on the formalism of pseudo-differential operators which is however not very
useful in specific calculations. During the past 
few years progress had been made
in a combination of conformal techniques and special case calculations. So for
Dirichlet and Robin\footnote{Robin \bc on a field $\varphi(x)$ are given by
  $(\pa/\pa n -h(x))\varphi(x)_{\mid_{x\in S}}=0$ where $h(x)$ is some
  function defined on the boundary. For $h=0$ they turn into Neumann \bcp} \bc
the coefficient $a_{\frac52}$ had been calculated for an arbitrarily 
shaped smooth
boundary. For boundaries with symmetries, a sphere or a generalized cone, the
coefficients up to quite high numbers are available.

Due to their properties as distributions, the structure of the coefficients can
be best expressed in the smeared form with a test function $f(x)$ as integral
over the manifold and  over its boundary
\be \label{smeared} a_{n}(f)=\int_{M} dx \ f(x)a_{n}(x,x)+
\int_{\partial M}dS_{x}
\ f(x)b_{n}(x) \,,
\ee
where $b_{n}(x)$ are the boundary dependent contributions. It is important
that these coefficients are all local in the sense that they may be
represented as integrals taking information from the background only at one
point.

In Eq. \Ref{smeared} the volume integrals contribute to coefficients with
integer number only whereas the surface integrals deliver coefficients with
integer and with half integer numbers. Another remarkable fact is  that the
coefficients can be expressed solely in terms of geometric characteristics of
the {\Mf}  like curvature and its derivatives and in this way they do not 
depend
on the dimension of the manifold. The first two coefficients are 
\be\label{exa}a_{0}(f)=\int_{M}dx   \ f \,, ~~~~~~~~~~
a_{\frac12}(f)=\frac{\sqrt{\pi}}{2}
\int_{\partial M}dS_{x} \ f \,.
\ee
So, $a_{0}(1)$ is the volume and 
$\frac{2}{\sqrt{\pi}}a_{\frac12}(1)$ is the surface area of the
manifold.  It follows that these two coefficients do not depend on the details
of the background and do not carry any information of  interest to us.
Especially in case of a flat manifold, $a_{0}$ is the so called Minkowski
space contribution. Mostly, it is of relevance only in curved space-time where
it enters the renormalization of the cosmological constant.

In order to give an impression on the general structure of the coeffecients
we note the next one, $a_{1}$, in the case of Dirichlet boundary conditions
\bea\label{a1}a_{1}(f)&=&\int_{M}dx f(x)(V(x)+\frac16 R(x)) \\
&&+\int_{\partial M}dS_{x}
( \frac13 f(x)L_{\al\al}+\frac12 f_{;N}(x)) \,, \nn
\eea
where $R(x)$ is the scalar curvature of \Mf, $L_{\al\al}$ is the trace of the
second fundamental form on {\dMf}  and $f_{;N}$ is the normal derivative of
$f$. For more details on these quantities the reader should consult a textbook
on differential geometry. 

It is important to notice that all higher coefficients after $a_{0}$ and
$a_{\frac12}$ are proportional to the background potential $V(x)$ or to the
mentioned geometrical quantities and their derivatives. It follows, for
instance, that for a flat {\Mf} without boundary or with flat boundary
(especially for plane parallel planes) all coefficients except for $a_{0}$
vanish. This is also the case in a field theory with temperature where the
corresponding manifold is simply $S^{1}$.
 
 To conclude the discussion of the general properties of the \hkks we note
 that the boundary dependent contributions $b_{n}$ to the coefficients inside
 and outside the boundary are connected by
\be\label{intext}{\mbox{$b_{n}$}}_{\bigl|_{inside}}=(-1)^{2n+1}
{\mbox{$b_{n}$}}_{\bigl|_{outside}}\,, \ee
i.e., the boundary dependent coefficients with half integer number are the
same on two sides of the boundary whereas that with integer numbers have
different signs.

In order to further 
illustrate the topic we note below the \hkks for some simple
configurations, such as 
for the Laplace operator with Dirichlet (D) and Neumann (N)
 \bc on a sphere and for a penetrable sphere (ps) given by a delta function
potential in three dimensions:
\be\label{hkksex}
\begin{array}{c|ccc}
&D&N&ps \\ \hline
a_{0}&\frac43 \pi R^{3}   & \frac43 \pi R^{3}  & \\
a_{\frac12}&- 2\pi^{3/2} R^{2}   & 2 \pi^{3/2} R^{2}  & 0 \\
a_{1}&\pm\frac83 \pi R   & \pm\frac{16}{9} \pi R  & -4\pi \al R\\
a_{\frac32}&- \frac16\pi^{3/2}    & \frac76 \pi^{3/2} & \pi^{3/2} \al^{2} \\
a_{2}&\mp\frac{16}{315}\frac{\pi}R   &  \pm\frac{16}{9}\frac{\pi}R  &
-\frac{2\pi}{3}\frac{\al^{3}}{R} \,.
\end{array}
\ee
Here the upper (lower) sign corresponds to the interior (exterior)
region (except for $a_0$ which is infinite in the exterior 
region) and $\al$ is the coefficient in front of the delta
function.  Note that for the penetrable sphere there is no
subdivision into interior and exterior region.

The calculation of the \hkks is  quite complicated but now a
large number of them are known.  For manifolds without boundary the
most complete calculation is done in \cite{Avramidi:1991je} and in
\cite{vandeVen:1997pf}, see also \cite{Barvinsky:1985an}. There is
even a computer program provided for that in \cite{Booth:1998xd}.  The
coefficients for Dirichlet and Neumann \bc can be found in
\cite{kenn78-11-173}. For a general surface the coefficients are given
in
\cite{bran90-15-245,Vassilevich:1995we,Branson:1997cm,Branson:1999jz,Kirsten:1997qd}. In
\cite{Bordag:1996gm} the coefficients for a d-dimensional sphere with
Dirichlet and Robin  \bc are given up to $n=10$. The coefficients
for the penetrable sphere where first calculated in
\cite{Bordag:1999vs} and generalized to an arbitrary penetrable
surface in \cite{Bordag:1999ed}. The coefficients for the dielectric
ball are given in \cite{Bordag:1999vs}. They don't have an explicit
expression. For a particular example see Eq. \Ref{a2diel} below.

\subsection{The divergent part of the \gse}\label{sec3.3}

We start with the representation 
\be\label{zreg}\E_{0}(s)=
\pm \frac{\mu^{2s}}{2}\sum_{J}(\la_{J}+m^{2})^{\frac12-s}
\ee
of the \gse in zeta-functional regularization.  Here we have shown the
dependence on the mass explicitly, 
and $\la_{J}$ are the eigenvalues of the
corresponding operator such as $\P$, Eq. \Ref{sevp1}. 
In parallel we consider the
frequency-cutoff regularization 
\be\label{dreg}\E_{0}(\delta)=\pm \frac12
\sum_{J}{(\la_{J}+m^{2})^{\frac12}}  \ e^{-\delta{(\la_{J}+
m^{2})^{\frac12}}}\,,
\ee
where the convergence is achieved by the exponential damping function and we
have to put $\delta=0$ in the end. Both \regs as well as many other have their
own advantages and limitations. The zeta-functional \reg is the most elegant
known \reg with pleasant mathematical properties. However, one has to take
care not to have modes with zero eigenvalue $\la_{J}=0$ for a massless
theory because $0^{-s}$ is ill defined. Here the mass can serve as an
intermediate infrared regulator. The cutoff \reg \Ref{dreg} has the physically
very intuitive meaning as to cut at frequencies $\sim1/\delta$ where any real
mirror becomes transparent.  With both \regs an arbitrariness comes in. In the
zeta-functional \reg there is the arbitrary parameter $\mu$ with dimension of a
mass which can be introduced (in the limit of $s\to 0$, i.e., of formally
removing the regularization it disappears). It gives the regularized \gse the
correct dimension. In the cutoff \reg the parameter $\delta$ (it has the
dimension of an inverse mass) can be multiplied by any finite, positive
number.

For physical reasons, all regularizations must deliver the same end result.
How to achieve this is subject to the normalization condition to be
discussed below. To conclude the general considerations, we would like to
note that in introducing a \reg one has to take care that it in fact removes
the divergencies, i.e., that it delivers a mathematically correct defined
expression. 

We want to study the divergent part of the \gse by making use of the \hkep 
For the \gse in zeta-functional \reg \Ref{zreg} we use formula \Ref{euga} 
and obtain
\be\label{zr1}\E_{0}(s)=\pm
\frac{\mu^{2s}}{2}\sum_{J}\int_{0}^{\infty}{dt\over
  t}{t^{s-\frac12}\over\Gamma(s-\frac12)} \ e^{-t(\la_{J}+m^{2})}\,.
\ee
Due to the convergence for $\Re s>\frac32$, the sum and the integral in the
\rhs may be interchanged and by means of the heat kernel
\be\label{hk}K(t)=\sum_{J}e^{-t\la_{J}} \,,
\ee
where the mass is not included into the heat kernel (in difference to
Eq. \Ref{ghk}), we obtain
\be\label{zr2}\E_{0}(s)=\pm
\frac{\mu^{2s}}{2}\int_{0}^{\infty}{dt\over
  t}{t^{s-\frac12}\over\Gamma(s-\frac12)} K(t) \ e^{-tm^{2}}\,.
\ee
Now, as we are interested in the divergent part of the  \gsek we can
insert the \hke into the \rhsp 
Then the $t$-integration can be carried out simply
by applying formula \Ref{euga} and we arrive at 
\be\label{}\E_{0}(s)=\pm
\frac{\mu^{2s}}{2}\sum_{n\ge0}{a_{n}\over (4\pi)^{3/2}}
{\Gamma(s+n-2)\over\Gamma(s-\frac12)}m^{2(2-s-n)} \,. 
\ee
From this expression it is seen by inspection that only the coefficients with
numbers $n\le 2$ contribute to divergencies at $s=0$. We take these
contributions and {\it define} the {\it divergent part} as sum 
of the non-vanishing terms
for $s\to0$  resulting from the \hkks with numbers $n\le2$:
\bea \label{Edivz}\E_{0}^{\rm div}  &=& -\frac{m^4}{64\pi^2}\left(\frac 1 {s} 
+\ln\frac
{4\mu^2}{m^2} -\frac 1 2 \right) a_0 -{m^{3}\over 24\pi^{3/2}}a_{1/2}\nn\\ 
& &+\frac{m^2}{32\pi^2}\left(\frac 1 {s} 
+\ln\frac{4\mu^2}{m^2} -1\right) a_1+{m\over 16 \pi^{3/2}}a_{3/2} \\ 
& &-\frac 1 {32\pi^2} \left(
\frac 1 {s} +\ln \frac{4\mu^2}{m^2}-2\right) a_2. \nn\eea
In fact, it contains some finite  contributions at $s=0$ also. 
Furthermore we
observe that for dimensional reasons $\E_{0}^{\rm div}$ contains only
nonnegative powers of the mass and that 
all terms of $\E_{0}^{\rm div}$ are of this type. 

A similar procedure can be applied to the cutoff \reg \Ref{dreg}. It is 
 technically a bit
more involved. First of all let us consider the integral
\be\label{Is1}I(s)\equiv\int_{0}^{\infty}d\delta \ \delta^{s}\E_{0}(\delta)\,.
\ee
Using Eq. \Ref{dreg} and formula \Ref{euga} we obtain
\be\label{Is2}I(s)=\frac12\Gamma(s+1)\sum_{J}
(\la_{J}+m^{2})^{-\frac{s}2}\,
=\frac12\Gamma(s+1)\zeta_{H}\left(\frac{s}2\right)\,.
\ee
So, $I(s)$ is proportional to the corresponding zeta function of the half
argument. Using a Mellin transform we obtain from Eqs. \Ref{Is1} and \Ref{Is2}
\be\label{}\E_{0}(\delta)=\frac12 \int_{-i\infty}^{i\infty}{ds\over 2\pi
  i}\delta^{-s-1}\Gamma(s+1)\zeta_{H}\left(\frac{s}2\right)\,, 
\ee
which is an expression of the \gse in cutoff \reg 
and given as an integral of the
Mellin-Barnes type of the zeta function. Here the integration path goes
parallel to the imaginary axis to the right of the poles of the integrand.
Being interested in the divergent part, i.e., in the behavior for $\delta\to
0$, we move the integration path to the left and collect the contributions
resulting from crossing the poles of the integrand. These poles are known to
result from the \hke  inserted into the representation \Ref{zausK} of
the zeta function. We obtain
\bea\label{}\E_{0}(\delta)&=& \frac12 \int_{-i\infty}^{i\infty}{ds\over 2\pi
  i}\delta^{-s-1}\Gamma(s+1)
 \int_{0}^{\infty}
{dt\over  t} {t^{\frac{s}2}\over\Gamma(\frac{s}2)} 
    \sum_{n\ge0} {a_{n}t^{n}\over (4\pi  t)^{3/2}} \ e^{-tm^{2}} \\
&=&  {1\over 16\pi^{3/2}}  
\int_{-i\infty}^{i\infty}{ds\over 2\pi  i} \ \delta^{-s-1} \
{\Gamma(s+1)\over\Gamma(s/2)}
  \sum_{n\ge0}a_{n}\Gamma\left(\frac{s-3}2+n\right) \ m^{3-s-2n} \,.\nn 
\eea
From this representation it is seen that there are poles at $s=3,2,1$. There
is no pole in $s=0$ due to the gamma function in the denominator. Furthermore
there is a double pole in $s=-1$. The poles to the left of $s=-1$ contribute
positive powers of $\delta$ and we do not need to consider them. Calculating
the contributions from the poles at $s=3,2,1,-1$ we obtain 
\bea\label{Edivd}\E_{0}^{\rm div}(\delta)&=& \frac{1}{16\pi^{2}}
\Bigg\{\left({24\over \delta^{4}}-2{m^{2}\over \delta^{2}}+{m^{4}\over
      2}\ln \delta\right) a_{0}
+{4\sqrt{\pi}\over \delta^{3}}a_{\frac12}   \nn \\
&&+\left({2\over \delta^{2}}-m^{2}\ln\delta\right)a_{1}
+ (\ln \delta) \ a_{2} \Bigg\} \,.
\eea
From this formula, it is seen that all contributions are divergent including
that from the coefficient $a_{\frac12}$. Moreover, these divergent
contributions are also present in the massless case so that their absence in
the zeta-functional regularization (except for $a_{2}$) is due to the specific
form of the regularization.

\subsection{Renormalization and normalization condition}\label{sec3.4}

In order to perform the renormalization we need a quantity to be
renormalized. In the framework of perturbative \qft there are the bare
constants (masses, couplings, etc.) which become renormalized by the
corresponding counter-terms -- a well known procedure. In case of the vacuum
energy (as well as other vacuum expectation values) the corresponding
quantities are the classical background fields 
that the \vevs depend on. The
simplest example to be considered is a scalar field theory with a classical
background field $\Phi(x)$ and a quantum field $\varphi(x)$ with the action 
\bea \label{scmod}S&=&\frac 1 2
  \int\d x~\left\{\Phi(x) (\Box +M^2 +\lambda \Phi^2 (x) )\Phi(x)
    +\nn\right.\\
  &&\left. ~~~~~~~~~~~~~~\varphi(x) (\Box +m^2 +\lambda ' \Phi^2(x)
    )\varphi(x)\right\},\eea
where $V(x)=\la'\Phi^{2}(x)$ is the background potential 
that is coupled to the quantum field. 
Now there is a classical energy associated with the background 
\be\label{3.4.2}\E^{\rm class}=\frac{1}{2} \int d^3 { x} ~\left(\left(\nabla
      \Phi(x)\right)^{2}+M^{2}\Phi^{2}(x)+\lambda \Phi^{4}(x)\right)
\ee
where we assumed the background to be static. Furthermore we have the \gse
of the quantum field which is given e.g. in zeta-functional \reg by
Eq. \Ref{zreg} and there is the complete energy of the system as a whole which
is the sum of these two. The renormalization procedure consists simply in
subtracting the divergent part of the \gsek e.g., given by Eq. \Ref{Edivz},
from  the \gse and adding it to the classical energy,
\bea \label{renallg}\E&=&\underbrace{\E^{\rm class}+\E_{0}^{\rm div}}+
 \underbrace{\E_{0}-\E^{\rm div}_{0}}\nn\\
&\equiv&~~~~~~\tilde{\E}^{\rm class}~~~+~~~~~\E^{\rm ren}_{0}.
\eea
The change from $\E^{\rm class}$ to $\tilde{\E}^{\rm class}$ can be
interpreted as a renormalization of the parameters of the classical system. In
the first model it reads:
\bea M^{2}&\to&M^{2}-{\lambda'm^{2}\over 16\pi^{2}}\left(\frac 1 s
  +\ln{4\mu\over m}-1\right),\nn\\
\lambda&\to&\lambda-{\lambda'm^{2}\over 64\pi^{2}}\left(\frac 1 s
  +\ln{4\mu\over m}-2\right).\label{ren}\eea

The divergence associated with $a_{0}$ from Eq. \Ref{Edivz} would lead
to a renormalization of a constant addendum to the classical
energy. As noted above, we drop such a contribution.
 
 This procedure works, in principle, for any background field. Some of the
 structures in the classical action may be missing 
for the  case of the corresponding
 \hkk being zero. Other coefficients  
may be present, for instance those with half
 integer numbers in case of a singular background field, say containing a
 delta function as considered e.g. in \cite{Bordag:1999vs}. 
The best known example is
 half-classical gravity. In this connection it is interesting to remark that in
 (3.82) we observe a renormalization of the self coupling $\la$ of the
 background field which in this sense is an inevitable part of the classical
 action. The terms to be included for renormalization are determined primarily
 by dimensional reasons. In the same manner one needs to include the
 contributions quadratic in the curvature into the left-hand side of the
 Einstein equation.

We would like to stress that it is very important
which interpretation we can give to the vacuum expectation values. In
general they must be viewed as a quantum correction to the classical
system, given e.g. by $\Phi(x)$ in \Ref{3.4.2}. This system must have its
own dynamics from which its characteristics like $M$ or $\la$ may be
determined or there must be additional, say experimental knowledge on
them. In any case they cannot be determined from the \gsek i.e., from
calculating quantum corrections to them. So we are left with the
question how to give a unique meaning to the \vevsp As we have seen with
introducing a regularization, some arbitrariness creeps in. In fact, this
is known from perturbative \qftp There are reasons, e.g., in QED, like
observable masses or couplings which have unique normalization
conditions. For \vevs viewed as quantum corrections to some
classical system there is also a natural normalization condition. One
has to require that in the limit of a large mass of the quantum field
its \vev must vanish,
\be\label{ncond}\Er \raisebox{-0.9ex}{${\to \atop m\to\infty}$} 0\,.
\ee
This condition is natural since a field in the limit of infinite mass should
not have quantum fluctuations. From the technical point of view this
condition removes completely the arbitrariness of the renormalization
procedure. The reason for this is that all divergent contributions come along
with non-negative powers of the mass. This follows from dimensional reasons
together with the well known fact that the heat kernel expansion is also an
adiabatic expansion in the limit of a large mass. 

There is no general normalization condition for a massless field, however. As
it is shown in \cite{Bordag:1999vs} a \gse normalized according to 
\Ref{ncond} doesn't
have a finite limit for $m\to0$ except for the case of a vanishing \hkk
$a_{2}$. Therefore, in the case of $a_{2}\ne0$, the \gse of a massless field
cannot be uniquely defined and, hence, it is physically meaningless. Below we
will discuss some examples of this case.  

It is necessary to stress the point  that it is just the coefficient $a_{2}$
which becomes problematic. 
This is because the non-uniqueness comes from the
logarithmic contributions. Moreover, in the zeta-functional \reg  for a
massless field it is only the contribution from $a_{2}$ which is present (see
Eq. \Ref{Edivz} for $m=0$). In another \regk e.g., the frequency cutoff,
Eq. \Ref{Edivd}, also at $m=0$, the only logarithmic contribution is the one
proportional to $a_{2}$. 

Some special considerations are required for the 
computation of \gse  in the case of boundaries, such as 
for the Casimir effect. Here we have two different
situations. First, let us consider a setup with two distinct bodies with \bc
on their surfaces and assume we are interested only in the dependence of the
\gse on the distance between these bodies. This is the typical situation for
measurements. In that case the \hkks (except for $a_{0}$ which is of no
interest here) have only surface contributions. Now, the surface {\dMf}
constitutes of two parts, 
$\partial M=\partial M_{1}\cup\partial M_{2}$ 
corresponding to the two
bodies and the \hkks become a sum of two contributions according to 
\be\label{sumcoeff}a_{n}=\int_{\partial M_{1}}d S \ b_{n}^{(1)}+
\int_{\partial M_{2}}d S \
b_{n}^{(2)} \,.
\ee
Due to the locality of the coefficients there is no dependence on the distance
between the bodies and, consequently, no distance dependent singularities or
ambiguities. 

Now we consider the vacuum energy for \bc given on one body so that the energy
itself has a certain meaning. One can think for example of a conducting
sphere where the dependence of the \gse on the radius defines a pressure. In
that case, as discussed, e.g., in \cite{Blau:1988kv}, one has to introduce
a classical energy associated with the geometrical configuration under
consideration. So for a spherical surface one has to consider 
\be \label{Eclass}\E^{\rm class}=pV+\sigma\Sigma+FR+k+\frac h R,
 \ee 
 in accordance with the dependence of the \hkks (3.66) on the radius
 $R$. Here $V=\frac 4 3 \pi R^{3}$ is the volume and 
$\Sigma=4 \pi R^{2}$ is the
 surface of the sphere. Correspondingly, $p$ is the pressure and $\sigma$ is
 the surface tension. The parameters $F$, $k$ and $h$ do not have a special
 meaning.  Now the addition of $\E^{\rm div}_{0}$ to $\E^{\rm class}$ in
 \Ref{renallg} can be reformulated as a redefinition of the corresponding
 constants $p,\sigma,\dots $. 
With regard to the normalization conditions the same
 consideration apply as above.

It is of particular interest to consider the Casimir effect for a conducting
sphere from this point of view. First of all we observe that the problem
initially can be divided into two, one for the interior and the other for the
exterior of the sphere. The quantum fields inside and outside are completely
independent from each other because the sphere with conductor (or Dirichlet
\resp Neumann) \bc is impenetrable. In case of a massive quantum field  its
vacuum energy can be uniquely defined (and calculated \cite{Bordag:1997ma}) 
in each region
independently thanks to the normalization condition \Ref{ncond}. However, for
a massless field this is not the case. Here we have $a_{2}\ne 0$
(e.g. $a_{2}=\pm {16\pi\over315R}$ for Dirichlet \bck see Eq. \Ref{hkksex}) 
and uniqueness cannot be
achieved. This can be seen in the given example by noticing that the vacuum
energy  for dimensional reasons is proportional to $1/R$. Therefore it is
impossible to formulate a condition of the type that the vacuum energy must
vanish for $R\to\infty$ because it doesn't remove the arbitrariness which
results from the logarithmic term which is, of course, also proportional to
$1/R$. The situation, however, changes when considering the interior and the
exterior of the sphere together. In that case the contributions 
to $a_2$ from inside
and from outside  cancel and an unique calculation of the \gse is
possible. For the same reason one can drop the contributions in the classical
energy which are generated by $a_{0}$, $a_{1}$ and $a_{2}$. 

Equipped with this knowledge it is interesting to consider the first
calculation \cite{16} of the Casimir effect for a conducting
sphere. The \elm field obeys conductor \bc and a frequency cutoff \reg was
taken. In that work a delicate cancellation of divergencies was observed. In
fact, the most divergent contribution is that delivered by $a_{0}$ in Eq.
\Ref{Edivd}. It vanishes when taking together the modes from inside 
and from outside. The same applies to $a_{1}$ and $a_{2}$. 
The remaining divergence is
that resulting from $a_{\frac12}$. It vanishes when taking both 
the TE and the TM
modes, i.e., the contributions from Dirichlet and Neumann \bcp It
was this coincidence of cancellations which made that important calculation
possible. Let's note that in zeta-functional \reg the only divergent
contribution is that of $a_{2}$ (see Eq. \Ref{Edivz} for $m=0$) which vanishes
when taking the   
inside and outside contributions together. Hence, in this \reg the
contributions from the TE and from the TM modes have 
individually a finite vacuum energy.

There is another example much discussed during the last few years where the
situation is not so pleasant, namely the dielectric ball. The \hkks have been
calculated in \cite{Bordag:1999vs} with the result that $a_{2}$ is in general
nonzero.  It vanishes only in the dilute approximation, i.e., for small
$\ep -1$ \resp for small difference in the speeds of light 
$c_{1}$ and $c_{2}$
inside and outside the ball. It holds
\be\label{a2diel}a_{2}=-{2656\pi\over 5005 R} \ {(c_{1}-c_{2})^{3}\over
  c_{2}^{2}} +O\left((c_{1}-c_{2})^{4}\right)\,.
\ee
From this fact two consequences follow. First, all calculations
in the dilute approximation (there are at least three different ones,
by mode summation \cite{11}, by summing the pairwise
Casimir-Polder forces \cite{11} and a perturbative approach
\cite{Barton1999c}) must deliver the same result. In fact they do. They are
performed in different regularizations and some divergent
contributions had to be dropped which is possible in a unique way
because there are no logarithmic contributions. The same applies to
calculation with equal speed of light inside and outside. The second
conclusion is that beyond the dilute approximation the \gse of the
\elm field is not unique. In fact this has the consequence that the
approximation of real matter in which a ball is characterized solely
by a dielectric constant $\ep$ constitutes an ill defined problem. At
the moment there is no satisfactory explanation for this. Let us
conclude with the remark that the same problem exists for a conducting
spherical shell of finite thickness as discussed in
\cite{Bordag:1999vs}.

The calculation of the Casimir energy and force for a ball and other
configurations will be considered in more detail in Sec. \ref{sec4}. 

\subsection{The photon propagator with \bc}\label{sec3.5}
In order to investigate the properties of the vacuum it is in most cases 
possible to investigate global quantities like the \gsep Here, a knowledge of
the spectrum is sufficient,  along with formulas discussed in the
preceding subsections like \Ref{Ereg1} or \Ref{EregIms}, to calculate (at
least in principle) the quantities of interest.
The eigenfunctions or propagators are not needed to this end.
We are presented 
with a somewhat different situation if  one is concerned 
with subjects such as
semiclassical gravity where there are difficulties with the definition of the
global energy and the investigation of the (local) Einstein equations is
preferred by most people.  The local vacuum energy density may be of interest
also in cases where the global energy doesn't exist because the background is
too singular. An example is the vacuum energy density in the background of an
infinitely thin magnetic string which is well defined at any point outside the
flux line but cannot be integrated in its vicinity. Yet another example is the
calculation of radiative corrections to the vacuum energy. Here, a higher loop
graph must be calculated where one of the  propagators (or all) obey \bcp 
A knowledge of only the spectrum is clearly insufficient and one needs
knowledge of the eigenfunctions also. 

The photon 
propagator in the presence of \bc had been widely used in the calculation
of the Casimir effect in various configurations. In the simplest case of plane
parallel plates it can be constructed using the reflection principle well
known from electrostatic problems.  It works also in some other geometries,
such as a wedge for example.  
There is a generalization to arbitrarily shaped
cavities in form of the multi-reflection expansion. This is a formal, infinite
series and may be useful in some specific examples. In general, it is a matter
of intuition combined with symmetry considerations to find a sufficiently
simple representation of the propagator for a given problem.
A number of examples is considered in  Sec. \ref{sec4}. Here we are going to
discuss a more systematic, general (and necessarily more formal) approach
which reflects the general properties of the propagator. Thereby we connect it
with the specific problem of which degrees of freedom of a gauge field
contribute to the \gsep We restrict ourselfs to QED, but the considerations
may be easily generalized to other gauge field theories.

\subsubsection{Quantization in the presence of \bc}\label{sec3.5.1}
The conductor \bc that the \elm field has to fulfil on some boundary 
${\S}$ are
\be n^{\mu}F^{*}_{\mu\nu}(x)_{|_{x\in \S}}=0\;,
\label{cbc}\ee
where $F^{*}_{\mu\nu}=\epsilon_{\mu\nu\alpha\beta}F^{\alpha\beta}$ is the dual
field strength tensor and $n^{\mu}$ is the (outer) normal of ${\S}$.
Being the idealization of a physical interaction (with the conductor), they
are formulated in terms of the field strengths and thus are gauge--invariant.
Now, for well known reasons, it is desirable to perform the quantization of
QED in terms of the gauge potentials $A_{\mu}(x)$. Obviously, the \bc
\Ref{cbc} do not unambiguously imply \bc for all components of $A_{\mu}(x)$ as
required in order to obtain a self-adjoint wave operator.
 
There are two ways to proceed. The first one is to impose boundary
conditions on the potentials in such a way that the conditions \Ref{cbc} are
satisfied and a self-adjoint wave operator is provided.  Local \bc of this
kind fix either the magnetic or the electric field on the boundary.  These
conditions are stronger than \Ref{cbc} and they are not gauge invariant. When
now requiring Becchi-Rouet-Stora-Tyutin (BRST) invariance the ghosts (and some
auxiliary fields) become boundary dependent too and contribute to physical
quantities like the \gsep This was probably first observed in the papers
\cite{ambjornhughes83,ambjornhughes83a} and has later also been noticed in
connection with some models in quantum cosmology \cite{kamench,kamencha}. The
common understanding is that the ghost contributions cancel those from the
unphysical photon polarizations, for recent discussions see
\cite{Moss1990v,Moss1997h,Esposito1999m,Bordag:1999vs}.

In a second approach one considers the boundary conditions \Ref{cbc} as 
constraints when quantizing the potentials $A_{\mu}(x)$ as it was first 
done in \cite{Bordag:1985zk}. In that case, explicit gauge invariance is kept. 
There is no need to impose any additional conditions. The conditions 
\Ref{cbc} appear to be incorporated in a 'minimal' manner. In that respect, 
this second approach resembles the so called dyadic formalism which has 
successfully been used in the calculation of the Casimir energy in 
a spherical geometry \cite{Milton:1978sf}.

There are two ways to put this approach into practice. The first way
is to solve the constraints explicitly. For this purpose one has to
introduce a basis of polarization vectors $E_{\mu}^{s}$ (instead of
the commonly used $e_{\mu}^{s}$) such that only two amplitudes
$a_{s}(x)$ ($s=1,2$) of the corresponding decomposition
\be
A_{\mu}(x)=\sum\limits_{s=0}^{3}E_{\mu}^{s}a_{s}(x)
\label{dec1}
\ee 
of the \elm potential have to satisfy boundary conditions. The other
two amplitudes ($s=0,3$) remain free. In the case where the surface
${\S}$ consists of two parallel planes such a polarization basis
was constructed explicitly in \cite{Bordag:1985zk}.  In this way,
roughly speaking half of the photon polarizations feel the boundary
(in \cite{Bordag:1984ht} they have been shown to be the physical ones
in the sense of the Gupta--Bleuler quantization procedure) and half of
them do not. However, such an explicit decomposition, which
simultaneously diagonalizes both, the action and the \bc can only be
found in the simplest case.  The problem is that for a non--planar
surface ${\S}$ the polarizations $E_{\mu}^{s}$ become position
dependent and no general expression can be given.

In the second way one realizes the boundary conditions as constraints
which is equivalent to restrictions on the integration space in the
functional integral approach.  Than the generating functional of the
Green functions in QED reads
\bea \label{gf}
Z(J,\overline{\eta},\eta )&=& C
\int{D}A~{D}\overline{\psi}~{D}\psi ~
\prod_\nu\prod _{x\in \S}\delta \left(n^\mu F^{*}_{\mu\nu}(x)\right)
\exp\left\{ iS\left[A_\mu, \psi,\overline{\psi}\right]\phantom{\int}\right.
\nn\\
&&\left. +i\int{\rm d}x~\left(A_\mu(x)J^\mu (x)+
    \overline{\psi}(x)\eta (x)+ \overline{\eta}(x)\psi (x)\right)\right\},
\eea 
where the integration runs over all fields with the usual asymptotic
behavior and $J_{\mu}(x)$, $\overline{\eta}(x)$ and $\eta(x)$ are the
corresponding sources. The functional delta function restricts the
integration space to such potentials $A_{\mu}$ that the corresponding
field strengths satisfy the \bc \Ref{cbc}. This approach had been used
in \cite{Bordag:1985zk} and in \cite{Bordag:1998sw}.  It was shown to
result in a new photon propagator and an otherwise unaltered covariant
perturbation theory of QED.  The \bc \Ref{cbc} appear to be
incorporated with a 'minimal' disturbance of the standard formalism,
completely preserving gauge invariance (as well as the gauge fixing
procedure) and Lorentz covariance as far as possible.

The spinor field deserves a special discussion with respect to the
boundary conditions.  We do not impose \bc on the electron and
consider the \elm field and the spinor field on the entire Minkowski
space with the conducting surface placed in it. In the case of ${\S}$ 
being a sphere we thus consider the interior and the exterior
region together. In general, the surface ${\S}$ need not be
closed. Only the electromagnetic field obeys boundary conditions on
the surface ${\S}$. The electron penetrates it freely, it does not
feel the surface. As for a physical model one can think of a very thin
metallic surface which does not scatter the electrons but reflects the
electromagnetic waves. If the thickness of the metallic surface
(e.g. $1\,\mu$m) is small compared to the radiation length of the
metal (e.g. $1.43\,$cm for copper), this approximation is well
justified. However, the radiative corrections are of order
$(\alpha\lambda_{c}/L)$ with respect to the Casimir force itself and
thus too small to be directly observable. A discussion of the validity
of this approximation seems therefore to be somewhat  academic
at this time.

We remark, that the situation is different in the case of the bag model 
of the hadrons in QCD (see
Sec.~4.2.3).  The boundary conditions of the gluon field and of the
spinor field are connected by means of the equation of motion (this is because
the field strength tensor enters the boundary conditions rather than the dual
field strength tensor in \Ref{cbc}).

\subsubsection{The photon propagator}\label{sec3.5.2}
After the quantization is given by the functional integral \Ref{gf} it
``remains'' the task to calculate it. We start from the standard 
representation of perturbation theory in QED  
\be \label{pert} Z(J,\overline{\eta},\eta)= 
  \exp\left[i S_{\rm int}\left({\delta\over \delta i J},
-{\delta\over\delta i\eta},
{\delta\over\delta i\overline{\eta}}\right)\right]
Z^{(0)}(J,\overline{\eta},\eta)\,,
\ee 
where $Z^{(0)}$ is the generating functional of the free Greens functions and
\[
S_{\rm int}(A,\overline{\psi},\psi)=
e\int{\rm d}x~\overline{\psi}(x)\hat{A}(x)\psi (x) 
\] 
is the interaction. In this way the problem is reduced to
that of a standard perturbative technique with the corresponding Feynman rules
and  a free theory which now depends on the boundary conditions.  

Before proceeding further we rewrite the boundary conditions \Ref{cbc} in the
following way. Let $E_{\mu}^{s}(x)$ ($s=1,2$) be the two polarization vectors
in \Ref{dec1} with the properties
\be
{\pa\over\pa x_{\mu}}E_{\mu}^{s}(x)=0\;,\label{transv} \quad
n^{\mu}E_{\mu}^{s}(x)=0\;\label{ort} 
\ee
for $x\in\S$ (assuming $\pa_{\mu}n_{\nu}(x)=\pa_{\nu}n_{\mu}(x)$). They span a
space of transversal vectors tangential to the surface ${\S}$.  Note that
due to \Ref{ort}, there is no derivative acting outside the tangential space,
i.e. no normal derivative in \Ref{transv}. Without loss of generality, we
assume the normalization ${E^{s}_{\mu}}^{\dag}g^{\mu\nu}E_{\nu}^{t}=
-\delta^{st}$. We remark that the transformations $A_{\mu}\to
A_{\mu}+\pa_{\mu}\varphi(x)$ and $A_{\mu}\to A_{\mu}+n_{\mu}\varphi(x)$
respect the \bc \Ref{cbc}, i.e. if $A_{\mu}(x)$ satisfies the boundary
conditions, then so does the transformed potential. 
The invariance under the
first transformation simply says that the \bc are gauge independent. The
second means that the projection of $A_{\mu}$ onto the normal $n^{\mu}$ of
${\S}$ is unaffected by the boundary conditions.  We therefore conclude
that the boundary conditions \Ref{cbc} can equivalently be expressed as
\be
E_{\mu}^sA^{\mu}(x)_{|_{x\in\S}}=0~~~~~(s=1,2)\;.\label{bc} 
\ee
Explicit examples for such polarization vectors $E_{\mu}^{s}$ are given in
\cite{Bordag:1985zk} for plane parallel plates (see below Eq. \Ref{polbas1})
and in \cite{Bordag:1998sw} for a sphere and for a cylinder. In
\cite{bali78-112-165} these polarization vectors had been used in connection
with the radial multiple scattering expansion for a sphere.
%
%

Now, with the \bc in the form \Ref{bc} it is possible represent the
delta functions in \Ref{gf} as functional Fourier integrals. Then the
functional integral is Gaussian and can be completed. With the notation
$K^{\mu\nu}=g^{\mu\nu}\partial^2-( 1-1/\alpha
)\partial^\mu\partial^\nu$ for the kernel of the free action of the
\elm field and its inverse, which is the free (i.e., without boundary
conditions) photon propagator $D^{\mu\nu}(x-y)$, we define a new
integral kernel on the boundary manifold $\S$ by
\begin{equation} {
\bar{K}^{st}(z,z')\equiv {E_\mu^{s}}^{\dag}(z)D^{\mu\nu}(z-z')E_\nu^t(z'),
~~~z,z'\in\S\,.
}\label{DefK}\end{equation}
This object is in fact the projection 
of the propagator $D_{\mu\nu}(z-z')$ on the surface $\S$ and, with 
respect to the Lorentz indices, into the tangential subspace spanned by the 
polarization vectors $E_\mu^s(x)$, $s=1,2$. Further we need to define the 
inversion $\mbox{$\bar{K}^{-1}$}^{st}(z,z')$ of this operation
\be 
{\int_{\S}{\rm d}z''~\mbox{$\bar{K}^{-1}$}^{st'}(z,z'')
\bar{K}^{t't}(z'',z')=\delta_{\S}(z-z')\delta_{st}
}\;,
\label{inv}
\ee
where $\delta_{\S}(z-z')$ is the delta function with respect to the 
integration over the surface $\S$, $\delta_{st}$ is the usual Kronecker 
symbol. Then the new photon propagator with \bc takes the form
\be\label{bprop}\begin{array}{l} ~^{\S}D_{\mu\nu}(x,y)\equiv D_{\mu\nu}(x-y) -
  \bar{D}_{\mu\nu}(x,y)\nn
  \\
  =D_{\mu\nu}(x-y) \\
  ~~~~-\int\limits_{\S}{\rm d}z
\int\limits_{\S}{\rm d}z'~D^{\phantom{\mu}\mu
    '}_{\mu}(x-z) E_{\mu '}^s(z)\bar{K}^{-1}_{st}(z,z') {E_{\nu
      '}^{t}}^{\dag}(z') D^{\nu '}_{\phantom{\nu '}\nu}(z'-y)  \nn
\end{array}\ee
and the generating functional of the free (in the sense of perturbation
theory) Greens functions obeying \bc takes the form
\bea
Z^{(0)}(J,\overline{\eta},\eta )&=&C\left(\det K\right)^{-\frac{1}{2}}
\left(\det \bar{K}\right)^{-\frac{1}{2}}\nn\\
&&\times\exp \left[ \frac{\i}{2}\int{\rm d}x\int {\rm d}y~
J^\mu (x)~^{\S}D_{\mu\nu}(x,y) J^\nu (y) \right.\nn\\
&&+\left.\frac{1}i\int{\rm d}x\int {\rm d}y~
\overline{\psi}(x)S(x-y)\psi (y)\right]\;.
\label{Z0}
\eea
Note the appearance   in addition to the $\det K$, 
which is known from the theory
without boundary conditions, 
of the determinant   $\det\overline{K}$ of the
operation $\overline{K}$ which is boundary dependent.

Representation \Ref{bprop} of the photon propagator is valid for an arbitrary
surface $\S$.  Its explicit form in the plane parallel
geometry   and for  \bc  on a sphere   (explicit in terms of Bessel
functions)  can be found in \cite{Bordag:1998sw}. In general, for
geometries allowing for a separation of variables it is possible to write down
the corresponding explicite expressions.  In this respect representation
\Ref{bprop} is equivalent to any other. Its main advantage consists in
separating the propagator into a free part 
and a boundary dependent one allowing at
least in one loop calculations for an easy subtraction of the Minkowski space
contribution. Another remarkable property is that the gauge dependence is only
in the free space part whereas the boundary dependent part does not contain 
the gauge parameter $\al$.

As an illustration of these general formulas we consider in the next
subsection the photon propagator in plane parallel geometry.

\subsubsection{The photon propagator in plane parallel geometry}
\label{sec3.5.3}
Here the surface ${\S}$ consists of two pieces. A
coordinization of the planes is given by $z=\{x_{\alpha}, x_{3}=a_{i}\}$,
where the subscript $i=1,2$ distinguishes the two planes and $\alpha =0,1,2$
labels the directions parallel to them (they are taken perpendicular to the
$x_{3}$-axis intersecting them at $x_{3}=a_{i}$, $|a_{1}-a_{2}|\equiv a$ is
the distance between them). The polarizations $E_{\mu}^{s}$ ($s=1,2$) 
can be chosen as
\cite{Bordag:1985zk}
\be
\begin{array}{rclrcl}
E_{\mu}^{1}&=&\left(\begin{array}{c} 0\\ i\pa_{x_{2}}\\ -i
    \pa_{x_{1}}\\0\end{array}\right){1\over\sqrt{-\pa^{2}_{x_{||}}}}, &
E_{\mu}^{2}&=&\left(\begin{array}{c} -\pa^{2}_{x_{||}}\\
    -\pa_{x_{0}}\pa_{x_{1}} \\ -\pa_{x_{0}}\pa_{x_{2}}\\0\end{array}
\right){1\over\sqrt{-\pa^{2}_{x_{||}}}\sqrt{-\pa^{2}_{x_{0}}+
    \pa^{2}_{x_{||}}}},\\
 E_{\mu}^{3}&=&\left(\begin{array}{c}0\\0\\0\\1\end{array}\right),&
E_{\mu}^{0}&=&\left(
\begin{array}{c}-i\pa_{x_{0}}\\-i\pa_{x_{1}}\\-i\pa_{x_{2}}\\0\end{array}
\right){1\over\sqrt{-\pa^{2}_{x_{0}}+\pa^{2}_{x_{||}}}}
\end{array}
\label{polbas1}
\ee 
(${E_{\mu}^{s}}^{\dag}g^{\mu\nu}E_{\nu}^{s}=g^{st}$). These polarization
vectors do not depend on $x_{\alpha}$ or $i$. Therefore they commute with the
free photon propagator
\be 
D_{\mu\nu}(x-y)=\left(g_{\mu\nu}-
(1-\alpha){\pa_{x_{\mu}}\pa_{x_{\nu}}\over\pa^{2}_{x}}\right)
~ \int{\d^{4}k\over (2\pi)^{4}}~{e^{ik(x-y)}\over
    -k^{2}-i\epsilon}
\;,\label{photprop}
\ee 
($\epsilon>0$). Inserting $E_{\mu}^{s}$ into \Ref{DefK} yields the operator
$\bar{K}^{st}$ in the form
\be 
\bar{K}^{st}(z,z')=-\delta_{st}~D(x-x')_{|_{x,x'\in\S}}\;.
\label{ksplane}
\ee
We proceed with deriving a special representation of the scalar propagator. 
It is obtained by performing the integration over $k_{3}$ in Eq. \Ref{photprop}
\be D(x-x')=\int{\d^{3}k_{\alpha}\over
  (2\pi)^{3}}~{e^{ik_{\alpha}(x^{\alpha}-{x'}^{\alpha})+i\Gamma
    |x^{3}-{x'}^{3}|}\over -2i\Gamma},\label{ksplane2}\ee

with $\Gamma=\sqrt{k_{0}^{2}-k_{1}^{2}-k_{2}^{2}+i\epsilon}$. Substituting 
$x_{3}=a_{i}$ and $x'_{3}=a_{j}$ we get
\be \bar{K}^{st}(z,z')=-\delta_{st}\int{\d^{3}k_{\alpha}\over
  (2\pi)^{3}}~{i\over
  2\Gamma}~h_{ij}~e^{ik_{\alpha}(x^{\alpha}-{x'}^{\alpha})}\,,
\label{kplane}
\ee
where the abbreviation 
\be 
h_{ij}=e^{i\Gamma |a_{i}-a_{j}|}~~~~~~~(i,j=1,2)\;
\label{h}
\ee
has been introduced. With \Ref{kplane} we have achieved a mode decomposition 
of the operator $\bar{K}^{st}$ on the surface ${\S}$. As an advantage of 
this representation the inversion of $\bar{K}^{st}$, defined by \Ref{inv}, 
is now reduced to the algebraic problem of inverting the (2x2)--matrix 
$h_{ij}$. With
\be 
h^{-1}_{ij}={i\over 2\sin\Gamma a}\left(
\begin{array}{cc}e^{-i\Gamma
      a}&-1\\-1&e^{-i\Gamma a}\end{array}\right)_{ij}
\label{invh}
\ee
we get
\be 
\mbox{$\bar{K}^{-1}$}^{st}(z,z')=-\delta_{st}\int{\d^{3}k_{\alpha}\over
  (2\pi)^{3}}~{2\Gamma\over i}~h^{-1}_{ij}~e^{ik_{\alpha}(x^{\alpha}-
{x'}^{\alpha})}\;.
\label{kiplane}
\ee 
After inserting this expression into \Ref{bprop} we find the photon
propagator for the \elm field in covariant gauge with conductor \bc on two
parallel planes which was first derived  in \cite{Bordag:1985zk}.  
The connection to different representations can be seen from the remark that
the zeros of the denominatior for $\sin\Gamma a=0$ correspond just to the
discrete momentum perpendicular to the plates. However, the photon
propagator \Ref{kiplane} is valid in the whole space, i.e., in the outside
region too. The connection to the reflection principle can be made by
expanding $1/\sin\Gamma a=-2i\sum_{n\ge0}\exp(2i(n+\frac12)\Gamma a)$. 

The photon propagator for plane parallel plates as given in this
subsection had been used in \cite{Bordag:1986vz} for the calculation
of boundary dependent contributions to the anomalous magnetic moment
of the electron, and in \cite{Bordag92,Cheon:1996bi} for the
calculation of boundary dependent level shifts of a hydrogen atom.
It will be applied to calculate the radiative corrections to the
Casimir force in Sec.4.5.


\setcounter{equation}{0}

\section{Casimir effect in various configurations}
\label{sec4}

In this section the Casimir energies and forces in various configurations
are calculated for flat and curved boundaries. To perform
this calculation different theoretical methods are applied.
For some configurations, such as the stratified media, wedge, sphere,
or a cylinder the exact calculational methods described above are
applicable. For other cases, e.g., for a sphere (lens) above
a disk, the approximate methods are developed. The application of
the obtained results are considered in Quantum Field Theory,
Condensed Matter Physics, and Cosmology. Certain of the results,
presented here, are basic for comparison of  theory and
experiment. They will be used in the subsequent sections.

\subsection{Flat boundaries}
\label{sec4.1}

Here two examples of flat boundaries are considered: semispaces
including stratified media and rectangular cavities. Both configurations
are of much current interest in connection with the experiments on the
measurement of the Casimir force and applications of the Casimir
effect in nanotechnologies. Only the simplest configurations are
considered below, i.e., empty cavities and gaps between
semispaces (see, e.g., \cite{add1,add2} where the influence of the
additional external fields onto the Casimir effect is calculated).
The role and size of different corrections to the Casimir force
important from the experimental point of view are discussed in Sec.5.

Flat boundaries play a distinguished role for the Casimir effect because they
allow for quite explicit formulas. In addition, due to the missing curvature
of the boundaries, most \hkks are zero which makes it much easier to extract
the finite part of the vacuum energy even if additional external factors
are included. In a series of papers
\cite{Actor:1996zj,Actor:1995mc} generalizations of flat boundaries to
rectangular regions (e.g., a half plane sticked to a plane), and in
\cite{Actor:1995vc} to softened boundaries (e.g. a background potential
growing to infinity at the position of the boundary) are considered. In this
connection also the penetrable plane mirrors (see, e.g., Sec. \ref{sec3.1.1},
Eq. \Ref{twodelta}) should be mentioned.

\subsubsection{Two semispaces and stratified media}
\label{sec4.1.1}

In the case of flat boundaries the method of separation of variables 
in the field equation can usually be applied, 
which permits the application of the exact
calculational methods. The best known example of flat boundaries is
the configuration of two semispaces filled in by two dielectric
materials and separated by a gap filled in by some other material.
This is the configuration investigated by E.M.~Lifshitz \cite{9}
for which he obtained the general representation of the van der Waals
and Casimir force in terms of the frequency dependent dielectric
permittivities of all three media (magnetic permeabilities
were suggested to be equal to unity). Actually, Lifshitz results
can be generalized for any stratified medium containing an
arbitrary number of plane-parallel layers of different materials.

The original Lifshitz derivation was based on the assumption that the
dielectric materials can be considered as continuous media
characterized by randomly fluctuating sources. The correlation function
of these sources, situated at different points, is proportional to
the $\delta$-function of the radius-vector joining these points.
The force per unit area acting upon one of the semispaces was calculated as 
the flux of incoming momentum into this semispace through the boundary plane.
This flux is given by the appropriate component of the stress tensor
($zz$-component if $xy$ is the boundary plane). Usual boundary conditions
on the boundary surfaces between different media were imposed on the
Green's functions. To exclude the divergences, the values of all the
Green's functions in vacuum were subtracted of their values in the 
dielectric media \cite{sIV1}.

Here we present another derivation of the Lifshitz results and their
generalization starting directly from the zero-point energy of
electromagnetic field. In doing so, the continuous media, characterized by
the frequency dependent dielectric permittivities, and  
appropriate boundary conditions on the photon states, can be considered as
some effective external field (which cannot be described, however,
by a potential added to the left-hand side of wave equation). The main
ideas of such a derivation were first formulated in Refs.\cite{sIV2,sIV3}
(see also \cite{7,28,sIV4} where they were generalized and
elaborated). We are interested here not only in the force values acting 
upon boundaries but also in the finite, renormalized values of the Casimir
energies for the purpose of future application to configurations
used in experiments. 

\begin{figure}[ht]
\vspace*{-2.5cm}
\epsfxsize=17cm\centerline{\epsffile{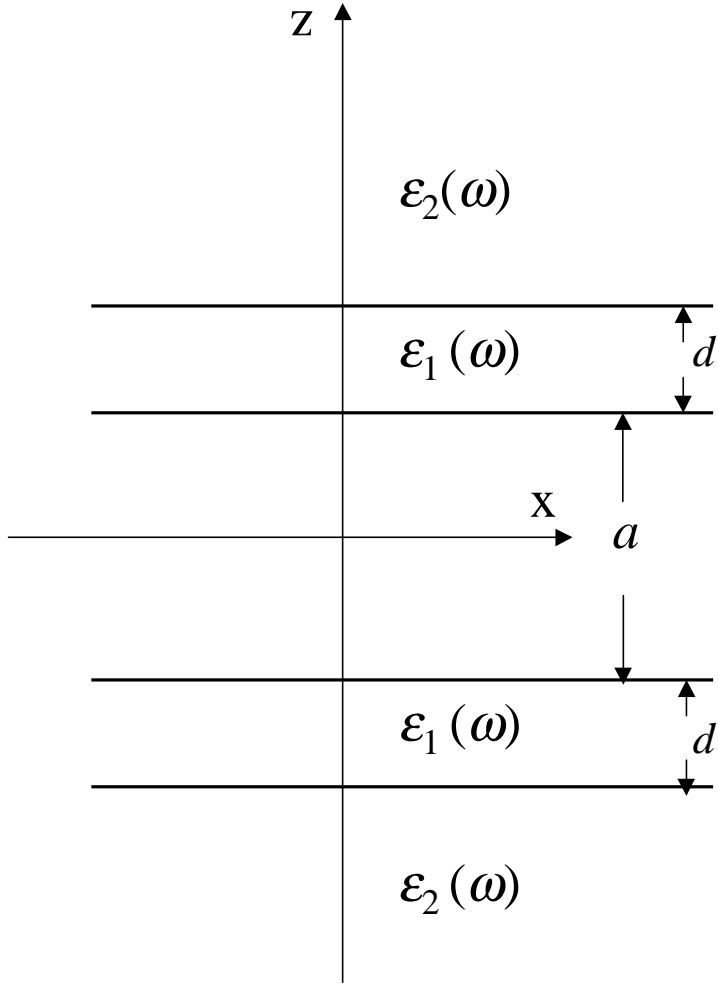} }
\vspace*{-5.5cm}
\caption{\label{fig4.1}
The configuration of two semispaces with a deilectric permittivity
$\varepsilon_2(\omega)$ covered by layers of thickness $d$ with
a permittivity $\varepsilon_1(\omega)$. The space separation between the
layers is $a$.
}
\end{figure}

In the experiments on the Casimir force measurements, symmetrical
configurations are usually used, i.e., both interacting bodies are made
of one and the same material which at times is covered by a thin layers
of another material \cite{32,33,34,35,36}. 
In line with this let us consider the
configuration presented in Fig.~\ref{fig4.1}. Here the main material of the
plates situated in ($x,y$) planes has the permittivity
$\varepsilon_2(\omega)$, and the covering layers (if any) ---
$\varepsilon_1(\omega)$. The empty space between the external surfaces
of the layers is of thickness $a$, and the layer thickness is $d$.

In line with Eq.~(2.29) from Section 2.2 the non-renormalized
vacuum energy density of electromagnetic field reads
\beq
{E_S}(a,d)=
\frac{E_0(a,d)}{S}=\frac{\hbar}{2}
\int\frac{dk_1\,dk_2}{(2\pi)^2}
\sum_{n}
\left(\omega_{{\bf k}_{\bot},n}^{(1)}+
\omega_{{\bf k}_{\bot},n}^{(2)}\right),
\label{sIV1}
\eeq
\noindent
where we have separated the proper frequencies of the modes with two
different polarizations of the electric field (parallel and perpendicular
to the plane formed by ${\bf k}_{\bot}$ 
and $z$-axis, respectively).
As in Sec.2.2, here ${\bf k}_{\bot}=(k_1,k_2)$ is the two-dimensional
propagation vector in the $xy$-plane. For simplicity $x$-axis is chosen
to be parallel to ${\bf k}_{\bot}$. However, it is more difficult than
for the perfectly conducting metallic planes to find the 
frequencies 
$\omega_{{\bf k}_{\bot},n}^{(1,2)}$.

In order to solve this problem we use the formalism of surface modes
\cite{sIV2,sIV3}, which are exponentially damping for
$z>\frac{a}{2}+d$ and $z<-\frac{a}{2}-d$.
These modes describe waves propagating parallel to the surface of the
walls \cite{evanescent}. 
They form a complete set of solutions 
and this approach is widely used. For another
approach using conventional scattering states see Sec. \ref{sec5.1} where it
is, in addition, generalized to nonzero temperature. 
%
%
To find these modes let us represent the orthonormalized set of
negative-frequency solutions to Maxwell equations in the form
\bes
&&
\mbox{\boldmath{$E$}}_{{\bf k}_{\bot},\alpha}^{(i)}(t,\rv )=
\mbox{\boldmath{$f$}}_{\alpha}^{\,(i)}({\bf k}_{\bot},z)\,
e^{i(k_x x+k_y y)-i\omega t},
\nonumber\\
&&
\mbox{\boldmath{$B$}}_{{\bf k}_{\bot},\alpha}^{(i)}(t,\rv )=
\mbox{\boldmath{$g$}}_{\alpha}^{(i)}({\bf k}_{\bot},z)\,
e^{i(k_x x+k_y y)-i\omega t},
\label{sIV2}
\ees
\noindent
where index $i$ numerates the same states of polarization as in
Eq.~(\ref{sIV1}), index $\alpha$ numerates the regions shown in 
Fig.~\ref{fig4.1}.

From Maxwell equations the wave equation for the $z$-dependent
vector functions follows
\beq
\frac{d^2\mbox{\boldmath{$f$}}_{\alpha}^{\,(i)}}{dz^2}-R_{\alpha}^2
\mbox{\boldmath{$f$}}_{\alpha}^{(i)}=0,\qquad
\frac{d^2\mbox{\boldmath{$g$}}_{\alpha}^{(i)}}{dz^2}-R_{\alpha}^2
\mbox{\boldmath{$g$}}_{\alpha}^{(i)}=0,
\label{sIV3}
\eeq
\noindent
where the notation is introduced
\beq
R_{\alpha}^2=k_{\bot}^2-\varepsilon_{\alpha}(\omega)\frac{\omega^2}{c^2},
\quad k^2=k_1^2+k_2^2,
\quad \varepsilon_0=1, \quad \alpha=0,\,1,\,2.
\label{sIV4}
\eeq
\noindent
In obtaining Eqs.~(\ref{sIV3}) we have assumed that the media are 
isotropic so that 
the electric displacement is 
{\boldmath{$D$}}${}_{\alpha}=
\varepsilon_{\alpha}${\boldmath{$E$}}${}_{\alpha}$.

According to the boundary conditions at the interface between two
dielectrics the normal component of {\boldmath{$D$}} and tangential
component of {\boldmath{$E$}} should be continuous. Also
{\boldmath{$B$}}${}_n$ and 
{\boldmath{$H$}}${}_t=${\boldmath{$B$}}${}_t$ (in our case of
non-magnetic media) are continuous. It is easy to verify that all these 
conditions are satisfied automatically if the quantities 
$\varepsilon_{\alpha}f_{z,\alpha}^{(1)}$ and
$df_{z,\alpha}^{(1)}/dz$ or
$f_{y,\alpha}^{(2)}$ and
$df_{y,\alpha}^{(2)}/dz$
are continuous. Let us consider detailly the first of these conditions.

According to Eq.~(\ref{sIV3}), the surface modes $f_z^{(1)}$ in
different regions of Fig.~4 can be represented as the following 
combinations of exponents:
\bes
f_z^{(1)}&=&
Ae^{R_2z}, \qquad z<-\frac{a}{2}-d,
\nonumber\\
{f_z^{(1)}}&=&
Be^{R_1z}+Ce^{-R_1z}, 
\qquad -\frac{a}{2}-d<z<-\frac{a}{2},
\nonumber\\
{f_z^{(1)}}&=&
De^{R_0z}+Ee^{-R_0z}, 
\qquad -\frac{a}{2}<z<\frac{a}{2},
\label{sIV5}\\
{f_z^{(1)}}&=&
Fe^{R_1z}+Ge^{-R_1z}, 
\qquad \frac{a}{2}<z<\frac{a}{2}+d,
\nonumber\\
{f_z^{(1)}}&=&
He^{-R_2z}, 
\qquad z>\frac{a}{2}+d.
\nonumber
\ees
\noindent
Imposing the continuity conditions on
$\varepsilon_{\alpha}f_{z,\alpha}^{(1)}$ and
$df_{z,\alpha}^{(1)}/dz$ at the points
$z=-\frac{a}{2}-d$, $-\frac{a}{2}$, $\frac{a}{2}$, and
$\frac{a}{2} +d$, and taking into account (\ref{sIV5}) we arrive at
the following system of equations
\bes
&&
A\varepsilon_2e^{R_2\left(-\frac{a}{2}-d\right)}=
B\varepsilon_1e^{R_1\left(-\frac{a}{2}-d\right)}+
C\varepsilon_1e^{-R_1\left(-\frac{a}{2}-d\right)},
\nonumber\\
&&
AR_2e^{R_2\left(-\frac{a}{2}-d\right)}=
BR_1e^{R_1\left(-\frac{a}{2}-d\right)}-
CR_1e^{-R_1\left(-\frac{a}{2}-d\right)},
\nonumber\\
&&
B\varepsilon_1e^{-R_1\frac{a}{2}}+
C\varepsilon_1e^{R_1\frac{a}{2}}=
De^{-R_0\frac{a}{2}}+
Ee^{R_0\frac{a}{2}},
\nonumber\\
&&
BR_1e^{-R_1\frac{a}{2}}-
CR_1e^{R_1\frac{a}{2}}=
DR_0e^{-R_0\frac{a}{2}}-
ER_0e^{R_0\frac{a}{2}},
\label{sIV6}\\
&&
De^{R_0\frac{a}{2}}+
Ee^{-R_0\frac{a}{2}}=
F\varepsilon_1e^{R_1\frac{a}{2}}+
G\varepsilon_1e^{-R_1\frac{a}{2}},
\nonumber\\
&&
DR_0e^{R_0\frac{a}{2}}-
ER_0e^{-R_0\frac{a}{2}}=
FR_1e^{R_1\frac{a}{2}}-
GR_1e^{-R_1\frac{a}{2}},
\nonumber\\
&&
F\varepsilon_1e^{R_1\left(\frac{a}{2}+d\right)}+
G\varepsilon_1e^{-R_1\left(\frac{a}{2}+d\right)}=
H\varepsilon_2e^{-R_2\left(\frac{a}{2}+d\right)},
\nonumber\\
&&
FR_1e^{R_1\left(\frac{a}{2}+d\right)}-
GR_1e^{-R_1\left(\frac{a}{2}+d\right)}=
-HR_2e^{-R_2\left(\frac{a}{2}+d\right)}.
\nonumber
\ees

This is a linear homogeneous system of algebraic equations relating 
the unknown coefficients $A,B,\ldots,H$. It has the non-trivial solutions
under the condition that the determinant of its coefficients is equal to 
zero. This condition is, accordingly, the equation for the determination
of the proper frequencies
$\omega_{{\bf k}_{\bot},n}^{(1)}$
of the modes with a parallel polarization \cite{28}
\bes
&&
\Delta^{(1)}\left(\omega_{{\bf k}_{\bot},n}^{(1)}
\right)
\equiv
e^{-R_2(a+2d)}\left\{\left(r_{10}^{+}r_{12}^{+}e^{R_1d}-
r_{10}^{-}r_{12}^{-}e^{-R_1d}\right)^2e^{R_0a}\right.
\nonumber\\
&&\phantom{aaaaaaaaaaa}\left.
-\left(r_{10}^{-}r_{12}^{+}e^{R_1d}-
r_{10}^{+}r_{12}^{-}e^{-R_1d}\right)^2e^{-R_0a}\right\}=0.
\label{sIV7}
\ees
\noindent
Here the following notations are introduced
\beq
r_{\alpha\beta}^{\pm}=R_{\alpha}\varepsilon_{\beta}\pm
R_{\beta}\varepsilon_{\alpha}, \qquad
q_{\alpha\beta}^{\pm}=R_{\alpha}\pm R_{\beta}.
\label{sIV8}
\eeq

Similarly, the requirement that the quantities
$f_{y,\alpha}^{(2)}$ and 
$df_{y,\alpha}^{(2)}/dz$ are continuous at boundary points results
in the equations for determination of the frequences
$\omega_{{\bf k}_{\bot},n}^{(2)}$
of the perpendicular polarized modes \cite{28}
\bes
&&
\Delta^{(2)}\left(\omega_{{\bf k}_{\bot},n}^{(2)}\right)
\equiv
e^{-R_2(a+2d)}\left\{\left(q_{10}^{+}q_{12}^{+}e^{R_1d}-
q_{10}^{-}q_{12}^{-}e^{-R_1d}\right)^2e^{R_0a}\right.
\nonumber\\
&&\phantom{aaaaaaaaaaa}\left.
-\left(q_{10}^{-}q_{12}^{+}e^{R_1d}-
q_{10}^{+}q_{12}^{-}e^{-R_1d}\right)^2e^{-R_0a}\right\}=0.
\label{sIV9}
\ees
\noindent
Note that to obtain Eqs.~(\ref{sIV7}), (\ref{sIV9}) 
we set the determinants of the linear
system of equations equal to zero and do not perform any additional 
transformations.
This is the reason why (\ref{sIV7}), (\ref{sIV9}) 
do not coincide with the corresponding
equations of \cite{7,sIV4} 
where some transformations were used which are
not equivalent in the limit $|\omega|\to\infty$ (see below).

Summation in Eq.~(\ref{sIV1}) over the solutions of 
Eqs.~(\ref{sIV7}), (\ref{sIV9}) can be performed
by applying the argument theorem which was 
applied for this purpose in
\cite{sIV2,sIV3}. According to this theorem
\beq
\sum_{n}
\omega_{{\bf k}_{\bot},n}^{(1,2)} =
\frac{1}{2\pi i}\left[
\int\limits_{i\infty}^{-i\infty}
\omega d\ln\Delta^{(1,2)}(\omega)
+
\int\limits_{C_{+}}
\omega d\ln\Delta^{(1,2)}(\omega)\right],
\label{sIV10}
\eeq
\noindent
where $C_{+}$ is a semicircle of infinite radius in the right 
one-half of the
complex $\omega$-plane with a center at the origin. Notice that the
functions $\Delta^{(1,2)}(\omega)$, defined in 
Eqs.~(\ref{sIV7}), (\ref{sIV9}), have no poles.
For this reason the sum over their poles is absent from (\ref{sIV10}).

The second integral in the right-hand side of (\ref{sIV10}) is simply
calculated with the natural supposition that
\beq
\lim\limits_{\omega\to\infty}\varepsilon_{\alpha}(\omega)=1,
\qquad
\lim\limits_{\omega\to\infty}
\frac{d\varepsilon_{\alpha}(\omega)}{d\omega}=0
\label{sIV11}
\eeq
\noindent
along any radial direction in complex $\omega$-plane. The result is
infinite, and does not depend on $a$:
\beq
\int\limits_{C_{+}}
\omega\, d\ln\Delta^{(1,2)}(\omega)
=4\int\limits_{C_{+}}
d\omega. 
\label{sIV12}
\eeq

Now we introduce a new variable $\xi=-i\omega$ in 
Eqs.~(\ref{sIV10}) and (\ref{sIV12}).
The result is
\beq
\sum_{n}
\omega_{{\bf k}_{\bot},n}^{(1,2)} =
\frac{1}{2\pi }
\int\limits_{\infty}^{-\infty}
\xi\, d\ln\Delta^{(1,2)}(i\xi)+
\frac{2}{\pi}
\int\limits_{C_{+}}
d\xi,
\label{sIV13}
\eeq
\noindent
where both contributions in the right-hand side diverge. To remove the
divergences we use a renormalization procedure which goes back to the
original Casimir paper \cite{1} 
(see also \cite{20,sIV3,sIV4}). 
The idea of this
procedure is that the renormalized physical vacuum energy density vanishes
for the infinitely separated interacting bodies. 
From Eqs.~(\ref{sIV7}), (\ref{sIV9}) and (\ref{sIV13})
it follows
\beq
\lim\limits_{a\to\infty}
\sum_{n}
\omega_{{\bf k}_{\bot},n}^{(1,2)} =
\frac{1}{2\pi }
\int\limits_{\infty}^{-\infty}
\xi\, d\ln\Delta_{\infty}^{(1,2)}(i\xi)+
\frac{2}{\pi}
\int\limits_{C_{+}}
d\xi,
\label{sIV14}
\eeq
\noindent
where the asymptotic behavior of $\Delta^{(1,2)}$ at $a\to\infty$
is given by
\bes
&&
\Delta_{\infty}^{(1)}=
e^{(R_0-R_2)a-2R_2d}
\left(r_{10}^{+}r_{12}^{+}e^{R_1d}-
r_{10}^{-}r_{12}^{-}e^{-R_1d}\right)^2,
\nonumber \\
&&
\Delta_{\infty}^{(2)}=
e^{(R_0-R_2)a-2R_2d}
\left(q_{10}^{+}q_{12}^{+}e^{R_1d}-
q_{10}^{-}q_{12}^{-}e^{-R_1d}\right)^2.
\label{sIV15}
\ees

Now the renormalized physical quantities are found with the help of
Eqs.~(\ref{sIV13})--(\ref{sIV15})
\bes
&&
\left(\sum_{n}
\omega_{{\bf k}_{\bot},n}^{(1,2)}\right)_{ren}
\equiv
\sum_{n}
\omega_{{\bf k}_{\bot},n}^{(1,2)}-
\lim\limits_{a\to\infty}
\sum_{n}
\omega_{{\bf k}_{\bot},n}^{(1,2)}
\nonumber\\
&&\phantom{\left(\sum_{n}
\omega_{{\bf k}_{\bot},n}^{(1,2)}\right)_{ren}}
=
\frac{1}{2\pi }
\int\limits_{\infty}^{-\infty}
\xi\, d\ln\frac{\Delta^{(1,2)}(i\xi)}{\Delta_{\infty}^{(1,2)}(i\xi)}.
\label{sIV16}
\ees
\noindent
They can be transformed to a more convenient form with the help of
integration by parts
\beq
\left(\sum_{n}
\omega_{{\bf k}_{\bot},n}^{(1,2)}\right)_{ren}=
\frac{1}{2\pi }
\int\limits_{-\infty}^{\infty}
d\xi\, \ln\frac{\Delta^{(1,2)}(i\xi)}{\Delta_{\infty}^{(1,2)}(i\xi)},
\label{sIV17}
\eeq
\noindent
where the term outside the integral vanishes.

To obtain the physical, renormalized Casimir energy density one should
substitute the renormalized quantities 
(\ref{sIV17}) into Eq.~(\ref{sIV1}) instead of
Eq.~(\ref{sIV13}) with the result
\beq
E_S^{ren}(a,d)=\frac{\hbar}{4\pi^2}
\int\limits_{0}^{\infty} k_{\bot}\,dk_{\bot}
\int\limits_{0}^{\infty} d\xi
\left[\ln Q_1(i\xi)+ \ln Q_2(i\xi)\right],
\label{sIV18}
\eeq
\noindent
where we introduced polar coordinates in $k_1,k_2$ plane, and
\bes
&&
Q_1(i\xi)\equiv\frac{\Delta^{(1)}(i\xi)}{\Delta_{\infty}^{(1)}(i\xi)}
=1-\left(\frac{r_{10}^{-}r_{12}^{+}e^{R_1d}-
r_{10}^{+}r_{12}^{-}e^{-R_1d}}{r_{10}^{+}r_{12}^{+}e^{R_1d}-
r_{10}^{-}r_{12}^{-}e^{-R_1d}}\right)^2e^{-2R_0a},
\nonumber\\
&&
Q_2(i\xi)\equiv\frac{\Delta^{(2)}(i\xi)}{\Delta_{\infty}^{(2)}(i\xi)}
=
1-\left(\frac{q_{10}^{-}q_{12}^{+}e^{R_1d}-
q_{10}^{+}q_{12}^{-}e^{-R_1d}}{q_{10}^{+}q_{12}^{+}e^{R_1d}-
q_{10}^{-}q_{12}^{-}e^{-R_1d}}\right)^2e^{-2R_0a}.
\label{sIV19}
\ees
\noindent
In Eq.~(\ref{sIV18}) the fact that $Q_{1,2}$ are even functions
of $\xi$ has been taken into account.

For the convenience of numerical calculations below we introduce the new
variable $p$ instead of $k_{\bot}$ defined by
\beq
k_{\bot}^2=\frac{\xi^2}{c^2}(p^2-1).
\label{sIV20}
\eeq
\noindent
In terms of $p,\,\xi$ the Casimir energy density 
(\ref{sIV18}) takes the form
\beq
E_S^{ren}(a,d)=\frac{\hbar}{4\pi^2c^2}
\int\limits_{1}^{\infty} p\,dp
\int\limits_{0}^{\infty} \xi^2\,d\xi
\left[\ln Q_1(i\xi)+ \ln Q_2(i\xi)\right],
\label{sIV21}
\eeq
\noindent
where a more detailed representation for the functions $Q_{1,2}$ from
(\ref{sIV19}) is
{\normalsize
\bes
&&
Q_1(i\xi)=1-\left[\frac{(K_1-\varepsilon_1p)(\varepsilon_2K_1+
\varepsilon_1K_2)-(K_1+\varepsilon_1p)(\varepsilon_2K_1-\varepsilon_1K_2)
e^{-2\frac{\xi}{c}K_1d}}{(K_1+\varepsilon_1p)(\varepsilon_2K_1+
\varepsilon_1K_2)-(K_1-\varepsilon_1p)(\varepsilon_2K_1-\varepsilon_1K_2)
e^{-2\frac{\xi}{c}K_1d}}\right]^2
e^{-2\frac{\xi}{c}pa},
\nonumber\\
&&
Q_2(i\xi)=1-\left[\frac{(K_1-p)(K_1+K_2)-(K_1+p)(K_1-K_2)
e^{-2\frac{\xi}{c}K_1d}}{(K_1+p)(K_1+K_2)-(K_1-p)(K_1-K_2)
e^{-2\frac{\xi}{c}K_1d}}\right]^2
e^{-2\frac{\xi}{c}pa}.
\label{sIV22}
\ees
\noindent}
Here all permittivities depend on $i\xi$ and
\beq
K_{\alpha}=K_{\alpha}(i\xi)\equiv\sqrt{p^2-1+\varepsilon_{\alpha}(i\xi)}=
\frac{c}{\xi}R_{\alpha}(i\xi),
\quad
 \alpha=1,\,2.
\label{sIV23}
\eeq
\noindent
For $\alpha=0$ one has $p=cR_0/\xi$ which is equivalent to 
Eq.~(\ref{sIV20}).

Notice that the expressions (\ref{sIV18}) and (\ref{sIV21}) 
give us finite values
of the Casimir energy density (which is in less common use than the force).
Thus in \cite{7} 
no finite expression for the energy density is presented
for two semi-spaces. 
In \cite{sIV4} the omission of infinities is performed
implicitly, namely instead of Eqs.~(\ref{sIV7}), (\ref{sIV9}) 
the result of their division
by the terms containing $\exp(R_0a)$ was presented. The coefficient near
$\exp(R_0a)$, however, turns into infinity on $C_{+}$. In other words the
Eqs.~(\ref{sIV7}), (\ref{sIV9}) 
are divided by infinity. As a result the integral along
$C_{+}$ is equal to zero in \cite{sIV4} 
and the quantity (\ref{sIV1}) would
seem to be finite. Fortunately, this implicit division is equivalent 
to the
renormalization procedure explicitly presented above. That is why the final
results obtained in \cite{sIV4} are indeed correct. 
In \cite{sIV1} the energy 
density is not considered at all.

From Eq.~(\ref{sIV21}) it is easy to obtain the Casimir force 
per unit area acting between
semispaces covered with layers:
\bes
&&
F_{ss}(a,d)=-\frac{\partial E_S^{ren}(a,d)}{\partial a}=
-\frac{\hbar}{2\pi^2c^3}
\int\limits_{1}^{\infty} p^2\,dp
\int\limits_{0}^{\infty} \xi^3\,d\xi
\label{sIV24}\\
&&\phantom{aaaaaaaaaaaaa}
\times
\left[\frac{1-Q_1(i\xi)}{Q_1(i\xi)}+ 
\frac{1-Q_2(i\xi)}{Q_2(i\xi)}\right].
\nonumber
\ees
\noindent
This expression coincides with Lifshitz result 
\cite{9,sIV1,sIV5} for the force 
per unit area between
semi-spaces with a dielectric permittivity $\varepsilon_2$ if the covering
layers are absent. To obtain this limiting case from 
Eq.~(\ref{sIV24}) one should
put $d=0$ and $\varepsilon_1=\varepsilon_2$
\bes
&&
F_{ss}(a)=
-\frac{\hbar}{2\pi^2c^3}
\int\limits_{1}^{\infty} p^2\,dp
\int\limits_{0}^{\infty} \xi^3\,d\xi
\left\{\left[\left(\frac{K_2+\varepsilon_2p}{K_2-\varepsilon_2p}\right)^2
e^{2\frac{\xi}{c}pa}-1\right]^{-1}\right.
\nonumber\\
&&
\phantom{aaaaaaaaaaaaaaaa}\left.
+\left[\left(\frac{K_2+p}{K_2-p}\right)^2
e^{2\frac{\xi}{c}pa}-1\right]^{-1}\right\}.
\label{sIV25}
\ees
\noindent
The corresponding quantity for the energy density follows from 
Eq.~(\ref{sIV21})
\bes
&&
E_S^{ren}(a)=
\frac{\hbar}{4\pi^2c^2}
\int\limits_{1}^{\infty} p\,dp
\int\limits_{0}^{\infty} \xi^2\,d\xi
\left\{\ln\left[1-\left(\frac{K_2-\varepsilon_2p}{K_2+\varepsilon_2p}\right)^2
e^{-2\frac{\xi}{c}pa}\right]\right.
\nonumber\\
&&\phantom{aaaaaaaaaaaaa}\left.
+\ln\left[1-\left(\frac{K_2-p}{K_2+p}\right)^2
e^{-2\frac{\xi}{c}pa}\right]\right\}.
\label{sIV26}
\ees 

It is well known \cite{sIV1} that Eqs.~(\ref{sIV25}), (\ref{sIV26})
contain the limiting cases of both van der Waals and Casimir forces
and energy densities. At small distances $a\ll\lambda_0$ (where
$\lambda_0$ is the characteristic absorption wavelength of the
semispace dielectric matter) these equations take the simplified form
\beq
F_{ss}(a)=-\frac{H}{6\pi a^3},
\qquad
E_S^{ren}(a)=-\frac{H}{12\pi a^2},
\label{sIV27}
\eeq
\noindent
where the Hamaker constant is introduced
\beq
H=\frac{3\hbar}{8\pi}
\int\limits_{0}^{\infty}x^2\,dx
\int\limits_{0}^{\infty}d\xi
\left[\left(\frac{\varepsilon_2+1}{\varepsilon_2-1}\right)^2
e^x-1\right]^{-1},
\label{sIV28}
\eeq
\noindent
and the new integration variable is $x=2p\xi a/c$.
This is the non-retarded van der Waals force per unit area and
the corresponding energy density between the semispaces.

In the opposite case of large distances $a\gg\lambda_0$ the
dielectric permittivities can be represented by their static values at $\xi=0$.
Introducing in Eq.~(\ref{sIV25}) the variable $x$ (now instead of
$\xi$) one obtains
\beq
F_{ss}(a)=-\frac{\hbar c\pi}{10 a^4}\Psi(\varepsilon_{20}),
\qquad
E_S^{ren}(a)=-\frac{\hbar c\pi}{30 a^3}\Psi(\varepsilon_{20}),
\label{sIV29}
\eeq
\noindent
where the function $\Psi$ is defined by
\bes
&&
\Psi(\varepsilon_{20})=\frac{5}{16\pi^3}
\int\limits_{1}^{\infty}\frac{dp}{p^2}
\int\limits_{0}^{\infty}x^3\,dx
\left\{
\left[\left(\frac{K_{20}+p}{K_{20}-p}\right)^2
e^x-1\right]^{-1}+\right.
\nonumber\\
&&\phantom{aaaaaaaaaaaaaaa}\left.
\left[\left(
\frac{K_{20}+p\varepsilon_{20}}{K_{20}-p\varepsilon_{20}}
\right)^2e^x-1\right]^{-1}\right\},
\label{sIV30}
\ees
\noindent
and  $K_{20}=(p^2-1+\varepsilon_{20})^{1/2}$, 
$\varepsilon_{20}=\varepsilon_2(0)$.

If both bodies are ideal metals the dielectric permittivity
$\varepsilon_2(i\xi)\to\infty$ for all $\xi$ including 
$\xi\to 0$.
Putting $\varepsilon_{20}\to\infty$, $\Psi(\varepsilon_{20})\to\pi/24$
we obtain the Casimir result for the force per unit area and
energy density
\beq
F_{ss}^{(0)}(a)=-\frac{\pi^2}{240}\frac{\hbar c}{a^4},
\qquad
E_S^{(0)ren}(a)=-\frac{\pi^2}{720}\frac{\hbar c}{a^3}.
\label{sIV31}
\eeq

The other method 
to obtain the force between semispaces (but with 
a permittivity
$\varepsilon_1$) is to consider limit $d\to\infty$ in (\ref{sIV24}).
In this limit we obtain once more the results 
(\ref{sIV25}), (\ref{sIV26})
where $K_2$, $\varepsilon_2$ are replaced by $K_1$, $\varepsilon_1$. Note
also that here we have not taken into account the effect of nonzero 
point temperature
which is negligible for $a\ll\hbar c/(k_BT)$.
The calculation of the Casimir force including the effect of non-zero
temperature is contained in Sec.5.1.

The above formulas can be applied also to describe the Casimir force
between two dielectric plates of finite thickness. For this purpose
it is enough to put $\varepsilon_2(\omega)=1$ in Eqs.~(\ref{sIV22}),
(\ref{sIV24}). This case was especially considered in \cite{addIV1-1}.
For anisotropic plates along with the vacuum force a torque can
appear which tends to change the mutual orientation of the bodies
(see \cite{21,addIV1-2,addIV1-3}).


\subsubsection{Rectangular cavities: attractive or repulsive force?}
\label{sec4.1.2}

As mentioned above, the Casimir energy may change its sign depending on
geometry and topology of the configuration. Probably the most evident
example of the dependence on the geometry is given by the Casimir
effect inside a rectangular box. As was noticed in \cite{sIV12-1},
the vacuum Casimir energy of electromagnetic field inside a
perfectly conducting box may change sign depending on the length of the 
sides. The detailed calculation of the Casimir energy inside a
rectangular box, when it is positive or negative, as a function of the 
box dimensions is contained in 
Refs.~\cite{17,sIV12-3}. In these references the analytical results for
two- and three-dimensional boxes were obtained by the repeated application 
of the Abel-Plana formula (\ref{2.25}). The results of 
\cite{17,sIV12-3} were later verified by other authors (see the
references below).

In \cite{sIV12-4} the Epstein zeta function was applied to calculate
the Casimir energy for a scalar and electromagnetic field in a
general hypercuboidal region with $p$ sides of finite length
$a_1,\ldots,a_p$ and $d-p$ sides with length $L\gg a_i$. Both the
periodic and perfect conductivity boundary conditions were considered,
and the contours were computed at which energy is zero.
The detailed examination of the attractive or repulsive nature of the 
Casimir force for a massless scalar field as a funciton of the dimensionality
of space and the number of compact sides $p$ was performed in
\cite{sIV12-5,addIV1-4} using the method of \cite{sIV12-4}. 
In \cite{sIV12-6}
the Casimir effect for electromagnetic field in three-dimensional
cavities was investigated by the use of Hertz potentials and their second
quantization. Particular attention has been given to the isolation of
divergent quantities and their interpretation which is 
in accordance with  \cite{17,sIV12-3}. The case of massless scalar
field in multidimensional rectangular cavity was reexamined in
\cite{sIV12-7}. The sign of the Casimir energy and its dependence on the
type of boundary conditions (periodic, Dirichlet, Neumann) was studied.

Let us apply the Epstein zeta function method to calculate the Casimir 
energy and force for the electromagnetic vacuum inside a rectangular
box with the side lengths $a_1$, $a_2$, and $a_3$. The box faces are 
assumed to be perfect conductors. Imposing the boundary conditions
of Eq.~(\ref{embc}) on the faces, the proper frequencies are
found to be
\beq
\omega_{n_1 n_2 n_3}^2=
\pi^2c^2\left(
\frac{n_1^2}{a_1^2}+
\frac{n_2^2}{a_2^2}+
\frac{n_3^2}{a_3^2}
\right).
\label{sIV12.1}
\eeq
\noindent
Here the oscillations for which all $n_i\neq 0$ 
and positive are doubly degenerate.
If one of the $n_i$ vanishes they are not degenerate. There are no
oscillations with two or three indices equal to zero because in such
cases electromagnetic field vanishes. As a consequence, the
nonrenormalized vacuum energy of electromagnetic field inside a box
takes the form
\bes
&&
E_0(a_1,a_2,a_3)=\frac{\hbar}{2}\left(
2\sum\limits_{n_1,n_2,n_3=1}^{\infty}
\omega_{n_1n_2n_3}+
\sum\limits_{n_2,n_3=1}^{\infty}
\omega_{0\,n_2n_3}\right.
\nonumber\\
&&\phantom{aaaaaaaaaaaaaaaaa}
\left.+
\sum\limits_{n_1,n_3=1}^{\infty}
\omega_{n_10\,n_3}+
\sum\limits_{n_1,n_2=1}^{\infty}
\omega_{n_1n_20}\right).
\label{sIV12.2}
\ees

We regularize this quantity with the help of Epstein zeta function
which for a simple case under consideration is defined by
\beq
Z_3\left(
\frac{1}{a_1},\frac{1}{a_2},\frac{1}{a_3};t\right)=
\sum\limits_{n_1,n_2,n_3=-\infty}^{\infty}
{\vphantom{\sum}}^{\!\!\!\!\!\!\!\!\!\!\!\!\!\!\prime}
\,\,\,\,\,\,\,\,\,
\left[\left(\frac{n_1}{a_1}\right)^2+
\left(\frac{n_2}{a_2}\right)^2+\left(\frac{n_3}{a_3}\right)^2
\right]^{-\frac{t}{2}}.
\label{sIV12.3}
\eeq
\noindent
This series is convergent if $t>3$. The prime near sum indicates 
that the term for which all $n_i=0$ is to be omitted.

At first, Eq.~(\ref{sIV12.2}) should be transformed identically to
\bes
&&
E_0(a_1,a_2,a_3)=\frac{\hbar}{8}
\sum\limits_{n_1,n_2,n_3=-\infty}^{\infty}
{\vphantom{\sum}}^{\!\!\!\!\!\!\!\!\!\!\!\!\!\!\!\prime}
\,\,\,\,
\omega_{n_1n_2n_3}\left(1-\delta_{n_1,0}\delta_{n_2,0}\right.
\label{sIV12.4}\\
&&\phantom{aaaaaaaaaaaaaaaaa}\left.
-
\delta_{n_1,0}\delta_{n_3,0}-\delta_{n_2,0}\delta_{n_3,0}\right).
\nonumber
\ees

Introducing the regularization parameter $s$ like in Sec.2.2 and
using the definitions (\ref{sIV12.3}), (\ref{Riezeta}) one obtains
\bes
&&
E_0(a_1,a_2,a_3;s)=\frac{\hbar}{8}
\sum\limits_{n_1,n_2,n_3=-\infty}^{\infty}
{\vphantom{\sum}}^{\!\!\!\!\!\!\!\!\!\!\!\!\!\!\!\prime}\,\,\,\,\,\,\,\,\,
\,\,\,\,
\omega_{n_1n_2n_3}^{1-2s}
\left(1-\delta_{n_1,0}\delta_{n_2,0}-
\delta_{n_1,0}\delta_{n_3,0}-\delta_{n_2,0}\delta_{n_3,0}\right)
\nonumber\\
&&\phantom{aaa}=
\frac{\hbar\pi c}{8}
\left[Z_3\left(\frac{1}{a_1},\frac{1}{a_2},\frac{1}{a_3};2s-1\right)
-2\zeta_R(2s-1)
\left(\frac{1}{a_1}+\frac{1}{a_2}+\frac{1}{a_3}\right)
\right].
\label{sIV12.5}
\ees
\noindent
To remove regularization $(s\to 0)$ we need the values 
of Epstein and Riemann zeta functions at $t=-1$. Both of them are
given by the analytic continuation of these functions.
As to $\zeta_R(t)$ the Eq.~(\ref{Riezeta,a}) should be used. For Epstein zeta
function the reflection formula analogical to (\ref{Riezeta,a}) is
\cite{sIV12-4}
\bes
&&
\Gamma\left(\frac{t}{2}\right)\pi^{-\frac{t}{2}}Z_3(a_1,a_2,a_3;t)=
(a_1a_2a_3)^{-1}\Gamma\left(\frac{3-t}{2}\right)\pi^{\frac{t-3}{2}}
\nonumber\\
&&\phantom{aaaaaaaaaaaaaaaaaaaaaaa}\times
Z_3\left(\frac{1}{a_1},\frac{1}{a_2},\frac{1}{a_3};3-t\right),
\label{sIV12.6}
\ees
\noindent
where $\Gamma(z)$ is gamma function. The results of their 
application are
\bes
&&
Z_3\left(\frac{1}{a_1},\frac{1}{a_2},\frac{1}{a_3};-1\right)
=-\frac{a_1a_2a_3}{2\pi^3}\,Z_3(a_1,a_2,a_3;4),
\nonumber\\
&&
\zeta_R(-1)=-\frac{1}{12}.
\label{sIV12.7}
\ees
\noindent
Substituting the obtained finite values into Eq.~(\ref{sIV12.5})
in the limit $s\to 0$ we obtain the renormalized (by means of
zeta function regularization) vacuum energy \cite{sIV12-4}
\bes
&&
E_0^{ren}(a_1,a_2,a_3)=
-\frac{\hbar c\,a_1a_2a_3}{16\pi^2}Z_3(a_1,a_2,a_3;4)
\nonumber\\
&&\phantom{aaaaaaaaaaaaaaaaaaa}
+\frac{\hbar c\pi}{48}
\left(\frac{1}{a_1}+\frac{1}{a_2}+\frac{1}{a_3}\right).
\label{sIV12.8}
\ees

It is clearly seen that the obtained result consists of the
difference of two positively defined terms and, therefore, can be both
positive and negative. Two renormalizations performed above
(associated with the analytical continuation of the Epstein and
Riemann zeta functions) may be interpreted as the omitting of terms
proportional to the volume $a_1a_2a_3$ and perimeter
$(a_1+a_2+a_3)$ of the box \cite{sIV12-6}.
For possible explanation of the Casimir repulsion in terms of vacuum 
radiation pressure see \cite{sIV12-7a}.

As usual, the forces acting upon the opposite pairs of faces and
directed normally to them are
\beq
F_i(a_1,a_2,a_3)=
-\frac{\partial E_0^{ren}(a_1,a_2,a_3)}{\partial a_i},
\label{sIV12.9}
\eeq
\noindent
so that the total vacuum force is
\beq
\mbox{\boldmath{$F$}}(a_1,a_2,a_3)=
-\nabla E_0^{ren}(a_1,a_2,a_3).
\label{sIV12.10}
\eeq

According to energy sign the forces (\ref{sIV12.9}) can be both repulsive 
or attractive depending on the relationship between the lengths of the sides
$a_1$, $a_2$, and $a_3$.

In Ref.\cite{sIV12-8} the detailed computations of the vacuum forces
and energies were performed numerically for boxes of different dimensions.
In particular, the zero-energy surfaces are presented, which separate
the positive-energy surfaces from the negative-energy ones, and the
surfaces of zero force. The analytical results of 
Refs.\cite{17,sIV12-1,sIV12-3,sIV12-6} were confirmed numerically in 
\cite{sIV12-8}.

Although the general result (\ref{sIV12.8}) settles the question of
the vacuum energy inside a perfectly conducting rectangular box,
we give some attention to the alternative calculation of the same
quantity based on the application of the Abel-Plana formula
\cite{17,sIV12-3}. The latter provides a simple way of obtaining
analytical results which is an advantage over the zeta function
method in application to a box. Let us start with a massless
scalar field in a two-dimensional box $0\leq x\leq a_1$,
$0\leq y\leq a_2$ for which the nonrenormalized vacuum energy is
expressed by
\beq
E_0(a_1,a_2)=\frac{\hbar}{2}
\sum\limits_{n_1,n_2=1}^{\infty}\omega_{n_1n_2},
\qquad
\omega_{n_1n_2}^2=\pi^2c^2
\left(\frac{n_1^2}{a_1^2}+\frac{n_2^2}{a_2^2}\right).
\label{sIV12.11}
\eeq
\noindent
To perform the summation in (\ref{sIV12.11}) we apply twice
the Abel-Plana formula (\ref{2.25}). As was noted in Sec.2.3,
the explicit introducion of the dumping function is not
necessary. After the first application one obtains
\bes
&&
S_{n_1}\equiv
\sum\limits_{n_2=1}^{\infty}
\left(\frac{n_1^2}{a_1^2}+\frac{n_2^2}{a_2^2}\right)^{1/2}=
-\frac{n_1}{2a_1}+
\int\limits_{0}^{\infty}dt
\left(\frac{n_1^2}{a_1^2}+\frac{t^2}{a_2^2}\right)
\nonumber\\
&&\phantom{aaaaa}
-2
\int\limits_{\frac{n_1a_2}{a_1}}^{\infty}
\left(\frac{t^2}{a_2^2}-\frac{n_1^2}{a_1^2}\right)^{1/2}
\frac{dt}{e^{2\pi t}-1},
\label{sIV12.12}
\ees
\noindent
where the last integral uses the fact that the difference of the radicals
is non-zero only above the branch point (see Eq.~(\ref{2.36})).
The result of the second application is
\bes
&&
\sum\limits_{n_1=1}^{\infty}S_{n_1}=
-\frac{1}{2}\left(\frac{1}{a_1}+\frac{1}{a_2}\right)
\int\limits_{0}^{\infty}t\,dt+
\int\limits_{0}^{\infty}dt
\int\limits_{0}^{\infty}dv
\left(\frac{v^2}{a_1^2}+\frac{t^2}{a_2^2}\right)^{1/2}
\nonumber\\
&&\phantom{aaaaa}
+\frac{1}{24a_1}-
\frac{a_2}{8\pi^2a_1^2}\zeta_R(3)+
\frac{2a_2}{a_1^2}\,G\left(\frac{a_2}{a_1}\right),
\label{sIV12.13}
\ees
\noindent
where
\beq
G(x)=
-\int\limits_{1}^{\infty}ds
\sqrt{s^2-1}
\sum\limits_{n_1=1}^{\infty}
\frac{n_1^2}{e^{2\pi x n_1 s}-1}.
\label{sIV12.14}
\eeq
\noindent
Renormalization of the quantity (\ref{sIV12.13}) is equivalent to the
omission of first two integrals in the right-hand side (the first is 
proportional to the perimeter, and the second --- to 
the volume, i.e., area of the box). As a result, the renormalized
vacuum energy is
\bes
&&
E_0^{ren}(a_1,a_2)=\frac{\hbar\pi c}{2}
\left(\sum\limits_{n_1=1}^{\infty}S_{n_1}\right)_{ren}
\nonumber\\
&&\phantom{aaaaa}
=\hbar c\left[\frac{\pi}{48a_1}-
\frac{\zeta_R(3)a_2}{16\pi a_1^2}+
\frac{\pi a_2}{a_1^2}\,G\left(\frac{a_2}{a_1}\right)\right].
\label{sIV12.15}
\ees

In the foregoing we performed the summation in $n_2$ first, followed by 
summation of $n_1$. This order is advantageous for $a_2\geq a_1$ because
$G(a_2/a_1)$ is small in this case. It can be easily shown from
Eq.~(\ref{sIV12.13}) that for $a_1=a_2$ the contribution of $G(1)$
to the vacuum energy is of order 1\% and for $a_2>a_1$ is
exponentially small. In such a manner one can neglect term
containing $G$ in (\ref{sIV12.15}) and get analytical expression for 
the vacuum energy which is valid with a high degree of accuracy. 
It is seen from this expression that the energy is positive if
\beq
1\leq\frac{a_2}{a_1}<\frac{\pi^2}{3\zeta_R(3)}\approx 2.74,
\label{sIV12.16}
\eeq
\noindent
and negative if $a_2>2.74a_1$ (we remind that 
$\zeta_R(3)\approx 1.202$).

The above derivation can be applied to the more realistic case of
electromagnetic vacuum confined in a three-dimensional box
$a_1\times a_2\times a_3$ with a perfectly conducting faces. After
three applications of the Abel-Plana formula and renormalization,
the result is the following \cite{17}
\bes
&&
E_0^{ren}(a_1,a_2,a_3)=
\hbar c\left[-\frac{\pi^2a_2a_3}{720a_1^3}-
\frac{\zeta_R(3)}{16\pi}\frac{a_3}{ a_2^2}\right.
\nonumber\\
&&\phantom{aaaaa}\left.
+\frac{\pi}{48}\left(\frac{1}{a_1}+\frac{1}{a_2}\right)+
H\left(\frac{a_2}{a_1},\frac{a_3}{a_1},\frac{a_3}{a_2}
\right)\right],
\label{sIV12.17}
\ees
\noindent
where function $H$ is exponentially small in all its arguments if
$a_1\leq a_2\leq a_3$ (the explicit expression for $H$ can be
found in \cite{17,20}).

If there is a square section $a_1=a_2\leq a_3$ and the small integral 
sums contained in $H$ are neglected the result is
\beq
E_0^{ren}(a_1,a_3)\approx
\frac{\hbar c}{a_1}
\left[\frac{\pi}{24}-
\left(\frac{\pi^2}{720}+\frac{\zeta_R(3)}{16\pi}\right)
\frac{a_3}{ a_1}\right].
\label{sIV12.18}
\eeq
\noindent
In the opposite case $a_1=a_2>a_3$ the vacuum energy is
\beq
E_0^{ren}(a_1,a_3)\approx
\frac{\hbar c}{a_1}
\left[\frac{\pi}{48}-\frac{\zeta_R(3)}{16\pi}
+\frac{\pi}{48}\frac{a_1}{a_3}-
\frac{\pi^2}{720}\left(\frac{a_1}{ a_3}\right)^3\right].
\label{sIV12.19}
\eeq

For the case $a_1=a_2$, one can easily obtain from (\ref{sIV12.17}) that
the energy is positive if
\beq
0.408<\frac{a_3}{a_1}<3.48
\label{sIV12.20}
\eeq
\noindent
and passes through zero at the ends of this interval.
Outside the interval (\ref{sIV12.20}) vacuum energy of electromagnetic
field inside a box is negative. For the cube $a_1=a_2=a_3$ the vacuum
energy takes the value
\beq
E_0^{ren}(a_1)\approx 0.0916\frac{\hbar c}{a_1}.
\label{sIV12.21}
\eeq
\noindent
Exactly the same results as in (\ref{sIV12.20}) and (\ref{sIV12.21})
are obtained from Eq.~(\ref{sIV12.8}) by numerical computation
\cite{sIV12-8}. If, instead of (\ref{sIV12.17}) one uses the 
approximate formulas (\ref{sIV12.18}), (\ref{sIV12.19}) the error is less
than 2\%. For example, for the case of the cube, where the error is largest, 
it follows
from (\ref{sIV12.18})that $E_0^{ren}(a_1)\approx 0.0933\hbar c/a_1$
instead of (\ref{sIV12.21}). This demonstrates the advantage of the
analytical representations (\ref{sIV12.17})--(\ref{sIV12.19}) for
the vacuum energies inside the rectangular boxes. The analogous results
to (\ref{sIV12.17}) can be obtained also for the periodic
boundary conditions imposed on the box faces (see, e.g., \cite{20}).

It is notable that the vacuum energy (\ref{sIV12.8}) preserves its value 
if two sides of the box are interchanged or the cyclic permutations
of all three sides is performed. The same property is implicitly present in 
Eqs.~(\ref{sIV12.15}), (\ref{sIV12.17}) even though it appears to be 
violated. The apparent violation of it is connected with the adopted condition
$a_1\leq a_2\leq a_3$ under which the additional contributions
$G$ and $H$ are small.

\subsection{Spherical and cylindrical boundaries}\label{sec4.2}
In this subsection we consider the \gse in spherical and 
cylindrical geometry.
The interest for these results comes 
from a number of sources. Historically the first
example emerged from Casimir's attempt to explain the stability of the
electron \cite{caselectron} stimulating Boyer to his work \cite{16} with
the surprising result of a repulsive force for the conducting sphere. In the
70ies there was the bag model in QCD and during the past decade there were
attempts to explain sonoluminescence stimulating the investigation of
spherical geometries. Technically quite closely related problems appear from
vacuum polarization in the background of black holes and in some 
models of
quantum cosmology.  Here we explain the methods suited to handle problems
given by local \bc and matching conditions on a sphere. In a separate
subsection we collect the corresponding results for a cylindrical surface.
These methods possess generalizations to smooth spherically symmetric
background potentials and to \bc on generalized cones as well as to higher
dimensions which will, however, not be discussed here.

We consider two types of \bcp The first one are ``hard'' 
\bc dividing the space
into two parts, the interior of the sphere ($0\le r\le R$) and the exterior
($R\le r <\infty$) so that the field is quantized independently in each
region. Examples are Dirichlet and Neumann \bc for a scalar field, conductor
\bc for the \elm field and bag \bc for the spinor field. The second type of
\bc can be rather called ``matching conditions'' connecting the field of both
sides of the boundary so that it must be quantized in the whole space as an
entity. Examples are the matching conditions which are equivalent to a delta
function potential and the conditions on the surface of a dielectric ball. 
The
fields which we consider here are in general massive ones except for the \elm
field. 


\subsubsection{Boundary conditions on a sphere}
\label{sec4.2.1}
In spherical geometry consider a   function which is after separation of  
variables 
 by means of Eq. \Ref{varsep} (see Sec. \ref{sec3.1.2})
subject to the equation \Ref{sevp2} which we rewrite here in the form
\be\label{lop}\left(-{\pa^{2}\over\pa r^{2}}+{l(l+1)\over
    r^{2}}+V(r)\right)\phi_{l}(r)=\lambda\phi_{l}(r) \,,
\ee
where we  kept the
background potential $V(r)$ for a moment. Dirichlet and Neumann \bc are
defined by
\be\label{Dbc}\begin{array}{lrcl}\mbox{Dirichlet:}&
  \phi_{l}(r)_{\mid_{r=R}}&=&0\,,  \\
\label{Nbc}\mbox{Neumann:}& {\pa\over\pa r}\phi_{l}(r)_{\mid_{r=R}}&=&0 \,.
\end{array}
\ee
Conductor \bc for the \elm field \Ref{cbc} turn into \bc for the two 
modes the
field can be decomposed in: Dirichlet \bc for the transverse electric (TE)
modes and Neumann\footnote{Note that it is a Neumann boundary condition for
  the function $\phi_{l}(r)$ defined by \Ref{varsep}, for the function
  $\psi(x)$ it is a Robin boundary condition.} \bc for the transverse 
magnetic
(TM) ones. In the bag model of QCD, the \bc on a surface $\S$ for the
gluon field reads $n^{\mu}F_{\mu\nu}(x)_{|_{x\in \S}}=0$ and can be treated 
by means
of duality in the same way as the conductor \bc \Ref{cbc} (at least 
at the one
loop level). For the spinor field the bag \bc read
\be\label{bbc} n^{\mu}\gamma_{\mu}\psi(x)_{|_{x\in\S}}=0 \,.
\ee
The bag \bc prevent  the color flux through the surface $\S$ simulating
confinement. Note that it is impossible to impose \bc on each 
component of the
spinor individually since this would result in an overdetermined problem
because the Dirac equation is of first order. 

These conditions are the most important local \bcp There are some other,
spectral \bc \cite{atiy75-77-43} or \bc containing tangential derivatives
\cite{Gilkey83,grub93-317-1123} which are discussed in some problems of
quantum cosmology or spectral geometry, see e.g. \cite{eath91-44-1713}.

Penetrable \bc interpolate to some extent between hard \bc and smooth
background potentials. There is a simple, although somewhat unphysical
example, given by a potential containing a delta function,
$V(r)=\frac{\al}{R}\delta(r-R)$. Here $\al$ is the dimensionless strength of
the potential.  
The problem with this potential in the equation \Ref{lop} can be reformulated
as a problem with no potential but matching condition
\be\label{mcondr}\phi_{l}(r)_{|_{r=R-0}}=\phi_{l}(r)_{|_{r=R+0}}, \quad {\pa
  \over\pa r}\phi_{l}(r)_{|_{r=R-0}}-{\pa \over\pa
  r}\phi_{l}(r)_{|_{r=R+0}}=\frac{\al}{R}\phi_{l}(r)_{|_{r=R}} \ee
in complete analogy to the one-dimensional case, Eq. \Ref{mcond}. 

Another example having an obvious physical meaning is that of the matching
condition for the \elm field on the surface between two bodies with different
dielectric constant requiring the continuity of the normal components of the
{\boldmath$D$} and {\boldmath$B$} and of the tangential components of
{\boldmath$E$} and {\boldmath$H$}. In this case we have different speeds of
light according to $c_{1,2}=c/\sqrt{{\ep}_{1,2}\mu_{1,2}}$ 
in the corresponding regions.

We are going to apply the methods explained in Sec.3 for
the calculation of the \gsep Consider first the field in the interior 
($0\le r\le R$) 
with hard \bcp In this case the spectrum is discrete and we need a
function whose zeros are just the discrete eigenvalues. For Dirichlet and
Neumann \bc these functions are  the solutions $\hat{j}_{l}(kr)$ to Eq.
\Ref{lop} with $V=0$  and its derivative 
$\hat{j}'_{l}(kr)$ (cf. the
discussion in Sec. \ref{sec3.1.2}). In order to use representation
\Ref{EregIms} we need these functions on the imaginary axis.  We remind the
reader of the notations
\be\label{slel}\begin{array}{rclrcl}
\hat{j}_{l}(iz)&=&i^{l} s_{l}(z)\,,  &s_{l}(z)&=&\sqrt{\pi z \over
  2}I_{l+\frac12}(z)\,, \\
\hat{h}^{+}_{l}(iz)&=&i^{-l} e_{l}(z)\,, & e_{l}(z)&=&\sqrt{2 z \over
  \pi}K_{l+\frac12}(z)\,.\end{array}
\ee
One should note that these functions depend on Bessel functions of
half-integer order so that they have explicit analytic expressions
\[s_{l}(z)=z^{l+1}\left(\frac1z\frac{\pa}{\pa z}\right)^{l}\frac{\sinh z}{z} 
\,, ~~~
e_{l}(z)=(-1)^{l}z^{l+1}\left(\frac1z\frac{\pa}{\pa z}\right)^{l}
\frac{e^{-z}}{{z}}
\,,  \]
which are, however, not of  much use with respect to the \gsep

As explained in Sec. \ref{sec3.1.2} we multiply these functions by a certain
power of their argument in order to make the Jost functions regular at $k=0$.
In this way we obtain for the Dirichlet problem in the interior
\be\label{Di}f^{\rm D,i}_{l}(ik)=(kR)^{-(l+1)}s_{l}(kR)
\ee
and for the Neumann problem
\be\label{Ni}f^{\rm N,i}_{l}(ik)=(kR)^{-l}s'_{l}(kR)\,.
\ee

Now we turn to the corresponding exterior problems. According to 
Sec. \ref{sec3.1.2}, the Jost function for the Dirichlet problem is given by
Eq. \Ref{kugela} which we note now in the form
\be\label{Da}f^{\rm D,e}_{l}(ik)=(kR)^{l}e_{l}(kR)\,.
\ee
In the same manner we obtain for the Neumann problem
\be\label{Na}f^{\rm N,e}_{l}(ik)=(kR)^{l+1}e'_{l}(kR)\,.
\ee

The additivity of the contributions from different modes to the regularized
\gse implies that the corresponding Jost functions  must
be multiplied. In this way we obtain for the Dirichlet problem on a thin
spherical shell by taking both the interior and the exterior problems together
\be\label{Dia}f^{\rm D}_{l}(ik)=(kR)^{-1}s_{l}(kR)e_{l}(kR)
\ee
and for the corresponding Neumann problem
\be\label{Nia}f^{\rm N}_{l}(ik)=kR \ s'_{l}(kR)e'_{l}(kR).
\ee
For the conductor \bc on the \elm field we have to take the product of
Dirichlet and Neumann \bcp For the interior problem we obtain in this way
\be\label{em,i}f^{\rm em,i}_{l}(ik)=(kR)^{-(2l+1)}s_{l}(kR)s'_{l}(kR)
\ee
and for the exterior problem
\be\label{em,a}f^{\rm em,e}_{l}(ik)=(kR)^{2l+1}e_{l}(kR)e'_{l}(kR)
\ee
and, finally, for the thin conducting sphere
\be\label{em}f^{\rm em}_{l}(ik)=s_{l}(kR)s'_{l}(kR)e_{l}(kR)e'_{l}(kR)\,.
\ee
This expression can be rewritten using the Wronskian
$s_{l}^{\prime}(z)e_{l}(z)-s_{l}(z)e_{l}^{\prime}(z)=1$ so that with
$s_{l}(z)s_{l}^{\prime}(z)e_{l}(z)e_{l}^{\prime}(z)=
-\frac14\left\{1-\left[(s_{l}(z)e_{l}(z))'\right]^{2}\right\}$ we can use
\[f^{\rm em}_{l}(ik)=1-\left\{\left[s_{l}(kR)e_{l}(kR)\right]'
\right\}^{2}
\]
instead of \Ref{em}, taking into account that a $k$-independent factor doesn't
influence the \gsep 

Note that for the \elm field the sum over the orbital momentum starts from
$l=1$, i.e., that the s-wave is missing which is in contrast to the 
scalar field.

Now we turn to the problems with matching conditions.  Let's note the general
form of the solution to Eq. \Ref{lop}
\be\label{msol}\phi_{l}(r)=\hat{j}_{l}(qr)\theta(R-r)+
\frac{i}2\left(f_{l}(k)\hat{h}^{-}_{l}(kr)-
f_{l}^{*}(k)\hat{h}^{+}_{l}(kr)\right) \,. 
\ee
With $q=k$ this is the solution for the delta function potential. Note that
in this case the asymptotic expressions \Ref{reg0} and \Ref{reginf} coincide
with the exact solutions in $0\le r\le R$ and $R\le r<\infty$ \resp as the
potential has a pointlike support in the radial coordinate. The matching
conditions applied to the solution \Ref{msol} can be solved and   the
Jost function follows to be
\be\label{delta}f^{\rm delta}_{l}(ik)=1+\frac{\al}{kR}s_{l}(kR)e_{l}(kR)\,.
\ee
Note that in the formal limit $\al\to\infty$ this function turns into $f^{\rm
  D}_{l}(ik)$ for the Dirichlet problem taken interior and exterior 
together. 

In the \elm case we consider a dielectric ball with $\ep_{1}$ and $\mu_{1}$
inside and  $\ep_{2}$ and $\mu_{2}$ outside,
$c_{1,2}=c/\sqrt{\ep_{1,2}\mu_{1,2}}$ are the corresponding speeds of
light. We insert solution \Ref{msol} into Eq. \Ref{lop} and obtain
$q^{2}={\om^{2}\over c_{1}^{2}}$ \resp $k^{2}={\om^{2}\over c_{2}^{2}}$ so
that we have $q={c_{2}\over c_{1}}k$. As in the case of the delta potential
the spectrum is completely continuous which can be understood in physical
terms as there are no bound states for the photons. The Jost functions
following from the matching conditions for the \elm field are well known in
classical electrodynamics and are denoted by $\Delta^{\rm TE}_{l}(k)$ and
$\Delta^{\rm TM}_{l}(k)$ for the two polarizations. They read
\bea\label{TE}f^{\rm TE}_{l}(ik)&\equiv&\Delta^{\rm TE}_{l}(ik)\\
&=& \sqrt{\ep_1 \mu_2} s _l ' (q R) e _l (k R) -\sqrt{\ep_2 \mu_1} s _l
(q R)
e _l ' (k R) \nn \,, \\[4pt]
\label{TM}     f^{\rm TM}_{l}(ik)&\equiv&\Delta^{\rm TM}_{l}(ik)\\
&=& \sqrt{\mu_1 \ep_2} s _l ' (q R) e _l (k R) -\sqrt{\mu_2 \ep_1} s _l
(q R) e _l ' (k R) \nn  \,.
\eea
The Jost function for the dielectric ball is the product of them:
\be\label{diel}f^{\rm diel}_{l}(ik)=\Delta^{\rm TE}_{l}(ik)\Delta^{\rm
  TM}_{l}(ik) \,.  \ee
A frequently discussed special case is that of equal speeds of light inside
and outside the sphere. In that case formulas simplify greatly and no \uv
divergences appear. After some trivial transformations making use of the
Wronskian one obtains
\be\label{em1}f^{\rm diel, c_{1}=c_{2}}_{l}(ik)=
1-\xi^{2}\left\{\left[s_{l}(kR)e_{l}(kR)\right]'\right\}^{2} \ee
with $\xi=(\ep_{1}-\ep_{2})/(\ep_{1}+\ep_{2})$. Note that in this case we
have $\mu_{1}/\mu_{2}=\ep_{2}/\ep_{1}\ne 1$ while for  unequal speeds it is
possible to have a pure dielectric ball, i.e., to have
$\mu_{1}=\mu_{2}=1$. For $\xi=1$ this Jost function coincides up to an
irrelevant factor with that of the conducting sphere, Eq. \Ref{em}. 

For completeness we note the Jost functions here for bag boundary
conditions. For the interior they read
\be\label{bagi}f_{l}^{\rm
  bag,i}(ik)=(kR)^{-2l-2}\left[s_{l}^{2}(kR)+s_{l+1}^{2}(kR)+ {2mc\over \hbar
    k}s_{l}(kR)s_{l+1}(kR)\right] \ee
and for the exterior  
\be\label{bage}f_{l}^{\rm
  bag,e}(ik)=(kR)^{2l+2}\left[e_{l}^{2}(kR)+e_{l+1}^{2}(kR)+ {2mc\over\hbar
    k}e_{l}(kR)e_{l+1}(kR)\right] \,.
\ee
Here $m$ is the mass of the spinor field. Its appearance in the Jost function
is a special feature of the \bc \Ref{bbc} and the corresponding equation of
motion has to be used. As discussed in \cite{Elizalde:1997hx} this dependence
on the mass causes some specific problems in the definition of the 
\gsek which
we do not however discuss here.

\subsubsection{Analytic continuation of the regularized \gse}
\label{sec4.2.2}
By means of Eq. \Ref{EregIms} we have a representation of the 
regularized \gse
well suited for analytic continuation in $s$. By means of Eq. \Ref{Edivz} we
have a representation of the singular at $s=0$ contributions, which are to be
subtracted in accordance with Eq. \Ref{renallg} in order to obtain the
renormalized ground state energy. 

Because it is impossible to put $s=0$ under the signs of summation and
integration in $\E^{\rm reg}$ we subtract and add in the integrand in Eq.
\Ref{EregIms} the first few terms of the uniform asymptotic expansion of the
logarithm of the Jost function, $\ln f_{l}^{\rm as}(ik)$, and split the
renormalized \gse accordingly. Then it takes the form
\be\label{Efas}\Er\equiv \E^{\rm reg}-\E_{0}^{\rm div}=\E^{f}+\E^{as}
\ee
with
\bea\label{Ef}\E^{f}&=& -\frac{\hbar c}{ 2\pi}\sum_{l=0}^{\infty}
\left(2l+1\right) 
\int_{mc/\hbar}^{\infty}{d k }
\left[k^{2}-\left({mc\over\hbar}\right)^{2}\right]^{1/2}\nn \\
&&~~~~~~\times{\pa\over\pa k}
\left[\ln f_{l}(ik)-\ln f_{l}^{\rm as}(ik)\right]
\eea
and 
\bea\label{Eas}\E^{as}&=&- \hbar c{\cos \pi s\over 2\pi}\sum_{l=0}^{\infty}
\left(2l+1\right) \int_{mc/\hbar}^{\infty} {d k
  }\left[k^{2}-\left({mc\over\hbar}\right)^{2}\right]^{1/2-s}\nn \\ &&~~~~~~
\times{\pa\over\pa k}\ln f_{l}^{\rm as}(ik) -\E_{0}^{\rm div}(s) \,.  \eea
We assume $\ln f_{l}^{\rm as}(ik)$ to be defined such that
\be\label{ascond}\ln f_{l}(ik)-\ln f_{l}^{\rm as}(ik) =O\left(l^{-4}\right)
\ee
in the limit $l\to\infty$, $k\to\infty$, uniform with respect to
$\frac{k}{l}$.  This allows us to put $s=0$ in the integrand of $\E^{f}$,
\Ref{Ef}, because of the convergence of both, the integral and the sum. Now
the computation of $\E^{f}$ for a given problem is left as a purely numerical
task.

The so called ``asymptotic contribution'', $\E^{as}$, can be analytically
continued to $s=0$ where it is finite because the pole term is subtracted. 
The
construction of this analytic continuation can be done analytically because
its structure is quite simple, see below for examples. In this way sometimes
explicit results for $\E^{as}$ can be reached, mainly for massless field.
Otherwise expressions in terms of well converging integrals or sums  can be
obtained. 

We note the asymptotic expansion in the form
\be\label{lnfas}\ln f_{l}^{\rm as}(ik)=
\sum_{i=-1}^{3}{X_{i}(t)\over \nu^{i}}.
\ee
This is an expansion in $\nu\equiv l+\frac12$ which is better than the
corresponding expansion in $l$. We use the notation
\be\label{deft}t={1\over\sqrt{1+\left({kR\over\nu}\right)^{2}}},
\ee
which is well known from the Debye polynomials appearing in the uniform
asymptotic expansion of the modified Bessel functions, see 
e.g. \cite{abra70b}.
For most problems the functions $X_{i}(t)$ ($i$=1,2,\dots) are polynomials in
$t$, for the dielectric ball they are more complicated.

Let us remark that in terms of the variable $t$ it becomes clear why it is
useful to work on the imaginary axis with respect to the variable $k$ in
representation \Ref{ci1} or \Ref{exam1}. On the real axis (by means of
$k\to-ik$ in \Ref{deft}) the expansion is not uniform for $k \sim \nu$ 
because
$t$ becomes large and  more complicated expansions should been used there.

The depth of the asymptotic expansion required for the procedure described
here is determined by the spatial dimension of the problem considered, 
up to 3
in our case. It is admissible to include higher terms into the definition of
$\ln f_{l}^{\rm as}(ik)$ which is some kind of over-subtraction not changing,
of course, $\Er$. In general one can expect for this to increase $\E^{as}$ on
the expense of $E^{f}$ diminishing the part left for pure numerical
calculation. This is, however, not useful for large background potentials
(e.g. large $\al$ in \Ref{delta}) as large compensations between $\E^{f}$ and
$\E^{as}$ will appear in that case. 

To obtain $\ln f_{l}^{\rm as}(ik)$ there are two possibilities. For smooth
background potentials one can use the Lippmann-Schwinger equation known in
potential scattering to obtain a recursion. Examples can be found in
\cite{Bordag:1996fv,Bordag:1998tg}. The other way is to have an explicit
expression for $\ln f_{l}(ik)$ in terms of special functions  like the
examples shown above and to expand them directly using the known expansions
for these functions. 
As all our examples are in terms of Bessel functions it is sufficient to 
note
their uniform asymptotic expansion which is well known, e.g., 
\be\label{}
I_{\nu} (\nu z)
\sim 
\frac 1 {\sqrt{2\pi \nu}}\frac{e^{\nu
\eta}}{(1+z^2)^{\frac 1 4}}\left[1+\sum_{k=1}^{\infty} \frac{u_k (t)}
{\nu ^k}\right],
\ee
with $t=1/\sqrt{1+z^2}$ and $\eta =\sqrt{1+z^2}+\ln
[z/(1+\sqrt{1+z^2})]$. The first few polynomials $u_{k}(t)$ and the recursion
relation for the higher ones can be found in \cite{abra70b}.

In this way one obtains for Dirichlet \bc on the sphere ($z=kR$)
\bea\label{XD}
X_{-1}&=&\eta(z)-\ln z, ~~~~ X_{0}=-\frac14 \ln (1+z^{2})=\frac12 \ln t, \\
X_{1}&=&\frac18 t-\frac5{24}t^{3}, ~~~~~ X_{2}= \frac 1 {16} t^2 -
\frac 3 8 t^4
+\frac 5 {16} t^6, \nn \\
X_{3}&=&\frac{25}{384}t^{3}-\frac{531}{640}t^{5}+\frac{221}{128}t^{7}-
\frac{1105}{1152}t^{9}. \nn
\eea
For the contribution from the exterior one has to change only signs, 
$X_{i}\to
(-1)^{i}X_{i}$. For a semi-transparent shell it holds
\cite{Bordag:1999vs,Scandurra:1998xa}
\be\label{Xdelta}
X_{1}=\frac{\al}{2}t, ~~ X_{2}=-\frac{\al^{2}}{8}t^{2}, ~~ X_{3}=
\left(\frac{\al}{16}+\frac{\al^{3}}{24}\right)t^{3}-
\frac{3\al}{8}t^{5}+\frac{5\al}{16}t^{7}.
\ee
For the dielectric sphere the first two coefficients are \cite{Bordag:1999vs}
\bea\label{Xdiel}
X_{-1}&=&\eta\left(\frac{z}{c_{1}}\right)-\eta\left(\frac{z}{c_{2}}\right),
\\
X_{0}&=&\ln {\sqrt{{\ep_{1}\mu_{2}\over \ep_{2}\mu_{1}}}c_{1}
  t_{2}+c_{2}t_{1}\over 2 \sqrt{c_{1}c_{2}t_{1}t_{2}}}+
\ln {\sqrt{{\ep_{2}\mu_{1}\over \ep_{1}\mu_{2}}}c_{1}
  t_{2}+c_{2}t_{1}\over 2 \sqrt{c_{1}c_{2}t_{1}t_{2}}} \nn 
\eea
with $t_{1,2}=1/\sqrt{1+(z/c_{1,2})^{2}}$. The higher $X_{i}$ are listed in
the appendix of \cite{Bordag:1999vs}. For bag \bc these functions can 
be found
in \cite{Elizalde:1997hx}.

Now, equipped with the explicit expression for the functions $X_{i}(t)$ in
\Ref{lnfas} we can construct the analytic continuation in $\E^{as}$
\Ref{Eas}. This had been done in all mentioned examples, sometimes
repeatedly by different authors. We demonstrate such calculations here 
on the simplest example of
the Casimir effect for a scalar field with Dirichlet \bc on a sphere. 

We introduce notations for the pieces of $\E^{as}$ as follows
\bea\label{Easdec}\E^{ as}&=&\sum_{i=-1}^{3}A_{i}-\E_{0}^{\rm div}(s)\\
&\equiv&-\hbar c \sum_{i=-1}^{3}{ \cos \pi s\over 2\pi}\sum_{l=0}^{\infty}
\left(2l+1\right) \int_{mc/\hbar}^{\infty}{d k }
\left[k^{2}-\left(\frac{mc}{\hbar}\right)^{2}\right]^{1/2-s}
{\pa\over\pa k}{X_{i}\over \nu^{i}}
-\E_{0}^{\rm div}(s)  \,.   \nn
\eea
For $X_{-1}$ and $X_{0}$  given by Eq. \Ref{XD} the $k$-integration delivers
hypergeometric functions which by means of their Mellin-Barnes integral
representation can be transformed into a sum in powers of the mass $m$. After
that the $l$-sum can be taken delivering a Hurwitz zeta function. We obtain
(for details see \cite{Bordag:1997ma})
\bea\label{}
A_{-1} &=&\frac{\hbar cR^{2s-1}}{4\sqrt{\pi}\Gamma (s-\frac12)}
\sum_{j=0}^{\infty} 
\frac{(-1)^j}{j!}
\left(\frac{mcR}{\hbar}\right)^{2j}
\frac{\Gamma\left( j+s-1\right)}{s+j-\frac12}\zeta_H (2j+2s
-3;1/2),\nn\\          
A_0  &=&-\frac{\hbar cR^{2s-1}}{4\Gamma (s-\frac12)}\sum_{j=0}^{\infty} 
\frac{(-1)^j}{j!} \left(\frac{mcR}{\hbar}\right)^{2j}
\Gamma (s+j-\frac12)
\zeta_H (2j+2s-2;1/2)\,.\nn
\eea
In the contributions from  $i=1,2,\dots$ the $X_{i}(t)$ are polynomials 
in $t$,
\be\label{Xia}X_{i}(t)=\sum_{a}x_{ia}{t^{a}\over\nu^{i}}\,,
\ee
and the $k$-integration can   be carried out explicitly 
\bea\label{}A_{i}  &=& -\frac{\hbar cR^{2s-1}}{\Gamma (s-\frac12)}
\sum_{j=0}^{\infty} {\frac{(-1)^j}{j!}} 
\left(\frac{mcR}{\hbar}\right)^{2j} \zeta_H
(2s-2+i+2j;1/2) \nn\\
& &\hspace{3cm} \times\sum_{a=0}^i x_{i,a} 
\frac{\Gamma\left( s+a+j+(i-1)/2\right)}{\Gamma\left( a+i/2\right)}.\nn
\eea
In these expressions the pole contributions are given explicitly in terms of
gamma and zeta functions. In fact they cancel exactly the pole 
contribution in
$\E^{\rm div}$, \Ref{Edivz}. In this way the analytic continuation 
to $s=0$ is
constructed. The sums over $j$ in these expressions converge for 
$mcR/\hbar<1$. A representation valid for all 
$mcR/\hbar$ had been derived in \cite{Bordag:1997ma} but
it is too complicated to be displayed here. 
In contrast to this, in the massless case the
expressions are simple:
\beq\label{dis}
A_{-1} &=& \frac{\hbar c}{2\pi R} \left\{ \frac 7 {1920} \left[ \frac 1 s 
           +\ln \left(\frac{\mu cR}{\hbar}\right)^2 
\right] +\frac 7 {1920} + \frac 1 {160} \ln 2 
           +\frac 7 8 \zeta_R ' (-3) \right \} +O(s)  ,\nn\\
A_0 &=& 0 ,\nn\\
A_1 &=& \frac{\hbar c}{2\pi R} \left\{ \frac 1 {192} 
                \left[ \frac 1 s
           +\ln\left(\frac{\mu cR}{\hbar}\right)^2 \right] 
-\frac 1 {36} -\frac 1 8 \zeta_R ' (-1)
              \right\} + O(s) ,\label{eq5.3a}\\
A_2 &=& 0 , \nn\\
A_3  &=& \frac{\hbar c}{2\pi R} \left\{-\frac{229}{40320} 
               \left[ \frac 1 s
           +\ln\left(\frac{\mu cR}{\hbar}\right)^2 \right] 
+\frac{269}{7560 } -\frac{229} {20160} 
         \gamma -\frac{229}{6720} \ln 2 \right\} +O(s) ,\nn
\eeq
where $\mu$ is the mass parameter introduced in (\ref{Ereg1}).
The pole contribution is 
\[
\sum_{i=-1}^{3}A_{i}=\frac{1}{630}
\frac{\hbar c}{\pi R  s}+O(1).
\] 
It cancels just $\E^{\rm div}=-a_{2}\hbar c/(32\pi^{2}s)$ 
(Eq.~\Ref{Edivz} 
for $m=0$) with $a_{2}=-16\pi/(315R)$ given in Eq.~\Ref{hkksex} 
as expected.

When adding the contributions from the exterior by means of the signs,
$X_{i}\to (-1)^{i}X_{i}$, the contributions from the $A_{i}$ cancel 
each other
and one is left with $\E^{f}$.

\subsubsection{Results on the Casimir effect on a sphere}\label{sec4.2.3}
As we have seen in the foregoing subsection, the Casimir energy is typically
represented as a sum of two parts. The first, $\E^{f}$, is a convergent
integral and sum which can be calculated only numerically. The second part is
sometimes explicit in terms of gamma and Hurwitz zeta functions, when this 
is not the case, 
then it has to be evaluated numerically also.  It is not meaningful to
describe the numerical procedures here. They are simple and can be done for
example using Mathematica. Therefore we restrict ourselves  
to a collection of most of 
the known results. 

The historically first result on a conducting sphere is that by Boyer
\cite{16}. It had been confirmed with higher precision in
\cite{bali78-112-165,Milton:1978sf} using Green function methods. In the bag
model the vacuum energy of spinors was calculated in
\cite{brow84-140-285,milt80-22-1441,milt80-22-1444,Milton:1983wy}
and more recently in \cite{DeFrancia:1992sf,Falomir:1991fy}. 
For massive
fields it was recognized quite early that there are additional divergencies
\cite{baac83-27-460}. The first complete calculation for the massive case was
done in \cite{Bordag:1997ma} for a scalar field and in \cite{Elizalde:1997hx}
for a spinor field with bag \bcp In \cite{lese96-250-448} the results for a
massless field in several dimensions are collected. A calculation of the
Casimir energy in any (even fractional) dimension is done in
\cite{Bender:1994zr,Milton:1997ri}. 

The Casimir energy  for the massless scalar field
inside the sphere of radius $R$ in $D=3$ dimensions 
with Dirichlet \bc 
is given by
\be\label{ECasml}
E_{Cas} = \frac{\hbar c}{R} \left( 0.0044 +\frac 1 {630 \pi} \left[
              \frac 1 s+ \ln\left(\frac{\mu cR}{\hbar}\right)^2 
\right] \right) .
\ee
It is proportional to $1/R$ for dimensional reasons. The pole part 
(in zeta-functional
regularization  with $s\to 0$) follows from \Ref{dis}, the logarithmic term
is a consequence of the divergence. It carries the mentioned  
arbitraryness in
terms of the parameter $\mu$ introduced in \Ref{Ereg1}. The finite part
results from the corresponding numerical evaluation of $E^{f}$. In the same
manner results for the exterior space and for different dimensions can be
obtained. Taking interior and  exterior space together, in odd dimensions the
divergencies cancel and so do the logarithmic terms. A finite and unique
result emerges. 

In the following Tables taken from \cite{Kirsten:2000xc}, where the known
results are collected and extended, the corresponding numbers are shown for
several spatial dimensions $D$.  In the first two columns the values of the
corresponding zeta function are given, the sums of which are twice the
coefficient of the Casimir energy given in the third column.  The pole parts
are given as real numbers for better comparison with the finite parts 
although
they all have explicit representations in terms of zeta functions.  The
Casimir energy for the whole space is finite in odd spatial dimension and
infinite in even.

It is interesting to note the changes in the signs for the Casimir 
energy for whole space. There is no general rule known for that. 


\newcommand\dati[3]{$#1$&$#2$&$#3$\\ \hline}

\def\capDir{{Coefficients for the zeta functions and the Casimir energy for a
    massless scalar field with Dirichlet boundary conditions.}  }
\def\twoDir{2&
     \dati{+0.010038 - 0.003906/s} 
          {-0.008693 - 0.003906/s} 
          {+0.000672 - 0.003906/s}
        } 
\def\threeDir{3&
     \dati{+0.008873 + 0.001010/s} 
          {-0.003234 - 0.001010/s} 
          {+0.002819}
        } 
\def\fourDir{4&
     \dati{-0.001782 + 0.000267/s} 
          {+0.000470 + 0.000267/s} 
          {-0.000655 + 0.000267/s}
        } 
\def\fiveDir{5&
     \dati{-0.000940 - 0.000134/s} 
          {+0.000364 + 0.000134/s} 
          {-0.000288}
        }
\def\sixDir{6&
     \dati{+0.000268 - 0.000033/s} 
          {-0.000062 - 0.000033/s} 
          {+0.000102 - 0.000033/s}
        }
\def\sevenDir{7&
     \dati{+0.000137 + 0.000021/s} 
          {-0.000055 - 0.000021/s} 
          {+0.000040}
        } 
\def\eightDir{8&
     \dati{-0.000045 + 5.228\times10^{-6}/s} 
          {+0.000010 + 5.228\times10^{-6}/s} 
          {-0.000017 + 5.228\times10^{-6}/s}
        } 
\def\nineDir{9&
     \dati{-0.000022 - 3.769\times10^{-6}/s} 
       {+9.399\times10^{-6} + 3.769\times10^{-6}/s} 
       {-6.798\times10^{-6}}
}

\def\capNeu{{Coefficients for the zeta functions and the
      Casimir energy for a massless scalar field with Neumann
      boundary conditions.}}
\def\twoNeu{2&
    \dati{-0.344916 - 0.019531/s} 
          {-0.021330 - 0.019531/s} 
           {-0.183123 - 0.019531/s}
 }
\def\threeNeu{3&
    \dati{-0.459240 - 0.035368/s} 
          {+0.012324 + 0.035368/s} 
           {-0.223458}
 }
\def\fourNeu{4&
    \dati{-0.512984 - 0.044716/s} 
          {-0.008760 - 0.044716/s} 
           {-0.260872 - 0.044716/s}
 }
\def\fiveNeu{5&
    \dati{-0.556588 - 0.048921/s} 
          {+0.016024 + 0.048921/s} 
          {-0.270281}
 }
\def\sixNeu{6&
    \dati{-0.677067 - 0.051373/s} 
          {-0.076351 - 0.051373/s} 
           {-0.376709 - 0.051373/s}
}

\def\capBag{{Coefficients for the zeta functions and the
    Casimir energy for a massless spinor field
with bag boundary conditions.}
 }
\def\twoBag{2&
     \dati{-0.00537  + 0.007812/s}
          {+0.02167  + 0.007812/s}
          {-0.008148 - 0.007812/s} 
               }
\def\threeBag{3&
     \dati{-0.060617 - 0.005052/s}
           {+0.019796 + 0.005052/s}
            {+0.020410}                       }
\def\fourBag{4&
    \dati{+0.005931 - 0.002838/s}
          {-0.010171 - 0.002838/s}
           {+0.002120 + 0.002838/s}    }
\def\fiveBag{5&
     \dati{+0.025059 + 0.002510/s}
           {-0.008981 - 0.002510/s}
            {-0.008039}                        }
\def\sixBag{6&
     \dati{-0.003039 + 0.001171/s}
           {+0.004602 + 0.001171/s}
            {-0.000781 - 0.001171/s}    }
\def\sevenBag{7&
     \dati{-0.010862 - 0.001174/s}
           {+0.004024 + 0.001174/s}
            {+0.003419}                        }

\def\capGS{{\bf Massless spinor field 
with global spectral boundary conditions.}
Values of the zeta function at $s=-1/2$ 
inside and outside a spherical shell and
values of the Casimir energy.} 
\def\twoGS{2&
    \dati{-0.008948 + 0.031976/s} 
         {+0.009622 + 0.031976/s} 
         {-0.000338 - 0.031976/s}
 }
\def\threeGS{3&
    \dati{-0.171226 - 0.003770/s} 
         {+0.001732 + 0.003770/s} 
         {+0.084747}
 }
\def\fourGS{4&
    \dati{+0.008194 - 0.011832/s} 
         {-0.003930 - 0.011832/s} 
         {-0.002132 + 0.011832/s}
 }
\def\fiveGS{5&
    \dati{-0.105971 + 0.001947/s} 
         {-0.000814 - 0.001947/s} 
         {+0.053393}
 }
\def\sixGS{6&
    \dati{-0.004186 + 0.004907/s} 
         {+0.001730 + 0.004907/s} 
         {+0.001228 - 0.004907/s}
 }
\def\sevenGS{7&
    \dati{-0.029105 - 0.000925/s} 
         {+0.000387 + 0.000925/s} 
         {+0.014359}
 }
\def\eightGS{8&
    \dati{+0.002027 - 0.002142/s} 
         {-0.000789 - 0.002142/s} 
         {-0.000619 + 0.002142/s}
 }
\def\nineGS{9&
    \dati{+0.012901 + 0.000435/s} 
         {-0.000177 - 0.000435/s} 
         {-0.006362}
 }
\def\tenGS{10&
    \dati{-0.000964  + 0.000962/s} 
         {+0.000362  + 0.000962/s} 
         {+0.000301 - 0.000962/s}
}

\def\capEF{{Coefficients for the zeta functions and the Casimir energy for the electromagnetic field in a perfectly 
conducting spherical shell.} 
It has to be noted that in even dimensions, 
in contrast with the scalar field,
the divergences between the inside and
outside energies are different for $D>2$. 
This is due to the fact that 
(only in even dimensions) the $l=0$
mode explicitly contributes to the poles of the
$\zeta$-function}
\def\twoEF{2&
    \dati{-0.344916 - 0.019531/s} 
          {-0.021330 - 0.019531/s} 
           {-0.183123 - 0.019531/s}
}
\def\threeEF{3&
    \dati{+0.167872 + 0.008084/s} 
          {-0.075471 - 0.008084/s} 
           {+0.046200}
}
\def\fourEF{4&
    \dati{-0.044006 - 0.073556/s} 
          {-0.351663 + 0.006021/s} 
           {-0.197834 - 0.033768/s}
}
\def\fiveEF{5&
    \dati{+0.580372 + 0.049692/s}                 
          {-0.593096 - 0.049692/s} 
           {-0.006362}
}
\def\DirBC{\begin{table}\label{Dir0}\begin{center}
\begin{tabular}{|r||p{4cm}|p{4cm}||p{4cm}|}\hline
D&Zeta Function  Inside&Zeta Function  Outside& Casimir Energy\\ 
\hline\hline
\twoDir\threeDir\fourDir\fiveDir\sixDir\sevenDir\eightDir\nineDir
\end{tabular}\caption{\protect\small\capDir}\end{center}\end{table}}
\def\NeuBC{\begin{table}\label{Neu0}\begin{center}
\begin{tabular}{|r||l|l||l|}\hline
D&Zeta Function   Inside&Zeta Function  Outside& Casimir Energy\\ 
\hline\hline
\twoNeu\threeNeu\fourNeu\fiveNeu\sixNeu  
\end{tabular}
\end{center}\caption{\protect\small\capNeu}\end{table}}
\def\BagBC{\begin{table} \label{Bag12}
\begin{center}\begin{tabular}{|r||l|l||l|}\hline
D& Zeta Function  Inside& Zeta Function  Outside& Casimir Energy\\
\hline\hline
\twoBag\threeBag\fourBag\fiveBag\sixBag\sevenBag
\end{tabular}\caption{\protect\small\capBag}\end{center}\end{table}}
\def\GSBC{\begin{table}\label{GS12}\begin{center}
\begin{tabular}{|r||l|l||l|}\hline
D& Zeta Function  Inside&Zeta Function   Outside& Casimir Energy\\
\hline\hline
\twoGS\threeGS\fourGS\fiveGS\sixGS\sevenGS\eightGS\nineGS
\end{tabular}\caption{\protect\small\capGS}
\end{center}\end{table}}
\def\EF{\begin{table}\label{EF1}\begin{center}
\begin{tabular}{|r||l|l||l|}\hline
D& Zeta Function  Inside& Zeta Function  Outside& Casimir Energy\\
\hline\hline
\twoEF\threeEF\fourEF\fiveEF
\end{tabular}\caption{\protect\small\capEF}
\end{center}\end{table}}

\def\alltables{\begin{table}\label{Dir0}\begin{center}
\begin{tabular}{|r||p{4cm}|p{4cm}||p{4cm}|}\hline
D&Zeta Function  Inside&Zeta Function  Outside& Casimir Energy\\ 
\hline\hline
\twoDir\threeDir\fourDir\fiveDir\sixDir\sevenDir\eightDir\nineDir
\end{tabular}\caption{\protect\small\capDir}\end{center}
\begin{center}
\begin{tabular}{|r||l|l||l|}\hline
D&Zeta Function   Inside&Zeta Function  Outside& Casimir Energy\\ 
\hline\hline
\twoNeu\threeNeu\fourNeu\fiveNeu\sixNeu  
\end{tabular}
\end{center}\caption{\protect\small\capNeu}
\begin{center}\begin{tabular}{|r||l|l||l|}\hline
D& Zeta Function  Inside& Zeta Function  Outside& Casimir Energy\\
\hline\hline
\twoBag\threeBag\fourBag\fiveBag\sixBag\sevenBag
\end{tabular}\caption{\protect\small\capBag}\end{center}
\label{GS12}\begin{center}
\begin{tabular}{|r||l|l||l|}\hline
D& Zeta Function  Inside&Zeta Function   Outside& Casimir Energy\\
\hline\hline
\twoGS\threeGS\fourGS\fiveGS\sixGS\sevenGS\eightGS\nineGS
\end{tabular}\caption{\protect\small\capGS}
\end{center}
\label{EF1}\begin{center}
\begin{tabular}{|r||l|l||l|}\hline
D& Zeta Function  Inside& Zeta Function  Outside& Casimir Energy\\
\hline\hline
\twoEF\threeEF\fourEF\fiveEF
\end{tabular}\caption{\protect\small\capEF}
\end{center}\end{table}}

\begin{figure}[h]
\setlength{\unitlength}{1cm}
\begin{picture}(13,9.2)
\put(0,0){\epsfig{figure=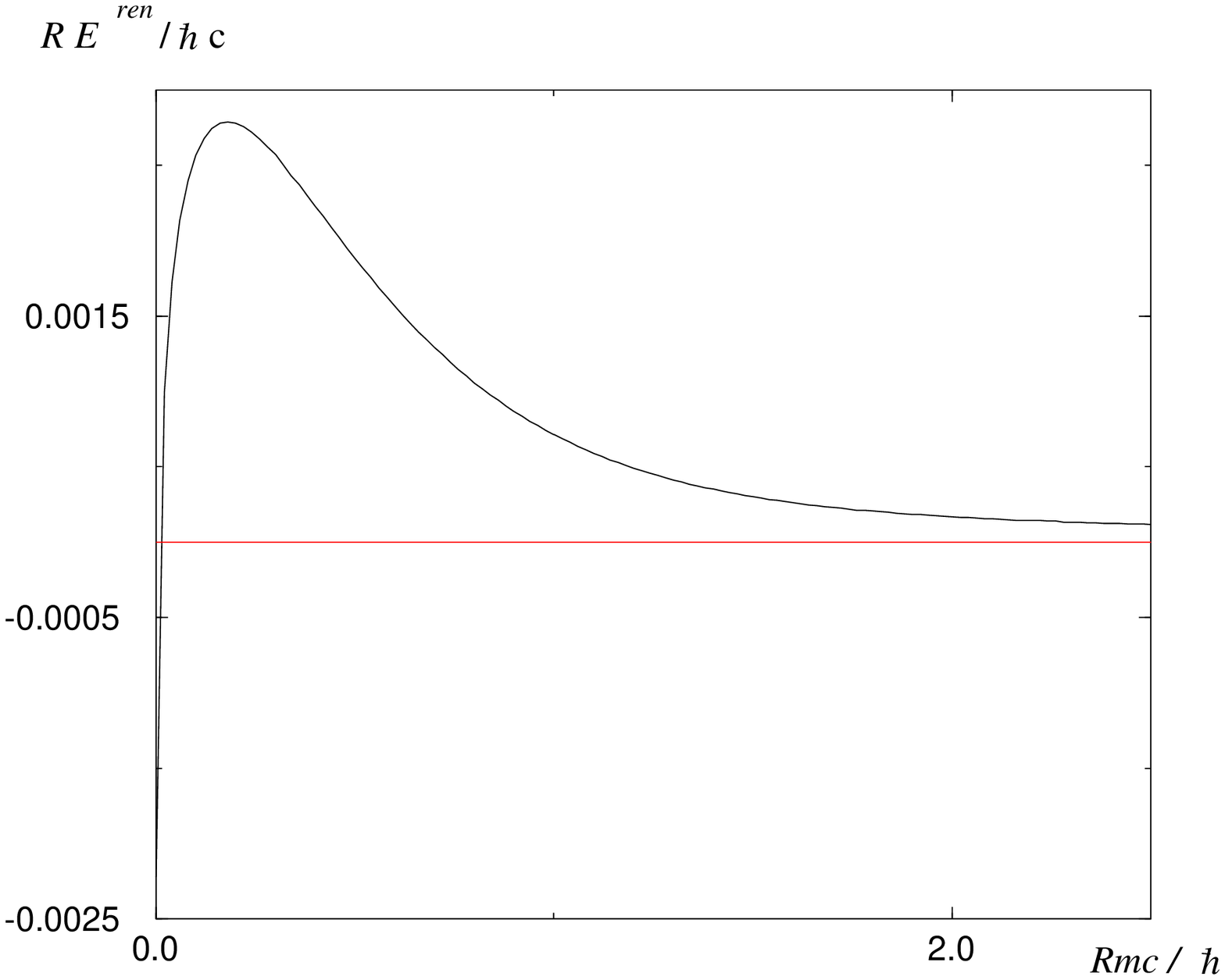,height=9cm,width=13cm}}
\end{picture}
\caption{\label{figint}The Casimir energy of a massive scalar field in the
  interior of a sphere of radius $R$.}
\setlength{\unitlength}{1cm}
\begin{picture}(13,9.5)
\put(0,0){\epsfig{figure=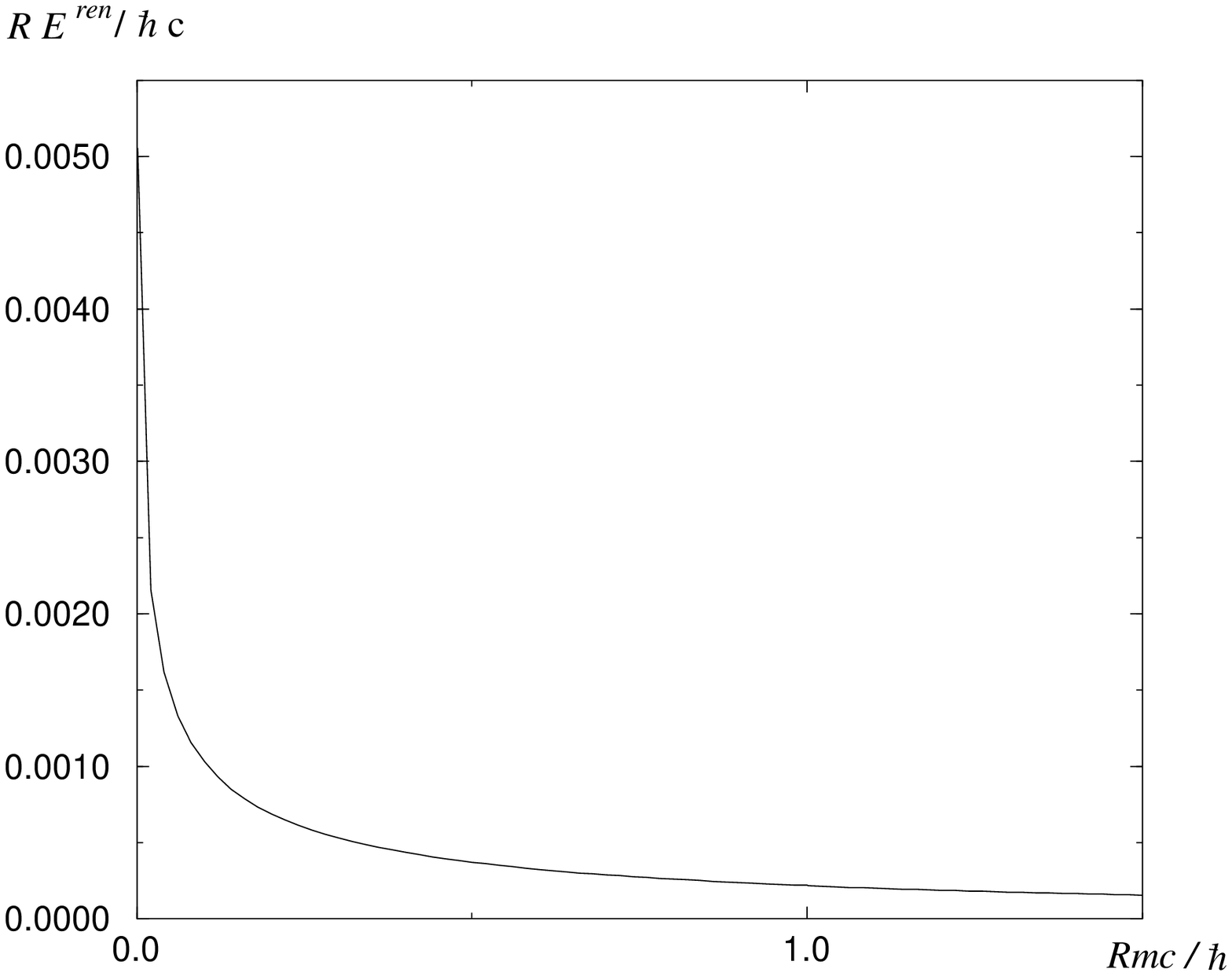,height=8.5cm,width=13cm}}
\end{picture}
\caption{\label{figext}The Casimir energy of a massive scalar field in the
  exterior of a sphere of radius $R$.}
\end{figure}
\begin{figure}[h]
\setlength{\unitlength}{1cm}
\begin{picture}(13,9)
\put(0,0){\epsfig{figure=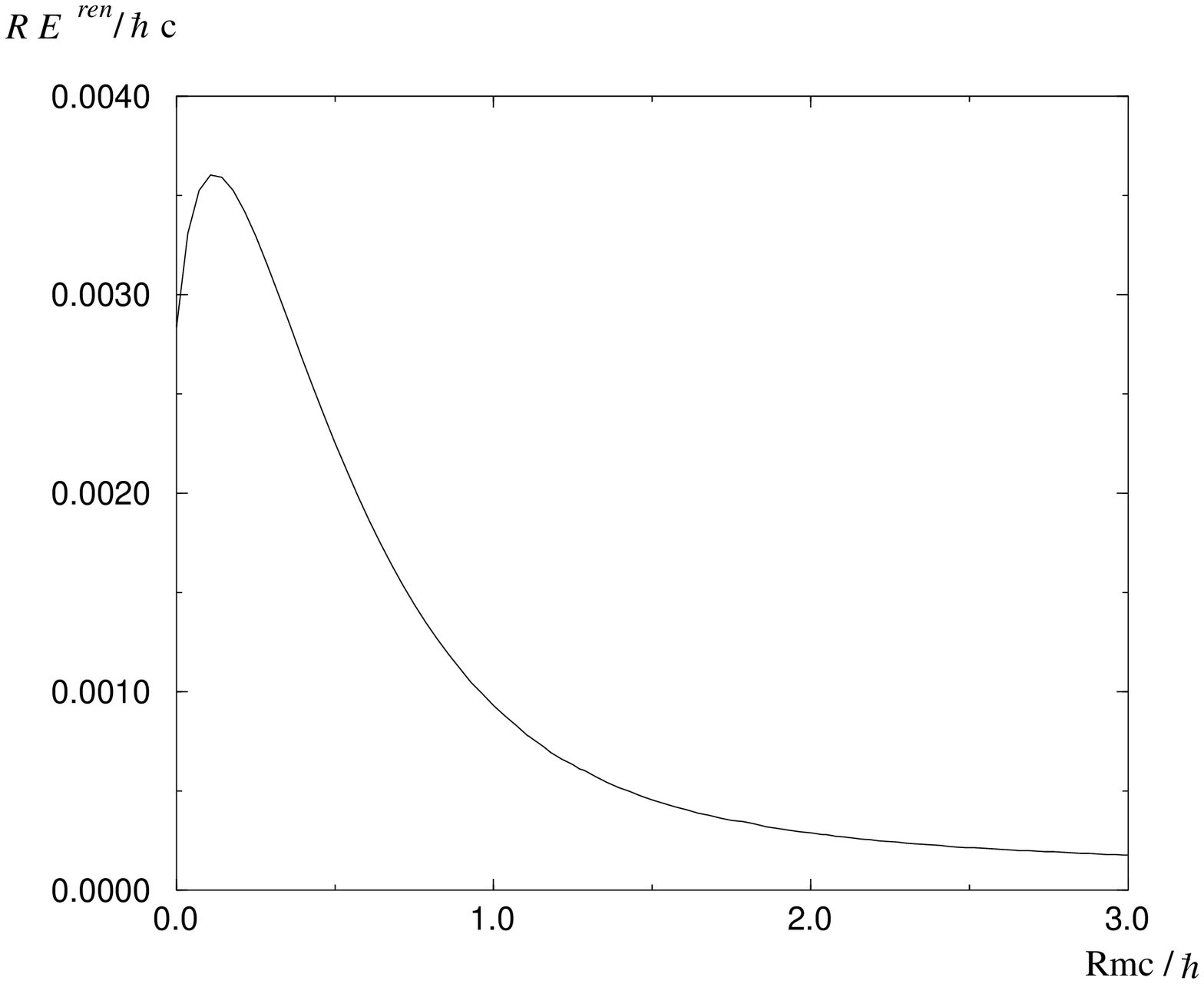,height=8.5cm,width=13cm}}
\end{picture}
\caption{\label{figboth}The Casimir energy of a massive scalar field in the
  whole space with Dirichlet \bc on a sphere of radius $R$.}
\setlength{\unitlength}{1cm}
\begin{picture}(13,9.5)(0,3.5)
\put(0,4){\epsfig{figure=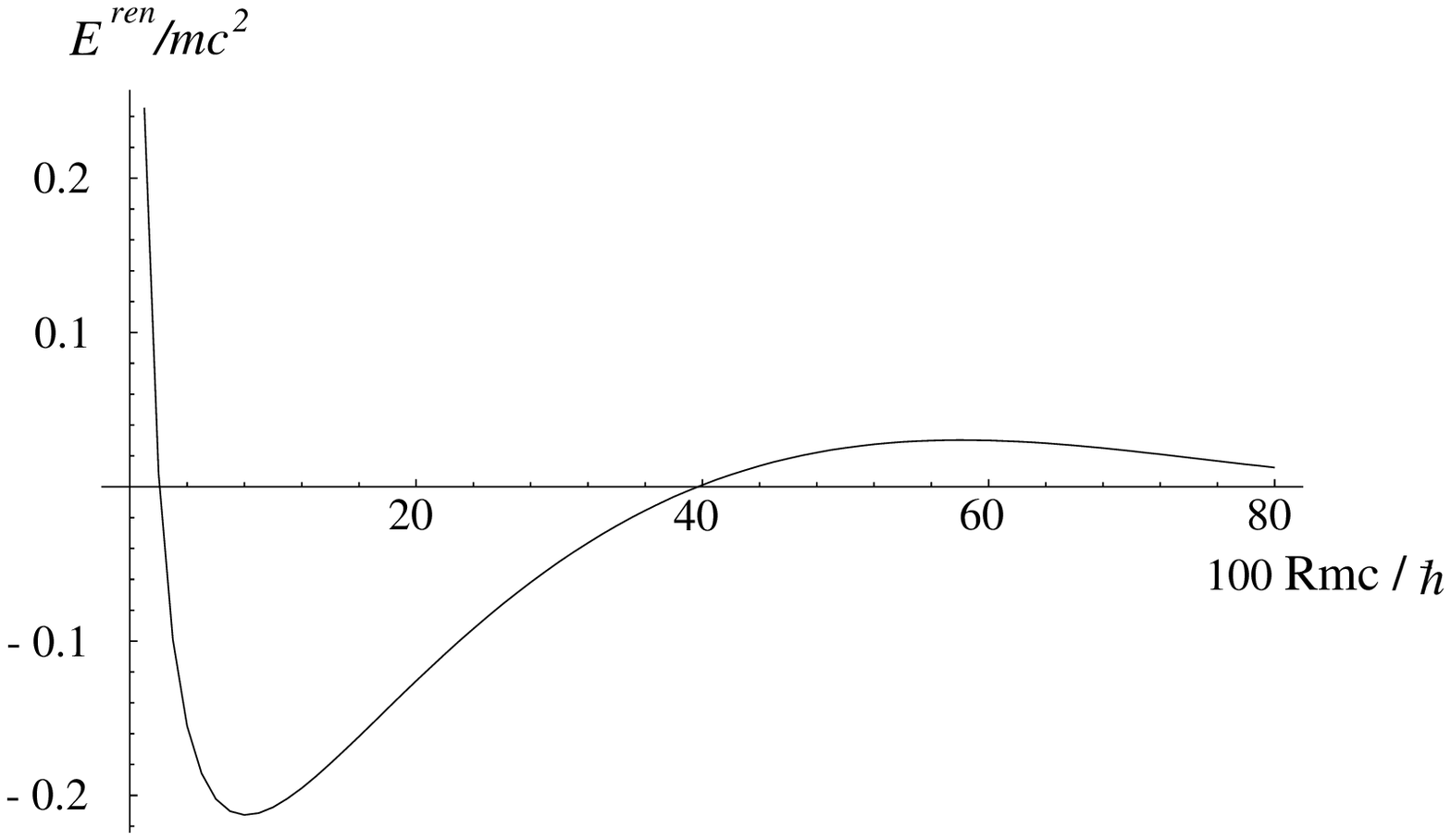,width=13cm}}
\end{picture}
\caption{\label{figbag}The Casimir energy for a spinor field obeying bag \bcp}
\end{figure}

\DirBC
\NeuBC
\BagBC
\EF
In the case of a massive field we have a renormalization prescription
delivering a unique result as described in Sec. \ref{sec3.4}.  
For dimensional
reasons the renormalized energy can be represented in the form
\be\label{darerenm}\E^{\rm ren}=
\hbar c{f\left(\frac{mcR}{ \hbar}\right)\over R},
\ee
where $f(mcR/\hbar)$ is some dimensionless function. 
This function is shown in the
figures, for the interior space in Fig.~\ref{figint}, for the exterior space
in Fig. \ref{figext}, and for the whole space in Fig. \ref{figboth}. This is
the Casimir energy measured in units of the inverse radius. In the
interior space the energy takes positive as well as negative values
demonstrating an essential dependence on the mass. It should be noted 
that for
large masses, the Casimir energy decreases as a power of the mass and it is not
exponentially damped as known from plane parallel geometry. This was probably
first noted in \cite{Bordag:1997ma}. This is due to the nonzero heat kernel
coefficient $a_{5/2}$ (in general the next nonvanishing coefficient after
$a_{2}$).  For small masses there is a logarithmic behavior in the interior
and exterior separately taken which is due to the normalization condition and
the nonzero \hkk $a_{2}$. For the whole space these contributions cancel and
the massless result \Ref{ECasml} is again obtained for $mcR/\hbar\to 0$. 
It should be
noted that the energy is positive for all values of the mass in contrast to
the contribution in the interior where the energy takes different signs as a 
function of the radius.

In the paper \cite{Elizalde:1997hx} the same had been done for the spinor
field obeying bag \bc on  a sphere. The result is shown in
Fig. \ref{figbag}. Again, a non trivial dependence on the radius is seen. 

In \cite{Scandurra:1998xa} the corresponding calculation had been done for the
semi-transparent spherical shell given by a delta function potential
corresponding to the matching condition \Ref{mcond} and the Jost function is
given by Eq. \Ref{delta}. The result is a \gse changing its sign in dependence
on the radius and the strength of the potential.

The vacuum energy of the \elm field in the presence of a dielectric body is of
general interest as an example for the interaction with real macroscopic
matter. Recently this was intensively discussed as a possible explanation of
sonoluminescence (there is  a number of  papers on this topic, see, e.g.,
\cite{11,Liberati:1998wg} and papers cited therein). The structure
of the \uv divergencies had been discussed in Section 3.3. For a
dielectric ball (permittivity $\ep_{1}$ ($\ep_{2}$) and permeability $\mu_{1}$
($\mu_{2}$) inside (outside)) these divergencies, represented by the \hkk
$a_{2}$, turned out to be present in general not allowing for a unique
definition of the vacuum energy of the \elm field due to the lack of a
normalization condition in a massless case. Only in the dilute approximation,
i.e., to order $(c_{1}-c_{2})^{2}$ for $c_{1}\sim c_{2}$ or for equal speeds
of light, $c_{1}=c_{2}$ (still allowing for $\ep_{1}\ne\ep_{2}$ by means of
$\mu_{1}\ep_{1}=\mu_{2}\ep_{2}$) a unique result can be obtained.

In general, the corresponding calculations can be done in a  similar 
manner to that
for the conducting sphere. One has to take  the corresponding Jost function,
Eq. \Ref{em1} for $c_{1}=c_{2}$ or Eq. \Ref{diel} which must be expanded in
powers of $c_{1}-c_{2}$ and to insert that into the regularized energy,
Eq. \Ref{EregIms}. The resulting expression which contains an infinite sum and
integral can be analytically continued in $s$ and has a finite continuation to
$s=0$. The first calculation for  $c_{1}=c_{2}$ had been done in
\cite{Brevik1982af} using a different representation. The vacuum energy turned
out to be positive. As a function of $\xi=(\ep_{1}-\ep_{2})/(\ep_{1}+\ep_{2})$
it interpolates smoothly between zero for $\xi=0$ and the known result for a
conducting sphere at $\xi=1$. For small $\xi$ it is possible to obtain an
analytical result
\be\label{Exi}\E =\frac{5}{16\pi}\frac{\hbar c \xi^{2}}{2R}+O(\xi^{4})\,.
\ee
It had been obtained in \cite{bali78-112-165} by the multiple reflection
expansion and in \cite{Klich:1999df} by performing the orbital momentum
summation prior to the radial momentum integration.

In the dilute approximation, i.e., for nonequal speeds of light, the result
in the first order in $c_{1}-c_{2}$ is zero. 
In second order it had been computed
by different methods. In \cite{Milton:1997ky} the Casimir-Polder potential
between pairs of molecules within the ball had been integrated over the
volume,
\be\label{CasPol}\E=-\frac{1}{2}\, 
\frac{23}{4\pi}\hbar c \al^{2}N^{2}\int_{V}
d{{\bf r}_{1}}d{{\bf r}_{2}} 
{1\over \mid {{\bf r}_{1}}- {{\bf r}_{2}}\mid^{7-\sigma}},
\ee
where $\al=(\ep-1)/( 4\pi N)$ is the polarizability and $N$ is the density
of molecules. The divergencies at ${\bf r}_{1}={\bf r}_{2}$ forced the
introduction of a regularization ($\sigma$ sufficiently large and $\sigma\to0$
in the end). After dropping some divergent contributions which can be
understood as particular examples (in a different regularization) of the
divergent terms in Eq. \Ref{Edivd}, as finite result
\be\label{Ediel}\E={23\over 1536\pi}{(\ep-1)^{2} \hbar c \over R}
\ee
emerges. Another approach was chosen in \cite{Barton1999c}. Here the
dielectric ball was taken as perturbation to the Minkowski space. Again, there
was a number of divergent contributions to be dropped in the same sense as
above and the remaining finite part coincides with \Ref{Ediel}. In this
calculation a wave number cut off was taken as regularization which is quite
close to the frequency cut off used in Eq. \Ref{dreg}.  These two papers
establish the equivalence of `Casimir-Polder summation' and ground state
energy which is known for distinct bodies since the fifties now for one body
which is quite important for the understanding of these phenomena. For
instance, it was demonstrated that the attractive force between individual
molecules in \Ref{CasPol} turns into a repulsive one after performing the
integration and dropping divergent contributions.

In \cite{Nesterenko:2000cf} the same result \Ref{Ediel} was obtained by mode
summation, i.e., starting from a formula equivalent to \Ref{EregIms}. However,
there no regularization was used. It would be of interest to get this result
using a mathematically correct regularization procedure.

Despite the mentioned 
results the situation with a dielectric body remains
unsatisfactory. Consider first the dilute approximation. There are 
divergent contributions
in some regularizations  
(in zeta-functional regularization we would expect
to have no divergent contributions according to Eq. \Ref{Edivz} with $m=0$
and $a_{2}=0$). So we are forced to introduce some classical model like
\Ref{Eclass}. There is no understanding how to do that as no classical energy
is associated with a dielectric body.  Beyond the dilute approximation the
situation is even worse. It is impossible to identify a unique quantum energy.
But on the other hand, we are confronted with real macroscopic bodies and
the clear existance of vacuum fluctuations of the \elm field constituting a
real physical situation so that no infinities or arbitraryness should occur.
At the moment the only conclusion that can be made
representing the properties of a
real body by dielectric constant  $\ep$ is not a good idealization.  
Dispersion and
other properties must be taken into account. A consequence would be that the
vacuum energy will depend on them. It should be remarked that spatial
dispersion, i.e., a position dependent $\ep({\bf x})$, makes the situation
even worse as then $a_{2}\ne 0$ even in the dilute approximation
\cite{Bordag:1997fh}.

\subsubsection{The Casimir effect for a cylinder}\label{sec4.2.4}
Using the methods explained above in this section it is possible to
investigate the Casimir effect for a large number of configurations which can
be characterized as generalized cones. Among them the conducting and
dielectric cylinder in three dimensions are of particular interest 
as intermediate situation between the parallel plates and the sphere where
the Casimir force is attractive and repulsive respectively.

The calculations go essentially the same way as in the spherical case and we
restrict ourselves  to pointing out the differences. Due to the translational
invariance along the axis of the cylinder, the quantum numbers labeling the
eigenvalues are $(k_{z},l,n)$ (corresponding to a cylindrical coordinate
system), where $k_{z}\in(-\infty,\infty)$ is the momentum corresponding to the
$z$-axis, $l=-\infty,\dots,\infty$ is the orbital momentum and $n$ is the
radial quantum number (assuming for the moment the presence of a 'large
cylinder'). In parallel to Eq. \Ref{EregIms} we obtain for the energy density
per unit length
\be\label{Eregcyl} \E_{0}^{\rm cyl}(s)=- \hbar c {\cos \pi s\over 2\pi}
\sum_{l=-\infty}^{\infty}\int_{-\infty}^{\infty}{\d k_{z}\over 2\pi}
\int_{mc/\hbar}^{\infty}{d k }\left(k^{2}-k_{z}^{2}\right)^{\frac12-s} 
{\pa\over\pa k}\ln
f_{l}(ik,k_{z}) \,.  \ee
The Jost function $f_{l}(ik,k_{z})$ is defined for the scattering problem in
the $(x,y)$-plane. Usually it does not depend on $k_{z}$ and the
corresponding integration can be carried out. But for the dielectric
cylinder it depends on $k_{z}$ (see below). The main difference with the
spherical case is that the corresponding Bessel functions are of integer order
now. Also, it is useful to take the notations of the modified Bessel
functions instead of the spherical ones, Eq. \Ref{slel}. For example, for
Dirichlet \bc on the cylinder we have for the interior problem
\be\label{cylDi}f^{\rm D,i \ {\rm cyl}}_{l}(ik)=(kR)^{-l}I_{l}(kR)\,.
\ee
A technical complication appears from the $l=0$ contribution. Instead of Eq.
\Ref{lnfas} we must define the asymptotic expansion in the form 
\be\label{}\ln f_{l}^{\rm as}(ik)=\sum_{i=-1}^{3}{X_{i}(t)\over l^{i}}
\ee
as an expansion in inverse powers of $l$. The $l=0$ contribution must be
treated separately using the corresponding asymptotic expansion for large
argument. Another difference is that in the electromagnetic case the $l=0$
mode must be included so that the Casimir energy for a conducting cylinder is
the sum of the corresponding energies for Dirichlet and Neumann \bcp The
results have been compiled in \cite{Gosdzinsky:1998qa}. For a massless scalar
field with Dirichlet \bc (taking interior and exterior regions together) the
Casimir energy is  
\be\label{cylDir}{\E}_{\rm Dir}^{\rm cyl}= \hbar c\frac{0.0006148}{R^{2}}\,,
\ee
for Neumann \bc it reads
\be\label{cylNeu}{\E}_{\rm Neum}^{\rm cyl}=- \hbar c\frac{0.014176}{R^{2}}
\ee
and the Casimir energy for a conducting cylinder is the sum of these two,
\be\label{cylem}{\E}_{\rm elm}^{\rm cyl}=- \hbar c\frac{0.01356}{R^{2}}\,,
\ee
and the force is attractive. The first calculation of this quantity had been
done in \cite{DeRaad:1981hb}.

The dielectric cylinder had been considered only recently. This problem
is more involved than the corresponding spherical case because the $TE$- and
$TM$-modes do not decouple in general. The corresponding Jost function reads
\be\label{fdielcyl} 
f_{l}(ik,k_{z})=\Delta^{\rm TE}\Delta^{\rm TM}
+l^{2}  {\left(k^{2}-k_{z}^{2}\right)k_{z}^{2}   \over
                         k^{2} q^{2}}  
{\left(c_{1}^{2}-c_{2}^{2}\right)^{2}c^2\over c_{1}^{4}c_{2}^{2}}
   \left[I_{l}(qR)K_{l}(kR)\right]^{2} 
\ee
with $q=\sqrt{\left({c_{2}\over
      c_{1}}\right)^{2}k^{2}+\left(1-\left({c_{2}\over
        c_{1}}\right)^{2}\right)k_{z}^{2}}$ and 
\be\label{TEcyl} \Delta^{\rm
  TE}=\mu_{1} \ kR \ I_{l}'(qR) \ K_{l}(kR)-\mu_{2} 
 \ qR \ I_{l}(qR) \  K_{l}'(kR) 
\ee
and
\be\label{TEcyl} \Delta^{\rm
  TM}=\ep_{1} \ kR \ I_{l}'(qR) \ K_{l}(kR)-\ep_{2}  
 \ qR \ I_{l}(qR) \ K_{l}'(kR)\,. 
\ee
Up to now an analysis of the \uv divergencies in terms of the \hkks is
missing. Calculations had been done only in special cases. For equal speeds of
light inside and outside the TE- and TM-modes decouple and the Casimir energy
is finite. As a function of the parameter 
$\xi=(\ep_{1}-\ep_{2})/( \ep_{1}+\ep_{2})$ 
it had been calculated for small $\xi$ in
\cite{Milton2000,Lambiase:1998tf,Nesterenko:1999bb,Klich:1999ip} and found to
be zero in order $\xi^{2}$ and nonzero in order $\xi^{4}$ and higher (for
$\xi=1$ the conducting cylinder, Eq. \Ref{cylem} is reobtained). For $c_{1}\ne
c_{2}$, but in dilute approximation the Casimir energy had been calculated in
the appendix of the paper \cite{Milton2000} summing the Casimir-Polder forces
and found to be zero too.  At the moment there is no explanation for these
somewhat unexpected results. As for the \hkks one may speculate that $a_{2}$
is zero in the dilute approximation like in the case of a dielectric
sphere. The reason is that in the Casimir Polder summation (where, of course,
a regularization had to be used) no polarization term and, hence, no 
logarithmic
singularity appeared. 

For a massive field the \gse had been calculated in \cite{Scandurra:2000qz}
for a semi-transparent cylinder using the same method as in the spherical case.
In contrast to the spherical case, the \gse turned out to be negative for 
all values of
the radius and the strength of the background potential.


\subsection{Sphere (lens) above a disk: additive methods and
proximity forces}
\label{sec4.3}

In the many calculations of the Casimir energies and forces
presented previously, the exact calculation methods were used, based
on the separation of variables in a field equation
and finding a frequency spectrum (in an explicit or implicit form).
Unfortunately, this can be done only for configurations possessing
high symmetry properties. Of particular interest is the configuration
of a sphere (lens) above a disk which is used in the modern
experiments on the measurement of the Casimir force 
\cite{32,33,34,35,36}.
For this configuration all the variables in the wave equation
cannot be separated which makes impossible to obtain the exact
expression for the Casimir force starting from the first principles.
Hence some approximate methods should be applied in this case. Such
methods are needed also to account for the surface roughness which is 
very important factor in the Casimir force
measurements (see below Sec.5.3).

There exist iterative procedures to
obtain the Green function for configurations where the variables
cannot be separated, starting from related configurations for which
the exact solution is known. Such procedure called the multiple
scattering expansion method has been proposed for the Casimir
problems with boundaries in \cite{sIV3-1}. It turns out to be rather 
complicated in applications to configurations
of experimental interest. Because of this, more phenomenological
approximate procedures which could be easily applied are
also of great interest.

One such procedure is called the Proximity Force Theorem
\cite{sIV3-2}. It has gained wide-spread acceptance in recent
years. The application range of this theorem extends from 
coagulations of aerosols (the subject where it was used
first \cite{sIV3-3}) to atomic nuclei \cite{sIV3-2}.
It is based on the expression for the proximity energy associated
with a curved gap of smoothly variable width $z$
\beq
V_p=
\int\!\!\!\int\limits_{\Sigma}E_S^{ren}(z)\,d\sigma,
\label{sIV3.1}
\eeq
\noindent
where ${E_S^{ren}}(z)$ is the known interaction energy per unit area of
the two parallel planes at the separation $z$, $\Sigma$ is one of the
two surfaces restricting a gap. It is apparent that in Eq.~(\ref{sIV3.1})
we neglect the non-parallelity of the area elements situated 
on the curved boundary surfaces restricting the gap. 
The corrections to (\ref{sIV3.1})
diminish as the curvatures of these surfaces become small.

After some steps of approximately the same accuracy as
 Eq.~(\ref{sIV3.1}) the expression for the force acting between the
gap boundaries follows \cite{sIV3-2}
\beq
F(a)=-\frac{\partial V_p}{\partial a}=
2\pi {\bar R}\,E_S^{ren}(a).
\label{sIV3.2}
\eeq 
\noindent
Here $a$ is the minimal value of the separation $z$ of boundary
surfaces, ${\bar R}=(R_1R_2)^{1/2}$ is the geometric mean of the
two principal radii of curvature $R_1$ and $R_2$, characterizing the gap.

Let us apply the Proximity Force Theorem expressed by the
Eq.~(\ref{sIV3.2}) to calculate the Casimir force in configuration
of a sphere (or a spherical lens) situated above a large disk.
Let the closest separation between the sphere and disk points be
$a\ll R$, where $R$ is a sphere (lens) radius. Under this condition the
force acting in both configurations is one and the same because the upper 
part of a sphere makes a negligible contribution 
to the force value. We put the
coordinate origin on a surface of a disk under a sphere center, which
is situated at a point $z=a+R$. Then the width of a gap is
\beq
z=R+a-\sqrt{R^2-x^2-y^2},
\label{sIV3.3}
\eeq
\noindent
and the principal radii of a gap curvature are simply calculated
\beq
R_1=\frac{1}{z_{xx}^{\prime\prime}(0,0)}=R,
\qquad
R_2=\frac{1}{z_{yy}^{\prime\prime}(0,0)}=R,
\label{sIV3.4}
\eeq
\noindent
which leads also to ${\bar R}=R$.

Now in accordance with Eq.~(\ref{sIV3.2}) the Casimir (van der Waals)
force acting between a disk and a sphere (lens) is
\beq
F_{dl}(a)=2\pi R\,E_S^{ren}(a),
\label{sIV3.5}
\eeq
\noindent
where the energy density ${E_S}^{ren}(a)$ between semispaces
(planes) is given by the Eq.~(\ref{sIV26}). This result as well as
(\ref{sIV26}) contains the limiting cases of both the van der Waals
and Casimir force. At small separations ($a\ll\lambda_0$, see
Sec.4.1.1) it follows from (\ref{sIV3.5})
and (\ref{sIV27}) for perfect conductors
\beq
F_{dl}(a)=-\frac{HR}{6a^2}.
\label{sIV3.6}
\eeq
\noindent
This is the van der Waals force between a disk and a lens. At large
separations $a\gg\lambda_0$ using Eq.~ (\ref{sIV3.5})
and (\ref{sIV31}) one obtains the Casimir force in configuration
of the perfectly conducting disk and lens \cite{sIV3-4}
\beq
F_{dl}^{(0)}(a)=-\frac{\pi^3\hbar cR}{360a^3}.
\label{sIV3.7}
\eeq
\noindent
If Eq.~(\ref{sIV18}) is substituted into the right-hand side of
(\ref{sIV3.5}) the expression for the Casimir (van der Waals) force
acting between a disk and a lens covered by the additional layers
follows. The question arises as to the accuracy of the results
(\ref{sIV3.5})--(\ref{sIV3.7}) given by the Proximity Force
Theorem. Before discussing this question we consider one more approximate
method which can be applied for the calculation of the Casimir force
in complicated configurations.

To calculate the Casimir and van der Waals force in configurations where
the exact methods do not work the additive summation of interatomic
pairwise interaction potentials is used widely. Let us consider
two atoms with electrical polarizability $\alpha(\omega)$ and let
$\omega_0$ be the characteristic frequency for transition between
the ground and excited states. If the distance $r_{12}$ between these
atoms is such that $\omega_0 r_{12}/c\gg 1$ the Casimir-Polder
interaction occurs \cite{2}
\beq
U(r_{12})=-\frac{C}{r_{12}^7},
\qquad
C\equiv\frac{23}{4\pi}\hbar c \alpha^2(0)
\label{sIV3.8}
\eeq
\noindent
(it is assumed that the magnetic polarizabilities of the atoms are
equal to zero). In the opposite case $\omega_0 r_{12}/c\ll 1$
the non-retarded van der Waals interaction with the inverse sixth
power of distance holds. Its coefficient does not depend on $c$.

The additive result for the Casimir interaction energy of two bodies
$V_1$ and $V_2$ separated by a distance $a$ is obtained by the
summation of the retarded interatomic potentials (\ref{sIV3.8})
over all atoms of the interacting bodies
\beq
U^{add}(a)=-CN^2
\int\limits_{V_1}d^3r_1
\int\limits_{V_2}d^3r_2
|{\rv}_1-{\rv}_2|^{-7},
\label{sIV3.9}
\eeq
\noindent
where $N$ is the number of atoms per unit volume.

It has been known that the additive result (\ref{sIV3.9})
reproduces correctly the dependence of $U$ on distance
(see, e.g., section 2.5 of \cite{sIV3-5} where a number of examples
and references on this point is contained). The coefficient near
this dependence is, however, overestimated in $U^{add}(a)$
compared to the true value. This is because Eq.~(\ref{sIV3.9}) does not
take into account the non-additivity effects connected with the
screening of more distant layers of material by closer
ones (the value of a coefficient in (\ref{sIV3.9}) tends to a true value
if $\varepsilon\to 1$, i.e. for a very rarefied medium).
In Ref.\cite{sIV3-6} a normalization procedure was suggested to take
approximately account for the non-additivity. According to this procedure the
additive expression (\ref{sIV3.9}) should be divided by a special
factor which is 
obtained by the division of
the additive result by the exact quantity found from
the plane-parallel configuration (see also \cite{21}).

For two semispaces separated by the distance $a$ the integration in
(\ref{sIV3.9}) yields additive energy per unit area
\beq
U^{add}(a)=-\frac{CN^2\pi}{30a^3}.
\label{sIV3.10}
\eeq
\noindent
The normalization factor is obtained by the division of (\ref{sIV3.10})
with the exact energy density from (\ref{sIV29})
\beq
K=\frac{U^{add}(a)}{E_S^{ren}(a)}=
\frac{CN^2}{\hbar c\Psi(\varepsilon_{20})}.
\label{sIV3.11}
\eeq
\noindent
Finally the expression for the Casimir energy with approximate
account of non-additivity is
\beq
U(a)\equiv\frac{U^{add}(a)}{K}=-\hbar c\Psi(\varepsilon_{20})
\int\limits_{V_1}d^3r_1
\int\limits_{V_2}d^3r_2
|{\rv}_2-{\rv}_1|^{-7}.
\label{sIV3.12}
\eeq

Applying Eq.~(\ref{sIV3.12}) to a sphere (spherical lens)
situated above a large disk at a minimum distance $a\ll R$
one easily obtains
\beq
U(a)=-\frac{\pi^2\hbar cR}{30a^2}\Psi(\varepsilon_{20}).
\label{sIV3.13}
\eeq
\noindent
Let us
calculate  the force as 
\beq
F_{dl}(a)=-\frac{\partial U(a)}{\partial a}=
-\frac{\pi^2\hbar cR}{15a^3}\Psi(\varepsilon_{20}).
\label{sIV3.13a}
\eeq
\noindent
Comparing 
the obtained result with Eq.~(\ref{sIV3.5}) 
for a sphere above a disk derived with the Proximity Force Theorem, 
taking into account the second equation of (\ref{sIV29}),
we can observe that both methods are in
agreement.

Now we start discussion of the accuracy of both methods. 
For two plane parallel plates the result (\ref{sIV3.12}) is
exact by construction. If to consider the arbitrary shaped body above
a conducting plane the maximal error occurs when this body is a
little sphere with a radius $R\ll a$ (the configuration which is
maximally distinct from two plane parallel plates). For this case the
independent result obtained from the first principles is
\cite{sIV3-7}
\beq
{E}^{ex}(a)=-\frac{9\hbar c R^3}{16\pi a^4}.
\label{sIV3.14}
\eeq
\noindent
For this configuration the additive method supplemented by
normalization (Eq.~(\ref{sIV3.13}) in the limit
$\varepsilon_{20}\to\infty$) gives the result
\beq
U(a)=-\frac{\hbar c\pi^3}{180}\frac{R^3}{a^4}.
\label{sIV3.15}
\eeq
\noindent
The comparison of Eqs.~(\ref{sIV3.14}) and (\ref{sIV3.15}) shows
that the maximum error of the above method is only 3.8\%.

It is necessary to stress that for such configurations like
a sphere (lens) above a disk under the condition $a\ll R$ the
actual accuracy of both Proximity Force Theorem and an additive
summation with normalization is much higher. For such configurations
the dominant contribution to the Casimir force comes from the
surface elements which are almost parallel.

In the papers \cite{sIV3-8,sIV3-9} the semiclassical approach was proposed 
for calculation of the Casimir energies and forces between curved 
conducting surfaces. In the framework of this approach the Casimir energy
is explained in terms of the classical periodic trajectories along which
the virtual photons are traveling between the walls. The contribution
from periodic trajectories decreases with their length (the contribution from each trajectory is inversely proportional
to the third power of its length). As shown in \cite{sIV3-8}
the semiclassical approximation reproduces the value of the Casimir
energy for a large class of configurations. It was applied to
configurations of two spheres of radii $R_1$, $R_2$, a distance
$a\ll R_1,\,R_2$ apart, and also of a sphere (lens) above a disk
under the condition $a\ll R$. In both cases the results are the same
as were obtained earlier by the application of the Proximity Force
Theorem or the additive summation with normalization (for a sphere
above a disk this result is given by Eq.~(\ref{sIV3.7})). Also,
several more complicated configurations were considered
semiclassically \cite{sIV3-9}. The semiclassical approach provides additional theoretical justification for the results
obtained by the Proximity Force Theorem and the additive method.
According to the semiclassical approach for a configuration of a sphere
(lens) above a disk the correction term to Eq.~(\ref{sIV3.7}) is
of order $a/R$ \cite{sIV3-9}. By this is meant for the typical
values used in experiment \cite{32} $a=1\,\mu$m, $R=10\,$cm
the accuracy of Eq.~(\ref{sIV3.7}) is of order $10^{-3}$\%.

Another possibility 
of checking and confirming the high accuracy of phenomenological
methods if there are the small deviations from
plane-parallelity is discussed in Sec.5.3.1. 
In this section the additive method
is applied for a plane plates inclined at a small angle to one
another. For this configuration the exact result is also obtainable.
The comparison of both results shows that the additive result
coincides with the exact one at least up to $10^{-2}$\% which is
in agreement with a semiclassical estimation of accuracy
(see Sec.5.3.1).

At the end of this section we discuss the application range of
the approximate methods mentioned above. The Proximity Force
Theorem is applicable for any material and distances between the
interacting bodies, i.e. in the range of the van der Waals forces,
Casimir forces, and also in the intermediate transition region.
It is, however, not applicable for stochastic and drastically
fluctuating surfaces, which is the case if the surface
distortions are taken into account (see Sec.5.3).

The additive method supplemented by the normalization procedure
starts from the summation of pairwise interatomic potentials.
These potentials decrease as $r_{12}^{-6}$ for the van der Waals
forces and as $r_{12}^{-7}$ for the Casimir ones. That is why
the transition region between the two types of forces is not
covered by the additive method. The advantage is that it can be
applied to calculate roughness corrections to the van der Waals
or Casimir force both for dielectrics and metals.

The more fundamental, semiclassical approach is applicable, however, 
only for the perfectly conducting bodies placed at a distance small
compared to the characteristic curvature radius of a boundary
surface. In the case of the large distances (e.g., sphere of a radius
$R\ll a$ or of the order $a$ where $a$ is a distance to the disk)
diffraction effects should be taken into account. What this means
is that trajectories which are going around a sphere make non-negligible contributions to the result. The inclusion of diffraction into the semiclassical
theory of the Casimir effect is an outstanding question to be
solved in the future \cite{sIV3-8,sIV3-9}.

Given that the modern experiments on the Casimir
force measurement \cite{32,33,34,35,36} 
use configurations such as a sphere (lens)
above a disk, the approximate methods discussed above are
gaining in importance. 

\subsection{Dynamical Casimir effect}
\label{sec4.4}

For the case of two dimensions (one-dimensional space and one-dimensional time)
the dynamical Casimir effect was already discussed in Sec.2.4.
It was emphasized there that the non-stationarity of the boundary
conditions leads to two different effects. The first one is the
dependency of the Casimir energy and force on the velocity of
a moving plate. The second effect is quite different and consists
in the creation of photons from vacuum such as happens in 
non-stationary external fields. Let us begin with the first effect
which was not discussed in Sec.2.4.

Consider the massless scalar field in the configuration of two
perfectly conducting planes one of which, $K$, lies in the plane
$x^3=0$, and the other, $K^{\prime}$, moves with a constant
velocity $v$ in a positive direction of the $x^3$-axis \cite{sIV4-1}.
The Green function of the field is the solution of the Dirichlet
boundary problem
\bes
&&
\Box_xG(x,x^{\prime})=-\delta(x-x^{\prime}),
\nonumber \\
&&
G(x,x^{\prime})|_{x,x^{\prime}\in K\ \mbox{\small{or}}\ K^{\prime}}
=0,
\label{sIV4.1}
\ees
\noindent
where $K,\,K^{\prime}$ are the planes at $x^3=0$ and at
$x^3=vt$ respectively. The Green's function has to be found separately 
for the three domains:
$x^3<0$, $0\leq x^3<vt$, $vt\leq x^3<\infty$. 
Given knowledge of Green functions it is not difficult to find
the non-renormalized vacuum energy density in all three domains
using the equalities
\bes
&&
\langle 0|T_{00}(x)|0\rangle=
\frac{\hbar c}{2}
\sum\limits_{k=0}^{3}
\langle 0|\partial_k\varphi(x)\partial_k\varphi(x)|0\rangle
\label{sIV4.2} \\
&& \phantom{aaa}
=\frac{\hbar c}{2}\lim\limits_{x^{\prime}\to x}
\sum\limits_{k=0}^{3}
\partial_k\partial_k^{\prime}
\langle 0|\varphi(x)\varphi(x^{\prime})|0\rangle=
\frac{i\hbar c}{2}\lim\limits_{x^{\prime}\to x}
\sum\limits_{k=0}^{3}
\partial_k\partial_k^{\prime}G(x,x^{\prime}).
\nonumber
\ees

Within the first domain the Green function is found by the use of
reflection principle with one reflection only
\beq
G^{<}(x,x^{\prime})=
\frac{i}{4\pi^2}\left[\frac{1}{(x-x^{\prime})^2}-
\frac{1}{(x-x_1^{\prime})^2}\right],
\quad
x^3,\,{x^3}^{\prime}<0,
\label{sIV4.3}
\eeq
\noindent
where
\beq
x_1^{\prime}=S_Kx^{\prime},
\qquad
S_K=\left(
\begin{array}{rrrr}
1&&&\\
&1&&\\
&&1&\\
&&&-1
\end{array}\right),
\label{sIV4.4}
\eeq
\noindent
and the diagonal operator $S_K$ describes the reflection in the plane
of a mirror $K$.

To find the Green function in the third domain we transform the point
under consideration into the reference system of a moving mirror 
$K^{\prime}$, find its reflection in $K^{\prime}$, and determine
the coordinates of this reflection relative to the mirror $K$ at rest
by an inverse Lorenz transformation. The result is
\beq
G^{>}(x,x^{\prime})=
\frac{i}{4\pi^2}\left[\frac{1}{(x-x^{\prime})^2}-
\frac{1}{(x-x_1^{\prime})^2}\right],
\quad
x^3,\,{x^3}^{\prime}>vt,
\label{sIV4.5}
\eeq
\noindent
where
\bes
&&
x_1^{\prime}=S_{K^{\prime}}x^{\prime},
\qquad
S_{K^{\prime}}=\left(
\begin{array}{cccc}
\cosh s&0&0&-\sinh s\\
0&1&0&0\\
0&0&1&0\\
\sinh s&0&0&-\cosh s
\end{array}\right),
\nonumber\\
&&
s\equiv\ln\frac{c+v}{c-v}.
\label{sIV4.6}
\ees

Now consider the case of the second domain where the point lies in
between the mirrors and experiences the multiple reflections in both 
of them. Here the Green function is given by the infinite sum over
all reflections
\beq
\bar{G}(x,x^{\prime})=
\frac{i}{4\pi^2}
\sum\limits_{m=-\infty}^{\infty}(-1)^m
\frac{1}{(x-x_m^{\prime})^2},
\quad
0\leq x^3,\,{x^3}^{\prime}\leq vt,
\label{sIV4.7}
\eeq
\noindent
where $x_0^{\prime}\equiv x^{\prime}$ and
\bes
&&
x_{2m}^{\prime}=(S_KS_{K^{\prime}})^mx^{\prime},
\qquad
x_{2m-1}^{\prime}=S_K(S_KS_{K^{\prime}})^mx^{\prime},
\nonumber \\
&&
x_{-2m}^{\prime}=(S_{K^{\prime}}S_K)^mx^{\prime},
\qquad
x_{-2m-1}^{\prime}=S_K(S_{K^{\prime}}S_K)^mx^{\prime}.
\label{IV4.8}
\ees

The Casimir energy density is calculated in all three regions
separately using Eqs.~(\ref{sIV4.2}), and (\ref{sIV4.3}), (\ref{sIV4.5}),
(\ref{sIV4.7}). To obtain the renormalized energy density, the
contributions of free Minkowski space is subtracted in each region.
This is equivalent to the descarding of the first terms in
  Eqs.~(\ref{sIV4.3}), (\ref{sIV4.5}), (\ref{sIV4.7}) which
coincide with Green function
$G_0=i/\left[4\pi^2(x-x^{\prime})^2\right]$ in free space without any
boundaries. The value of a force per unit area is calculated by
differentiation of the obtained energy per unit area with respect to
the time dependent distance $a(t)=vt$ between the plates
\beq
F\left(a(t)\right)=-\frac{d}{d(vt)}
\int\limits_{-\infty}^{\infty}dx^3\,E_0^{ren}(x^3,s).
\label{sIV4.9}
\eeq
\noindent
The result is \cite{sIV4-1}
\beq
F\left(a(t)\right)=-\frac{\pi^2\hbar c}{480a^4(t)}
\left[1+\frac{8}{3}\left(\frac{v}{c}\right)^2+
O\left(\frac{v^4}{c^4}\right)\right].
\label{sIV4.10}
\eeq

It is seen that in the first approximation the Casimir force between
moving boundaries coincides with the well known result for the
massless scalar field between plates separated by a distance $a(t)$
(which is one half of the electromagnetic force from
Eq . (\ref{2.38})).
%
%

The same method was applied in \cite{sIV4-2} to calculate the velocity
dependent correction to the Casimir force for electromagnetic field.
Both the cases of small and large velocities of plate were investigated.
For $v\ll c$ the following result is obtained \cite{sIV4-2}
\beq
F\left(a(t)\right)=-\frac{\pi^2\hbar c}{240a^4(t)}
\left[1-\left(\frac{10}{\pi^2}-\frac{2}{3}\right)
\left(\frac{v}{c}\right)^2+
O\left(\frac{v^4}{c^4}\right)\right].
\label{sIV4.11}
\eeq
\noindent
In the limiting case $c-v\ll c$ the force is given by
\beq
F\left(a(t)\right)=-\frac{3\hbar c}{8\pi^2a^4(t)}
\left[1+\frac{(c^2-v^2)^2}{16c^4}+
O\left(\frac{(c^2-v^2)^4}{c^8}\right)\right].
\label{sIV4.12}
\eeq
\noindent
It is seen from Eqs.~(\ref{sIV4.10}) and (\ref{sIV4.11}) that
the velocity dependent correction to the Casimir force has
different sign in the scalar and electromagnetic cases. Also,
the velocity dependence of the Casimir force for the electromagnetic
field appears to be very slight (less than 8\% of the static
value for the same separation).

Now let us discuss the effect of photon creation in case of a
three-dimensional non-stationary cavity and the possibility of the 
experimental observation of this effect. As was told in Sec.2.4 the 
method suggested in \cite{46} cannot be used in four-dimensional
space-time (the generalization of this method to the resonant
oscillations of one-dimensional cavity was given in \cite{sIV4-3}).
A perturbation method applicable to a single boundary moving along 
some prescribed trajectory was developed in \cite{sIV4-4}. It was
used to calculate the radiated energy for a plane mirror in
four-dimensional space-time. The analytical method of Ref.~\cite{50}
used in Sec.2.4 leaves room for the application to three-dimensional
resonant cavities.

Let a rectangular cavity have dimensions $a_1,\,a_2,\,a_3$ respectively. If these
dimensions do not depend on time the eigenfrequencies
$\omega_{n_1n_2n_3}$ are given by Eq.~(\ref{sIV12.1}). For
simplicity consider the case $a_3\ll a_1\sim a_2$ \cite{50}.
With this condition the frequencies with $n_3\neq 0$ are much greater
than the frequencies with $n_3=0$. Studying the excitation of the
lowest modes one may put $n_3=0$ when taking into account the fact that the
interaction between low- and high-frequency modes is weak.
As a consequence the vector-potential of the electromagnetic field
is directed along the $z$-axis and depends only on two space
coordinates $x$ and $y$. For $t\leq 0$ the field operator is
given by
\beq
A_z(t\leq 0,x,y)=
\sum\limits_{n_1,n_2}
\psi_{n_1n_2}(x,y)\left[
e^{-i\omega_{n_1n_20}t}a_{n_1n_2}+ 
e^{i\omega_{n_1n_20}t}a_{n_1n_2}^{+}\right],
\label{sIV4.13}
\eeq
\noindent
where
\beq
\psi_{n_1n_2}(x,y)=\frac{\sqrt{2c}}{\sqrt{a_1a_2a_3\omega_{n_1n_20}}}
\,\sin\frac{\pi n_1x}{a_1}\,\sin\frac{\pi n_2y}{a_2}.
\label{sIV4.14}
\eeq

For $t>0$ dimension $a_1$ depends on time given by
$a_1=a(t)$. The boundary conditions for the operator of
the vector-potential are
\beq
A_z|_{x=0}=A_z|_{x=a(t)}=A_z|_{y=0}=A_z|_{y=a_2}=0.
\label{sIV4.15}
\eeq
\noindent
For any time $t$ it can be found in the form
\beq
A_z(t,x,y)=
\sum\limits_{n_1,n_2}
{\tilde{\psi}}_{n_1n_2}(x,y)Q_{n_1n_2}(t),
\label{sIV4.16}
\eeq
\noindent
where the functions ${\tilde{\psi}}_{n_1n_2}$ are obtained from
(\ref{sIV4.14}) by replacing $a_1=$const for $a_1=a(t)$.
The operators $Q_{n_1n_2}(t)$ for $t>0$ are unknown. Substituting
(\ref{sIV4.16}) into the wave equation
\beq
\frac{\partial^2A_z(t,x,y)}{\partial t^2}-\Delta A_z(t,x,y)=0
\label{sIV4.17}
\eeq
\noindent
one obtains the coupled system similar to (\ref{2.74}) for their
determination. The coefficients of this system are given by
\beq
h_{n_1n_2,k_1k_2}=a
\int\limits_{0}^{a}dx
\int\limits_{0}^{a_2}dy
\int\limits_{0}^{a_3}dz
{\tilde{\psi}}_{k_1k_2}(x,y)
\frac{\partial{\tilde{\psi}}_{n_1n_2}(x,y)}{\partial a}
\label{sIV4.18}
\eeq
\noindent
(compare with (\ref{2.75})).

Now let us assume that the wall oscillates according
 to (\ref{2.82})
\beq
a(t)=a_1\left[1-\varepsilon\cos(2\omega_{n_1n_20}t)\right]
\label{sIV4.19}
\eeq
\noindent
with $\varepsilon\ll 1$. Using considerations analogous to those
of Sec.2.4 in one-dimensional case it is possible to find
the total number of photons created by the time $t$ \cite{50}
\beq
n(t)=\sinh^2(\omega_{n_1n_20}\gamma t),
\label{sIV4.20}
\eeq
\noindent
where the frequency modulation depth $\gamma$ is defined by
\beq
\gamma=\frac{1}{2}\varepsilon\left[1+
\left(\frac{n_1a_1}{n_2a_2}\right)^2\right]^{-1/2}.
\label{sIV4.21}
\eeq
\noindent
It is notable that in a three-dimensional case both the total
energy of the cavity and also the number of photons grows exponentially
with time.

Let us discuss the possibility of detecting the photons created by virtue 
of the dynamical Casimir effect. Instead of oscillations of a wall 
as a whole it is more realistic to consider the oscillations of its
surface due to the strong acoustic waves excited inside a wall.
The maximal possible value of the dimensionless displacement
$\varepsilon$ from the Eqs.~(\ref{2.82}), (\ref{sIV4.19}), which
a wall material can endure, is estimated as
$\varepsilon_{\max}\sim 3\times 10^{-8}$ for the lowest mode
$\omega_1\sim c\pi/a_0$ \cite{50}. Considering the separation
between plates of order several centimeters the frequency of the
lowest mode is $\omega_1\sim 60\,$GHz. Substituting these numbers
into Eq.~(\ref{2.88}) one obtains the photon creation rate for the
one-dimensional cavity
\beq
\frac{dn_1(t)}{dt}\approx 
\frac{4}{\pi^2}\varepsilon_{\max}\omega_1
\sim 700\,{\mbox{s}}^{-1}.
\label{sIV4.22}\eeq

In the case of three-dimensional cavity the number of created photons
can be even larger due to the exponential dependence on time in 
Eq.~(\ref{sIV4.20}). Using the same frequency value as in
one-dimensional case $\omega_{n_1n_20}\sim\omega_1\sim 60\,$GHz,
and the displacement parameter 
$\varepsilon=\varepsilon_{\max}/100\sim 3\times 10^{-10}$,
the frequency modulation depth of Eq.~(\ref{sIV4.21}) takes the
value $\gamma\sim 10^{-10}$. As a result it follows from
Eq.~(\ref{sIV4.20})
\beq
n(t)\approx\sinh^2(6t),
\label{sIV4.23}
\eeq
\noindent
$t$ being measured in seconds, which results in approximatelly 
$4\times 10^4$ photons created in
the cavity during one second due to the wall oscillations. 
So large number of photons can be observed experimentally. 
The main problem is, however, how to excite the high-frequency
surface vibrations in GHz range of sufficient amplitude.

There are many important factors which should be taken into account
in future experiments on the dynamical Casimir effect.
It should be noted that the above derivations were performed for the perfectly
reflecting walls. The case of the wall with any finite (but
nondispersive) refractive index was considered in \cite{sIV4-5}
for the scalar field in one-dimensional space. The methods,
elaborated in \cite{sIV4-5} can be generalized, however, for the
electromagnetic field in four-dimensional space-time.

The important point is the type of detector and its influence on
the photon creation process. In \cite{50} two types of detectors were
discussed. The first one proposes that the beam of Rydberg atoms be
injected into the cavity after the creation of sufficient number of
photons. The typical frequencies of order 10\,GHz mentioned above
correspond to the transitions between the atomic levels with
$n\sim 100$ ($n$ being the principal quantum number). These transitions
can be observed by the well known methods. The second type of detector
suggests that a harmonic oscillator tuned to the frequency of
resonant mode be placed into the cavity from the start, so that 
a quantum system consisting of two subsystems is built up.
The interaction between the radiated resonant modes and detector can
be described by the quadratic Hamiltonian with time dependent
coefficients. As a result, the squeezed state of electromagnetic
field is generated, and the number of photons in each subsystem is
equal to one half of the result given by Eq.~(\ref{sIV4.20}) in the
absence of detector.

Another point which affects the possibility of experimental
observation of the dynamical Casimir effect is the back reaction
of the radiated photons upon a mirror. In Ref.~\cite{sIV4-6}
not only the total energy of radiated photons but also 
the dissipative force exerted on a single mirror moving
nonrelativistically is considered. The effect, however, turns out to be very
small (creation rate there is of order $10^{-5}$ photon per second
only, to compare with much larger above rates in the case of 
a cavity). In Ref.~\cite{sIV4-7} a master equation is derived describing
one-dimensional nonrelativistic mirror interacting with vacuum via
radiation pressure (fluctuations of a position of the dispersive
mirror driven by the vacuum radiation pressure were considered
earlier in\cite{sIV4-8}). The other part of the radiative reaction force
exerted on a free mirror, which is not dissipative but of reactive
nature, was considered in \cite{sIV4-5,sIV4-9}. The existence of this 
force leads to corrections of the inertial mass of the mirrors.

All the above considerations of the dynamical Casimir effect deals with
the case of zero temperature. The Casimir effect at non-zero temperature
will be considered in Sec.5.1. Here we touch only on the influence of
non-zero temperature upon the photon creation rate in the dynamical
Casimir effect. In Ref.~\cite{sIV4-10} the temperature correction
to the number of photons created by a moving mirror is derived
in the framework of response theory. It was shown that for a
resonantly vibrating cavity of a typical size of about 1\,cm at room
temperature a thermal factor of order $10^3$ should be considered along with 
the zero-temperature result
(\ref{sIV4.20}).  Due to this, at room temperature, there will 3 orders of magnitude larger number of photons created in comparison to the 
$4\times 10^4$ photons created during 1 second, for a displacement
parameter $\varepsilon\sim 3\times 10^{-10}$ (see Eq.~(\ref{sIV4.23})).
This provides the possibility to decrease the displacement
parameter and makes more probable the experimental observation of
the dynamical Casimir effect in near future.

\subsection{Radiative corrections to the Casimir effect}\label{sec4.5}
{}From the point of view of \qft the Casimir effect is a one loop radiative
correction to an external classical background given by some \bc 
(or background fields if in
a broader understanding). The question of higher loop corrections naturally 
appears. So the first radiative correction to the Casimir effect is 
in fact a two
loop contribution. In general, the expected effects are very small. First, they
are suppressed by the corresponding coupling constant. Second, for \bc they
are suppressed by the ratio of the Compton wavelength $\la_{c}$ of the
corresponding quantum field and the characteristic macroscopic geometrical
size, the plate separation $a$, for instance. Nevertheless there is general
interest in the consideration of radiative corrections, mainly as a test of
the applicability of the general methods of perturbative \qftp
For 
example, the covariant formulation in the presence of explicitly 
non--covariant boundaries and 
the generalization of the perturbation expansion of a gauge theory are 
among the important issues to be addressed in this context. Boundary 
conditions necessitate a modification of the renormalization procedure as 
well. The interaction of the quantum fields with the boundary leads to
ultraviolet divergencies in vacuum graphs that cannot simply be discarded
by normal ordering.  Last but not least the question for the order of the
geometrical effects is raised, i.e. the question for the leading power
of $\lambda_{c}/a$ contributing to the effective action. Additional 
attention to radiative corrections arises in the bag model of QCD where 
they are not a priori negligible.

In the framework of general perturbative QED the radiative corrections to the
\gse can be obtained from the effective action. We start from the 
representation
\Ref{pert} of the generating functional of the Greens functions with
$Z^{(0)}$ being the corresponding functional of the free Greens functions,
Eq. \Ref{Z0}. Then the effective action is given 
by\footnote{As we restored in this section the dimensional constants
$\hbar$ and $c$ we note that $K,\>\overline{K}$ do have dimensions.
However, as the functional determinants here and below are defined up
to a constant it would be an unnecessary complication to introduce factors
making the arguments of the logarithms dimensionless.}
\be\label{effac}\Gamma=\frac{i\hbar}{2}\Tr \ln K +
\frac{i\hbar}{2}\Tr \ln \overline{K}  -  
i\hbar\sum\{\mbox{\rm 1PI graphs}\}\,.
\ee
Here, in addition to Eq. \Ref{effaction} we have the trace of the operation
$\overline{K}$ according to the representation Eq. \Ref{Z0} of the generating
functional and the sum is over all one-particle irreducible (1PI) graphs with
no external legs (vacuum graphs) resulting from Eq. \Ref{pert}.  Since the \bc
are static, the effective action is proportional to the total time $T$ and the
ground state energy is given by
\be 
E_0=-{1 \over T}\Gamma\;.
\label{E01}
\ee
If there are translation invariant directions (e.g., parallel to the
plates), the relevant physical quantity is the energy density and we have to
divide by the corresponding volume also.

In this representation, the first
contribution, $\frac{i}{2}\hbar\Tr \ln K$ does not depend on the boundary and
delivers just the Minkowski space contribution. The second contribution,
$\frac{i}{2}\hbar\Tr \ln\overline{K}$, delivers the boundary dependent 
part and the third
contribution contains the higher loops. We consider the first of
them. Graphically it is given by \raisebox{-7mm}{\epsffile{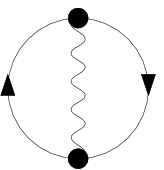}}. Here,
the solid line is the spinor propagator and the wavy line represents the
photon propagator. 
So up to the two loop order we find the 
following expression for the ground state energy  
\bea\label{E02}
E_0\equiv E_{0}^{(0)}+E_0^{(1)}&=&-{\frac{i\hbar}{ 2T}}{\rm Tr}\ln K 
-{\frac{i\hbar}{2T}}{\rm Tr}\ln \bar{K}\\
&&+{\frac{i\hbar}{2T}}
\int {\rm d}x\int {\rm d}y~\left[D_{\nu\mu}(x-y)-
\bar{D}_{\nu\mu}(x,y)\right]\Pi^{\mu\nu}(y-x)\nn\; ,
\eea
where 
\be\label{polten}
\Pi^{\mu\nu}(x-y)=-i \alpha{\rm Tr}\gamma^\mu S^c(x-y)\gamma^\nu S^c(y-x)\; 
\ee
is the known polarization tensor, $S^c(x-y)$ is the spinor propagator
and $\alpha=e^2/\hbar c$ is the fine structure constant.

Here we briefly pause from the calculation of the radiative correction and
instead recalculate the one loop contribution in order to illustrate the given
representation. Using the explicit expression \Ref{kplane}, the boundary
dependent part takes the form 
\[E_{0}^{(0)}\equiv 
-{{i\hbar}\over 2TV_{||}}{\rm Tr}\ln \bar{K}
=-{{i\hbar}\over 2TV_{||}}\int \d^{3}x_{\alpha}\int{\d^{3}k_{\alpha}\over
  (2\pi)^{3}}\tr\ln\left(-\delta_{st}{i\over 2\Gamma}h_{ij}\right)\,,
\]
where $\tr$ is related to two-dimensional indices,
$V_{||}$ is the volume of the translational invariant directions
parallel to the plates and $E_{0}^{(0)}$ is the energy density per unit area
of the plates.  We perform the Wick rotation $k_{0}\to ik_{0}$ (thereby
$\Gamma\to i\gamma\equiv i\sqrt{k_{0}^{2}+k_{1}^{2}+k_{2}^{2}}$) and calculate
the ``$\tr\ln$''
\[
\tr\ln\left(-\delta_{st}{1\over 2\gamma}h_{ij}\right)=
\ln\det\left(-\delta_{st}{1\over 2\gamma}h_{ij}\right)=
2\ln\left(1-e^{-2\gamma a}\right)-4\ln(2\gamma) \;.
\]
The term $-4\ln(2\gamma)$ yields a distance independent contribution
and will be dropped. In this way the known result for the Casimir energy is
reproduced
\[ 
E_{0}^{(0)}=\hbar c\int{\d^{3}k_{\alpha}\over
  (2\pi)^{3}}\ln \left(1-e^{-2\gamma a}\right)=
-{{\pi^{2}\hbar c}\over 720 a^{3}}\;.
\]

Now we turn back to the radiative correction.  Again, to do the separation 
of the
photon propagator into two parts, we see that the radiative correction in
\Ref{E02} separates into corresponding two parts also. That proportional to
the free space photon propagator $D_{\nu\mu}(x-y)$ is just the same as without
\bcp It deliveres a constant contribution not depending on the boundary and
we drop it as part of the Minkowski space contribution. Hence, we can
restrict ourselves to consider the contribution from 
$\overline{D}_{\nu\mu}(x-y)$. We
rewrite it using \Ref{bprop} and \Ref{polten} in the form
\be 
E_0^{(1)}=-{{i \hbar}\over 2T V_{||}}
\int\limits_{\S}\d^3 z\int\limits_{\S}\d^3 z'~
\mbox{$\bar{K}^{-1}$}^{st}(z,z')~\bar{\Pi}^{ts}(z',z)\; ,
\label{E03}
\ee
where the abbreviation
\bea \label{pist} 
& & \bar{\Pi}^{ts}(z',z)=    \\
& & ~\int d x \int d y~ {E^{t}}^{\dag}_{\nu'}(z')D^{\nu'}_{\phantom{\nu}\nu}
  (z'-y) \Pi^{\nu\mu}(y-x)D_\mu^{\phantom{\mu}\mu'}(x-z)E_{\mu'}^{s}
(z)_{|_{z,z'\in\S}}\; \nn
\eea
has been introduced.
Using the transversality of $\Pi_{\mu\nu}(x)$,
\[ 
\Pi_{\mu\nu}(x)=(g_{\mu\nu}\pa^{2}_{x}-\pa_{x_{\mu}}\pa_{x_{\nu}})
{\Pi}(x^{2})\; ,
\]
this expression simplifies to
\beao&&
{\bar{\Pi}}^{ts}(z',z)= \\
&&\int\d x\int\d y~{E^{t}}_{\mu}^{\dag}(z')
D(z'-y)g^{\mu\nu}\pa_{y}^{2}\Pi((y-x)^{2})D(x-z)E^s_{\nu}(z)_{|_{z,z'\in\S}}\;.
\eeao
Introducing the Fourier transform of $\Pi(x)$
\be 
{\Pi}(x^{2})=\int{\d^{4}q\over (2\pi)^{4}}~e^{iqx}~\tilde{\Pi}
(q^{2})
\label{Pq}
\ee
we rewrite it in the final form 
\be 
{\bar{\Pi}}^{ts}(z',z)={E_\mu^{t}}^{\dag}(z')\int{\d^{4}q\over (2\pi)^{4}}
e^{iq(z-z')}{E^{t}}_{\mu}^{\dag}(z')g^{\mu\nu}{\tilde{\Pi}
(q^2)\over -q^2}E_\nu^s(z')_{|_{z,z'\in\S}}\; .
\label{Pz}
\ee 
This representation of the radiative correction is still valid for a
general surface $\S$. Here, we observe another important conclusion. The scalar
part $\tilde{\Pi} (q^2)$ of the polarization tensor is known to possess a
logarithmic divergence which is independent of $q$ (for example, in
Pauli--Villars regularization it is of the form $-(2\alpha/3\pi)\ln(M/m)$
where $M$ is the regularizing mass, $m$ --- the mass of the electron). 
However, if only a $q$--independent
constant is added to $\tilde{\Pi} $, i.e. $\tilde{\Pi} (q^2)\to \tilde{\Pi}
(q^2)+C$, the quantity ${\bar{\Pi}}^{ts}$ changes according to
${\bar{\Pi}}^{ts}(z',z)\to {\bar{\Pi}}^{ts}(z',z)+C~\bar{K}^{ts}(z',z)$.  Then
it may be verified with the help of \Ref{inv} that the corresponding change in
the ground state energy is a simple constant which is independent of the
geometry. Consequently, the removal of the divergence of the polarization
tensor can be interpreted as a renormalization of the cosmological 
constant which is 
analogous to the free Minkowski space contributions discussed above.
As a result only the finite, renormalized part of the polarization tensor
needs to be taken into account when calculating the boundary dependent part of
the radiative correction $E_0^{(1)}$.

Now we rewrite the radiative correction \Ref{Pz} in the specific geometry of
two parallel planes. Using the polarization vectors $E^{s}_{\mu}$ ($s=1,2$),
\Ref{polbas1}, which commute with the polarization tensor, we obtain
\be
{\bar{\Pi}}^{st}(z-z')=\delta_{st}~\int{\d^{4}q\over (2\pi)^{4}}
e^{iq(z-z')}{\tilde{\Pi}
(q^2)\over q^2}_{|_{z,z'\in\S}}\;.
\label{er2}
\ee
We substitute this expression into the radiative correction \Ref{E03} and find
\be \label{150}
E_0^{(1)}=\hbar c \sum\limits_{i,j=1}^{2}\int{\d^{4}k\over
    (2\pi)^{4}}~\Gamma~h_{ij}^{-1}~e^{-ik_{3}(a_{j}-a_{i})}~
{\tilde{\Pi}
(k^{2})\over k^{2}}
\ee
with $\Gamma=\sqrt{k_{\alpha}k^{\alpha}+i\epsilon}$ and
$k^{2}=k_{\mu}k^{\mu}$. 
Using \Ref{invh} the last equation takes the form
\be \label{208}
E_0^{(1)}=2i\hbar c\int{\d^{4}k\over (2\pi)^{4}}~
{\Gamma\over\sin\Gamma a}~\left(e^{-i\Gamma a}-\cos k_{3}a\right)~
{\tilde{\Pi}
(k^{2})\over k^{2}}\;.
\ee
By means of the trivial relation 
$\exp(-i\Gamma a)=\exp(i\Gamma a)-2i\sin\Gamma a$ 
we separate again a distance independent contribution that will be omitted.
Now we perform the Wick rotation and obtain the final result for the 
radiative correction to the ground state energy to order $\alpha$ in the 
geometry of two parallel conducting planes
\be  
E_0^{(1)}=2\hbar c\int{\d^{4}k\over (2\pi)^{4}} {\gamma\over\sinh \gamma a}
\left(e^{-\gamma a}-\cos k_{3}a\right) {\tilde{\Pi}
(k^{2})\over k^{2}}\;.
\label{radcorr}
\ee

It is interesting to study the radiative correction \Ref{radcorr} 
in the limit $\lambda_{c}/a<<1$. For this purpose we transform the 
integration path of the $k_3$--integration. Due to $\tilde{\Pi}
(k_3^2+\gamma^2)$ there is a cut
with branch point $k_{3}=i\sqrt{4(mc/\hbar)^{2}+\gamma^{2}}$ 
in the upper half of the
complex $k_3$--plane. The discontinuity of the one loop vacuum polarization 
$\tilde{\Pi}
(k^2)$ across the cut, 
${\rm disc}\tilde{\Pi}
(k^{2})=\tilde{\Pi}
(k^{2}+i\epsilon)
-\tilde{\Pi}
(k^{2}-i\epsilon)$, is well known \cite{6}
\be  
{\rm disc}\tilde{\Pi}
(k^{2})=-\frac{2i}{3} \alpha
\sqrt{1-\frac{4m^{2}c^2}{\hbar^2k^{2}}}\left(1+
\frac{2m^{2}c^2}{\hbar^2k^{2}}\right)\;.
\label{disc}
\ee
We move the integration contour towards the imaginary axis so
that the cut is enclosed. The result can be written in the form
\be\label{255}  
E_0^{(1)}=\frac{i\hbar c}{\pi }\int{\d^{3}k\over (2\pi)^{3}}
{\gamma\over\sinh \gamma a}
\int\limits_{2mc/\hbar}^{\infty}{\d k_{3}\over k_3}
\ {{\rm disc}\tilde{\Pi}
\left(k_3^2
\right) \over \sqrt{k_3^2+\gamma^2}}
\left(e^{-\gamma a}-
e^{-\sqrt{k_3^2+\gamma^2}a}\right).
\ee
In the limit $mca>>\hbar$, the second term in the round brackets 
is exponentially  
suppressed and can be neglected. The desired series in inverse 
powers of $mca/\hbar$
is now simply achieved by expanding the square root in the denominator
\be 
E_0^{(1)}=\frac{i\hbar c}{2\pi}\int{\d^{3}k\over (2\pi)^{3}}~
{\gamma~e^{-\gamma a}\over\sinh \gamma a}~
\int\limits_{2mc/\hbar}^{\infty}\d k_{3}
{{\rm disc}\tilde{\Pi}
\left(k_3^2\right) \over k_{3}^{2}}
\left(1+O\left(\frac{\gamma^2}{k_3^2}\right)\right)\;.
\label{rkp2}
\ee
After some elementary integrations and with the help of 
\be 
\int\limits_{2mc/\hbar}^{\infty}\d k_{3}
{{\rm disc}\tilde{\Pi}
\left(k_3^2\right) \over k_{3}^{2}}= 
-\frac{3i\pi}{16}{\alpha\hbar\over2mc}
\label{discint}
\ee
we find the leading order contribution to the radiative correction
\be  
\mbox{$E_0^{(1)}$}=
\frac{\pi^2\alpha\hbar^2}{2560ma^4}+
O\left(\frac{\hbar^3}{m^2ca^5}\right),
\label{rkorrp1}
\ee
in agreement with
\cite{Bordag:1985zk} and \cite{add1}.

The calculation of the radiative correction to the Casimir effect for a
conducting sphere had been carried out in \cite{Bordag:1998sw}. It follows
essentially the lines shown here but is technically more involved. 
Therefore we
discuss only  the results here. First, the \uv divergencies which could be
simply dropped in the case of plane parallel plates (as they were distance
independent) deserve special consideration in the spherical case. In zeta
functional regularization there are divergent contributions
\be\label{rkdiv}E_{0}^{(1) {\rm div}}=
-{16 \over9\pi}\alpha m^{3}R^{2}\frac{c^4}{\hbar^2}-{4\over 15\pi}
\alpha m c^2
\ee
to $\E_{0}^{(1)}$ 
(here we use the same definition of the divergent part as in
subsection \ref{sec3.3} as those from contributions with non-negative powers of
the mass). In this case the mass is that of the spinor field. The physical
argument is the same, the vacuum energy must vanish when the fluctuating field
becomes too massive. We note that there is actually no divergence in
\Ref{rkdiv}. This is due to the specific regularization used. One can expect
that the same quantity calculated in another regularization shows up as a real
divergence like those in the corresponding contributions in Eq. \Ref{Edivd}. 
Finally
we note that in the (one loop) expression for the divergent part of the \gse
in zeta functional regularization, Eq. \Ref{Edivz}, for a massless field only
the contribution of $a_{2}$ is present. Thus we may consider the divergent
contributions given by Eq. \Ref{rkdiv} as radiative corrections to the
corresponding \hkks the $a_{\frac12}$ and $a_{\frac32}$. In order to perform
the renormalization we must assume a procedure as described in Sec.
\ref{sec3.4}.  We have to define a classical system with energy $E^{\rm
  class}=\sigma\S +K$ as special case of Eq. \Ref{Eclass} because there are
only two \hkks different from zero.

We present the finite parts,  
the renormalized \gse 
for parallel plates and that for the sphere together with the corresponding one
loop contribution respectively as:
\bea\label{rkorrc} E_{0}^{\rm plates}&=& -{\pi^{2}\over 720}{{\hbar c}\over
  a^{3}}+{\pi^{2}\over
  2560}{{\alpha\hbar^2}\over ma^{4}},     \\[4pt]
E_{0}^{\rm sphere}&=&{{0.092353\hbar c}\over 2R}- 
\left(7.5788\cdot 10^{-4}\ln\frac{mcR}{\hbar}
  +6.4833\cdot 10^{-3}\right) \ {{\alpha\hbar^2}\over mR^{2}} \,. \nn \eea
The numbers in $E_{0}^{\rm sphere}$ are a result of numerical computation in
\cite{Bordag:1998sw} whereby the one loop contribution is in fact that of
Boyer, the number being taken from \cite{Milton:1978sf}. The common features
of both equations is that the radiative correction is the largest 
possible, i.e.,
it is proportional to the first power of the fine structure constant and the
first power of the ratio of the two geometric quantities, namely the 
distance $a$ 
the radius $R$ respectively to the
Compton wavelength of the electron $\la_{c}=\hbar/(mc)$. 
The appearance of the logarithmic contribution for the sphere
can be understood as a consequence of the radius dependent \uv divergence,
Eq. \Ref{rkdiv}.  In both cases, i.e., for the plates as well as for the
sphere, the radiative correction can be interpreted as a renormalization of
the distance by the spinor loop such as
\bea
a&\to& a\left( 1+{3\over 32}{\alpha\lambda_{c}\over a}\right)   \nn \\
R&\to& R\left[ 1+\left(1.4040\cdot 10^{-1} +1.6413\cdot
    10^{-2}\ln\frac{mcR}{\hbar}\right){\alpha\lambda_{c}\over R}\right]
\label{radren}
\eea 
thus making them larger. This is the same in both cases regardless of the
different overall sign in the energy.

An extension of the above calculations is given in \cite{Bordag:1998yf} where
the radiative correction to the Casimir force between partly transmitting
mirrors was calculated. These mirrors are given by delta function potentials
on parallel planes which is equivalent to the matching conditions
\Ref{mcond}with the transmission coefficient \Ref{twodelta}. 
It was demonstrated that
the calculation can be performed in straightforward generalization of the
above formulas. As a result the relative weight of the radiative corrections
$\Delta E_{0}$ can be represented as
\be\label{relweight}{\Delta E_{0}\over E_{0}} =
\frac{\alpha\hbar}{mca} \ f(\beta a)\,,
\ee
where $\beta$ is now the strength of the delta potential 
(see Eq. \Ref{mcond}) and
$f(\beta a)$ is a smooth function with limiting values
\be\label{hart}f(\beta a)=-\frac{9}{32}\left(1+{2.92\over {\beta a}}
+\dots\right)
~~~\mbox{\rm for} ~~ \beta a\to\infty \,,
\ee
i.e., in the limit of impenetrable mirrors \Ref{rkorrc} is  again obtained, and
\be\label{weich}f(\beta a)=\frac{3}{16}\beta a
~~~\mbox{\rm for} ~~ \beta a\to0
\ee
for nearly completely penetrable mirrors. 

It is worth noting that in this case the ``largest possible''
radiative correction, proportional to the first powers of $\alpha$ and
$\lambda_{c}/a$, is present. In \cite{add1} the radiative
correction was calculated at finite temperature. 
For intermediate temperatures, obeying $k_BT\gg mc^2$, 
it turned out to be given by the 
first line in
Eq. \Ref{radren}, i.e., by the same substitution as in the zero
temperature case.


\subsection{Spaces with non-Euclidean topology}
\label{sec4.6}

As was mentioned in Introduction, the Casimir effect arises not
only in the presence of material boundaries but also in spaces
with non-trivial topology. The latter causes the boundary 
conditions, or, more exactly, periodicity conditions, imposed
on the wave functions which are of the same kind as those
due to boundaries. In Sec.~2.3 the simple examples in one- and
two-dimensional spaces were already given. Here we consider
in detail the four-dimensional space-time of a closed Friedmann model
which is physically important.
Also the vacuum polarization in the space-time of cosmic strings is
discussed and vacuum interaction between two parallel strings is
calculated. The other examples of the Casimir effect in spaces with
non-Euclidean topology is given by Kaluza-Klein theories. There the local
Casimir energy density and pressure can lead to spontaneous
compactification of extra spatial dimensions with a definite value of
compactification parameter.

\subsubsection{Cosmological models}
\label{sec4.6.1}

According to the concept of standard cosmology, the space of our
Universe is homogeneous and isotropic. Depending on the value of
the mean density of matter the cosmological models can be open
(curved hyperbolic space of infinite volume), quasi-Euclidean
(flat infinite space in 3-dimensions), and closed (curved
spherical space of finite volume). At present the precise value of the mean 
density of matter is not known. Thus it does not offer 
a unique means to determine among the three possible models
which one (closed) possesses non-Euclidean topology and,
consequently, Casimir effect. Note, that the non-trivial topology
may be introduced in any model, including the quasi-Euclidean case,
by imposing some point identification (see below). 
But the closed Friedmann model is 
topologically non-trivial and naturally incorporates the 
Casimir effect 
as compared to the other models mentioned above.

It is well known that in gravitational theory, it is the energy density 
and the other components of the
stress-energy tensor and not the total energy which are the physical 
quantities of importance. They are present in the right-hand side of the
Einstein equations and determine the space-time geometry. Because of this,
we are interested here (and in the next section also) not by the
global characteristics of a vacuum but by the local ones:
$\langle 0|T_i^k(x)|0\rangle^{ren}$.
Due to the symmetry properties of the space-time only diagonal
components survive. In addition the space components are equal
to each other and can be expressed through the energy density by
means of the conservation condition. Thus we are interested here only 
in the energy density
\beq
\varepsilon(\eta)=\langle 0|T_0^0(x)|0\rangle,
\label{sIV62.1}
\eeq
\noindent
where $\eta$ is a conformal time variable, and the stress-energy
tensor is defined in Eq.~(\ref{t11}).

Metric of the closed Friedmann model has the form
\bes
&&
ds^2=a^2(\eta)\left(d\eta^2-dl^2\right),
\label{sIV62.2}\\
&&
dl^2=d\chi^2+\sin^2\chi\left(d\theta^2+\sin^2\theta d\varphi^2\right),
\nonumber
\ees
\noindent
where $a(\eta)$ is a scale factor with dimensions of length,
$0\leq\chi\leq\pi$, $\theta$ and $\varphi$ are the usual spherical
angles.

It is seen that the space section of a closed model is the surface 
of a 3-sphere with topology $S^3$. The dependence of the radius of 
curvature $a(\eta)$ on time complicates the issue. With this dependence
the case under consideration can be likened to a dynamical Casimir
effect with a moving boundaries (see Secs.~2.4 and 4.4). Because of
this, except of the usual Casimir vacuum polarization depending on
a sphere radius the other vacuum quantum effects may be expected also.

Taking into account that $N=3$, $\xi=1/6$ and the scalar curvature
$R=6(a^{\prime\prime}+a)/a^3$, where prime denotes differentiation
with respect to conformal time $\eta$, we rewrite Eq.~(\ref{t10}) in the
form
\beq
\varphi^{\prime\prime}(x)+2\frac{a^{\prime}}{a}
\varphi^{\prime}(x)-\Delta^{\!(3)}\varphi(x)+
\left(\frac{m^2c^2a^2}{\hbar^2}+\frac{a^{\prime\prime}}{a}
+1\right)\varphi(x)=0,
\label{sIV62.3}
\eeq
\noindent
where $\Delta^{\!(3)}$ is the angular part of the Laplacian operator
on a 3-sphere, $x=(\eta,\chi,\theta,\varphi)$.

The orthonormal set of solutions to Eq.~(\ref{sIV62.3}) can be 
represented as
\bes
&&
\varphi_{\lambda lM}^{(+)}(x)=\frac{1}{\sqrt{2}\,a(\eta)}
g_{\lambda}(\eta)\Phi_{\lambda lM}^{*}(\chi,\theta,\varphi),
\nonumber\\
&&
\varphi_{\lambda lM}^{(-)}(x)=
\left[\varphi_{\lambda lM}^{(+)}(x)\right]^{*},
\label{sIV62.4}
\ees
\noindent
where the eigenfunctions of the Laplacian operator are defined
according to
\beq
\Phi_{\lambda lM}(\chi,\theta,\varphi)=
\frac{1}{\sqrt{\sin\chi}}
\sqrt{\frac{\lambda(\lambda+l)!}{(\lambda-l+1)!}}
\,P_{\lambda-\frac{1}{2}}^{-l-\frac{1}{2}}(\cos\chi)\,
Y_{lM}(\theta,\varphi),
\label{sIV62.5}
\eeq
\noindent
$\lambda=1,2,\ldots$, $l=0,1,\ldots ,\lambda-1$, $Y_{lM}$ are the
spherical harmonics and $P_{\mu}^{\nu}(z)$ are the adjoint Legendre 
functions on the cut. The discrete quantity $\lambda$ has the sense
of a dimensionless momentum, the physical momentum being
$\hbar\lambda/a$. The time-dependent function $g_{\lambda}$
satisfies the oscillatory equation
\beq 
g_{\lambda}^{\prime\prime}(\eta)+
\omega_{\lambda}^2(\eta)g_{\lambda}(\eta)=0,
\qquad
\omega_{\lambda}^2(\eta)=\lambda^2+\frac{m^2c^2a^2(\eta)}{\hbar^2}
\label{sIV62.6}
\eeq
\noindent
with the time-dependent frequency and initial conditions fixing the
frequency sign at the initial time
\beq
g_{\lambda}(\eta_0)=\frac{1}{\sqrt{\omega_{\lambda}(\eta_0)}},
\qquad
g_{\lambda}^{\prime}(\eta_0)=i\sqrt{\omega_{\lambda}(\eta_0)}.
\label{sIV62.7}
\eeq
\noindent
Eigenfunctions (\ref{sIV62.4})--(\ref{sIV62.7}) define the vacuum
state at a moment $\eta_0$. In the homogeneous isotropic case one may put
$\eta_0=0$.

Substituting the field operator expanded in terms of the functions
(\ref{sIV62.4}), (\ref{sIV62.5}) into the $00$-component of
(\ref{t11}) and calculating the mean value in the initial vacuum
state according to (\ref{sIV62.1}) one obtains the nonrenormalized
vacuum energy density
\bes
&&
\varepsilon^{(0)}(\eta)=\frac{\hbar c}{4\pi^2a^4(\eta)}
\sum\limits_{\lambda=1}^{\infty}
\lambda^2\omega_{\lambda}(\eta)
\left[2s_{\lambda}(\eta)+1\right],
\label{sIV62.8}\\
&&
s_{\lambda}(\eta)=\frac{1}{4\omega_{\lambda}}
\left(|g_{\lambda}^{\prime}|^2+\omega_{\lambda}^2|g_{\lambda}|^2
-2\omega_{\lambda}\right).
\nonumber
\ees
\noindent
The corresponding vacuum energy density in tangential Minkowski
space at a given point is
\beq
\varepsilon_{M}^{(0)}(\eta)=\frac{\hbar c}{4\pi^2a^4(\eta)}
\int\limits_{0}^{\infty}\lambda^2\,d\lambda\,\omega_{\lambda}(\eta).
\label{sIV62.9}
\eeq
\noindent
Subtracting (\ref{sIV62.9}) from (\ref{sIV62.8}) with the help of
Abel-Plana formula (\ref{2.25}) we come to the result
\beq
\varepsilon_{ren}^{(0)}(\eta)={E}^{(0)}(\eta)+
\frac{\hbar c}{2\pi^2a^4(\eta)}
\sum\limits_{\lambda=1}^{\infty}
\lambda^2\omega_{\lambda}(\eta)s_{\lambda}(\eta),
\label{sIV62.10}
\eeq
\noindent
where the Casimir energy density of a scalar field in a closed
Friedmann model is \cite{15}
\beq
{E}^{(0)}(\eta)=\frac{\hbar c}{2\pi^2a^4(\eta)}
\int\limits_{mca(\eta)/\hbar}^{\infty}
\frac{\lambda^2\,d\lambda}{e^{2\pi\lambda}-1}
\left[\lambda^2-
\frac{m^2c^2a^2(\eta)}{\hbar^2}\right]^{1/2}.
\label{sIV62.11}
\eeq
\noindent
Note that under subtraction of two infinite quantities
(\ref{sIV62.8}) and (\ref{sIV62.9})
the damping function was introduced implicitly (see Sec.~2.2
for details). Also, Eq.~(\ref{2.36}) was taken into account to
obtain Eq.~(\ref{sIV62.11}). The index $ren$ near the energy density
from (\ref{sIV62.10}) is put by convention. However, as it will 
become clear soon, additional renormalizations are needed
in (\ref{sIV62.10}).
In the frame of dimensional regularization the quantity
$\langle 0_M|T_{ik}|0_M\rangle$ is demonstrated to be proportional
to a metrical tensor of space-time $g_{ik}$. What this means is
the subtraction of the quantity (\ref{sIV62.9}) performed above
is equivalent to the renormalization of a cosmological constant
in the effective action of the gravitational field \cite{42,DeWitt}.
The same interpretation is valid also when one subtracts the
contribution of free Minkowski space in the problem with material
boundaries (if to put the renormalized value of cosmological
constant be equal to zero).

Up to this point we have considered the closed Friedmann model
which is non-stationary. If we consider Eq.~(\ref{sIV62.2})
with $a^{\prime}(\eta)=0$ we obtain the metric of the static
Einstein model $R^1\times S^3$. In that case 
$s_{\lambda}(\eta)=s_{\lambda}(\eta_0)=0$ as is evident from
Eq.~(\ref{sIV62.8}), and the total vacuum energy density is given by
the Casimir term ${E}^{(0)}$ from the Eq.~(\ref{sIV62.11})
with $a=$const.

It follows from (\ref{sIV62.6}), (\ref{sIV62.8}) that
$s_{\lambda}(\eta)=s_{\lambda}(\eta_0)=0$ for massless field
($m=0$) even if $a^{\prime}(\eta)\neq 0$, i.e. metric is
non-stationary. In that case, however, the total vacuum energy
density does not reduce to the static-like Casimir term
${E}^{(0)}$ only. The point to note is that the second term in
the right-hand side of Eq.~(\ref{sIV62.10}) is the subject of two
additional renormalizations in accordance with the general
structure of infinities of Quantum Field Theory in curved
space-time \cite{42,43,DeWitt}. This is the renormalization of
the gravitational constant and of constants near the terms that are 
quadratic in the curvature in the effective gravitational action.
Both these renormalizations are accidentally finite for the
conformal scalar field in isotropic homogeneous space.
As a result of these renormalizations the total vacuum energy
of massless scalar field in closed Friedmann model takes the form
\beq
\varepsilon_{ren}^{(0)}(\eta)={E}_0^{(0)}(\eta)+
\frac{\hbar c}{960\pi^2a^4(\eta)}
\left(2b^{\prime\prime}b-{b^{\prime}}^2-2b^4\right),
\label{sIV62.12}
\eeq
\noindent
where $b=b(\eta)\equiv a^{\prime}(\eta)/a(\eta)$.
The Casimir energy density of a massless field which appears
in this expression is obtained from Eq.~(\ref{sIV62.11}) for
both constant and variable $a$ \cite{14,15} as:
\beq
{ E}_0^{(0)}(\eta)=\frac{\hbar c}{2\pi^2a^4(\eta)}
\int\limits_{0}^{\infty}
\frac{d\lambda\,\lambda^3}{e^{2\pi\lambda}-1}=
\frac{\hbar c}{480\pi^2a^4(\eta)}.
\label{sIV62.13}
\eeq

The second term in the right-hand side of Eq.~(\ref{sIV62.12}) may
be interpreted in a two ways. It can be viewed as a part of the dymanical
Casimir effect on one hand, as it disappears when
$a^{\prime}(\eta)\to 0$. Closer examination of this term makes it more 
reasonable to take the second interpretation, according to which it is the
usual vacuum polarization by the external gravitational field
having no connection with the periodicity conditions.
Actually this term which leads to the conformal anomaly, is
present in the open Friedmann model also,  where  the periodic
boundary conditions (and therefore the Casimir effect) are
absent.

In the case of massive fields Eq.~(\ref{sIV62.11}) can be integrated 
analytically in the limit $mca(\eta)/\hbar\gg 1$
both for the constant and variable $a$ with the result \cite{15}
\beq
{ E}^{(0)}(\eta)\approx
\frac{\left(mca(\eta)/\hbar\right)^{5/2}\,\hbar c}{8\pi^3a^4(\eta)}
e^{-2\pi mca(\eta)/\hbar}.
\label{sIV62.14}
\eeq
\noindent
It is seen that the Casimir energy density of a massive field on
$S^3$ is exponentially small which is not typical for the
configurations with nonzero curvature.

For the non-stationary metric and massive field the quantity
$s_{\lambda}(\eta)$ from Eqs.~(\ref{sIV62.8}), (\ref{sIV62.10}) 
is not equal to zero except at initial moment. As a consequence,
the total vacuum energy density consists of the Casimir contribution
(\ref{sIV62.11}), the vacuum polarization contribution given by
the second term in the right-hand side of Eq.~(\ref{sIV62.12})
(it is the same for both massless and massive fields), and the contribution of particles created from vacuum by the non-stationary
gravitational field. The latter effect owes its origin to the
variable gravitational field, not to the periodicity conditions due
to non-Euclidean topology. It exists in the open and quasi-Euclidean
models which are topologically trivial. In this regard cosmological 
particle production is different from the production of photons
in the dymanical Casimir effect with boundaries.
There the material boundaries themselves play the role of a
non-stationary external field, so that without boundaries the effect
is absent. Because of this, we do not discuss the effect of particle
production here in further detail (see, e.g., \cite{42,43}).

All the above considerations on the Casimir energy density of scalar
field in the closed Friedmann model can be extended for the case of
quantized spinor field. The details of quantization procedure can
be found in \cite{21,22}. Here we dwell on the results only.
After the calculation of the vacuum energy density and subtraction
of the tangential Minkowski space contribution the result similar
to Eq.~(\ref{sIV62.10}) is
\beq
\varepsilon_{ren}^{(1/2)}(\eta)={ E}^{(1/2)}(\eta)+
\frac{2\hbar c}{\pi^2a^4(\eta)}
\sum\limits_{\lambda=3/2}^{\infty}
\left(\lambda^2-\frac{1}{2}\right)\omega_{\lambda}(\eta)s_{\lambda}(\eta),
\label{sIV62.15}
\eeq
\noindent
where $s_{\lambda}(\eta)$ is expressed in terms of the solution
to the oscillatory equation with a complex frequency obtained from
the Dirac equation after the separation of variables.
The Casimir contribution in the right-hand side is \cite{42}
\beq
{ E}^{(1/2)}(\eta)=\frac{2\hbar c}{\pi^2a^4(\eta)}
\int\limits_{mca(\eta)/\hbar}^{\infty}
\frac{d\lambda}{e^{2\pi\lambda}+1}
\left(\lambda^2+\frac{1}{4}\right)
\left[\lambda^2-\frac{m^2c^2a^2(\eta)}{\hbar^2}\right]^{1/2}.
\label{sIV62.16}
\eeq
\noindent
To obtain (\ref{sIV62.16}) the Abel-Plana formula (\ref{2.26}) for
the summation over the half-integers was used. In the static
Einstein model $s_{\lambda}(\eta)=0$ once more and the total
vacuum energy density is given by the Casimir term
${ E}^{(1/2)}(\eta)$ from Eq.~(\ref{sIV62.16}) with
$a=$const.

In the non-stationary case $a^{\prime}(\eta)\neq 0$ two additional
renormalizations mentioned above are needed to obtain the total
physical energy density of a vacuum (note that for the spinor field
the sum in Eq.~(\ref{sIV62.15}) is divergent). The result in
a massless case is given by
\beq
\varepsilon_{ren}^{(1/2)}(\eta)={ E}_0^{(1/2)}(\eta)+
\frac{\hbar c}{480\pi^2a^4(\eta)}
\left(6b^{\prime\prime}b-3{b^{\prime}}^2-
\frac{7}{2}b^4+5b^2\right),
\label{sIV62.17}
\eeq
\noindent
where the Casimir energy density of a massless spinor field
is 
\beq
{ E}_0^{(1/2)}(\eta)=\frac{2\hbar c}{\pi^2a^4(\eta)}
\int\limits_{0}^{\infty}
\left(\lambda^2+\frac{1}{4}\right)
\frac{\lambda\,d\lambda}{e^{2\pi\lambda}+1}
=\frac{17\hbar c}{960\pi^2a^4(\eta)}.
\label{sIV62.18}
\eeq
\noindent
The second contribution in the right-hand side of (\ref{sIV62.17}),
as well as in (\ref{sIV62.12}) for a scalar case, is interpreted as
a vacuum polarization by the non-stationary gravitational field.
In the case of a massive field, the effect of fermion pair creation in vacuum
by the gravitational field is possible \cite{42,43}. It has no direct
connection, however, to the periodicity conditions and the Casimir
effect. As to the Casimir energy density of massive spinor field,
it decreases exponentially for $mca(\eta)/\hbar\gg 1$, and
${ E}^{(1/2)}(\eta)\approx 4{E}^{(0)}(\eta)$,
where the scalar result is given by Eq.~(\ref{sIV62.14}).

At last for the quantized electromagnetic field the Casimir
energy density in the closed Friedmann model is
\beq
{ E}_0^{(1)}(\eta)=\frac{11\hbar c}{240 \pi^2a^4(\eta)}.
\label{sIV62.19}
\eeq

Notice that although the Casimir energy densities (\ref{sIV62.13}),
(\ref{sIV62.18}), (\ref{sIV62.19}) are very small at the present stage
of the evolution of the Universe their existence provides
an important means to determine the global topological
structure of space-time from the results of local measurements.

Theoretically different topological structures of space-time
on cosmological scale are possible. It is well known that one and the same
metric, which is a solution to Einstein equations, may correspond to
different spatial topologies. For example in \cite{sIV62-2} the
quasi-Euclidean space-time was considered with a 3-torus topology
of space. The latter means that the points $x+kL$, $y+mL$, $z+nL$,
where $k,\,m,\,n$ are integers and both negative and positive values, are
identified with a characteristic identification scale $L$.
As was shown in \cite{sIV62-2} the Casimir energy density in such 
a model can drive the inflation process. A number of
multi-connected cosmological models have been investigated in the
literature (see the review \cite{sIV62-3}). If the identification
scale is smaller than the horizon, some observational effects of
non-trivial topology are possible. This was first discussed
in \cite{sIV62-4,sIV62-5}. In particular, due to non-trivial
topologies, the multiple images of a given cosmic source may
exist (in some specific cosmological models the detailed analyses
of this possibility was performed in \cite{sIV62-6}). The
observable effects of the various topological properties of space-time 
in the various cosmological models make the calculation of the Casimir 
energy densities important (see, e.g., the monograph
\cite{sIV62-7} and review \cite{sIV62-8} where it was calculated
in spherical and cylindrical Universes of different dimensionality).


\subsubsection{Vacuum interaction between cosmic strings}\label{sec4.6.2}
Cosmic strings are classical objects which may have been produced in the early
universe as byproducts of cosmological phase transitions. They are interesting
due to their cosmological implications such as their influence on primordial 
density fluctuations
\cite{VilenkinShellard}. 

Straight cosmic strings are solutions to the Einstein equations whose metric
is flat everywhere except for the string axis where it is conical, i.e., it 
has an
angle deficit. As a consequence, in classical theory
there are gravitational forces between test
bodies and cosmic strings but there is no interaction between the strings
themselves.

The interaction of a matter field with a string can be described by \bc
requiring the field to be periodic after rounding the string with an angle of
$2\pi\alpha$:
\be\label{bccs}\Phi(t,z,r,\varphi+2\pi\alpha)=\Phi(t,z,r,\varphi)
\ee
where $0<\alpha\le 1$ and $(r,\varphi)$ 
are polar coordinates in the $(x,y)$-plane. 

In this way we have a situation analogous to the Casimir effect for
plane conductors -- no interaction between the classical objects and \bc on
the matter fields. Indeed as one might expect, the \gse appearing from 
the quantization of these
fields may result in vacuum polarization and in a Casimir force between two
(or more) cosmic strings. 
Vacuum polarization in the background of one string was first considered in
\cite{Helliwell1986n}, later on  in more general space-times containing
conical flux tube singularities, for example, in
\cite{Dowker1987ar,Dowker1987as}.  For purely dimensional reasons, the local
energy density  $\langle 0\mid T_{00}(x)\mid 0 \rangle$ has a singularity
proportional to $r^{-4}$ near the string where $r$ is the distance from the
string. Therefore the global vacuum energy is not integrable near $r=0$.
One way
to avoid this problem is to consider a string of finite thickness. 
Here a complete analysis is
still missing. A  calculation  for small angle deficit showed a vanishing
vacuum energy in the first nontrivial order \cite{Khusnutdinov:1999tf}. 

The situation is different if one considers two (or more) cosmic strings. Here
one would expect a force acting between them due to the vacuum fluctuations in
analogy with the force between conducting planes whereby there should be no
distance dependent singularities. This was done in
\cite{Bordag:1990if,Galtsov} for a scalar field, in \cite{Aliev:1997qm} for a
spinor and in \cite{Aliev:1997va} for the \elm field. Let us note that the
same holds for magnetic strings, \cite{Bordag:1991pv}, and can easily be
generalized to cosmic strings carrying magnetic flux.

From the technical point of view, however, the problem with two or more
strings is complicated because in that background the variables do not
sepa\-rate and at present it is not known how to calculate the vacuum energy
in such situation. So one is left with perturbative methods. Fortunately, for
cosmic strings this is reasonable because the corresponding couplings are
small, see below.

The background containing parallel cosmic strings is given by the interval
\cite{Letelier}
\be\label{interval}ds^{2}=c^2dt^{2}-dz^{2}-
e^{-2\Lambda({\bf x_{\perp}})}(dx^{2}+dy^{2}) 
\ee
with $\Lambda({\bf x_{\perp}})=\sum_{k}4\lambda_{k}\ln (\mid {\bf
  x_{\perp}}-{\mbox{\boldmath$a$}}_{k}\mid/\rho_0)$ , 
where ${\bf x_{\perp}}=(x,y)$ is a vector in
the $(x,y)$-plane and {\boldmath{${a}$}}${}_{k}$ are 
the positions of the strings in that
plane, $\rho_0$ is the unit of length.  
The coupling to the background, $\lambda_{k}=G\mu_{k}/c^{2}$, where
  $\mu_{i}$ are the linear mass densities and $G$ is the gravitational
  constant, is connected with the angle deficit in Eq. \Ref{bccs} by means of
  $\alpha_k=1-4\lambda_k$.  These couplings are small
  quantities of  typical order 
$\lambda_k\sim\left({M_{GUT}\over m_{pl}}\right)\sim10^{-6}$.

In order to perform the calculations it is useful to start from the 
local energy density, from Eq. \Ref{sIV62.1} for example, and then consider
energy per unit length
\be\label{}\E_{0}=\int d{\bf x_{\perp}} \>\sqrt{-g}\>
\langle 0\mid T_{00}(x)\mid 0 \rangle
,\ee
where
\be\label{}i\langle 0\mid T_{00}(x)\mid 0 \rangle=-\frac{\hbar c}{2}
\pa_{x_{\rho}}\pa_{y_{\rho}}
D(x,y)_{|_{y=x}}
\ee
is the vacuum expectation value of the $(00)$-component of the
energy-momentum tensor, given here for the massless scalar real field.

The propagator $D(x,y)$ obeys the equation
\be\label{}\left[e^{-2\Lambda(x)}
\left(c^{-2}\pa_{t}^{2}-\pa_{z}^{2}\right)-\pa_{x}^{2}
-\pa_{y}^{2}\right] D(x,y)=-\delta(x,y).
\ee
The perturbative setup is obtained from rewriting this equation in the form
\be\label{}\Box  D(x,y)=-\delta(x,y)-V(x)D(x,y),
\ee
where
\be\label{}V(x)=-\left(1-e^{-2\Lambda(x)}\right)
\left(c^{-2}\pa_{t}^{2}-\pa_{z}^{2}\right)
\ee
is the perturbation. Iteration yields
\bea\label{D0VD0} D(x,y)&=&D^{(0)}(x,y)+
\int dz D^{(0)}(x,z) V(z)D^{(0)}(z,y)+\\&&
\int dz \int dz' D^{(0)}(x,z) V(z)D^{(0)}(z,z') V(z')D^{(0)}(z',y)+\dots \,.
\nn\eea

To first order in $V(z)$ the propagator $D^{(1)}(x,y)$ can be calculated quite
easily, see for example \cite{Aliev:1997va}, with the result (for $y=x$)
\be\label{prop1orng}D^{(1)}(x,x)=-\frac{i}{6\pi^{2}}\sum_{k}
\frac{\lambda_{k}}{r_{k}^{2}},
\ee
where $r_{k}=|{\bf x_{\perp}}-{\mbox{\boldmath{$a$}}}{}_{k}|$ 
and the sum goes over the
strings. In the same way the vacuum expectation value of the energy momentum
tensor  can be calculated, and one obtains
\[
\langle 0\mid T_{00}(x)\mid 0 \rangle =-\frac{2\hbar
  c}{15\pi^{2}}\sum_{k}\frac{\lambda_{k}}{r_{k}^{4}}
\]
for the \elm field. The singularity $\sim r^{-4}$ near each string is clearly
observed. 

In order to obtain the corresponding contribution to the interaction of two
strings one has to take the second order  term in \Ref{D0VD0}. There one has
to expand $V(z)$ and $V(z')$ for small $\lambda_k$ and to pick up the term
proportional to $\lambda_{1}\lambda_{2}$. In this way the interaction energy
and force
per unit length of two parallel strings  can be obtained. 
They have the form
\be\label{Eint2st}
\E_{0}=-\sigma \hbar c {\lambda_{1}\lambda_{2}\over a^{2}},
\qquad
F_{0}=-2\sigma \hbar c {\lambda_{1}\lambda_{2}\over a^{3}},
\ee
where $a$ is the distance between the strings, and $\sigma$ is a number.
Eqs.~\Ref{Eint2st} follow for dimensional reasons and the number $\sigma$ 
has to
be calculated.  In \cite{Galtsov} this had been done for the 
massless scalar field with
the result $\sigma =4/(15\pi)$, in \cite{Aliev:1997qm} for the 
massless spinor field
with the same result and in \cite{Aliev:1997va} for the \elm field with the
result of $\sigma$ being twice the value of the scalar field.  The force acting
between the strings is attractive.

We presented here the typical setup of the vacuum polarization and of the
Casimir force between cosmic strings and would like to mention that more work
had been done in the past years for the vacuum polarization in space-times
with conical singularities.  The corresponding \hkks have been calculated in
\cite{Fursaev1994y,Cognola:1994qg,Bordag:1996fw} and further papers.  Similar
methods have been applied to other topological singularities, to monopoles 
\cite{Bezerra} or
black holes for example.

\subsubsection{Kaluza-Klein compactification of extra dimensions}
\label{sec4.6.3}

The previous discussion has shown that in the spaces with non-Euclidean
topology there is a non-zero vacuum energy density of a Casimir nature.
This fact is widely used in multi-dimensional Kaluza-Klein
theories which provide the basis for the modern extended studies
of unified descriptions of all the fundamental interactions
including gravitational. The main idea of Kaluza-Klein
approach is that the true dimensionality of space-time is $d=4+N$,
where the additional $N$ dimensions are compactified and form
a compact space with geometrical size of order of Planck length
$l_{Pl}=\sqrt{G\hbar/c^3}\sim 10^{-33}\,$cm ($G$ is the
gravitational constant). Originally this idea arose in the twenties
through investigation of the possibility of unifying gravitation and
electromagnetism. For this purpose a five-dimensional space was
used with a topology $M^4\times S^1$. A review of the modern
applications of Kaluza-Klein theories in the framework of
supersymmetric field models can be found in \cite{sIV63-1}.

According to present concepts, the most promising theory of
fundamental interactions is the Superstring Theory \cite{sIV63-2}.
The dimensionality of space-time in string theory is fixed as
$(9+1)$, which results from the Lorenz invariance and unitarity.
There is no way to establish a link between this theory and the real
world except by suggesting the real space-time is of the form
$M^4\times K^6$, where $K^6$ is a six-dimensional compact space in
the spirit of Kaluza-Klein theories. The superstring theories attract 
so much attention because they do not contain ultra-violet divergencies
and some of them incorporate both non-abelian gauge interactions and
gravitational interaction.

The important problem of string theories is to find the mechanism of
dynamical compactification that is responsible for the stable existence 
of six-dimensional compact manifold $K^6$. 
Probably the most popular mechanism of this kind is based on the
Casimir effect; in so doing the Casimir energy density of different
quintized fields defined on $M^4\times K^6$ is substituted
into the right-hand side of the multi-dimensional Einstein equations
to look for the self-consistent solutions. The values of the
geometrical parameters of a compact space $K^6$ are determined in
the process.

In four-dimensional space-time in the right-hand side of the classical 
Einstein equations,
a vacuum stress-energy tensor of the quantized fields is widely used 
\cite{42,43}. Such equations are, in fact,
approximate in the framework of a one-loop approximation in comparison 
to a fully
quantized theory of gravitation. This imposes the evident restrictions
on their application range: the characteristic geometrical values of
obtained solutions must be larger than the Planck length. In cosmology
the self-consistent non-singular solutions to Einstein equations
with the energy densities (\ref{sIV62.12}) and (\ref{sIV62.17}) as
the sources in the closed, open and quasi-Euclidean Friedmann
models were first performed in \cite{sIV63-3,sIV63-3a,sIV63-4}.
Among the obtained solutions there are the famous ones describing
inflation. For the closed model, the Casimir effect makes nonzero 
contribution to
vacuum energy density. However, a crucial role is played by the
vacuum polarization due to the non-stationary nature of the gravitational 
field.

In Kaluza-Klein compactification the manifold $K$ is presumed to be
stationary. Because of this, the Casimir vacuum energy density is
important for the determination of the parameters of $K$ and its
stability. Representing the Casimir stress-energy tensor by
$T_{AB}^{ren}$, the multi-dimensional Einstein equations take the form
\beq
R_{AB}-\frac{1}{2}R_dg_{AB}+\Lambda_d g_{AB}=
-\frac{8\pi G_d}{c^4}\,T_{AB}^{ren},
\label{sIV63.1}
\eeq
\noindent
where $A,B=0,1,\ldots,d-1$ (we leave $d$ arbitrary, not necessarily
equal to ten as in string theory), $G_d$ and $\Lambda_d$ are the
gravitational and cosmological constants in $d$ dimensions.

In the literature a great number of different compact manifolds were 
considered and the corresponding Casimir stress-energy tensors
$T_{AB}^{ren}$ were calculated. In many cases the self-consistent
solutions to the equations (\ref{sIV63.1}) were found. In \cite{sIV63-5},
as an example, the Casimir energies for scalar and spinor fields were
computed in even-dimensional Kaluza-Klein spaces of the form
$M^4\times S^{N_1}\times S^{N_2}\times\ldots$.
For the massless scalar and spinor fields defined on
$M^4\times S^2\times S^2$ (four-dimensional compact internal space)
the stable self-consistent solution was found. The Casimir energies
on the background $M^4\times T^N$, where $T^N$ is $N$-dimensional
torus are considered and reviewed in \cite{sIV63-6}.
This includes the case $N=6$ which is of interest for the superstring
theory. In \cite{sIV63-7,sIV63-8} the Casimir energies were computed
on the background of $M^4\times S^N$. The self-consistent solutions
to Eqs.~(\ref{sIV63.1}) were found and the problem of their stability was 
discussed. The case of $M^4\times B$, where $B$ is the Klein bottle,
was considered in \cite{sIV63-9}. The self-consistent solutions to
the Einstein equations for a static space-time with spatial section
$S^3\times S^3$ at finite temperature were examined in \cite{sIV63-10}. 
It was shown that the Casimir stress-energy tensor of massless Dirac field
determines the self-consistent value of sphere radii for all $T$. Many
more complicated Kaluza-Klein geometries were studied also. To take
one example, in \cite{sIV63-11} the Casimir effect for a free
massless scalar field defined on a space-time $R^2\times H^{d-1}/\Gamma$
was investigated, where $\Gamma$ is a torsion free subgroup of
isometries of $(d-1)$-dimensional Lobachevsky space $H^{d-1}$
(see also \cite{sIV62-7}).

All the above research is aimed at finding the genuine structure of
the internal space compatible with the experimental knowledge of
High Energy Physics. Unfortunately, this objective has not been
met up to the present time. Because of this, there is no point in going 
through here all the details of the numerous multi-dimensional models where
the Casimir energies are calculated and serve as a mechanism
of spontaneous compactification of extra dimensions. Instead, we give
below one example only \cite{sIV63-8} based on space-time of
the form $M^4\times S^N$, which illustrates the main ideas in this field
of research (note that six-sphere does not satisfy the consistency
requirements of string theory \cite{sIV63-12} but presents a clear
typical case).

We are looking for the solution of Eqs.~(\ref{sIV63.1}) which are
Poincar\'{e} invariant in four dimensions. What this means is
the metrical tensor $g_{AB}$ and Ricci tensor $R_{AB}$ have the
block structure
\beq
g_{AB}=\left(
\begin{array}{cc}
\eta_{nm} & 0 \\
0 & h_{ab}(\mbox{\boldmath$u$})
\end{array}\right),
\qquad
R_{AB}=\left(
\begin{array}{cc}
0  & 0 \\
0 & R_{ab}(\mbox{\boldmath$u$})
\end{array}\right),
\label{sIV63.2}
\eeq
\noindent
where $\eta_{mn}$ ($m,n=0,\,1,\,2,\,3$) is the metric tensor in
Minkowski space-time $M^4$, and $h_{ab}(\mbox{\boldmath$u$})$
is the metric tensor on a manifold $S^N$ with coordinates
{\boldmath$u$} ($a,b=4,\,5,\ldots,\,d-1$). It is clear that the scalar
curvature $R_d$ coincides with the one calculated from the metrical
tensor $h_{ab}(\mbox{\boldmath$u$})$.

The Casimir stress-energy tensor also has the block structure
\beq
T_{mn}^{ren}=T_1\eta_{mn},
\qquad
T_{ab}^{ren}(\mbox{\boldmath$u$})=
T_2h_{ab}(\mbox{\boldmath$u$}).
\label{sIV63.3}
\eeq
\noindent
Note that $T_{1,2}$ do not depend on {\boldmath$u$} due to space
homogeneity.

The Ricci tensor on an $N$-dimensional sphere is
\beq 
R_{ab}(\mbox{\boldmath$u$})=
-\frac{N-1}{a^2}h_{ab}(\mbox{\boldmath$u$}),
\label{sIV63.4}
\eeq
\noindent
where $a$ is a sphere radius.

To find $T_1$ and $T_2$ we remind the reader that the Casimir 
stress-energy tensor
$T_{ab}^{ren}$ can be expressed in terms of the effective potential $V$ by
variation with respect to the metric
\beq
T_{ab}^{ren}=-\frac{2}{\sqrt{|\mbox{det}h_{ab}|}}
\frac{\delta V(h)}{\delta h^{ab}}.
\label{sIV63.5}
\eeq
\noindent
The variation of the metric tensor $h_{ab}$ can be considered as a change
in the sphere radius $a$. Multiplying both sides of (\ref{sIV63.5})
by $h^{ab}$, summing over $a$ and $b$, and integrating over the
volume of $S^N$, one obtains with the use of (\ref{sIV63.3})
\cite{sIV63-8}
\beq
\frac{1}{2}T_2N\Omega_N=
-\int d^Nu\,h^{ab}\frac{\delta V(h)}{\delta h^{ab}}
=-a^2\frac{dV}{da^2},
\label{sIV63.6}
\eeq
\noindent
where the volume of the sphere is
\beq
\Omega_N=\int d^Nu\,\sqrt{|\mbox{det}h_{ab}|}.
\label{sIV63.7}
\eeq

To express $T_1$ in terms of the effective potential a similar trick is
used. It is provisionally assumed that the Minkowski tensor is of
the form $g_{mn}=\lambda^2\eta_{mn}$, where $\lambda$ is varied.
The result is \cite{sIV63-8}
\beq
T_1\Omega_N=-V.
\label{sIV63.8}
\eeq

Now we rewrite the Einstein equations (\ref{sIV63.1}) separately
for the subspaces $M^4$ and $S^N$ using 
Eqs.~(\ref{sIV63.2})--(\ref{sIV63.4}) and (\ref{sIV63.6}),
(\ref{sIV63.8}). The result is
\bes
&&
\frac{N(N-1)}{2a^2}+\Lambda_d=
\frac{8\pi G}{c^4}\,V(a),
\label{sIV63.9}\\
&&
-\frac{N-1}{a^2}+\frac{N(N-1)}{2a^2}+\Lambda_d=
\frac{8\pi G}{c^4N}\,a\frac{dV(a)}{da}.
\nonumber
\ees
\noindent
Subtracting the second equation from the first, one obtains
\beq
\frac{c^4(N-1)}{8\pi Ga^2}=V(a)-\frac{a}{N}\frac{dV(a)}{da},
\label{sIV63.10}
\eeq
\noindent
where the usual gravitational constant $G$ is connected with
the $d$-dimensional one by the equality $G_d=\Omega_NG$.

From dimensional considerations for the massless field we have
$T_{1,2}\sim a^{-d}$ in a $d$-dimensional space-time. With account
of Eqs.~(\ref{sIV63.6}), (\ref{sIV63.8}) and $\Omega_N\sim a^N$
this leads to
\beq
V(a)=\frac{\hbar c C_N}{a^4}.
\label{sIV63.11}
\eeq
\noindent
Here $C_N$ is a constant whose values depend on the dimensionality
of a compact manifold. Substituting this into Eq.~(\ref{sIV63.10})
we find the self-consistent value of the radius of the sphere
\beq
a^2=\frac{8\pi C_N(N+4)G\hbar}{N(N-1)c^3}.
\label{sIV63.12}
\eeq
\noindent
Then from (\ref{sIV63.9}) the cosmological constant is
\beq
\Lambda_d=-\frac{N^2(N-1)^2(N+2)c^3}{16\pi C_N(N+4)^2G\hbar}.
\label{sIV63.13}
\eeq
\noindent
Thus, the self-consistent radii are possible when $N>1$, and
$C_N>0$. In that case the multi-dimensional cosmological
constant is negative.

It is seen from Eq.~(\ref{sIV63.12}) that $a\sim l_{Pl}$, and the value of
the coefficient in this dependence is determined by the value of $C_N$.
Generally speaking one should take account not of one field but of
all kinds of boson and fermion fields contributing to the Casimir
energy. From this point of view the self-consistent radii are
expressed by
\beq
a=\left[\frac{8\pi (N+4)}{N(N-1)}C_N\right]^{\frac{1}{2}}\,l_{Pl},
\qquad
C_N=n_BC_B^N+n_FC_F^N,
\label{sIV63.14}
\eeq
\noindent
where $C_B^N$ and $C_F^N$ are the dimensionless constants in the
Eq.~(\ref{sIV63.11}) written for each field separately, $n_B$ and
$n_F$ are the numbers of boson and fermion massless fields. It is 
important that $C_N\geq 1$. In other case the one-loop approximation
in frames of which the self-consistent solutions are found is not
valid and one should take into account the corrections to it due
to the quantization of gravity. 

To determine the values of $C_B^N$ and $C_F^N$ it is necessary to
calculate the effective potential (\ref{sIV63.11}) explicitly. This was
done in \cite{sIV63-8} for odd values of $N$ using the method of
dimensional regularization and generalized zeta functions. In the scalar
case both the conformally and minimally coupled fields were
considered (see Sec.~2.3). Some of the obtained results are
represented in Table~\ref{tab-kk}.
\begin{table}[ht]
\caption{\label{tab-kk}
The coefficients of the effective potential for the minimally 
coupled scalar field, conformally coupled scalar field, and
fermion field.}
\vspace{1cm}
\begin{tabular}{|r|r|r|r|}
\hline
  N & $10^5\,C_B^N\,$ (min.) & $10^5\,C_B^N\,$ (conf.) &
 $10^5\,C_F^N\,$ 
\\ \hline
 3 & 7.56870 & 0.714589 & 19.45058 \\
 5 & 42.8304$\phantom{0}$ & --0.078571 & --11.40405 \\
 7 & 81.5883$\phantom{0}$ & 0.007049 & 5.95874 \\
 9 & 113.389$\phantom{00}$ & --0.000182 & --2.99172 \\
11 & 132.932$\phantom{00}$ & --0.000157 & 1.47771 \\ \hline
\end{tabular}
\end{table}

It is seen from the Table~\ref{tab-kk} that the values of all coefficients 
are rather small, being especially small for the conformally coupled
scalar field. The values of $C_N$ from Eq.~(\ref{sIV63.14}) are
positive, e.g., for $N=3$ or 7 regardless of the number of
different fields. For $N=5,\,9$ the number of fermions and
conformally coupled scalar fields should not be too large comparing
the number of minimally coupled scalar fields in order to assure the
condition $C_N\geq 0$. In all this cases Eq.~(\ref{sIV63.14})
provides the self-consistent values of a compactification radius.
To reach the value $C_N\approx 1$ (which is minimally permissable
for the validity of a one-loop approximation) a large number of
fields is needed, however. For example, even for $N=11$ where the
value of $C_B^N\approx 133\times 10^{-5}$ is achieved, one needs to have 752
scalar fields with minimal coupling to have $C_N\approx 1$.
The enormous number of light matter fields required to get the
reasonable size of the compactification radius (no smaller than a
Planckean one) is the general characteristic feature of the
spontaneous compactification mechanisms based on the Casimir effect.
Although such mechanisms are of great interest from a theoretical
point of view, only with future development can we recognize if 
they have
any relationship to the real world.

\setcounter{equation}{0}
\section{Casimir effect for real media}\label{sec5}
In this section the Casimir effect for real media is considered which is
to say that the realistic properties of the boundary surfaces
are taken into account. In the previous sections the highly
symmetrical configurations were mostly  restricted by the 
geometrically perfect boundaries on which the idealized boundary
conditions were imposed. Contrary to this, here we concentrate 
on the distinguishing features of all physically realizable situations 
where the impact of test conditions like nonzero temperature, surface 
roughness or finite conductivity of the boundary metal should be
carefully taken into account in order to obtain highly accurate
results. As indicated below, these conditions separately, and also
their combined effect, have a dramatic influence on the value of
the Casimir force. Thus, the present section forms the basis for
experimental investigation of the Casimir effect presented in Sec.6.

\subsection{The Casimir effect at nonzero temperature}\label{sec5.1}
The Casimir effect at nonzero temperature is interesting for the description
of the present experimental situation and as an example for 
nonzero temperature quantum field theory
as well.  We consider the static Casimir effect. This is a system in
thermal equilibrium and can be described by the Matsubara formulation. One has
to take the Euclidean version of the field theory with fields periodic
(antiperiodic) in the Euclidean time variable for bosons (fermions) within the
interval of time $\beta=\hbar /k_{B}T$ where $k_{B}$ is the Boltzmann
constant.  This procedure is well known and we restrict ourselves here to some
questions which are specific to the Casimir effect. Its temperature dependence
was first investigated in \cite{Fierz60} and \cite{Mehra67} for conducting
planes with the result that the influence of temperature is just below what
was measured in the experiment by Sparnaay \cite{18}. Later on the temperature
dependence had been intensively investigated theoretically, see for example
\cite{bali78-112-165,BrownMaclay,SRM,Dowker1978bl}. It is still an active area
of research, see for instance a recent investigation of the thermodynamics of
the Casimir effect with different temperatures in between and outside the
plates, \cite{Mitter:1999hu}. It is impossible to cover the whole area in this
review. So we focus on two specific moments.

From the mathematical point of view, the inclusion of nonzero temperature is
nothing other than the addition of another pair of boundaries, just in the
time coordinate with periodic boundary conditions (strictly speaking, this is
a compactification with no boundary).  The spectrum in the corresponding
momentum variable is equally spaced in both cases and one problem can be
transformed into another. So, for example, the simple Casimir effect (two
parallel conducting planes) can be reduced to the same Riemann zeta function
(see Sec. \ref{sec2.2}, Eq.  \Ref{Riezeta}), as it appears in the black body
radiation in empty space. Yet another example is the Casimir effect for a
rectangular domain say with two discrete frequencies ${\pi n_{x}\over a_{x}}$
and ${\pi n_{y}\over a_{y}}$ which can be expressed in terms of an Epstein zeta
function in the same way as the Casimir effect for a pair of plates at nonzero
temperature.  Thereby one has of course to take into account that there are
different boundary conditions, periodic ones in the imaginary time and
Dirichlet ones in the spatial directions.

Recently the precision of the Casimir force measurements increased in a way
that there is a hope to measure the temperature effects.  Therefore they must
be calculated with sufficient accuracy, see Sec. \ref{sec5.4}. Of particular
interest is the force between dielectric bodies at finite temperature. This
problem had been first solved by Lifshitz \cite{9}. Because of its complicated
form it had been reconsidered later on, see for example \cite{Spruch}. In view
of its actual importance we give below another derivation which is based on
field theoretical methods, especially the representation of the \gse for a
background depending on one coordinate developed in Sec. \ref{sec3.1}.

\subsubsection{Two semispaces}\label{sec5.1.1}
We start from the representation of the free energy $F_{E}$ in a field theory
at nonzero temperature which is on the one loop level given by
\be\label{FE}F_{E}=\frac{\hbar}{2} \Tr \ln \left[\Box_{E}+
V({\bf x})+\left(\frac{mc}{\hbar}\right)^{2}\right]\,,
\ee
where the trace and the wave operator are taken in Euclidean space. This
formula can be obtained from the effective action \Ref{effaction} by Wick
rotation, $x_{0}\to ix_{4}$, with $x_{4}\in [0,\beta]$ and periodic \bc in
$x_{4}$ on the field. 

\begin{figure}[h]
\setlength{\unitlength}{1cm}
\begin{picture}(13,5.5)
\put(0,0){\epsfig{figure=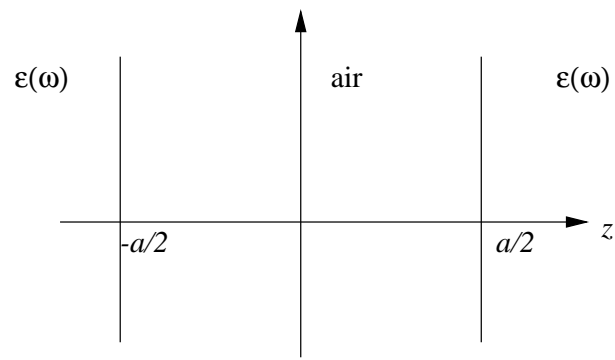
}}
\end{picture}
\caption{\label{air}{The configuration of two dielectric bodies 
separated by air.}}
\end{figure}

To be specific we consider two dielectric bodies with frequency dependent 
permittivity 
$\ep(\om)$, restricted by two
plane parallel surfaces $z=\pm a/2$ and separated by air with
distance $a$ between them, see Fig. \ref{air}. Then Eq. \Ref{FE} can be
rewritten in the more specific form
\be\label{FE1}F_{E}=-\frac{\hbar}{2}\frac{\pa}{\pa s}\frac1{\beta}
  \sum_{l}\int{ d {\bf  k_{\perp}}\over (2\pi)^{2}}
    \sum_{n}\left(\varepsilon(i\omega_l)\frac{\om_{l}^2}{c^2}+
k_{\perp}^{2}+\la_{n}\right)^{-s} \,,
\ee
where ${\bf k_{\perp}}=(k_{x},k_{y})$ are the momenta in the translational
invariant directions perpendicular to the $z$-axis and $\om_{l}={2\pi
  l\over\beta}$ ($l=-\infty,\dots,\infty$) are the Matsubara frequencies
(we do not need to introduce the factor $\mu^{2s}$ into this formula
because it drops out from the distance dependent contributions to free
energy like (5.5) below).
Following the discussion in Sec. \ref{sec3.1} we assume for a moment in the
$z$-direction the presence of a ``large box'' in order to have the 
corresponding
eigenvalues $\la_{n}$ discrete. As usual we use zeta functional regularization
with $s\to0$ in the end and the logarithm is restored by the derivative with
respect to $s$.

As a next step we assume the scattering problem on the $z$-axis 
to be formulated
in the same manner as in Sec. \ref{sec3.1} for a background depending on one
Cartesian coordinate. This is a reasonable setup for the problem under
consideration. An \elm wave coming from the left (or from the right) in the
dielectric will be scattered on  the air strip and there will be a
transmitted and a reflected wave. Note that there are no bound states because
the photon cannot be confined in between the two dielectric bodies. Also note
that in this formulation there is no need to consider evanescent waves
separately as the asymptotic states given by the incident waves from the left
and from the right are complete. Now we use again the same discussion as in
Sec. \ref{sec3.1} to pass to the representation
\be\label{FEd}F_{E}=-\frac{\hbar}{2\beta}{\pa\over\pa s}\sum_{l}\int{ d {\bf
    k_{\perp}}\over (2\pi)^{2}} \int_{0}^{\infty}{d k\over \pi}
\left(\varepsilon(i\omega_l)\frac{\om_{l}^{2}}{c^2}+
k_{\perp}^{2}+k^{2}\right)^{-s} 
{\pa\over\pa k} \delta(k,{\bf k_{\perp}})\,,
\ee
which is analogous to Eq. \Ref{EregRe}. Here, however, having in mind the
properties of a dielectric we allow the scattering phase $\delta(k,{\bf
  k_{\perp}})$ to depend on ${\bf k_{\perp}}$. In passing to Eq. \Ref{FEd} we
dropped a distance independent contribution and took into account that there
are no bound states. Now we turn the integration over $k$ towards the imaginary
axis and obtain in parallel to Eq. \Ref{EregIm}
\bea\label{feo}F_{E}&=&\frac{\hbar}{2\pi\beta}{\pa\over\pa s}\sin\pi s
\sum_{l}\int{ d {\bf  k_{\perp}}\over
  (2\pi)^{2}} 
\int_{\sqrt{\frac{\varepsilon(i\omega_l)\om_{l}^{2}}{c^2}+
k_{\perp}^{2}}}^{\infty}d k \
  \left(k^{2}-\varepsilon(i\omega_l)\frac{\om_{l}^{2}}{c^2}-
{\bf k_{\perp}}^{2}
\right)^{-s} \nn \\
&&  ~~~~~~~~~~  ~~~~~~~~~~  ~~~~~~~~~~  ~~~~~~~~~~ ~~~~~~~~~~ 
\times{\pa\over\pa k}
  \ln s_{11}(ik,{\bf k_{\perp}})\,.
\eea

To proceed we assume that the integral over $k$ converges for $s=0$, i.e., for
removed regularization. This is justified for the problem   we are
interested in, namely for the force between two separated bodies where
divergent distance independent contribution can be dropped. In the limit 
$s\to 0$ 
the derivative with respect to $s$ can be carried out by means of
$\pa_{s}sf(s)_{|_{s=0}}=f(0)$ for any function $f(s)$ regular in $s=0$. Then
the integral over $k$ is over a total derivative and can be carried out with
the result
\be\label{fes}F_{E}=-\frac{\hbar}{2\beta}\sum_{l}\int{ d {\bf k_{\perp}}\over
  (2\pi)^{2}} \ln s_{11}
\left(i\sqrt{\varepsilon(i\omega_l)\frac{\om_{l}^{2}}{c^2}+
{\bf k_{\perp}}^{2}},{\bf
  k_{\perp}}\right) \,.  \ee
This is the generalization of Eq. \Ref{EregIm} to nonzero temperature. For
$T\to 0$ in the sense that $k_BT\ll\hbar c/a$
the sum over $l$ turns into an integral by means of
$\frac{1}{\beta}\sum_{l}\to \int{d\om\over 2\pi}$ with $\om_{l}\to\om$ and
after a change of variables to 
$\frac{\om^{2}}{c^2}+k_{\perp}^{2}=k^{2}$ Eq. \Ref{EregIm}
is reobtained. 

So it remains to obtain an expression for the scattering coefficient
$s_{11}(ik,{\bf k_{\perp}})$ for the problem with two dielectric bodies 
as shown in
Fig. \ref{air}. In fact this scattering problem for an \elm wave is quite
simple. We consider the electric field strength ${\bf E}(t,{\bf x})$ subject
to the Maxwell equations
\bea\label{Max1}\left({\ep\over c^{2}}{\pa^{2}\over\pa t^{2}}-\Delta\right){\bf
  E}(t,{\bf x})&=&0,\nn \\
{\nabla} \ {\bf E}(t,{\bf x})&=&0 \,.\eea
We turn to Fourier space in the translational invariant variables $t$ and
${\bf k_{\perp}}$ by means of
\be\label{four}{\bf E}(t,{\bf x})=e^{i\om t-i{\bf k_{\perp}{\bf x_{\perp}}}}
{\bf E}(\om,{\bf k_{\perp}},z)
\ee
and the equations become
\bea\label{Max2}\left(-{\ep \om^{2}\over c^{2}} +
  {\bf k_{\perp}}^{2}-
{\pa^{2}\over  \pa z^{2}}\right)
{\bf  E}(\om,{\bf k_{\perp}},z)&=&0,\nn \\
 i{\bf k_{\perp}} {\bf E_{\perp}}(\om,{\bf k_{\perp}},z)-
{\pa\over\pa z}{   E}_{z}(\om,{\bf k_{\perp}},z)&=&0 \,.
\eea
Now we allow the permittivity $\ep$ to be a function of the frequency $\om$.
Although the formalism described here works also in the case of an
$\ep(\om,z)$ which is a general function of $z$ we consider the specific form
\be\label{epp}\ep(\om,z)=\left\{\begin{array}{lcc}1 & ~~ {\rm for} ~~& 
-\frac{a}{2}\le z\le
    \frac{a}{2}, \\ \ep(\om) & {\rm for} ~~&\frac{a}{2}\le \mid z \mid 
\end{array}\right.
\ee
for the dependence of $\ep$ on $z$ in accordance with Fig.~\ref{air}, where
$\ep(\om)$ is  assumed to be a function of the frequency $\om$ with the
necessary analytic properties.  

As a  next step we have to separate the polarizations. 
This can be done using the
standard polarization vectors known from  the transverse photons in Coulomb
gauge. We introduce the decomposition
\be\label{decomp}{\bf  E}(\om,{\bf
  k_{\perp}},z)=N_1 \left(\begin{array}{c}
-ik_{1}{\pa\over\pa z}\\
-ik_{2}{\pa\over\pa z}\\
k_{1}^{2}+k_{2}^{2}\end{array}\right) \ e_{1}({\bf k_{\perp}},z) +
N_2 \left(\begin{array}{c}
-ik_{2}\\
ik_{1}\\0\end{array}\right) \ e_{2}({\bf k_{\perp}},z)
\ee
so that the second of the Eqs. (\ref{Max2}) is satisfied
($N_{1,2}$ are the normalization factors). 

The matching conditions on the
surface of the dielectric demanding ${\bf  E}_{\perp}(\om,{\bf
  k_{\perp}},z)$ and $\ep(\om,z) {{\bf  E}_{z}}(\om,{\bf
  k_{\perp}},z)$ to be continuous  are satisfied if
\be\label{mconde1} \ep(\om,z)e_{1}({\bf k_{\perp}},z)\,,  
\ \ \ \ {\pa\over\pa
  z}e_{1}({\bf k_{\perp}},z)\,,  
\ee
\noindent
and
\be
e_{2}({\bf k_{\perp}},z)\,, \ \ \ \ 
{\pa\over\pa  z}e_{2}({\bf k_{\perp}},z)
\label{mconde2}
\ee
are continuous in $z=\pm{a}/2$. Note that the second condition differs
from the first one simply by the absence of $\ep$. By the first of the Eqs.
\Ref{Max2} and the matching conditions 
(\ref{mconde1}), (\ref{mconde2}) we have for the
functions $e_{1}({\bf k_{\perp}},z)$ and $e_{2}({\bf k_{\perp}},z)$ a one
dimensional scattering problem. So we can find a solution of the form
\be\label{eisol}e_{i}({\bf k_{\perp}},z)=\left\{ \begin{array}{cc}
e^{ikz}+s_{12}e^{-ikz},  &  z\le -\frac{a}{2},   \\
\al e^{iqz}+\beta e^{-iqz},  &  -\frac{a}{2}\le z\le \frac{a}{2},   \\
s_{11} e^{ikz},   &  \frac{a}{2}\le z   \\
\end{array}\right.
\ee
for each of the polarizations $i=1,2$.  From Eq. \Ref{Max2} it follows
\be\label{kundq}-\ep(\om){\om^{2}\over c^{2}}+{\bf k_{\perp}}^{2}+k^{2}=0 \ \
{\rm and} \ \ \ -{\om^{2}\over c^{2}}+{\bf k_{\perp}}^{2}+q^{2}=0 \,.
\ee
The coefficients can be determined from the matching conditions. Consider
first the polarization with $i=1$. In $z=\frac{a}{2}$ we have from Eq.
\Ref{mconde1}
\bea\label{}\al e^{i\frac{qa}{2}} +\beta e^{-i\frac{qa}{2} }
&=& \ep(\om)s_{11}e^{i\frac{ka}{2}}, \nn \\
iq\left(\al e^{i\frac{qa}{2}} -\beta e^{-i\frac{qa}{2}}\right) 
&=&ik s_{11}e^{i\frac{ka}{2}}, \nn
\eea
wherefrom
\be\label{albet}
\alpha=
\frac{s_{11}}{2}\left(\ep(\om)+\frac{k}{q}\right) e^{i\frac{(k- q)a}{2}},
\quad
\alpha=
\frac{s_{11}}{2}\left(\ep(\om)-\frac{k}{q}\right) e^{i\frac{(k+ q)a}{2}}
\ee
follows. In $z=-\frac{a}{2}$ we have
\bea\label{}
\ep(\om)\left(\al e^{-i\frac{ka}{2}} +s_{12} e^{i\frac{ka}{2} }\right)
&=&
\al e^{-i\frac{qa}{2}} +\beta e^{i\frac{qa}{2} }, \nn \\
ik\left(\al e^{-i\frac{ka}{2}} -s_{12} e^{i\frac{ka}{2} }\right)
&=&
iq\left(\alpha e^{-i\frac{qa}{2}} -\beta e^{i\frac{qa}{2} }\right) \,. \nn
\eea
Adding the first equation divided by $\ep(\om)$ to the second one divided by
$ik$ we eliminate $s_{12}$. After inserting $\al$ and $\beta$ from
Eq. \Ref{albet} we arrive at
\be\label{s11diel}s_{11}(k, {\bf k_{\perp}}) = {4\ep(\om)kqe^{-ika}
\over
\left[\ep(\om)q+k\right]^{2} e^{-iqa}-\left[\ep(\om)q-k\right]^{2} e^{iqa}
} .
\ee
For representation \Ref{fes} of the free energy we need $s_{11}$ on the
imaginary $k$-axis,
\be\label{s11dieli}s_{11}(ik, {\bf k_{\perp}}) = {4\ep(i\xi)kqe^{ka}
\over
\left[\ep(i\xi)q+k\right]^{2} e^{qa}-\left[\ep(i\xi)q-k\right]^{2} e^{-qa}
} ,
\ee
where we have substituted $k\to ik$, $q\to iq$ and $\om\to i\xi$ so that these
quantities are now related by
\[ \ep(i\xi){\xi^{2}\over c^{2}}+{\bf k_{\perp}}^{2}-k^{2}=0 \ \ \ {\rm and} \
\ {\xi^{2}\over c^{2}}+{\bf k_{\perp}}^{2}-q^{2}=0\,.
\]
For the second polarization the same formulas hold by using $\ep=1$ where
it explicitly appears in Eq. \Ref{s11diel} and \Ref{s11dieli}. By means of
Eq. \Ref{s11diel} we have the coefficient $s_{11}$ ($s_{12}$ can be obtained
easily) of the scattering problem. It matches all usually expected analytic properties. 
For instance it is an analytic function in the upper half plane of
$k$. There are no poles on the positive imaginary axis in accordance with the
absence of bound states in this problem. The coefficient $s_{11}(k, {\bf
  k_{\perp}})$ has poles in the lower half plane which correspond to
resonance states the photon has between the two dielectric bodies.

Next we are going to insert $s_{11}(ik, {\bf k_{\perp}})$, Eq. \Ref{s11dieli}
into the free energy, Eq. \Ref{fes}. As previously we rewrite it in the form
\be\label{s11umg}s_{11}(ik, {\bf k_{\perp}})=\left\{ \left
    [ {\ep(i\xi)q+k\over\ep(i\xi)q-k}\right]^{2}-e^{-2aq}\right\}^{-1} 
{4\ep(i\xi)kq\over \left[\ep(i\xi)q-k\right]^{2}}\ e^{(k-q)a} .
\ee
Now we remark that the second factor is distance independent and that the
third factor delivers a contribution linear in the distance $a$ between the
dielectrics. Both deliver contributions to the energy which are not relevant
for the force $F_{ss}^{T}=-\partial F_{E}/\partial a$ 
and we drop them. So we are left with the
first factor in the \rhs of Eq. \Ref{s11umg}. Here we note that it
provides a convergent contribution to the free energy, Eq. \Ref{feo} so that
the transition to Eq. \Ref{fes} is justified. Inserting this first factor into
Eq. \Ref{fes} and taking the derivative with respect to $a$ we arrive at
\bea\label{foo}
F_{ss}^{T}&=&-{\hbar \over\beta }\sum_{l}\int{ d {\bf k_{\perp}}\over
  (2\pi)^{2}} q_l 
\left\{ \left[
\left
  ( {\ep(i\xi_{l})q_l+k_l\over\ep(i\xi_{l})q_l-k_l}\right)^{2}e^{2aq_l}-
1\right]^{-1}\right. \nn
  \\
&& ~~~~~~~~~~~~~~~~~~~~~~~~~ \left.
+
\left[
\left( {q_l+k_l\over q_l-k_l}\right)^{2}e^{2aq_l}-1\right]^{-1}
\right\} 
\eea
with $q_l=\sqrt{{\xi_{l}^{2}\over c^{2}}+{k_{\perp}^{2}}}$ and
$k_l=\sqrt{\ep(i\xi_{l}){\xi_{l}^{2}\over c^{2}}+{ k_{\perp}^{2}}}$,
$\xi_l=2\pi l/\beta$. 

Introducing a new variable $p$ according to
\be
k_{\bot}^2=\frac{\xi_l^2}{c^2}(p^2-1)
\label{sV1.20}
\ee
we rewrite Eq.~(\ref{foo}) in the original Lifshitz form \cite{9,sIV1}
\bea
&&
F_{ss}^{T}(a)=-\frac{k_BT}{\pi c^3}
\sum\limits_{l=0}^{\infty}{\vphantom{\sum}}^{\prime}\xi_l^3
\int\limits_{1}^{\infty}p^2\,dp\,
\left\{\left[\left(
\frac{K(i\xi_l)+\varepsilon(i\xi_l)p}{K(i\xi_l)-\varepsilon(i\xi_l)p}
\right)^2e^{2a\frac{\xi_l}{c}p}-1\right]^{-1}\right.
\nonumber\\
&&
\phantom{F_{ss}^{T}(a)}+\left.
\left[\left(
\frac{K(i\xi_l)+p}{K(i\xi_l)-p}
\right)^2e^{2a\frac{\xi_l}{c}p}-1\right]^{-1}\right\},
\label{sV1.21}
\eea
\noindent
where the notation
\be
K(i\xi_l)\equiv\left[p^2-1+\varepsilon(i\xi_l)\right]^{1/2}
\label{sV1.22}
\ee
\noindent
is introduced in analogy with the one from Eq.~(\ref{sIV23}), and the 
prime near
summation sign means that the zeroth term is taken with the
coefficient 1/2.

As was already mentioned above, in the limit of low temperatures
$k_BT\ll\hbar c/a$ the summation in $l$ in Eq.~(\ref{sV1.21}) can be changed
for integration with respect to $dl=\hbar d\xi/(2\pi k_BT)$. As a result
we are returning back to Eq.~(\ref{sIV25}) which is the Casimir force
between two semispaces at zero temperature.

Note that the representation of Eq.~(\ref{sV1.21}) for the temperature
Casimir force has a disadvantage as the $l=0$ term in it is the product
of zero by a divergent integral. This is usually eliminated
\cite{sIV1} by one more change of variables $z=2a\xi_lp/c$.  Both this
change and also (\ref{sV1.20}) are, however, singular at $l=0$.
Because of this, the representation of Eq.~(\ref{foo}) is preferable
as compared to (\ref{sV1.21}) and other representations obtained
from it by change of variables at the singularity at $l=0$ 
(in Sec.5.4.2 the additional difficulties are discussed connected with 
the zeroth term
of Lifshitz formula in application to real metals).

Now let us apply Eq.~(\ref{foo}) to ideal metals of infinitely high 
conductivity in order to find the temperature correction to the
Casimir force $F_{ss}^{(0)}(a)$ between perfect conductors (see, e.g.,
Eq.~(\ref{FCas})). To do this we use the prescription
by Schwinger, DeRaad and Milton that the limit $\varepsilon\to\infty$
should be taken before setting $l=0$ \cite{SRM}. Then, introducing a new
variable $y=2aq_l$ in Eq.~(\ref{foo}) instead of $|{\bf k_{\perp}}|=k$
(note that $y$ is regular at any $l$) we arrive at
\be
F_{ss}^{T}(a)=-\frac{k_BT}{4\pi a^3}
\sum\limits_{l=0}^{\infty}{\vphantom{\sum}}^{\prime}
\int\limits_{\frac{2a\xi_l}{c}}^{\infty}
\frac{y^2\,dy}{e^{y}-1}.
\label{sV1.23}
\ee
\noindent
This expression can be put in a form \cite{BrownMaclay,SRM}
\bea
&&
F_{ss}^{T}(a)=F_{ss}^{(0)}(a)\left\{1+\frac{30}{\pi^4}
\sum\limits_{n=1}^{\infty}\left[
\left(\frac{T}{T_{eff}}\right)^4\frac{1}{n^4}\right.\right.
\label{sV1.24}\\
&&
\phantom{F_{ss}^{T}(a)=}\left.\left.
-\pi^3\,\frac{T}{T_{eff}}\,\frac{1}{n}
\cosh\left(\pi n\frac{T_{eff}}{T}\right)\,
\sinh^{-3}\left(\pi n\frac{T_{eff}}{T}\right)
\vphantom{\left(\frac{T}{T_{eff}}\right)^4}\right]\right\},
\nonumber
\eea
\noindent
where the effective temperature is defined as $k_BT_{eff}=\hbar c/(2a)$.
Note that the quantity in square brackets is always positive.

At low temperatures $T\ll T_{eff}$ it follows from (\ref{sV1.24})
\be
F_{ss}^{T}(a)\approx F_{ss}^{(0)}(a)\left[1+\frac{1}{3}
\left(\frac{T}{T_{eff}}\right)^4\right].
\label{sV1.25}
\ee
\noindent
At high temperature limit $T\gg T_{eff}$
\be
F_{ss}^{T}(a)\approx 
-\frac{k_BT}{4\pi a^3}\zeta_R(3).
\label{sV1.26}
\ee
\noindent
Note that the corrections to the above asymptotic results are
exponentially small as $\exp(-2\pi T_{eff}/T)$ at low temperatures and
as $\exp(-2\pi T/T_{eff})$ at high temperatures. As a consequence, the 
asymptotic regime is even achieved when the temperature is only two 
times lower (higher) that the effective temperature value.
In \cite{SaTTort} the analogical results are obtained for the so called
unusual pair of parallel plates, i.e. the first be a perfectly
conducting and the second --- infinitely permeable one.

Other new fascinating problems arising from the study of the nonzero
temperature Casimir force for the real metals of finite conductivity are
discussed in Secs.5.4.2 and 5.4.3.

\subsubsection{A sphere (lens) above a disk}
\label{sec5.1.2}

Here we deal with the nonzero temperature Casimir force for the 
configuration of
a spherical lens (sphere) above a plate (disk) which is used 
in most experiments. The sphere will be
considered to have large radius comparing the space separation to
a plate. As was noted in Sec.4.3 in this situation Proximity Force
Theorem produces high accuracy results.

To apply the Proximity Force Theorem one should derive first the
expression for the free energy density in the configuration of two dielectric
plates at a temperature $T$. It obtained by the integration of
$-F_{ss}^{T}(a)$ from Eq.~(\ref{foo}) with respect to $a$ or simply
by the substitution of Eq.~(\ref{s11umg}) into Eq.~(\ref{fes}).
The result is
\bea
&&
F_E(a)=\frac{k_BT}{4\pi}
\sum\limits_{l=-\infty}^{\infty}
\int\limits_{0}^{\infty}k_{\bot}\,dk_{\bot}
\left\{\ln\left[1-\left(
\frac{\varepsilon(i\xi_l)q_l-k_l}{\varepsilon(i\xi_l)q_l+k_l}\right)^2
\,e^{-2aq_l}\right]\right.
\nonumber\\
&&
\phantom{aaaaaa}
\left.+\ln\left[1-\left(\frac{q_l-k_l}{q_l+k_l}\right)^2
\,e^{-2aq_l}\right]\right\}.
\label{sV13.1}
\eea
\noindent
Once more, the contribution to the free energy is omitted which does
not depend on $a$ in order that $F_E(a)\to 0$ when $a\to\infty$.

The nonzero temperature Casimir force in configuration of a sphere (lens)
above a disk (plate) is given by Eq.~(\ref{sIV3.5}) as follows
\be
F_{dl}^{T}=2\pi RF_E(a).
\label{sV13.2}
\ee

Let us apply the results (\ref{sV13.1}), (\ref{sV13.2}) in the case
of a disk and a sphere made of ideal metals. By the use of the same
change of variable $y=2aq_l$ as in Sec.5.1.1 in the limit
$\varepsilon\to\infty$ one arrives at
\be
F_{dl}^{T}(a)=\frac{k_BTR}{2a^2}
\sum\limits_{l=0}^{\infty}{\vphantom{\sum}}^{\prime}
\int\limits_{\frac{2a\xi_l}{c}}^{\infty}y\,dy
\ln\left(1-e^{-y}\right).
\label{sV13.3}
\ee
\noindent
This expression is in a direct analogy with Eq.~(\ref{sV1.23})
for two semispaces. It can be put in an equivalent form
\cite{BrownMaclay}
\bea
&&
F_{dl}^{T}(a)=F_{dl}^{(0)}(a)\left\{1+
\frac{45}{\pi^3}
\sum\limits_{n=1}^{\infty}\left[\left(\frac{T}{T_{eff}}\right)^3
\frac{1}{n^3}\coth\left(\pi n\frac{T_{eff}}{T}\right)
\right.\right.
\nonumber\\
&&\phantom{aaaaaaaa}\left.\left.
+\pi\,\left(\frac{T}{T_{eff}}\right)^2\frac{1}{n^2}
\sinh^{-2}\left(\pi n\frac{T_{eff}}{T}\right)\right]
-\left(\frac{T}{T_{eff}}\right)^4\right\}.
\label{sV13.4}
\eea
\noindent
Note that as in (\ref{sV1.24}) the temperature correction to the
Casimir force is always
positive. It is approximately 2.7\% of $F_{dl}^{(0)}$ at $a=1\,\mu$m.
But, e.g., at $a=6\,\mu$m the temperature correction is equal to
$1.74\,F_{dl}^{(0)}$, i.e. is already larger than the zero temperature 
force.

At low temperature $T\ll T_{eff}$ the hyperbolic functions behave
like exponents of large arguments. As a result the Casimir force
is approximately equal to
\be
F_{dl}^{T}(a)\approx F_{dl}^{(0)}(a)\left[1+
\frac{45\zeta_R(3)}{\pi^3} \left(\frac{T}{T_{eff}}\right)^3
-\left(\frac{T}{T_{eff}}\right)^4\right].
\label{sV13.5}
\ee
\noindent
In the opposite case of high temperatures $T\gg T_{eff}$ all the 
terms of series (\ref{sV13.3}) are exponentially small. As a result
the zeroth term alone determines the Casimir force value \cite{32}
\be
F_{dl}^{T}(a)\approx \frac{k_BTR}{4a^2}
\int\limits_{0}^{\infty}y\,dy
\ln\left(1-e^{-y}\right)=
-\frac{k_BTR\zeta_R(3)}{4a^2}.
\label{sV13.6}
\ee
\noindent
The corrections to Eq.~(\ref{sV13.5}) are exponentially small in
$\exp(-2\pi T_{eff}/T)$, and to  Eq.~(\ref{sV13.6}) --- in
$\exp(-2\pi T/T_{eff})$.  For this reason the transition region between the
two asymptotic regimes is actually very narrow.

\subsubsection{The asymptotics of the  Casimir force at high and low
temperature}\label{sec5.1.3}
As explained in the beginning of Sec.5.1 the Casimir effect at nonzero
temperature can be described in \qft by the Matsubara formalism. This has a
number of important consequences. For example one can easily
understand that all \uv divergencies at finite temperature are the same as at
zero temperature. Hence the temperature dependent part is a finite expression
which can be calculated without facing major problems, at least
numerically.  A second important consequence is that for high and low
temperatures asymptotic expansions can be obtained in quite general terms
whereby the remainders are exponentially small. These asymptotics had been
obtained first in \cite{Dowker1978bl} and later reconsidered and
generalized, see
for example \cite{Actor:1989mj}. 

Let $P$ be some three dimensional operator describing the spatial part of the
considered system in the same sense as in Eqs. \Ref{levp1} or \Ref{sevp1} and
take Eq. \Ref{evp2} as its eigenvalue problem. The free energy as the
quantity to be considered can be written following Eqs. \Ref{FE} and \Ref{FE1}
as 
\be\label{FEa}F_{E}=-\frac{\hbar}{2 \beta}{\pa\over\pa s}
\mu^{2s}\sum_{l=-\infty}^{\infty}\sum_{J}\left({\om_{l}^{2}\over
    c^{2}}+\la_{J}\right)^{-s} \ee
($\om_{l}=2\pi l/\beta, \ \beta=\hbar/(k_{B}T)$) 
with $s\to0$ in the end. In
this formula, $\mu$ is the arbitrary parameter (here with dimension of an
inverse length) entering zeta-functional regularization, see Sec.
\ref{sec3.1}. Using Eq. \Ref{euga} we rewrite this expression in the form
\be\label{FEk}F_{E}=-{\hbar c\over 2}{\pa\over\pa
  s}\mu^{2s}\int_{0}^{\infty}{d t\over t}{t^{s}\over\Gamma (s)} \
  K_{T}(t) \ K_{P}(t), \ee
where
\be\label{sre}K_{T}(t)=\frac{1}{\beta c}\sum_{l=-\infty}^{\infty} 
e^{-t\om_{l}^{2}/c^{2}} 
\ee
is the temperature dependent heat kernel and
\be\label{ghk1}
K_{P}(t)=\sum_{J}e^{-t\lambda_{J}}\ee
is the same heat kernel of the operator $P$
as given by Eq. \Ref{ghk}. 

In order to obtain the low temperature expansion it is useful to employ the
Poisson formula 
\be\label{Poisson}K_{T}(t)=\frac{1}{\sqrt{4\pi t}}+\frac{2}{\sqrt{4\pi
    t}}\sum_{l=1}^{\infty}\exp \left(-{l^{2}\beta^{2}c^{2}\over 4 t}\right)
.  \ee
Being inserted into $F_{E}$, Eq. \Ref{FEk}, the first term in the {\rhs}
of Eq.~(\ref{Poisson})
delivers just the zero temperature ($\beta=\infty$) contribution 
\be\label{FE0}F_{0}\equiv {F_{E}}_{|_{T=0}},
\ee
where we split
\be\label{}F_{E}=F_{0}+F_{T}.
\ee
The second term in the {\rhs} of Eq.~(\ref{Poisson})
makes the $t$-integral in Eq. \Ref{FEk}
exponentially converging for $t\to0$ so that there are no \uv
divergencies. The limit $s\to0$ can be done trivially and we obtain
\be\label{5.8}
F_{T}=-\frac{\hbar c}{2}\sum_{l=1}^{\infty} \int_{0}^{\infty}{d
  t\over t} {2\over\sqrt{4\pi t}} \ e^{-{l^{2}\beta^{2}c^{2}\over 4 t}} 
\ K_{P}(t), 
\ee
where sum and integral are absolutely convergent. Now we return to the sum
representation \Ref{ghk1} of $K_{P}(t)$. The $t$-integration 
in (\ref{5.8}) can be done
explicitly and after that the sum over $l$ also, resulting in 
\beq\label{FT1}
F_{T}=\frac{\hbar}{\beta}\sum_{J}\ln \left[1-\exp\left(-\beta c
    \sqrt{\lambda_{J}}\right)\right] .
\eeq
Note that this expression can be obtained by applying the Abel-Plana
formula \Ref{2.25} to Eq.~\Ref{FEa}. 

Representation \Ref{FT1} is well suited to discuss the behavior for $T\to0$,
see \cite{Dowker1978bl}. For a purely discrete spectrum, $F_{T}$ is
exponentially decreasing as 
$\exp(-\beta c \sqrt{\lambda_{0}})$ where $\lambda_{0}$ is
the smallest eigenvalue of the operator $P$. If the spectrum of $P$ is partly
(or completely) continuous going down to $\lambda=0$ then powers of $T$ may be
present. Consider as an example the Casimir effect bet\-ween conducting
planes. In this case for the temperature dependent part of free energy 
per unit area we have
\be\label{FT2}F_{T}={\hbar\over\beta} \int{d k_{1} d k_{2}\over (2\pi)^{2}}
\sum_{n=-\infty}^{\infty} 
\ln \left[1-\exp\left(-\beta c 
\sqrt{k_{1}^{2}+k_{2}^{2}+\left({\pi n\over a}\right)^{2}}\right)\right] .
\ee
The power-like contribution at $T\to0$ comes from $n=0$,
\bea\label{FT3}
F_{T}&=&{\hbar\over\beta} \int{d k_{1}d k_{2}\over
(2\pi)^{2}} \ln \left[1-\exp\left(-\beta c \sqrt{
k_{1}^{2}+k_{2}^{2}}\right)\right] +O\left(e^{-{\beta c \pi\over
a}}\right)\nn\\ &=&-{\hbar \zeta_{R}(3)\over
2\pi\beta^{3}c^{2}}+O\left(e^{-{\beta c \pi\over a}}\right), \eea
which is just the result first found by Mehra \cite{Mehra67}.  In
order to compare with \Ref{sV1.25} we remark that Eq. \Ref{FT2} gives
the temperature contribution to the 
free energy confined between the two plates. From the expansion of
Eq.~\Ref{FT3} for small $T$ there is no contribution to the force.  In
Eq.~\Ref{sV1.25}, however, the contribution from the exterior region
is taken into account (it was derived as a special case from extended
dielectric media). 
It is not difficult to write down the temperature dependent part of the free 
energy per unit volume from Eq.~(\ref{FT1})  for the exterior region of
one plate
\bea\label{blackbody}
f_{T}^{ext}
&=&\frac{\hbar}{\beta}
\int{d k_{1}\,d k_{2}\,dk_{3} \over (2\pi)^{3}}
\ln \left[1-\exp\left(-\beta c\sqrt{
      k_{1}^{2}+k_{2}^{2}+k_{3}^{2}}\right)\right] \nn \\
&=&-{\pi^{2}\hbar \over 90\beta^{4}c^{3}}.
\eea

Multiplying this by 2 to account for the photon polarizations one obtains
the familiar free energy density of the black body radiation
in empty space
\beq
f_{BB}=2f_{T}^{ext}.
\label{5.12a}
\eeq
\noindent
We now recall that at zero temperature to obtain the renormalized energy 
between plates
we subtracted the energy of zero-point fluctuations of empty space in
the same volume (interval $a$ in this case). To obtain the renormalized
free energy we do the same i.e. subtract from it the free energy of free
space in  the
line interval $a$ at the given temperature with
the result
\beq
F_{T}^{ren}=F_{T}-af_{BB}=F_{T}-2af_{T}^{ext}.
\label{5.12b}
\eeq
\noindent
It is evident that calculation of a force in accordance with
$-\partial F_{T}^{ren}/\partial a$ gives us the well known result
in the second term of Eq.~(\ref{sV1.25}).
Alternatively, one can say that there is equilibrium thermal radiation
on the outside whose pressure is equivalent to the force which is
attracting the plate.

{}From the structure of these expressions one can, for instance,
conclude $F_{T}\sim T^{2}$ for the conducting cylinder. Let us
consider as one more example the exterior of a sphere or of a
spherically symmetric background potential as in
Sec. \ref{sec3.1.2}. The discussion leading there from Eq. \Ref{sevp3}
to Eq. \Ref{EregRes} can be applied without change to $F_{T}$,
Eq. \Ref{FT1}. Note that it is not useful to turn the integration path
towards the imaginary axis, i.e., to pass to a formula in analogy with
\Ref{EregIms}.  In this case we obtain
\be\label{FT4}F_{T}={\hbar\over \beta}\sum_{l}(2l+1)\int_{0}^{\infty}{d
  k\over\pi} \ln\left(1-e^{-\beta c k}\right){\pa\over\pa k}\delta_{l}(k), 
\ee
where $l$ is the sum over the orbital momentum and $\delta_{l}(k)$ are the
scattering phase shifts. After the substitution $k\to k/\beta$ we see that for
$\beta\to\infty$ the phase shifts can be expanded at $k=0$ and the expansion
\be\label{FT5}F_{T}=-\frac{\hbar }{\pi \beta} \sum_{n\ge
  0}{\zeta_{R}(n+1)\over(\beta c)^{n}}
  \sum_{l}(2l+1)\delta^{(n)}_{l}(0) \ee
emerges relating the power-like contributions for $T\to0$ to the derivatives
of the phase shifts at zero momentum. 

In order to obtain the high temperature expansion it is useful to separate in
$K_{T}(t)$ \Ref{sre} the zeroth Matsubara frequency
\be\label{ZMat}K_{T}(t)=\frac{1}{\beta c}+\frac{2}{\beta c}\sum_{l=1}^{\infty}
e^{-t\om_{l}^{2}/c^{2}} \ .
\ee
Being inserted into Eq. \Ref{FEk}, the first contribution delivers by means of
Eq. \Ref{zausK} the derivative of the zeta function of the operator $P$.
Because of 
\[\ln\det P=\Tr\ln P=-\frac{\pa}{\pa s} \Tr {P^{-s}}_{|s=0}=
-\frac{\pa}{\pa s} \zeta_{P}(s)_{|s=0}=-\zeta'_{P}(0)
\]
 it is directly related
to the determinant of the operator $P$ (in three dimensions). This is an
example of the well known dimensional reduction connected with the zeroth
Matsubara frequency. Consider now the second contribution in the \rhs of Eq.
\Ref{ZMat} being inserted into Eq. \Ref{FEk}. The $t$-integration is 
exponentially convergent for $t\to\infty$. Hence, we can use the \hke
\Ref{hke1} for $K_{P}(t)$ neglecting possible contributions which are
exponentially small for $t\to0$ resulting in exponentially small for
$T\to\infty$ contributions to $F_{E}$. After that, the integration over $t$
can
be carried out and we arrive at
\bea\label{FEh0}F_{E}&=&-{\hbar\over2\beta}\left[\zeta'_{P}(0)
+\zeta_P(0)\,\ln\mu^2\right] \\
&&-{\hbar  \over\beta}{\pa\over\pa
s}\mu^{2s}\sum_{n=0,\frac12,\dots}{a_{n}\over
  (4\pi)^{3/2}} {\Gamma(s+n-\frac32)\over\Gamma(s)} \left({2\pi
    \over\beta c}\right)^{3-2(s+n)} \zeta_{R}(2(s+n)-3). \nn \eea
The limit $s\to0$ can be performed, now. One has to take into account that
there are poles from the gamma function in the numerator for $n=\frac12$ and
for $n=\frac32$. For $n=\frac12$ there is a compensation by $\zeta_{R}(-2)=0$.
For $n=2$ there is a pole from the zeta function so that for $n=\frac32$ and
for $n=2$ there are logarithmic contributions.  Finally (using
\Ref{Riezeta,a})
one obtains for $T\to\infty$
\bea\label{FEh}{F_{E}\over\hbar c}&=&
-{1\over2\beta c}\zeta_{P}'(0)-{a_{0}\over(\beta c)^{4}} {\pi^{2}\over
90}-{a_{1/2}\over(\beta c)^{3}}{\zeta_{R}(3)\over
4\pi^{3/2}}-{a_{1}\over(\beta c)^{2}}{1\over 24}   \\
&&
+{a_{3/2}\over
(\beta c)}{\ln\beta c\mu \over (4\pi)^{3/2}}  -{a_{2}\over 16\pi^{2}}
\left(\gamma+\ln {{\beta c\mu}\over{4\pi}}\right) \nn\\
&&-\sum_{n>2}
{a_{n}\over(\beta c)^{4-2n}} {(2\pi)^{3/2-2n}\over
2\sqrt{2}}\Gamma\left(n-\frac32\right) \zeta_{R}(2n-3) \nn \eea
in agreement with Eq. (41) in \cite{Dowker1978bl} up to a temperature
independent contribution due to the \uv subtraction done there. The result is
remarkable since the high-$T$ expansion is expressed completely
in terms of the \hkks and the determinant of the operator $P$ and in
this way in terms of quantities usually depending only locally on  the
background. 
On the other hand, this is just what is generally
expected from a high energy expansion. Sometimes the high-$T$ expansion is
called the classical limit. For instanse, the contributions from the
determinant and that from $a_{3/2}$ to $F_{E}$ do not contain $\hbar$.

Consider again the simplest example of two conducting plates. All distance
dependent \hkks are zero. The determinant of $P$ which leads to 
nonzero contribution  can  be obtained from the zeta function 
\be\label{}\zeta_{P}(s)=\sum_{n=1}^{\infty}\int{ d k_{1} d
  k_{2}\over(2\pi)^{2}}
\left[k_{1}^{2}+k_{2}^{2}+\left({\pi n\over a}\right)^{2}\right]^{-s}
\ee
for a real scalar field with $\zeta'_{P}(0)=\zeta_{R}(3)/(8\pi a^{2})$ (see
Eqs. \Ref{Espar} and \Ref{Riezeta}).
Substituting this into the first term in the right-hand side of 
(\ref{FEh}) one obtains \Ref{sV1.26} after
multiplication by 2 for the photon polarizations. Another easy example is a
sphere with various \bcp 
In this case the \hkks and the determinant of $P$ which is
the Laplace operator are known (see \cite{Dowker1995j} for Dirichlet and
\cite{Bordag:1995zc} for Robin \bc where also further references 
can be found).

\subsection{Finite conductivity corrections}\label{sec5.2}
The original Casimir result (\ref{FCas}) is valid only for perfect
conductors of infinitely high conductivity, i.e. for the planes
which fully reflect the electromagnetic oscillations of
arbitrary frequency. In fact, for sufficiently high frequency
any metal becomes transparent. This is connected with the finiteness
of its conductivity.
Because of this, there are finite conductivity corrections to
the results like (\ref{FCas}) which are derived for the perfect metal.
These corrections may contribute of order (10--20)\% of the net
result at separations $a\sim1\,\mu$m. Thus, they are very important
in the modern high precision experiments on the Casimir force. In this section we present different approaches to the calculation 
of the finite conductivity corrections in configurations of two
semispaces and a sphere (lens) above a plate.

\subsubsection{Plasma model approach for two semispaces}
\label{sec5.2.1}

The general expression for the Casimir and van der Waals force between
two semispaces made of material with a frequency dependent dielectric
permittivity $\varepsilon_2$ is given by Eq.~(4.25).
It is convenient now to use the notation $\varepsilon$ instead of
$\varepsilon_2$, introduce the new variable $x=2\xi pa/c$ instead of
$\xi$, and change the order of integration. As a result the force
equation takes a form
\bes
&&
F_{ss}^{C}(a)=
-\frac{\hbar c}{32\pi^2a^4}
\int\limits_{0}^{\infty}\! x^3dx
\int\limits_{1}^{\infty}\! \frac{dp}{p^2}\left\{\left[
\frac{(K+p\varepsilon)^2}{(K
-p\varepsilon)^2}e^{x}-1\right]^{-1}\right.
\nonumber\\
&&\phantom{aaaaaaaaaaaaaaaaa}\left.
+\left[
\frac{(K+ p)^2}{(K
- p)^2}e^{x}-1\right]^{-1}\right\},
\label{sV2.1}
\ees
\noindent
where the quantity $K=K(i\xi)=K\left(icx/(2pa)\right)$ is defined
by Eq.~(\ref{sIV23}), and index $C$ is introduced to underline that
the nonideality of the boundary metal is taken into account.

It is common knowledge that the dominant contribution to
the Casimir force comes from frequencies 
$\xi\sim c/a$. We consider the micrometer domain with
$a$ from a few tenths of a micrometer to around a
hundred micrometers. Here the dominant frequencies are
of visible light and infrared optics. In this domain, the
free electron plasma model works well. In the framework of this 
model the dielectric
permittivity of a metal can be represented as
\beq
\varepsilon(\omega)=1-\frac{\omega_p^2}{\omega^2},
\qquad
\varepsilon(i\xi)=1+\frac{\omega_p^2}{\xi^2},
\label{sV2.2}
\eeq
\noindent
where the plasma frequency $\omega_p$ is different for
different metals. It is given by the formula
\beq
\omega_p^2=\frac{4\pi Ne^2}{m^{*}},
\label{sV2.3}
\eeq
\noindent
where $N$ is the density of conduction electrons, $m^{*}$ is
their effective mass. Note that the plasma model does not
take into account relaxation processes. The relaxation
parameter, however, is much smaller than the plasma
frequency. That is why relaxation can play some role only
for large distances between the metal surfaces 
$a\gg \lambda_p=2\pi c/\omega_p$,
where the corrections to the Casimir force due to finite
conductivity are very small.

Let us expand the expression under the integral with 
respect
to $p$ in Eq.~(\ref{sV2.1}) in powers of a small parameter
\beq
\alpha\equiv\frac{\xi}{\omega_p}=
\frac{c}{2\omega_p a}\cdot \frac{x}{p}=
\frac{\delta_0}{a}\cdot\frac{x}{2p},
\label{sV2.4}
\eeq
\noindent
where $\delta_0=\lambda_p/(2\pi)$ is the effective
penetration depth of the electromagnetic 
zero-point oscillations into
the metal. Note that in terms of this parameter
$\varepsilon(\omega)= 1+ (1/\alpha^2)$.
The perturbative expansion of the Casimir force in powers
of the relative penetration depth $\delta_0/a$ can then be
obtained.

Such formulation of the problem was given first in \cite{sIV5},
where the first order finite conductivity correction to the
Casimir force was calculated (with an error in a numerical
coefficient which was corrected in \cite{sV2-1}, see also
\cite{7,SRM}). In \cite{sV2-3} the second order finite
conductivity correction was found in frames of the Leontovich
impedance approach which has seen rapid progress recently
\cite{sV2-4,sV2-5}. Here we present the perturbative results
up to fourth order \cite{31} which give the possibility
to take accurate account of finite conductivity corrections
in a wide separation range. Comparison of the perturbation results
up to the fourth order with the numerical computations using the 
optical tabulated data for the complex refractive index (see below 
Sec.5.2.3) shows that perturbation theory works well for separations 
$a\geq\lambda_p$ (not
$a\gg\lambda_p$ as one would expect from general considerations).

After straightforward calculations one obtains the expansion
of the first term in the integrand of Eq.~(\ref{sV2.1})
\bes
&&
\left[
\frac{(K+p\varepsilon)^2}{(K-p\varepsilon )^2}
e^{x}-1\right]^{-1}=
\frac{1}{e^x-1}\left[\vphantom{\frac{A}{p^4}}
1-\frac{4A}{p}\alpha +
\frac{8A}{p^2}(2A-1)\alpha^2\right.
\nonumber\\
&&\phantom{aaaaaaa}
+\frac{2A}{p^3}(-6+32A-32A^2+2p^2-p^4)\alpha^3
\label{sV2.5}\\
&&\phantom{aaaaaaa}\left.
+\frac{8A}{p^4}(2A-1)(2-16A+16A^2-2p^2+p^4)\alpha^4
+O(\alpha^5)\right],
\nonumber
\ees
\noindent
where $A\equiv e^x/(e^x-1)$, $\alpha$ is defined in Eq.~(\ref{sV2.4}).

In perfect analogy, the other contribution from Eq.~(\ref{sV2.1})
is 
\bes
&&
\left[
\frac{(K+p)^2}{(K-p)^2}
e^{x}-1\right]^{-1}=
\frac{1}{e^x-1}\left[
1-4Ap\alpha +
8A(2A-1)p^2\alpha^2\right.
\nonumber\\
&&\phantom{aaaaaaa}
+2A(-5+32A-32A^2)p^3\alpha^3
\label{sV2.6}\\
&&\phantom{aaaa}\left.
+8A(1+18A-48A^2+32A^3)p^4\alpha^4
+O(\alpha^5)\right]
\nonumber
\ees
\noindent 
(note that this expression actually does not depend on
$p$ due to (\ref{sV2.4})).

After substitution of (\ref{sV2.5}) and (\ref{sV2.6}) into
(\ref{sV2.1}) all integrals with respest to $p$ have the
form $\int_{0}^{\infty}dpp^{-k}$ with $k\geq 2$ and are
calculated next. The integrals with respect to
$x$ have the form
\beq
\int\limits_{0}^{\infty}dx
\frac{x^ne^{mx}}{(e^x-1)^{m+1}}
\label{sV2.7}
\eeq
\noindent
and can be easily calculated with the help of \cite{sV2-6}.
Substituting their values into (\ref{sV2.1}) we obtain
after some transformations the Casimir force between
metallic plates with finite conductivity corrections
up to the fourth power in relative penetration depth
\cite{31}
\bes
&&
F_{ss}^{C}(a)=F_{ss}^{(0)}(a)\left[
1-\frac{16}{3}\frac{\delta_0}{a}+
24\frac{\delta_0^2}{a^2}-
\frac{640}{7}\left(1-\frac{\pi^2}{210}\right)
\frac{\delta_0^3}{a^3}\right.
\nonumber\\
&&
\phantom{aaaaaaaaaaaaa}\left.+
\frac{2800}{9}\left(1-\frac{163\pi^2}{7350}\right)
\frac{\delta_0^4}{a^4}\right],
\label{sV2.8}
\ees
\noindent
where $F_{ss}^{(0)}(a)$ is defined in Eq.~(\ref{sIV31}).
The first order term of this expansion was obtained first
in \cite{sIV5,sV2-1}, and the second order one ---
in \cite{sV2-3}.

In Sec.5.2.3 the dependence of Eq.~(\ref{sV2.8}) is displayed
graphically for aluminium and gold in comparison with
numerical computations demonstrating high accuracy of the
perturbation result for all separations between semispaces
larger than the plasma wave length of a corresponding
metal. 

\subsubsection{Plasma model approach for a sphere (lens)
above a disk} 
\label{sec5.2.2}

We consider now the plasma model perturbation approach in
configuration of a sphere (lens) above a disk. The sphere (lens)
radius $R$ is suggested to be much larger than the
sphere-disk separation $a$. Thanks to this the Proximity
Force Theorem is valid (see Sec.4.3) and the Casimir force
is given by Eqs.~(\ref{sIV3.5}) and (\ref{sIV26}).
Introducing once more the variable $x=2\xi pa/c$ the following
result is obtained
\bes
&&
F_{dl}^{C}(a)=
\frac{\hbar cR}{16\pi a^3}
\int\limits_{0}^{\infty}\! x^2dx
\int\limits_{1}^{\infty}\! \frac{dp}{p^2}
\left\{\ln\left[1-
\frac{(K-p\varepsilon)^2}{(K+p\varepsilon )^2}
e^{-x}\right]\right.
\nonumber\\
&&\phantom{aaaaaaaaaaaaaa}\left.
+\ln\left[1-
\frac{(K-p)^2}{(K+p)^2}e^{-x}\right]\right\}.
\label{sV2.9}
\ees

Bearing in mind the need to do perturbative expansions it is convenient
to perform in (\ref{sV2.9}) an integration by parts with respect
to $x$. The result is
\bes
&&
F_{dl}^{C}(a)=
-\frac{\hbar cR}{48\pi a^3}
\int\limits_{0}^{\infty}\! x^3dx
\int\limits_{1}^{\infty}\! \frac{dp}{p^2}
\left[
\frac{(K-p\varepsilon)^2-(K+p\varepsilon )^2
\frac{\partial}{\partial x}\frac{(K-
p\varepsilon)^2}{(K+p\varepsilon )^2}}{(K+
p\varepsilon)^2 e^x-(K-p\varepsilon )^2}
\right.\nonumber \\
&&\phantom{aaaa}\left.+
\frac{(K-p)^2-(K+p)^2\frac{\partial}{\partial x}
\frac{(K-p)^2}{(K+p)^2}}{(K+p)^2 e^x-(K-p)^2}
\right].
\label{sV2.10}
\ees

The expansion of the first term under the integral in
powers of the parameter $\alpha$ 
introduced in (\ref{sV2.4}) is
\bes
&&
\frac{(K-p\varepsilon)^2-(K+p\varepsilon )^2
\frac{\partial}{\partial x}\frac{(K-
p\varepsilon)^2}{(K+p\varepsilon )^2}}{(K+
p\varepsilon)^2 e^x-(K-p\varepsilon )^2}
=
\frac{1}{e^x-1}\left\{\vphantom{\frac{A}{p^4}}
1+\frac{4}{px}(1-Ax)\alpha \right.
\nonumber\\
&&\phantom{aaaaa}
+
\frac{8A}{p^2x}(-2-x+2Ax)\alpha^2
+\frac{2}{p^3x}\left[\vphantom{A^2}
2-6p^2+3p^4\right.
\nonumber\\
&&\phantom{aaaaa}\left.
+Ax(-6+32A-32A^2+2p^2-p^4)
+16A(2A-1)\right]\alpha^3
\nonumber\\
&&\phantom{aaaaa}
+\frac{8A}{p^4x}\left[-8+32A-32A^2+8p^2-4p^4\right.
\label{sV2.11}\\
&&\phantom{aaaaaaaa}\left.\left.+
x(2A-1)(2-16A+16A^2-2p^2+p^4)\right]\alpha^4
+O(\alpha^5)
\vphantom{\frac{A}{p^4}}\right\}.
\nonumber
\ees 

In the same way for the second term under the integral
of (\ref{sV2.10}) one obtains
\bes
&&
\frac{(K-p)^2-(K+p)^2\frac{\partial}{\partial x}
\frac{(K-p)^2}{(K+p)^2}}{(K+p)^2 e^x-(K-p)^2}
=
\frac{1}{e^x-1}\left[
1+\frac{4}{x}(1-Ax)p\alpha \right.
\label{sV2.12}\\
&&\phantom{aaaaa}
+
\frac{8A}{x}(-2-x+2Ax)p^2\alpha^2
+\frac{2}{x}\left(\vphantom{A^2}
-1-16A+32A^2-5Ax\right.
\nonumber\\
&&\phantom{aaaaa}
\left.+32A^2x-32A^3x\right)
p^3\alpha^3
+\frac{8A}{x}\left(-4+32A-32A^2-x\right.
\nonumber\\
&&\phantom{aaaaaaaa}\left.\left.
+18Ax-48A^2x+32A^3x\right)p^4\alpha^4
+O(\alpha^5)
\right].
\nonumber
\ees 

Substituting Eqs.~(\ref{sV2.11}) and (\ref{sV2.12}) 
into Eq.~(\ref{sV2.10})
we first calculate integrals with respect to $p$.
All integrals with respect to $x$ are of the form
(\ref{sV2.7}). Calculating these we come to the
following result after some tedious algebra \cite{31}
\bes
&&
F_{dl}^{C}(a)=F_{dl}^{(0)}(a)\left[
1-4\frac{\delta_0}{a}+
\frac{72}{5}\frac{\delta_0^2}{a^2}-
\frac{320}{7}\left(1-\frac{\pi^2}{210}\right)
\frac{\delta_0^3}{a^3}\right.
\nonumber\\
&&\phantom{aaaaaaaaaaaaaaaa}
\left.+
\frac{400}{3}\left(1-\frac{163\pi^2}{7350}\right)
\frac{\delta_0^4}{a^4}\right],
\label{sV2.13}
\ees
\noindent
where $F_{dl}^{(0)}(a)$ is defined in Eq.~(\ref{sIV3.7}).

Although the results (\ref{sV2.8}) and (\ref{sV2.13}) for the two
configurations were obtained
independently they can be related by the use of Proximity
Force Theorem.
By way of example, the energy density
associated with the fourth order contribution 
in (\ref{sV2.8}) is
\beq
E_{ss}^{C(4)}(a)=
\int\limits_{a}^{\infty}F_{ss}^{C,(4)}(a)da=
-\frac{5\pi^2\hbar c }{27}
\left(1-\frac{163\pi^2}{7350}\right)
\frac{\delta_0^4}{a^7}.
\label{sV2.14}
\eeq
\noindent
Then the fourth order contribution to the force between
a disk and a lens given by
\beq
F_{dl}^{C(4)}(a)=2\pi RE_{ss}^{C,(4)}(a)=
-\frac{10\pi^3\hbar c R}{27a^3}
\left(1-\frac{163\pi^2}{7350}\right)
\frac{\delta_0^4}{a^4}
\label{sV2.15}
\eeq
\noindent
 is in agreement with (\ref{sV2.13}). The other coefficients
of (\ref{sV2.13}) can be verified in the same way.

Note that the linear and quadratic corrections from 
Eq.~(\ref{sV2.13}) for the configuration of a sphere (lens)
above a disk were first obtained in 
\cite{32,sV2-7}, respectively. 
In Sec.5.2.3 Eq.~(\ref{sV2.13}) is displayed 
graphically and the comparison with numerical computations is
made showing the excellent agreement for all $a\geq\lambda_p$.

\subsubsection{Computational results using the optical tabulated data}
\label{sec5.2.3}

The plasma model representation for the dielectric permittivity
(\ref{sV2.2}) was applied above to calculate the finite conductivity
corrections to the Casimir force. The obtained perturbation results
are adequate in some distance range. The plasma model does not take
into account, however, the absorption bands of the boundary metal
and the relaxation of conduction electrons. In addition, the plasma
frequency of the metal under consideration (e.g. aluminium or gold)
is not known very precisely. Because of this in \cite{sV23-1} the
Lifshitz formalism was applied numerically to different metals.
For this purpose the tabulated data for the frequency dependent
complex refractive index of that metals were used together with the
dispersion relation to calculate the values of dielectric permittivity
on the imaginary frequency axis. Thereupon the Casimir force was
calculated numerically for configurations of two plates and a spherical
lens above a plate. As shown in \cite{27} (see also \cite{sV23-2})
computations of \cite{sV23-1} contain errors in the interpolation and
extrapolation procedures which resulted in the deviations of the
obtained results from the correct values. The same computations were
performed in \cite{28} in a wider range of space separations and with
account of thin layers covering the metallic surfaces. The results of
\cite{27} and \cite{28} are in agreement. Let us discuss them in more
detail.

We begin from the force per unit area for the configuration of two
semispaces or the force for a sphere (lens) above a semispace given by
Eq.~(\ref{sIV25}), and Eqs.~(\ref{sIV26}) and (\ref{sIV3.5}).
To calculate numerically the corrections to the results for ideal
metal $F_{ss}^{(0)}$, $F_{dl}^{(0)}$ 
due to the finite
conductivity we use the tabulated data for the complex index
of refraction $n+ik$ as a function of frequency \cite{sV23-3}. 
The values of
dielectric permittivity along the imaginary axes can be expressed through 
Im$\varepsilon(\omega)=2nk$ with the help of dispersion relation 
\cite{sIV1}
\beq
\varepsilon(i\xi)=1+\frac{2}{\pi}
\int\limits_{0}^{\infty}
\frac{\omega\,\mbox{Im}\varepsilon(\omega)}{\omega^2+\xi^2}d\omega.
\label{sV23.1}
\eeq
\noindent

All calculations were performed in \cite{28}
for $Al$ and $Au$ surfaces because these
metals were used in the recent experiments on the Casimir force
measurements (see Sec.6).
The complete tabulated refractive indices extending from 0.04\,eV 
to 10000\,eV for Al and from 0.1\,eV to 10000\,eV for Au from 
\cite {sV23-3} are 
used to calculate Im$\varepsilon(\omega)$. For frequencies below 0.04 eV 
in the case of Al and below 0.1 eV in the case of Au, the table values 
can be extrapolated using the Drude model, which is more exact than 
the plasma one because it takes relaxation into account. 
In this case, the dielectric
permittivity along the imaginary axis is represented as:
\beq
\varepsilon(i\xi)=1+\frac{\omega_{p}^2}{\xi(\xi+\gamma)},
\label{sV23.2}
\eeq
\noindent
where $\omega_{p}=(2\pi c)/\lambda_{p}$ is the plasma frequency 
and $\gamma$ is the relaxation frequency.  
The values $\omega_{p}$=12.5\,eV and 
$\gamma$=0.063\,eV were used for the case of $Al$ based on the last 
results in Table XI 
on p.394 of \cite{sV23-3}. Note that the Drude model was used 
to describe the dielectric
function of a sphere which undergoes the Casimir attraction to a perfectly
reflecting wall \cite{sV23-3a}.

\begin{figure}[ht]
\setlength{\unitlength}{1cm}
\begin{flushleft}
\hspace*{-3cm}
\epsfig{figure=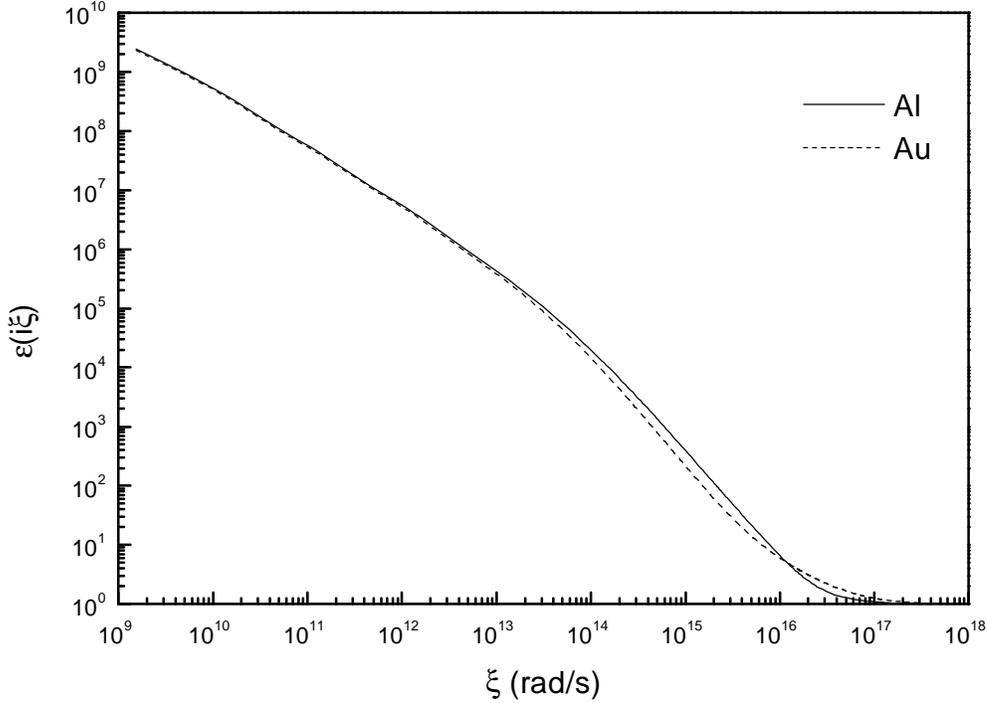}
\end{flushleft}
\vspace*{-12cm}
\caption{\label{figlay1}
The dielectric permittivity as a function of imaginary frequency for
$Al$ (solid line) and $Au$ (dashed line).
}
\end{figure}

In the case of $Au$ the analysis is not as straightforward, 
but proceeding in the manner outlined in \cite{27} we obtain 
$\omega_{p}$=9.0\,eV and $\gamma$=0.035\,eV. While the values of $\omega_{p}$ 
and $\gamma$ based on the optical data of various sources might differ 
slightly we have found that the resulting numerically computed Casimir 
forces to differ by less than 1\%. In fact, if for $Al$ metal, a 
$\omega_p$=11.5\,eV and $\gamma$=0.05\,eV
as in \cite{27} is used, the differences are extremely small. Of the values 
tabulated 
below, only the value of the 
force in the case of a sphere and a semispace at 0.5\,$\mu$m separation is 
increased by 0.1\%, which on round-off to the second significant figure 
leads to an increase of 1\%.
The results of numerical integration 
by Eq.~(\ref{sV23.1}) for $Al$ (solid
curve) and $Au$ (dashed curve) are presented in Fig.~\ref{figlay1} 
on a logarithmic
scale. 
As is seen from Fig.~\ref{figlay1}, the dielectric permittivity along the
imaginary axis decreases monotonically with increasing frequency (in
distinction to Im$\varepsilon(\omega)$ which possesses peaks corresponding
to interband absorption).

\begin{figure}[ht]
\setlength{\unitlength}{1cm}
\begin{flushleft}
\hspace*{-3cm}
\epsfig{figure=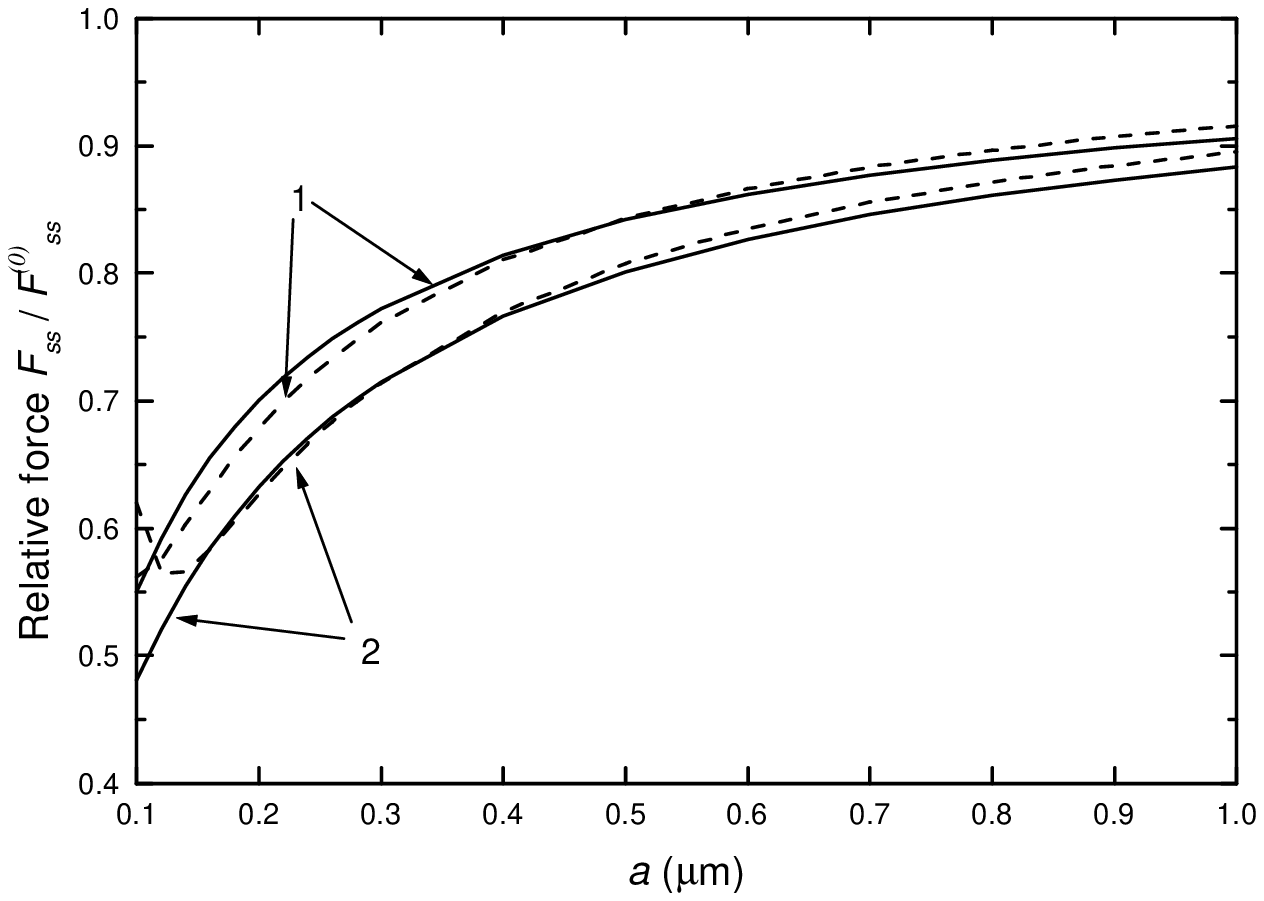}
\end{flushleft}
\vspace*{-12cm}
\caption{\label{figlay2}
The correction factor to the Casimir force due to finite conductivity of
the metal as a function of the surface separation in the
configuration of two semispaces.
The solid lines 1 and 2 represent the computational results 
for $Al$ and $Au$ respectively.
The dashed lines 1 and 2
represent
the perturbation correction factor up to the 4th order for the
same metals.
}
\end{figure}

\begin{figure}[ht]
\setlength{\unitlength}{1cm}
\begin{flushleft}
\hspace*{-3cm}
\epsfig{figure=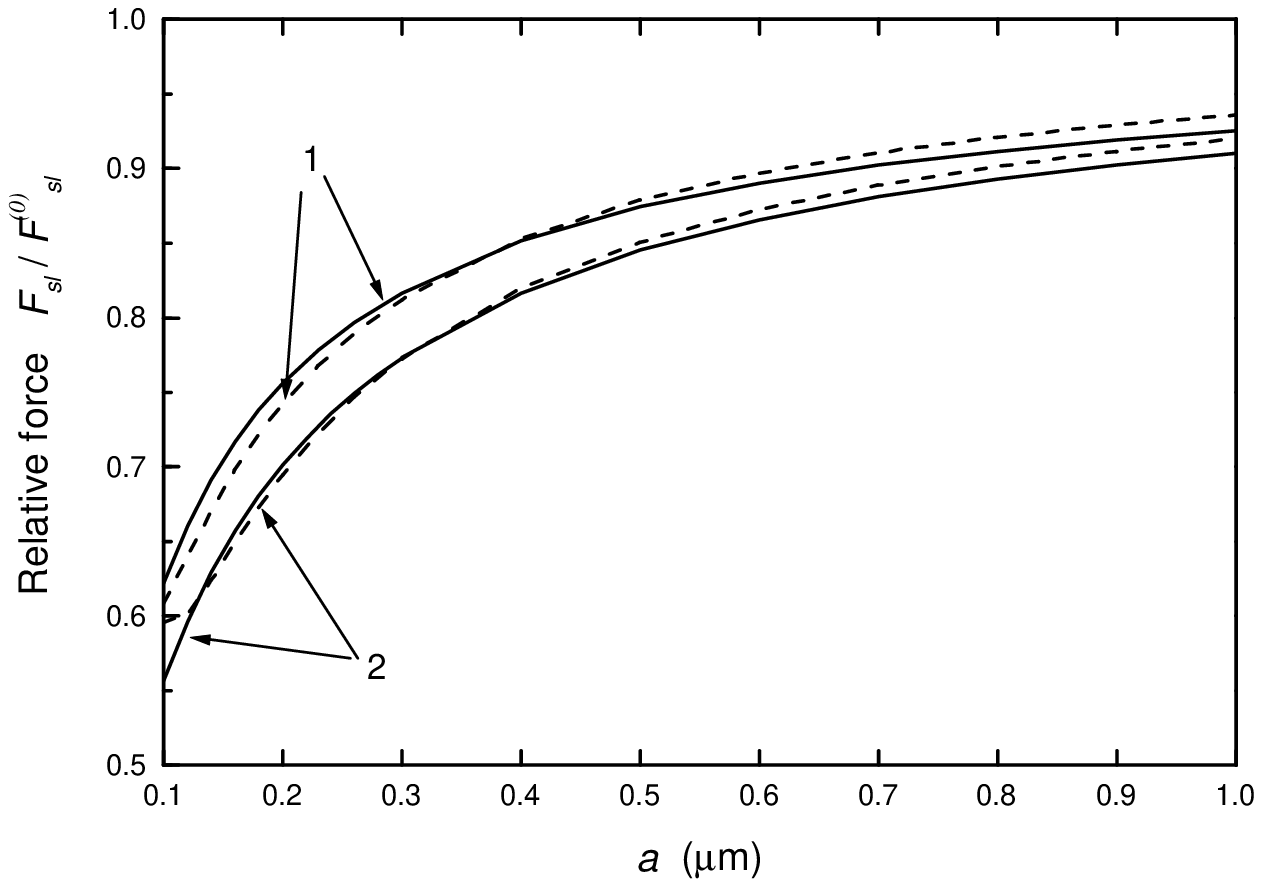}
\end{flushleft}
\vspace*{-12cm}
\caption{\label{figlay3}
The correction factor to the Casimir force due to finite conductivity of
the metal as a function of the surface separation
for a sphere (lens) above a disk.
The solid lines 1 and 2 represent the computational results 
for $Al$ and $Au$ respectively.
The dashed lines 1 and 2 represent
the perturbation correction factor up to the 4th order for 
the same metals.
}
\end{figure}

The obtained values of the dielectric permittivity along the imaginary axis 
were substituted into Eqs.~(\ref{sIV25}) and (\ref{sIV26}),
(\ref{sIV3.5}) 
to calculate the Casimir force acting between real metals in
configurations of two semispaces (ss) and a sphere (lens) above 
a disk (dl).  Numerical integration was done from an upper limit 
of $10^{4}\,$eV to a lower limit of $10^{-6}\,$eV.  Changes in the 
upper limit or 
lower limit by a factor of 10 lead to changes of less than 0.25\% in 
the Casimir force.  If the trapezoidal rule is used in the numerical 
integration of Eq.~(\ref{sV23.1}) the corresponding Casimir force decreases 
by a factor less than 0.5\%.   The results are presented 
in Fig.~\ref{figlay2} 
(two semispaces) 
and in Fig.~\ref{figlay3} for a sphere
above a disk by the solid lines 1 (material of the test bodies is
aluminium) and 2 (material is gold). 
In the vertical axis the relative
force $F_{ss}^{C}/F_{ss}^{(0)}$ is plotted in Fig.~\ref{figlay2} 
and $F_{dl}^{C}/F_{dl}^{(0)}$ in Fig.~\ref{figlay3}.
These quantities provide a sense of the correction factors to the Casimir
force due to the effect of finite conductivity. In the horizontal axis
the space separation is plotted in the range 0.1--1$\,\mu$m. We do not
present the results for larger distances because then the temperature
corrections to the Casimir force become significant.
 At room temperature
the temperature corrections contribute only 2.6\% of $F_{dl}^{(0)}$ at
$a=1\,\mu$m, but at $a=3\,\mu$m they contribute 47\% of $F_{dl}^{(0)}$,
 and
at $a=5\,\mu$m --- 129\% of $F_{dl}^{(0)}$ \cite{37}. It is seen that the
relative force for $Al$ is larger than for $Au$ at the same separations
as it should be because of better reflectivity properties of $Al$.
At the same time the relative force for $Cu$ is almost the same as
for $Au$ \cite{27}.

The computational results presented here are in good agreement with 
analytical perturbation expansions of the Casimir force in powers of
relative penetration depth of the zero-point oscillations into the
metal (see the two above sections).
In Fig.~\ref{figlay2} (two semispaces) 
the dashed line 1 represents the results
obtained by (\ref{sV2.8}) for $Al$ with $\lambda_p=107\,$nm
(which corresponds to $\omega_p=11.5\,$eV), and the dashed
line 2 --- the results obtained by (\ref{sV2.8}) for $Au$ with
$\lambda_p=136\,$nm ($\omega_p=9\,$eV) \cite{27}. 
In Fig.~\ref{figlay3} the dashed lines 1 and 2 
represent the
perturbation results obtained for $Al$ and $Au$ by (\ref{sV2.13}) for a lens
above a disk. As one can see from the figure,
the perturbation results are in good (up to 0.01) agreement with
computations for all distances larger than $\lambda_p$. Only at
$a=0.1\,\mu$m for $Au$ there are larger deviations because
$\lambda_{p1}\equiv\lambda_p^{Au}>0.1\,\mu$m.
This proves the fact that the perturbation expansions (\ref{sV2.8})
and (\ref{sV2.13}) are applicable with rather high accuracy not
for $a\gg\lambda_p$, as could be expected from general considerations,
but for all $a\geq\lambda_p$.

The same formalism gives the possibility to consider the influence of
thin outer
 metallic layers
on the Casimir force value \cite{28}. 
Let the semispace made of $Al$ ($\varepsilon_2$)
be covered by $Au$ ($\varepsilon_1$) layers as shown in Fig.~{\ref{fig4.1}.
For a configuration of a sphere above a plate such covering made of $Au/Pd$
was used in experiments \cite{33,34,35} with different values of layer
thickness $d$. In this case the Casimir force is given by the 
Eqs.~(\ref{sIV24}), and (\ref{sIV21}), (\ref{sIV3.5}),
where the quantities $Q_{1,2}(i\xi)$ are expressed by 
Eqs.~(\ref{sIV22}), (\ref{sIV23}). The computational results for
$\varepsilon(i\xi)$ are obtained from
Eq.~(\ref{sV23.1}). Substituting them into (\ref{sIV24})
and (\ref{sIV21}), (\ref{sIV3.5}) and
performing  a numerical integration in the same way as above
one obtains the Casimir force including the effect of covering layers.

The numerical computations described above show that
a  $Au$ layer of $d=20\,$nm thickness
significantly decreases the relative Casimir force between $Al$ surfaces.
With this layer the force approaches the value for pure $Au$
semispaces. For a thicker $Au$ layer of $d=30\,$nm thickness the relative
Casimir force is scarcely affected by the underlying $Al$. For example, 
at a space
separation $a=300\,$nm in the configuration of two semispaces we have
$F_{ss}^{C}/F_{ss}^{(0)}=0.773$ for pure $Al$,
$F_{ss}^{C}/F_{ss}^{(0)}=0.727$ for $Al$ with 20\,nm $Au$ layer,
$F_{ss}^{C}/F_{ss}^{(0)}=0.723$ for $Al$ with 30\,nm $Au$ layer,
and $F_{ss}^{C}/F_{ss}^{(0)}=0.720$ for pure $Au$.
In the same way for the configuration of a sphere above a disk 
the results
are:
$F_{dl}^{C}/F_{dl}^{(0)}=0.817$ (pure $Al$),
0.780 ($Al$ with 20\,nm $Au$ layer),
0.776 ($Al$ with 30\,nm $Au$ layer),
and 0.774 (for pure $Au$). 

Let us now discuss the application range of the obtained results for 
the case of covering layers \cite{28}. 
First, from a theoretical standpoint, the main
question concerns the layer thicknesses to which the obtained formulas 
(\ref{sIV24}), and (\ref{sIV21}), (\ref{sIV3.5})
and the above computations can be applied. In the derivation of
Sec.~4.1.1 the spatial dispersion is neglected and, as a consequence, the
dielectric permittivities $\varepsilon_{\alpha}$ depend only on $\omega$
not on the wave vector {\boldmath$k$}. In other words the field of vacuum
oscillations is considered as time dependent but space homogeneous. Except
for the thickness of a skin layer $\delta_0$ the main parameters of our problem
are the velocity of the electrons on the Fermi surface, $v_F$, the
characteristic frequency of the oscillation field, $\omega$, and the mean free 
path of the electrons, $l$. For the considered region of high
frequencies (micrometer distances between the test bodies) the following
conditions are valid \cite{sV23-4}
\beq
\frac{v_F}{\omega}<\delta_0\ll l.
\label{sV23.3}
\eeq
\noindent
Note that the quantity $v_F/\omega$ in the left-hand side of 
Eq.~(\ref{sV23.3})
is the distance travelled by an electron during one period of the field, so
that the first inequality is equivalent to the assumption of spatial
homogeneity of the oscillating field. Usually the corresponding frequencies
start from the far infrared part of spectrum, which means the space
separation $a\sim 100\,\mu$m \cite{20}. The region of high frequencies is
restricted by the short-wave optical or near ultraviolet parts of the spectrum
which correspond to the surface separations of several hundred nanometers.
For smaller distances absorption bands, photoelectric effect and other physical
phenomena should be taken into account. For these phenomena, the general 
Eqs.~(\ref{sIV24}) and  (\ref{sIV21}), (\ref{sIV3.5}), 
however, are still valid if one substitutes the experimental
tabulated data for the dielectric permittivity along the imaginary axis
incorporating all these phenomena.

Now let us include one more physical parameter --- the thickness $d$ of the
additional, i.e. $Au$, covering layer.
It is evident that Eqs.~(\ref{sIV24})  and (\ref{sIV21}), (\ref{sIV3.5}) 
are applicable only for
layers of such thickness that
\beq
\frac{v_F}{\omega}<d.
\label{sV23.4}
\eeq
\noindent
Otherwise an electron goes out of the thin layer during one period
of the oscillating field and the approximation of space homogeneity is 
not valid. If $d$ is so small that the inequality 
(\ref{sV23.4}) is violated, 
the spatial
dispersion should be taken into account which means that the dielectric
permittivity would depend not only on frequency but on a wave vector also:
$\varepsilon_1=\varepsilon_1(\omega,{\mbox{\boldmath$k$}})$. 
So, if (\ref{sV23.4})
is violated, the situation is analogous to the anomalous skin effect where
only space dispersion is important and the inequalities below are valid
\beq
\delta_0(\omega)<\frac{v_F}{\omega}, 
\qquad \delta_0(\omega)<l.
\label{sV23.5}
\eeq
\noindent
In our case, however, the role of $\delta_0$ is played by the layer
thickness $d$ (the influence of nonlocality effects on van der Waals force
is discussed in \cite{sV23-5,sV23-6}).

From (\ref{sV23.3}) and (\ref{sV23.4}) it follows that for pure $Au$ layers
($\lambda_p\approx 136\,$nm) the space dispersion can be neglected only if
$d\geq(25-30)\,$nm. For thinner layers a more general theory taking into
account nonlocal effects should be developed to calculate the Casimir force.
Thus for such thin layers the bulk tabulated data of the dielectric 
permittivity
depending only on frequency cannot be used (see experimental investigation
\cite{sV23-7} demonstrating that for $Au$ the bulk values of dielectric
constants can only be obtained from films whose thickness is about 30\,nm
or more). That is why the above
calculated results for the case of $d=20\,$nm
are subject to corrections due to the influence of spatial dispersion.
From an experimental 
standpoint thin layers of order a few nm grown by evaporation or sputtering 
techniques are 
highly porous. This is particularly so in the case of sputtered 
coatings as shown in \cite{sV23-8}. 
The nature of porosity is a function of the material and 
the underlying substrate. Thus it should be noted that the theory presented 
here which 
used the bulk tabulated data for
$\varepsilon_1$ cannot be applied to calculate the influence of thin
covering layers of $d=20\,$nm like those used in \cite{26,33} 
and of $d=8\,$nm used in \cite{34,35}
on the Casimir force. The measured high transparency of such layers for the
characteristic frequencies \cite{26,33} corresponds to a larger change of 
the force
than what follows from Eqs.~(\ref{sIV24}) and (\ref{sIV21}). 
This is in agreement with the above qualitative analyses.

Note that the role of spatial dispersion was neglected in \cite{sV23-9}
where the van der Waals interaction was considered between the metallic 
films of several nanometer thickness. According to the above 
considerations such neglect is unjustified. With account of spatial
dispersion the Casimir attraction between a bulk conductor and a
metal film deposited on a dielectric substrate was studied in
\cite{sV23-10}.

As is seen from Figs.~\ref{figlay2},\,12 at room temperature the Casimir 
force does not follow its ideal field-theoretical expressions 
$F_{ss}^{(0)}$, $F_{dl}^{(0)}$.
 For the
space separations less than $a=1\,\mu$m the corrections due to finite
conductivity of the metal are rather large 
(thus, at $a=1\,\mu$m they are
around 7--9\% for a lens above a disk, and 10--12\% for two
semispaces; at $a=0.1\,\mu$m, around 38--44\% (dl), and 45--52\% (ss)).
For $a>1\,\mu$m the temperature corrections increase very quickly (see
Sec.~5.1). Actually, the range presented in 
Figs.~\ref{figlay2},\,\ref{figlay3} 
is the beginning 
of a transition with decreasing $a$ from the Casimir force to the van 
der Waals force. 
In \cite{28} the intermediate region is investigated in more
detail for smaller $a$ and the  values of $a$ are found where the pure
(nonretarded) van der Waals regime described by Eq.~(\ref{sIV27}) starts.
In doing so, more exact values of the Hamaker constants for $Al$ and
$Au$ were also calculated
\beq
H^{Al}=(3.6\pm 0.1)\times 10^{-19}\,\mbox{J}, \qquad
H^{Au}=(4.4\pm 0.2)\times 10^{-19}\,\mbox{J}
\label{sV23.6}
\eeq
\noindent
(see Eq.~(\ref{sIV28}) for the determination of $H$).

\subsection{Roughness corrections}
\label{sec5.3}

The next point characterizing the real media is the imperfectness
of their boundaries. In reality there are always small deviations 
from the perfect geometry, whether it be the two plane parallel plates 
or a spherical lens (sphere) situated above a plate. These deviations
can be of different types. For example, plates can be under some
nonzero angle to each other. In a more general way, the boundary
surfaces may have a point dependent small deviations from the perfect
plane, spherical or cylindrical shape. In all cases ``small'' means
that the characteristic distortion amplitude $A$ is much less than the
space separation $a$ between the two bodies. The distortion period $T$ 
can be much larger, of order of or much smaller compared to $a$.
In the latter case one may speak about short scale distortions
describing surface roughness. Some kinds of roughness can be
described also by large scale distortions. Roughness is necessarily
present on any real surface and contribute to the value of the
Casimir force. Its contribution is rather large at $a\sim 1\,\mu$m and
should be taken into account when comparing theory with experiment.

The problem of roughness corrections has long attracted the attention
of researchers (see, e.g., \cite{sV3-1,sV3-2} where the roughness
corrections to the nonretarded van der Waals force were found).
In principle they can be calculated with perturbation theory based on
the Green function method \cite{bali78-112-165,sIV3-1}.
In doing so a small parameter characterizes the deviation from the 
basic geometry. Also the formalism based on functional integration
can be applied for this purpose \cite{sV3-3,sV3-4}. However, the
resulting expressions turn out to be quite complicated and not very
effective for specific applications. Because of this, we use here
phenomenological methods of Sec.4.3 to calculate corrections to the
Casimir force due to imperfectness of the boundary geometry
\cite{10,sV2-7,sV3-5,sV3-7,sV3-8,sV3-8a,sV3-10}. As is shown below, for
small deviations from the plane parallel geometry the accuracy of
these methods is very high, so that obtained results are quite
reliable and can be used for the interpretation of the modern
precision experiments on Casimir force measuring. We consider first the
configuration of two plane parallel plates and next a sphere (spherical 
lens) above a plate which are the two cases of experimental interest.

\subsubsection{Expansion in powers of relative distortion amplitude:
two semispaces}
\label{sec5.3.1}

Consider two semispaces modeled by plates which are made of a
material with a static dielectric permittivity $\varepsilon_0$ bounded
by  surfaces with small deviations from plane geometry. The
approximate expression for the Casimir energy in this configuration
is given by Eq.~(\ref{sIV3.12}), where the function $\Psi$ was defined
in (\ref{sIV30}), and we now change the notation $\varepsilon_{20}$
used in Sec.4.1.1 for $\varepsilon_0$.

As zeroth approximation in the perturbation theory we consider the
Casimir force between the square plane plate $P_1$ with the sidelength
$2L$ and the thickness $D$ and the other plane plate $P_2$, which is
parallel to $P_1$ and has the same length and thickness. Our aim is
to calculate the Casimir force between plates whose surfaces possess
some small deviations from the plane geometry. Let us describe the
surface of the first plate by the equation
\beq
z_1^{(s)}=A_1f_1(x_1,y_1)
\label{sV3.1}
\eeq
and the surface of the second plate by
\beq
z_2^{(s)}=a+A_2f_2(x_2,y_2),
\label{sV3.2}
\eeq
\noindent
where $a$ is the mean value of the distance between the plates.
The values of the amplitudes are chosen in such a way that
$\max\left| f_i(x_i,y_i))\right| =1$.
It is suitable to choose the zero point in the $z$-axis so that
\bes
&&
\langle z_1^{(s)}\rangle\equiv A_1\langle f_1(x_1,y_1)\rangle
\equiv\frac{A_1}{(2L)^2}\int\limits_{-L}^{L}dx_1
\int\limits_{-L}^{L}dy_1\,f_1(x_1,y_1)=0,
\nonumber\\
&&
\langle z_2^{(s)}\rangle\equiv a+A_2\langle f_2(x_2,y_2)\rangle
=a.
\label{sV3.3}
\ees

We assume in our perturbation expansion that $A_i\ll a$,
$a\ll D$, and $a\ll L$. At the same time, in all real situations,
we have $a/D,\,a/L\ll A_i/a$, so that we are looking for the
perturbation expansion in the powers of $A_i/a$ and in the zeroth
orders in $a/D$ and $a/L$.

The non-normalized potential of one atom at a height $z_2$
over the plate $P_1$ is given by
\bes
&&
U_A(x_2,y_2,z_2)=-CN\int\limits_{-L}^{L}dx_1
\int\limits_{-L}^{L}dy_1
\int\limits_{-D}^{A_1f_1(x_1,y_1)}dz_1
\nonumber\\
&&\phantom{aaaaaaaaaaaa}
\times\left[(x_2-x_1)^2+(y_2-y_1)^2+(z_2-z_1)^2\right]^{-7/2},
\label{sV3.4}
\ees
\noindent
where $C$ is an interaction constant from Eq.~(\ref{sIV3.8}),
$N$ is the number of atoms per unit volume of plates $P_1$ and $P_2$.

Let us calculate this expression as a series with respect to the
parameter $A_1/z_2$, which is small due to $z_2\geq a\gg A_1$.
In carrying out these calculations we can neglect the corrections of
order of $z_2/L$ and $z_2/D$, i.e. assume that the thickness and
length of the sides of the first plate are infinitely large.

The result of the expansion up to the fourth order with respect to
the small parameter $A_1/z_2$ may be written in the form
\bes
&&
U_A(x_2,y_2,z_2)=-CN\left\{\frac{\pi}{10z_2^4}+
\int\limits_{-L}^{L}dx_1\int\limits_{-L}^{L}dy_1\right.
\nonumber\\
&&\phantom{U_A(x_2,y_2,z_2)=}
\times\left[\frac{z_2f_1(x_1,y_1)}{X^{7/2}}
\left(\frac{A_1}{z_2}\right)+
\frac{7z_2^3f_1^2(x_1,y_1)}{2X^{9/2}}
\left(\frac{A_1}{z_2}\right)^2\right.
\nonumber\\
&&\phantom{U_A(x_2,y_2,z_2)=}
+\frac{7z_2^3}{6X^{9/2}}\left(\frac{9z_2^2}{X}-1\right)
f_1^3(x_1,y_1)\left(\frac{A_1}{z_2}\right)^3
\nonumber\\
&&\phantom{U_A(x_2,y_2,z_2)=}
\left.\left.
+\frac{21z_2^5}{8X^{11/2}}\left(\frac{11z_2^2}{X}-3\right)
f_1^4(x_1,y_1)\left(\frac{A_1}{z_2}\right)^4\right]
\vphantom{\int\limits_{-L}^{L}}\right\},
\label{sV3.5}
\ees
\noindent
with $X=(x_1-x_2)^2+(y_1-y_2)^2+z_2^2$. Here, the limit
$L\to\infty$ is performed in the first item, which describes the
perfect plates without deviations from planar case.

The normalized potential of the Casimir force between the plates may 
be obtained by the integration of Eq.~(\ref{sV3.5}) over the volume 
$V_2$ of the second plate using the boundary function $A_2f_2(x_2,y_2)$
and by division of the obtained result by the normalization constant
$K$ from Eq.~(\ref{sIV3.11})
\beq
U^R(a)=\frac{N}{K}
\int\limits_{-L}^{L}dx_2\int\limits_{-L}^{L}dy_2
\int\limits_{a+A_2f_2(x_2,y_2)}^{a+D}dz_2\,
U_A(x_2,y_2,z_2).
\label{sV3.6}
\eeq
\noindent
The Casimir force between the plates per unit area is given by
\beq
F_{ss}^R(a)=-\frac{1}{(2L)^2}
\frac{\partial U^R(a)}{\partial a}.
\label{sV3.7}
\eeq
\noindent
Substituting (\ref{sV3.6}) into (\ref{sV3.7}) we can write
\beq
F_{ss}^R(a)=-\frac{1}{(2L)^2}
\frac{N}{K}
\int\limits_{-L}^{L}dx_2\int\limits_{-L}^{L}dy_2
U_A\left[x_2,y_2,a+A_2f_2(x_2,y_2)\right].
\label{sV3.8}
\eeq

Let us now represent the quantity $U_A$ defined in Eq.~(\ref{sV3.5}) as
a series up to the fourth order in the small parameter $A_2/a$.
Then we substitute this series into Eq.~(\ref{sV3.8}). The result
may be written in the form
\beq
F_{ss}^R(a)=F_{ss}(a)
\sum\limits_{k=0}^{4}\sum\limits_{l=0}^{4-k}
c_{kl}\left(\frac{A_1}{a}\right)^k \left(\frac{A_2}{a}\right)^l,
\label{sV3.9}
\eeq
\noindent
where $F_{ss}(a)$ is as defined in Eq.~(\ref{sIV29}) in the case
of perfect plates. For the coefficients in (\ref{sV3.9}) we note that
\beq
c_{00}=1,\qquad c_{01}=c_{10}=0.
\label{sV3.10}
\eeq
\noindent
The last two equalities follow from the choice (\ref{sV3.3}).

The coefficients whose first index is zero are
\beq
c_{02}=10\,\langle f_2^2\rangle , \qquad
c_{03}=-20\,\langle f_2^3\rangle , \qquad
c_{04}=35\,\langle f_2^4\rangle , 
\label{sV3.11}
\eeq
\noindent
where the notation for the averaged values are used as 
in Eq.~(\ref{sV3.3}).

The remaining coefficients $c_{kl}$ in Eq.~(\ref{sV3.9}) are more
complicated. They read
\bes
&&
c_{20}=\frac{35}{\pi}a^7\langle f_1^2Y^{-9}\rangle , \qquad
c_{30}=\frac{35}{\pi}a^7\langle f_1^3\varphi_1(Y)\rangle ,
\nonumber\\
&&
c_{40}=\frac{105}{4\pi}a^9\langle f_1^4\varphi_2(Y)\rangle , \qquad
c_{11}=-\frac{70}{\pi}a^7\langle f_1f_2Y^{-9}\rangle , 
\nonumber\\
&&
c_{12}=\frac{35}{\pi}a^7\langle f_1f_2^2\varphi_1(Y)\rangle , \qquad
c_{21}=-\frac{35}{\pi}a^7\langle f_1^2f_2\varphi_1(Y)\rangle ,
\label{sV3.12}\\
&&
c_{13}=-\frac{105}{\pi}a^9\langle f_1f_2^3\varphi_2(Y)\rangle , \qquad
c_{31}=-\frac{105}{\pi}a^9\langle f_1^3f_2\varphi_2(Y)\rangle ,
\nonumber\\
&&
c_{22}=\frac{315}{\pi}a^9\langle f_1^2f_2^2\varphi_2(Y)\rangle .
\nonumber
\ees
\noindent
In (\ref{sV3.12}) the following notations are used
\bes
&&
Y=\left[(x_1-x_2)^2+(y_1-y_2)^2+a^2\right]^{1/2},
\label{sV3.13}\\
&&
\varphi_1(Y)=9a^2Y^{-11}-Y^{-9}, \quad
\varphi_2(Y)=11a^2Y^{-13}-3Y^{-11},
\nonumber
\ees
and the following averaging procedure for a function depending on
four variables
\beq
\langle \Phi(x_1,y_1;x_2,y_2)\rangle =\frac{1}{(2L)^2}
\int\limits_{-L}^{L}dx_2\int\limits_{-L}^{L}dy_2
\int\limits_{-L}^{L}dx_1\int\limits_{-L}^{L}dy_1
\Phi(x_1,y_1;x_2,y_2).
\label{sV3.14}
\eeq
\noindent
We have also used the notations $f_1\equiv f_1(x_1,y_1)$,
$f_2\equiv f_2(x_2,y_2)$.

To obtain the general result for corrections to the Casimir force, we need
to calculate the values of the coefficients $c_{lk}$ defined in
(\ref{sV3.12}). We start with the coefficients $c_{l0}$ which depend on 
one distortion function only. Integrating over $x_2$ and $y_2$
according to (\ref{sV3.14}) from $-\infty$ to $\infty$, we obtain
to zeroth order in $a/L$
\beq
c_{20}=10\,\langle f_1^2\rangle , \qquad
c_{30}=20\,\langle f_1^3\rangle , \qquad
c_{40}=35\,\langle f_1^4\rangle .
\label{sV3.15}
\eeq
\noindent
These results are in agreement with Eq.~(\ref{sV3.11}) and may be obtained 
also from symmetry considerations.

The calculation of the mixed coefficients (depending on the deviation
functions of both plates) is quite complicated. The results read
\cite{10}
\bes
&&
c_{11}=-\frac{4}{3}\sqrt{\frac{2}{\pi}}
\sum\limits_{m=0}^{\infty}\sum\limits_{n=0}^{\infty}
G_{mn}^{(1,1)}z_{mn}^{7/2}K_{7/2}(z_{mn}),
\nonumber\\
&&
c_{12}=\frac{2}{3}\sqrt{\frac{2}{\pi}}
\sum\limits_{m=0}^{\infty}\sum\limits_{n=0}^{\infty}
G_{mn}^{(1,2)}z_{mn}^{7/2}\left[z_{mn}K_{9/2}(z_{mn})-
K_{7/2}(z_{mn})\right],
\label{sV3.16}\\
&&
c_{13}=-\frac{2}{9}\sqrt{\frac{2}{\pi}}
\sum\limits_{m=0}^{\infty}\sum\limits_{n=0}^{\infty}
G_{mn}^{(1,3)}z_{mn}^{9/2}\left[z_{mn}K_{11/2}(z_{mn})-
K_{9/2}(z_{mn})\right],
\nonumber\\
&&
c_{22}=210g_{00}^{(2)}h_{00}^{(2)}
+\frac{1}{3}\sqrt{\frac{2}{\pi}}
\sum\limits_{m=0}^{\infty}\sum\limits_{n=0}^{\infty}
G_{mn}^{(2,2)}z_{mn}^{9/2}\left[z_{mn}K_{11/2}(z_{mn})-
K_{9/2}(z_{mn})\right].
\nonumber
\ees
\noindent
Let us explain the notations in Eqs.~(\ref{sV3.16}). The functions
$K_{\nu}(z)$ are the modified Bessel functions,
\beq
z_{mn}\equiv \pi\frac{a}{L}\sqrt{n^2+m^2}.
\label{sV3.17}
\eeq
\noindent
The quantities $G_{mn}^{(i,k)}$ are given by
\beq
G_{mn}^{(i,k)}=\frac{1}{4}\left(1+\delta_{m0}+\delta_{n0}\right)
\sum\limits_{l=1}^{4}
g_{l,mn}^{(i)}h_{l,mn}^{(k)},
\label{sV3.18}
\eeq
\noindent
where $\delta_{m0}$ is the Kronecker symbol and 
$g_{l,mn}^{(i)}$ and $h_{l,mn}^{(k)}$ are the Fourier coefficients
of the functions $f_1^i$ and $f_2^k$, considered as periodic
functions with the period $2L$ (for details see \cite{10}).
The quantities $g_{00}^{(2)}$ and $h_{00}^{(2)}$ are the zeroth
terms of the Fourier expansions of the functions $f_1^2$ and $f_2^2$
respectively. The coefficient $c_{21}$ differs from $c_{12}$ by its
sign and by the sequence of upper indices of $G$. For obtaining of
$c_{31}$ it is enough just to change the sequence of upper indices of $G$
in $c_{13}$.

So, the perturbation formalism developed allows one to obtain the
Casimir force for configurations with deviations from plane
parallel geometry in the form of the series, given by Eq.~(\ref{sV3.9}),
with coefficients defined in Eqs.~(\ref{sV3.10}), (\ref{sV3.11}),
(\ref{sV3.15}) and (\ref{sV3.16}). These coefficients can be
calculated explicitly for various kinds of distortions. However, the
simple case of nonparallel plates is of prime interest here because
an exact Casimir force value is also known for it. This gives the possibility
to estimate the accuracy of the above phenomenological approach.

\subsubsection{Casimir force between nonparallel plates and plates covered
by large scale distortions}
\label{sec5.3.2}

Let us consider the Casimir force for the configuration of two plane
plates with angle $\alpha$ between them. This angle
is assumed to be small, so that the inequality $\alpha L\ll a$ holds.
This configuration is a particular example for the deviation from plane
parallel geometry which a characteristic length scale is much larger than $a$.
For all deviations of such kind the coefficients (\ref{sV3.16}) can
be calculated in a general form. To do this we use the explicit
expressions of the modified Bessel functions \cite{sV2-6} in
(\ref{sV3.16}) and obtain \cite{10} 
\bes
&&
c_{11}=-20
\sum\limits_{m=0}^{\infty}\sum\limits_{n=0}^{\infty}
G_{mn}^{(1,1)}e^{-z_{mn}}\left(1+z_{mn}+
\frac{2}{5}z_{mn}^2+\frac{1}{15}z_{mn}^3\right),
\nonumber\\
&&
c_{12}=60
\sum\limits_{m=0}^{\infty}\sum\limits_{n=0}^{\infty}
G_{mn}^{(1,2)}e^{-z_{mn}}\left(1+z_{mn}+
\frac{13}{30}z_{mn}^2+\frac{1}{10}z_{mn}^3
+\frac{1}{90}z_{mn}^4\right),
\nonumber\\
&&
c_{13}=-140
\sum\limits_{m=0}^{\infty}\sum\limits_{n=0}^{\infty}
G_{mn}^{(1,3)}e^{-z_{mn}}\left[1+z_{mn}\Pi(z_{mn})\right],
\label{sV3.19}\\
&&
c_{22}=210\left\{g_{00}^{(2)}h_{00}^{(2)}+
\sum\limits_{m=0}^{\infty}\sum\limits_{n=0}^{\infty}
G_{mn}^{(2,2)}e^{-z_{mn}}\left[1+z_{mn}\Pi(z_{mn})\right]\right\},
\nonumber
\ees
\noindent
where 
$\Pi(z)=1+\frac{19}{42}z+\frac{5}{42}z^2
+\frac{2}{105}z^3+\frac{1}{630}z^4$.

The Fourier coefficients of the distortion functions $f_{1,2}$ and 
their powers decrease quickly with the number of harmonics. Due to
Eq.~(\ref{sV3.18}) the quantities $G_{mn}^{(i,k)}$ also decrease with the
increase in $m$ and $n$. The quantity $z_{mn}$, defined in
(\ref{sV3.17}), is of order of $a/L\ll 1$ for all harmonics
which give a significant contribution to the coefficients $c_{ik}$.
So, we can put in Eq.~(\ref{sV3.19}) $z_{mn}=0$ without loss of
accuracy:
\bes
&&
c_{11}=-20
\sum\limits_{m=0}^{\infty}\sum\limits_{n=0}^{\infty}
G_{mn}^{(1,1)},
\qquad
c_{12}=60
\sum\limits_{m=0}^{\infty}\sum\limits_{n=0}^{\infty}
G_{mn}^{(1,2)},
\nonumber\\
&&
c_{13}=-140
\sum\limits_{m=0}^{\infty}\sum\limits_{n=0}^{\infty}
G_{mn}^{(1,3)},
\label{sV3.20}\\
&&
c_{22}=210\left(g_{00}^{(2)}h_{00}^{(2)}+
\sum\limits_{m=0}^{\infty}\sum\limits_{n=0}^{\infty}
G_{mn}^{(2,2)}\right).
\nonumber
\ees
\noindent
Taking into account Eq.~(\ref{sV3.18}) and the elementary properties
of Fourier expansions, it is possible to rewrite Eqs.~(\ref{sV3.20}) 
in the form
\bes
&&
c_{11}=-20\langle f_1f_2\rangle ,\qquad
c_{12}=60\langle f_1f_2^2\rangle ,
\nonumber\\
&&
c_{13}=-140\langle f_1f_2^3\rangle ,\qquad
c_{22}=210\langle f_1^2f_2^2\rangle ,
\label{sV3.21}\\
&&
c_{21}=-60\langle f_1^2f_2\rangle ,\qquad
c_{31}=-140\langle f_1^3f_2\rangle 
\nonumber
\ees
\noindent
(in the last line the correlations between $c_{ik}$ and $c_{ki}$ have 
been used --- see Sec.5.3.1).

By the use of Eqs.~(\ref{sV3.10}), (\ref{sV3.11}), (\ref{sV3.15})
and (\ref{sV3.21}) the final result for the Casimir force from
Eq.~(\ref{sV3.9}) can be represented in the form \cite{10}
\bes
&&
F_{ss}^R(a)=F_{ss}(a)\left\{1+10\left[
\langle f_1^2\rangle\left(\frac{A_1}{a}\right)^2-
2\langle f_1f_2\rangle\left(\frac{A_1}{a}\right)\left(\frac{A_2}{a}\right)+
\langle f_2^2\rangle\left(\frac{A_2}{a}\right)^2\right]\right.
\nonumber\\
&&
\phantom{F_{ss}^R(a)=}
+20\left[
\langle f_1^3\rangle\left(\frac{A_1}{a}\right)^3-
3\langle f_1^2f_2\rangle\left(\frac{A_1}{a}\right)^2
\left(\frac{A_2}{a}\right)\right.
\nonumber\\
&&
\phantom{F_{ss}^R(a)=}
+\left.
3\langle f_1f_2^2\rangle\left(\frac{A_1}{a}\right)
\left(\frac{A_2}{a}\right)^2
-\langle f_2^3\rangle\left(\frac{A_2}{a}\right)^3\right]
\nonumber\\
&&
\phantom{F_{ss}^R(a)=}
+35\left[
\langle f_1^4\rangle\left(\frac{A_1}{a}\right)^4-
4\langle f_1^3f_2\rangle\left(\frac{A_1}{a}\right)^3
\left(\frac{A_2}{a}\right)
+6\langle f_1^2f_2^2\rangle\left(\frac{A_1}{a}\right)^2
\left(\frac{A_2}{a}\right)^2
\right.
\nonumber\\
&&
\phantom{F_{ss}^R(a)=}
-\left.\left.
4\langle f_1f_2^3\rangle\left(\frac{A_1}{a}\right)
\left(\frac{A_2}{a}\right)^3
+\langle f_2^4\rangle\left(\frac{A_2}{a}\right)^4\right]\right\}.
\label{sV3.22}
\ees

As can be seen from Eq.~(\ref{sV3.22}), the mixed terms have an 
evident interference character. For example, in the particular case
$f_2=\mp f_1$ we have
\bes
&&
F_{ss}^R(a)=F_{ss}(a)\left[1+10
\langle f_1^2\rangle\left(\frac{A_1}{a}\pm\frac{A_2}{a}\right)^2
+20\langle f_1^3\rangle\left(\frac{A_1}{a}\pm\frac{A_2}{a}\right)^3
\right.
\nonumber\\
&&\phantom{aaaaaaaaaaa}
\left.
+35\langle f_1^4\rangle\left(\frac{A_1}{a}\pm\frac{A_2}{a}\right)^4
\right].
\label{sV3.23}
\ees

Now, let us apply the results (\ref{sV3.22}) and (\ref{sV3.23})
to the case of two plane plates with a small angle $\alpha$ between them. 
This configuration can be realized in three ways. For example, the
upper plate may be left flat and the lower plate allowed to deviate from
the parallel position as described by the function
\beq
f_1(x_1,y_1)=\frac{x_1}{l}
\label{sV3.24}
\eeq
\noindent
with the amplitude $A_1=\alpha L$. Substituting (\ref{sV3.24}) into
the expression (\ref{sV3.22}) for the Casimir force and putting
$A_2=0$, one obtains
 \beq
F_{ss}^R(a)=F_{ss}(a)\left[1+10
\langle f_1^2\rangle\left(\frac{A_1}{a}\right)^2
+20\langle f_1^3\rangle\left(\frac{A_1}{a}\right)^3
+35\langle f_1^4\rangle\left(\frac{A_1}{a}\right)^4
\right].
\label{sV3.25}
\eeq
\noindent
Using the averaged values calculated from the function (\ref{sV3.24}),
one further obtains 
 \beq
F_{ss}^R(a)=F_{ss}(a)\left[1+\frac{10}{3}
\left(\frac{\alpha L}{a}\right)^2
+7\left(\frac{\alpha L}{a}\right)^4
\right].
\label{sV3.26}
\eeq

The same configuration can be obtained in a way when both plates
are perturbed with the amplitudes $A_1=\alpha_1L$ and $A_2=\alpha_2L$.
The angle between them is, thereby, $\alpha=\alpha_1+\alpha_2$ or
$\alpha=|\alpha_1-\alpha_2|$ and the deviation functions are given
by Eq.~(\ref{sV3.24}) supplemented by $f_2=\mp f_1$ respectively.

It is easy to see that the calculation of the Casimir force according to
Eq.~(\ref{sV3.23}) repeats the result (\ref{sV3.26}). This example is
interesting, because it gives the possibility of observing the role
of the interference terms in (\ref{sV3.22}). Such terms must be taken
into account in order to obtain the proper result.

The configuration of two nonparallel plates can also be considered by the
method of Green functions. For this purpose let us use the exact expression 
for the Casimir energy density of electromagnetic field inside a
wedge obtained by this method in \cite{Dowker1978bl,sV3-13}
\beq
{{E}}(\rho)=-\frac{(n^2-1)(n^2+11)}{720\pi^2\rho^4},
\label{sV3.27}
\eeq
\noindent
where $\rho$ is the distance from the line of intersection of the
wedge faces, and $n=\pi/\alpha$.

The total energy in the space between two square plates of side
length $2L=\rho_2-\rho_1$ at an angle $\alpha$ to each other is
\beq
U^R=\int\limits_{0}^{\alpha}d\varphi
\int\limits_{0}^{2L}dz\int\limits_{\rho_1}^{\rho_2}
d\rho\,\rho {E}(\rho).
\label{sV3.28}
\eeq
\noindent
Substituting Eq.~(\ref{sV3.27}) into Eq.~(\ref{sV3.28}) and
integrating, keeping $\alpha\ll 1$, one obtains the result
\beq
U^R=U^R(a)=-\frac{\pi^2L}{720\alpha}\left[
\frac{1}{(a-\alpha L)^2}-\frac{1}{(a+\alpha L)^2}\right],
\label{sV3.29}
\eeq
\noindent
where $a$ is the mean distance between the two plates.

Then the Casimir force can be obtained from Eq.~(\ref{sV3.7}) as
\beq
F_{ss}^R(a)=-\frac{\pi^2}{1440\alpha L}\left[
\frac{1}{(a-\alpha L)^3}-\frac{1}{(a+\alpha L)^3}\right].
\label{sV3.30}
\eeq

Expanding this expression for the force as a series with respect to
powers of the small parameter $\alpha L/a$ up to fourth order, one exactly 
obtains Eq.~(\ref{sV3.26}).

So it has been shown that, at least up to the fourth order inclusive
in the parameter $\alpha L/a$, the approximate approach based on the
additive summation with a subsequent normalization yields exactly the
same result as the Green function method if the slope angle between
plates is small. This allows to make an estimate of the relative
error of the above approach used for configurations with small deviations
from the plane parallel geometry. Taking the realistic estimation
of $\alpha L/a\approx 10^{-1}$, the conclusion is obtained that the relative
error of the result (\ref{sV3.22}) for the Casimir force is much less 
than $10^{-2}$\%. Therefore, the application of the approach under
consideration to configurations with small deviations from plane
parallel geometry can be expected to give reliable results up to the 
fourth order in the parameter $A_{1,2}/a$.

As was noted above, the Eq.~(\ref{sV3.22}) is valid for any large scale 
distortions whose characteristic length scale $T\gg a$. Such
distortions may describe large scale surface roughness.
Let us consider the longitudinal distortions with amplitudes $A_{1,2}$
described by the functions
\beq
f_1(x_1,y_1)=\sin\omega x_1, \qquad
f_2(x_2,y_2)=\sin(\omega x_2+\delta).
\label{sV3.31}
\eeq
\noindent
In the case when $\omega^{-1}\ll L$ the mean values of the functions 
$f_{1,2}^k$ in Eq.~(\ref{sV3.22}) can be calculated as the mean values
for one period. Substituting these results into Eq.~(\ref{sV3.22}),
one obtains \cite{10}  
\bes
&&
F_{ss}^R(a)=F_{ss}(a)\left\{1+5\left[
\left(\frac{A_1}{a}\right)^2-
2\cos\delta
\left(\frac{A_1}{a}\right)\left(\frac{A_2}{a}\right)+
\left(\frac{A_2}{a}\right)^2\right]\right.
\nonumber\\
&&
\phantom{F_{ss}^R(a)=}
+\frac{105}{8}\left[
\left(\frac{A_1}{a}\right)^4-
4\cos\delta
\left(\frac{A_1}{a}\right)^3
\left(\frac{A_2}{a}\right)
\right.
\label{sV3.32}\\
&&
\phantom{F_{ss}^R(a)=}
+2(2+\cos 2\delta)
\left(\frac{A_1}{a}\right)^2
\left(\frac{A_2}{a}\right)^2
-\left.\left.
4\cos\delta
\left(\frac{A_1}{a}\right)
\left(\frac{A_2}{a}\right)^3
+\left(\frac{A_2}{a}\right)^4\right]\right\}.
\nonumber
\ees
\noindent
It is seen that the result depends significantly on the value of
phase shift $\delta$.

If the periods of the distortion functions in Eq.~(\ref{sV3.31}) are 
different, it is necessary to calculate the mean values in the mixed 
terms over the whole plate according to Eq.~(\ref{sV3.3}). It can be
seen that in this case only the contribution of 
$\langle f_1^2f_2^2\rangle =1/4$ to the mixed terms is not equal to zero. 
So, the Casimir force is 
\bes
&&
F_{ss}^R(a)=F_{ss}(a)\left\{1+5\left[
\left(\frac{A_1}{a}\right)^2
+\left(\frac{A_2}{a}\right)^2\right]\right.
\label{sV3.33}\\
&&
\phantom{F_{ss}^R(a)=}
+\frac{105}{8}\left.\left[
\left(\frac{A_1}{a}\right)^4
+4\left(\frac{A_1}{a}\right)^2
\left(\frac{A_2}{a}\right)^2
+\left(\frac{A_2}{a}\right)^4\right]\right\}.
\nonumber
\ees
\noindent
The force from Eq.~(\ref{sV3.33}) is evidently larger than the force
from Eq.~(\ref{sV3.32}) for $\delta=0$, but smaller than that for
$\delta=\pi$.

\subsubsection{Casimir force between plates covered by short-scale
roughness}
\label{sec5.3.3}

The surfaces of real plates are always covered by some short scale
distortions or short scale roughness of different types. It is necessary 
to take into account the contribution of such distortions in
precision experiments on Casimir force measurements. The characteristic
longitudinal scales of such distortions are of order (or less than) the
distance $a$ between plates.

The distortions of the plate surfaces may be periodic or nonperiodic.
In both cases the general result for the Casimir force (see
Eq.~(\ref{sV3.9}) with the coefficients (\ref{sV3.10}), (\ref{sV3.11}),
(\ref{sV3.15}) and (\ref{sV3.19})) acquires one and the same,
simpler form than those presented by Eq.~(\ref{sV3.22}). As we shall
see below, most of the mixed terms in Eq.~(\ref{sV3.9}) turn into zero
for the short scale distortions.
The reasons for such simplifications are different for the cases of
periodic and nonperiodic distortions. Let us start with the
nonperiodic case, which is the more general.

If the functions $f_{1,2}$ are nonperiodic, the result (\ref{sV3.22})
for the Casimir force applies. Because the characteristic scale of the
distortions is of the order $a\ll L$, and by means of Eq.~(\ref{sV3.3}),
we obtain for the odd numbers ($i,k$)
\beq
\langle f_1^if_2^k\rangle =0.
\label{sV3.34}
\eeq
\noindent
By the use of this, Eq.~(\ref{sV3.22}) takes the form
\bes
&&
F_{ss}^R(a)=F_{ss}(a)\left\{1+10\left[
\langle f_1^2\rangle
\left(\frac{A_1}{a}\right)^2
+\langle f_2^2\rangle
\left(\frac{A_2}{a}\right)^2\right]\right.
\label{sV3.35}\\
&&
\phantom{F_{ss}^R(a)=}
+35\left.\left[
\langle f_1^4\rangle
\left(\frac{A_1}{a}\right)^4
+6\langle f_1^2f_2^2\rangle
\left(\frac{A_1}{a}\right)^2
\left(\frac{A_2}{a}\right)^2
+\langle f_2^4\rangle
\left(\frac{A_2}{a}\right)^4\right]\right\}.
\nonumber
\ees
\noindent
The same result occurs for large scale periodic distortions with 
different periods.

Now, let the functions $f_{1,2}$ be periodic. If the periods of
$f_1$ and $f_2$ are different (at least in one coordinate), 
Eq.~(\ref{sV3.34}) is valid once more and we return to the expression
(\ref{sV3.35}) for the Casimir force.

A different formula appears only in the case when periods $T_x$ and $T_y$
of  $f_1$ and $f_2$ are equal in both coordinates respectively. It is
evident in our case that $2L\approx mT_x$, $2L\approx nT_y$ holds
where $m,\,n\gg 1$. In the Fourier expansions of the functions
 $f_1^i$ and $f_2^k$ the coefficients of the modes with the corresponding
large numbers are large. So, in Eq.~(\ref{sV3.19}) the main contributions
are given by the terms containing the coefficients $G_{mn}^{(i,k)}$,
which were defined in Eq.~(\ref{sV3.18}), with $m,\,n$ large. The
parameter $z_{mn}$ defined in Eq.~(\ref{sV3.17}) is of the order of
$a/T_{x,y}$ in that case.

For extremely short scale distortions $T_{x,y}\ll a$ we have
$e^{-z_{mn}}\to 0$ and only one coefficient among all the mixed ones
survives:
\beq
c_{22}=210g_{00}^{(2)}h_{00}^{(2)}\equiv
210\langle f_1^2\rangle\langle f_2^2\rangle.
\label{sV3.36}
\eeq

As a consequence the Casimir force for the extremely short scale
distortions has the form \cite{10}
\bes
&&
F_{ss}^R(a)=F_{ss}(a)\left\{1+10\left[
\langle f_1^2\rangle
\left(\frac{A_1}{a}\right)^2
+\langle f_2^2\rangle
\left(\frac{A_2}{a}\right)^2\right]\right.
\label{sV3.37}\\
&&
\phantom{F_{ss}^R(a)=}
+35\left.\left[
\langle f_1^4\rangle
\left(\frac{A_1}{a}\right)^4
+6\langle f_1^2\rangle\langle f_2^2\rangle
\left(\frac{A_1}{a}\right)^2
\left(\frac{A_2}{a}\right)^2
+\langle f_2^4\rangle
\left(\frac{A_2}{a}\right)^4\right]\right\}.
\nonumber
\ees
\noindent
For the case $T_{x,y}\sim a$ one should calculate the coefficients of
Eq.~(\ref{sV3.19}) according to nonsimplified formulas.

As an example of periodic distortions let us examine the longitudinal
ones described by  Eq.~(\ref{sV3.31}) with $\omega^{-1}\sim a$.
Calculating the mixed coefficients according to Eq.~(\ref{sV3.19}) and
taking into account that only $G_{m,0}=1$ with $m=\omega L/\pi$ is
not equal to zero, one obtains
\bes
&&
c_{11}=-20\cos\delta e^{-z_{m0}}\left[1+z_{m0}+
\frac{2}{5}(z_{m0})^2+\frac{1}{15}(z_{m0})^3\right],
\nonumber\\
&&
c_{12}=c_{21}=0,
\label{sV3.38}\\
&&
c_{13}=c_{31}=-\frac{105}{2}
\cos\delta\, e^{-z_{m0}}\left[1+z_{m0}\Pi(z_{m0})\right],
\nonumber\\
&&
c_{22}=\frac{105}{2}\left\{1+
\cos2\delta\, e^{-z_{m0}}\left[1+z_{m0}\Pi(z_{m0})\right]\right\},
\nonumber
\ees
\noindent
where $\Pi(z)$ is as defined in Eq.~(\ref{sV3.19}) and
$z_{m0}=\omega a$.

If $\omega a\gg 1$, the terms of Eq.~(\ref{sV3.38}) containing the
exponential factors can be neglected and we return to the result
(\ref{sV3.37}) with
\beq
\langle f_1^2\rangle =\langle f_2^2\rangle =\frac{1}{2}, \qquad
\langle f_1^4\rangle =\langle f_2^4\rangle =\frac{3}{8}.
\label{sV3.39}
\eeq
\noindent
It is seen that this result coincides with (\ref{sV3.33}) and does not
depend on the phase shift $\delta$ as would be expected in this case.

For $\omega^{-1}\sim a$ one ought to make computations. For example, for
$z_{m0}=1$ the results are
\bes
&&
c_{11}=-18.08\cos\delta, \qquad
c_{13}=-49.98\cos\delta,
\nonumber\\
&&
c_{22}=\frac{105}{2}(1+0.953\cos 2\delta).
\label{sV3.40}
\ees
\noindent
The other coefficients coincide with those obtained in Sec.5.3.2 for this
example. They are contained in Eq.~(\ref{sV3.32}) and do not depend on the
period. The detailed investigation of the transition region between
the large scale and short scale roughness is contained in \cite{sV3-8a}.

Substituting all these coefficients into Eq.~(\ref{sV3.9}), one obtains
the Casimir force which is analogous to  Eq.~(\ref{sV3.32}) for larger
periods. It is seen that under a decrease of the period the absolute
values of the coefficients of the interference terms, i.e. the terms,
containing $\delta$, decrease \cite{10}.

In the above manner, the Casimir force between the plates covered by all
types of small distortions can be described perturbatively with a required 
accuracy (note that the case of an atom near the cavity wall covered
by roughness is examined in detail in \cite{sV3-10}).

\subsubsection{Expansion in powers of relative distortion amplitude:
a spherical lens above a plate}
\label{sec5.3.4}

The configuration of a spherical lens (or a sphere) above a plate (or 
a disk) is the most preferable from the experimental point of view
(see Sec.6). Because of this a knowledge of roughness corrections in 
this configuration is necessary to compare theoretical predictions with 
the results of the Casimir force measurements. Let us start from the
same approximate expression for the Casimir force potential as for two
plates given by Eqs.~(\ref{sIV3.12}), (\ref{sIV30}). Let the lens
curvature radius be $R$, thickness $h$, diameter $2r$. The parameters
of a plate are the same as in Sec.5.3.1. The coordinate plane ($x,\,y$)
is chosen to coincide with the plate surface.

As above, the surface of the plate with some small distortions
can be described by Eq.~(\ref{sV3.1}) with a condition given by the
first equality of (\ref{sV3.3}). Let us consider the surface of a lens
with small deviations from the ideal spherical shape described by the
equation
\beq
z_2^{(s)}=a+R-\sqrt{R^2-\rho^2}+A_2f_2(\rho,\varphi),
\label{sV3.41}
\eeq
\noindent
where $\rho,\,\varphi$ are the polar coordinates. The perturbation
theory can be developed in the same way as in Sec.5.3.1.
As a result the normalized potential of the Casimir force acting between
the lens and the plate is given by
\beq
U^R(a)=-\hbar c\Psi(\varepsilon)
\int\limits_{0}^{2\pi}d\varphi
\int\limits_{0}^{r}\rho d\rho
\int\limits_{z_2^{(s)}}^{h+a}dz_2
U_A(\rho,\varphi,z_2),
\label{sV3.42}
\eeq
\noindent
where
\bes
&&
U_A(\rho,\varphi,z_2)=
\int\limits_{-L}^{L}dx_1
\int\limits_{-L}^{L}dy_1
\int\limits_{-D}^{z_1^{(s)}}dz_1
\label{sV3.43}\\
&&
\phantom{U_A(\rho,\varphi,z_2)=}
\times\left[(x_1-\rho\cos\varphi)^2+(y_1-\rho\sin\varphi)^2+
(z_1-z_2)^2\right]^{-7/2}.
\nonumber
\ees

The function (\ref{sV3.43}) was calculated in Sec.5.3.1 as a series with
respect to the parameter $A_1/z_2$, which is small due to $z_2>a\gg A_1$.
Neglecting the corrections of the order of $z_2/L$ and  $z_2/D$
one has
\bes
&&
U_A(\rho,\varphi,z_2)=\frac{\pi}{10z_2^4}+
\int\limits_{-L}^{L}dx_1
\int\limits_{-L}^{L}dy_1
\label{sV3.44}\\
&&
\phantom{U_A(\rho,\varphi,z_2)=}
\times\left[
z_2f_1(x_1,y_1)X^{-7/2}\left(\frac{A_1}{z_2}\right)+
\frac{7}{2}z_2^3f_1^2(x_1,y_1)X^{-9/2}\left(\frac{A_1}{z_2}\right)^2
\right],
\nonumber
\ees
\noindent
with $X=(x_1-\rho\cos\varphi)^2+(y_1-\rho\sin\varphi)^2+z_2^2$.
Here the limiting transition $L\to\infty$ is performed in the first
item, which describes the contribution of the perfect plate without
deviations from the plane. For brevity we restrict ourselves to second 
order perturbation.

Substituting Eq.~(\ref{sV3.44}) in  Eq.~(\ref{sV3.42}) and calculating
force as $-\partial U^R/\partial a$, we arrive to the result in zeroth
order with respect to the small parameter $a/h$
\beq
F_{dl}^R(a)=-\hbar c\Psi(\varepsilon)
\int\limits_{0}^{2\pi}d\varphi
\int\limits_{0}^{r}\rho\,d\rho
U_A(\rho,\varphi,z_2^{(s)}).
\label{sV3.45}
\eeq
\noindent
Here $z_2^{(s)}$ is as defined by Eq.~(\ref{sV3.41}).

Let us now represent the quantity $U_A(\rho,\varphi,z_2^{(s)})$ in
Eq.~(\ref{sV3.45}) as a series up to the second order in the small 
parameter $A_2/a$. After integration, the result can be written in
the form
\beq
F_{dl}^R(a)=F_{dl}(a)
\sum\limits_{k=0}^{2}\sum\limits_{l=0}^{2-k}
C_{kl}\left(\frac{A_1}{a}\right)^k \left(\frac{A_2}{a}\right)^l,
\label{sV3.46}
\eeq
\noindent
where $F_{dl}(a)$ is defined by Eq.~(\ref{sIV3.13a}). 
For the first coefficient in Eq.~(\ref{sV3.46}) we note that
\beq
C_{00}=1.
\label{sV3.47}
\eeq

The other coefficients in Eq.~(\ref{sV3.46}) in  zeroth order with 
respect to small parameters $a/h$, $a/r$, $r/R$, $a/L$ are \cite{sV3-8}
\bes
&&
C_{01}=-\frac{6}{\pi Ra}
\int\limits_{0}^{2\pi}d\varphi
\int\limits_{0}^{\infty}\rho d\rho
f_2(\rho,\varphi)\,\left(1+\frac{\rho^2}{2aR}\right)^{-5},
\nonumber\\
&&
C_{02}=\frac{15}{\pi Ra}
\int\limits_{0}^{2\pi}d\varphi
\int\limits_{0}^{\infty}\rho d\rho
f_2^2(\rho,\varphi)\,\left(1+\frac{\rho^2}{2aR}\right)^{-6},
\nonumber\\
&&
C_{10}=\frac{15a^4}{\pi^2 R}
\langle f_1Y^{-7}\rangle,
\label{sV3.48}\\
&&
C_{20}=\frac{105a^5}{2\pi^2 R}
\langle f_1^2\left(a+\frac{\rho^2}{2R}\right)Y^{-9}\rangle,
\nonumber\\
&&
C_{11}=-\frac{105a^5}{2\pi^2 R}
\langle f_1f_2\left(a+\frac{\rho^2}{2R}\right)Y^{-9}\rangle.
\nonumber
\ees

In Eqs.~(\ref{sV3.48}) we use the notation 
\beq
Y=\left[(x_1-\rho\cos\varphi)^2+(y_1-\rho\sin\varphi)^2+
\left(a+\frac{\rho^2}{2R}\right)^2\right]^{1/2}
\label{sV3.49}
\eeq
\noindent 
and the following avaraging procedure for a function
depending on four variables
\beq
\langle \Phi(\rho,\varphi;x_1,y_1)\rangle =
\int\limits_{0}^{2\pi}d\varphi
\int\limits_{0}^{\infty}\rho d\rho
\int\limits_{-\infty}^{\infty}dx_1
\int\limits_{-\infty}^{\infty}dy_1
\Phi(\rho,\varphi;x_1,y_1).
\label{sV3.50}
\eeq
\noindent
Besides, we have used in Eqs.~(\ref{sV3.48}) the notations
$f_1\equiv f_1(x_1,y_1)$, $f_2\equiv f_2(\rho,\varphi)$. 
A point that should be mentioned is that the coefficients 
$C_{01}$, $C_{10}$
in Eq.~(\ref{sV3.46}) in general differ from zero. 
For the configuration of two parallel plates with small distortions, 
the analogical perturbative expansion starts from the second order 
(see Sec.5.3.1).

The coefficients given by Eqs.~(\ref{sV3.48}), and also the ones of 
higher orders, can be calculated in the same way as illustrated in 
Secs.5.3.1, 5.3.2 for the example of two plates. For this purpose the 
distortion function $f_1$ is considered as a periodic one with a period
$2L$ in both variables and the function $f_2(\rho,\varphi)$ is
considered as a periodic function of $\rho$ with a period $r$.
All the details can be found in Ref.~\cite{sV3-8}. Instead of presenting
the detailed algebra, here we only discuss the specific results which are 
applicable to experiments.

\subsubsection{Corrections to the Casimir force between a plate and
a lens due to different kinds of roughness}
\label{sec5.3.5}

Let us consider first the short scale roughness on the plate (disk) and 
on the lens with characteristic longitudinal scales $T_d$, $T_l$ 
which are the subject to the inequality
\beq
T_d,\,T_l\ll\sqrt{aR}.
\label{sV3.51}
\eeq
\noindent
In this case only the terms of the Fourier expansions of the
functions $f_1^i$, $f_2^k$ with sufficiently large numbers are
significant. As a result all the expansion coefficients from
Eq.~(\ref{sV3.46}) can be calculated in a closed form. The result
up to the fourth order in the relative roughness amplitude is

\bes
&&
F_{dl}^R(a)=F_{dl}(a)\left\{1+6\left[
\langle f_1^2\rangle\left(\frac{A_1}{a}\right)^2-
2\langle f_1f_2\rangle\left(\frac{A_1}{a}\right)\left(\frac{A_2}{a}\right)+
\langle f_2^2\rangle\left(\frac{A_2}{a}\right)^2\right]\right.
\nonumber\\
&&
\phantom{F_{ss}^R(a)=}
+10\left[
\langle f_1^3\rangle\left(\frac{A_1}{a}\right)^3-
3\langle f_1^2f_2\rangle\left(\frac{A_1}{a}\right)^2
\left(\frac{A_2}{a}\right)\right.
\nonumber\\
&&
\phantom{F_{ss}^R(a)=}
+\left.
3\langle f_1f_2^2\rangle\left(\frac{A_1}{a}\right)
\left(\frac{A_2}{a}\right)^2
-\langle f_2^3\rangle\left(\frac{A_2}{a}\right)^3\right]
\nonumber\\
&&
\phantom{F_{ss}^R(a)=}
+15\left[
\langle f_1^4\rangle\left(\frac{A_1}{a}\right)^4-
4\langle f_1^3f_2\rangle\left(\frac{A_1}{a}\right)^3
\left(\frac{A_2}{a}\right)
+6\langle f_1^2f_2^2\rangle\left(\frac{A_1}{a}\right)^2
\left(\frac{A_2}{a}\right)^2
\right.
\nonumber\\
&&
\phantom{F_{ss}^R(a)=}
-\left.\left.
4\langle f_1f_2^3\rangle\left(\frac{A_1}{a}\right)
\left(\frac{A_2}{a}\right)^3
+\langle f_2^4\rangle\left(\frac{A_2}{a}\right)^4\right]\right\}.
\label{sV3.52}
\ees

For the extremely short scale roughness which is given by the condition
$T_d,\,T_l\ll a$ one obtains $\langle f_1f_2\rangle =0$ and the
result (up to the second order in the relative roughness amplitude) 
is \cite{sV3-8}
\beq
F_{dl}^R(a)=F_{dl}(a)\left\{1+6
\langle f_1^2\rangle\left(\frac{A_1}{a}\right)^2
+6\langle f_2^2\rangle\left(\frac{A_2}{a}\right)^2\right].
\label{sV3.53}
\eeq

As is seen from Eqs.~(\ref{sV3.52}), (\ref{sV3.53}), there is no
contribution to the Casimir force from the linear terms in relative
distortion amplitude. This is analogous with the case of two
plates. However, the large, linear, corrections to the Casimir
force between a plate and a lens may appear in the case of large
scale distortions. If the characteristic longitudinal distortion
scales $T_d,\,T_l$ are much larger than $a$, the Fourier modes
with rather small numbers contribute into the result. The corresponding
corrections to the Casimir force can be important. Their magnitude and 
sign are affected by the position of the lens. By way of illustration
let us consider the longitudinal periodic distortions of the plate
(disk) described by
\beq
f_1(x)=\cos\left(\frac{2\pi x}{T_d}+\delta_1\right)
\label{sV3.54}
\eeq
\noindent
and the concentric distortions of the lens
\beq
f_2(\rho)=\cos\left(\frac{2\pi\rho}{T_l}+\delta_2\right).
\label{sV3.55}
\eeq

For the large scale distortions there are $T_d\sim L$, $T_l\sim r$.
The parameter $\delta_2$ in Eq.(\ref{sV3.55}) defines the type of
distortion in the lens center: convex or concave, smooth ($\delta_2=0$
or $\delta_2=\pi$) or sharp. The parameter $\delta_1$ in Eq.(\ref{sV3.54})
fixes the position of the lens above the plate.

Calculating the expansion coefficients with the functions 
(\ref{sV3.54}) and (\ref{sV3.55}), one obtains the Casimir force 
(\ref{sV3.46}) in the form
\bes
&&
F_{dl}^R(a)=F_{dl}(a)\left[1+
3\cos\delta_1
\left(\frac{A_1}{a}\right)-
3\cos\delta_2
\left(\frac{A_2}{a}\right)\right.
\label{sV3.56}\\
&&
\phantom{F_{ss}^R(a)=}
+\left.
3\left(\frac{A_1}{a}\right)^2 +3\left(\frac{A_2}{a}\right)^2-
12\cos\delta_1
\left(\frac{A_1}{a}\right)
\left(\frac{A_2}{a}\right)\right].
\nonumber
\ees
\noindent
Here the dependence of the Casimir force on the parameters 
$\delta_1,\,\delta_2$ is seen in explicit form. At the same time the 
Casimir force for the extremely short scale distortions of the
form (\ref{sV3.54}) and (\ref{sV3.55}) with $T_d,\,T_l\ll a$ is
defined by Eq.~(\ref{sV3.53}) and naturally does not depend on the
parameters $\delta_1,\,\delta_2$.

In Ref.~\cite{sV3-8} the smooth transition of the Casimir force
described by Eq.~(\ref{sV3.56}) to Eq.~(\ref{sV3.53}) was followed
with decreasing of the scales $T_d,\,T_l$ in
Eqs.~(\ref{sV3.54}), (\ref{sV3.55}). Also the case of two crossed
cylinders was considered there and the roughness corrections to the
Casimir force were calculated.

It is notable that the result (\ref{sV3.22}) for two plane plates and
(\ref{sV3.52}) for a lens above a plate are connected by the Proximity
Force Theorem (see Sec.4.3). This means that one may obtain the
perturbation expansion for the energy density between the plates by
integration of Eq.~(\ref{sV3.22}) and then determine the force between 
a plate and a lens by multiplying the result by $2\pi R$. But in the
case of large scale distortions it is impossible to derive equation
like (\ref{sV3.56}) using this theorem. Indeed, the distance between the
interacting bodies was defined above as the distance between such ideal 
surfaces relatively to which the average values of all distortions
are equal to zero. As this takes place, for the configuration of
two parallel plates there are no first order corrections for all
distortion types considered above. However, for the configuration
of a lens above a plate there exist first order corrections in the
case of large scale distortions. It is easy to see that the
origin of such corrections is connected with the definition
of distance between the interacting bodies, i.e. they disappear if one
defines $a$ as a distance between the nearest points of the distorted 
surfaces bounding the lens and the plate \cite{sV2-7}. Thus, the choice 
of the theoretical formula for the Casimir force to be compared with the
experimental results depends on how the distance is measured experimentally.

The results of above calculations do not depend on $L$, i.e. the plate (disk)
was effectively taken to be infinitely large. In some experiments,
however, the size of a lens can be even larger than of the plate
(see, e.g., \cite{32}). This is the reason why the boundary effects
due to finite sizes of the plate may be interesting. According to
analyses of this problem performed in \cite{sV2-7} the Casimir force
with account of finiteness of the plate is given by
\bes
&&
F_{dl}^L(a)\approx
F_{dl}(a)\left[1-\frac{a^3}{R^3}\frac{1}{(1-T)^3}\right],
\label{sV3.57}\\
&&
T\equiv\max\left(\frac{R}{\sqrt{R^2+L^2}},\frac{R-h}{R}\right).
\nonumber
\ees
\noindent
For the parameters of experiment \cite{32} (see Sec.6.3) this results in
a correction which does not exceed $6\times 10^{-7}$ in
the complete measurement range. Even smaller contribution is given by the boundary
effects in experiments using the atomic force microscope to
measure the Casimir force. This is the reason why we limit our discussion of 
these effects here.

\subsubsection{Stochastic roughness}
\label{sec5.3.6}

In the preceding subsections the surface roughness was 
described by regular functions. Here we discuss the
case of extremely irregular roughness which can be modeled in a better way
by stochastic functions. Let us discuss the case of two
parallel plates first \cite{sV3-7} and a lens (sphere) above a plate
next \cite{sV3-8}.

Now it is assumed that the surfaces of both plates with the dimensions
$2L\times 2L$ and of thickness $D$ are described by the stochastic
functions $\{\delta_if_i(x_i,y_i)\}$, $i=1,\,2$, with dispersions
$\delta_i$ and mean values
\beq
\langle\delta_if_i(x_i,y_i)\rangle_{\! i}=0.
\label{sV3.58}
\eeq
\noindent
Here, $\langle\ \rangle_{\! i}$ denotes the averaging over the ensembles of
all particular realizations $\delta_if_i(x_i,y_i)$ of the corresponding
stochastic functions. The factor $\delta_i$ is written in front of $f_i$
to have the dispersion of the functions $\{f_i(x_i,y_i)\}$ be equal
to unity. So the surfaces of the plates under consideration are given
by the functions
\beq
z_1^{(s)}=\delta_1f_1(x_1,y_1), \qquad
z_2^{(s)}=a+\delta_2f_2(x_2,y_2).
\label{sV3.59}
\eeq
\noindent
For test bodies with the surfaces (\ref{sV3.59}), the potential $U^R$ in
Eq.~(\ref{sIV3.12}) has to be substituted by 
$\langle\langle U^R(a)\rangle_{\! 1}\rangle_{\! 2}$. Then the Casimir
force is given by
\beq
F_{ss}(a)=-\frac{1}{(2L)^2}\,\frac{\partial}{\partial a}
\langle\langle U^R(a)\rangle_{\! 1}\rangle_{\! 2}.
\label{sV3.60}
\eeq
\noindent
Naturally we assume the absence of any correlation between the stochastic
functions describing the deviations from plane parallel geometry on both
plates.

In our perturbation treatment we assume $\delta_i\ll a$. In direct
analogy with the above treatment $a/L\ll \delta_i/a$, so that
a perturbative expansion with respect to powers $\delta_i/a$ is
considered, taken in zeroth order with respect to $a/D$ and $a/L$.

The non-normalized potential of one atom at the height $z_2$ above the
first plate is given by Eq.~(\ref{sV3.4}) where dispersion $\delta_1$
should be substituted instead of $A_1$. As a result of the perturbation
expansion up to the fourth order with respect to the small parameter
$\delta_1/z_2$ Eq.~(\ref{sV3.5}) follows once more with the same
substitution.

In accordance with Eq.~(\ref{sV3.60}) it is necessary to calculate the 
mean value over all realizations of the stochastic functions
$\{f_{1,2}\}$. For the first function it is suitable to do this
directly in Eq.~(\ref{sV3.5}). Using the normal distribution at each point
of the surfaces and the corresponding mean values
\beq
\langle f_i\rangle_{\! i}=\langle f_i^3\rangle_{\! i}=0, \quad
\langle f_i^2\rangle_{\! i}=1, \quad
\langle f_i^4\rangle_{\! i}=3
\label{sV3.61}
\eeq
\noindent
one obtains from (\ref{sV3.5})
\bes
&&
\langle U_A(x_2,y_2,z_2)\rangle_{\! 1}
=-CN\left\{\vphantom{\int\limits_{-L}^{L}}
\frac{\pi}{10z_2^4}\right.
\label{sV3.62}\\
&&
\phantom{aa}\left.
+
\int\limits_{-L}^{L}dx_1
\int\limits_{-L}^{L}dy_1
\left[\frac{7z_2^3}{2X^{9/2}}\left(\frac{\delta_1}{z_2}\right)^2+
\frac{63z_2^5}{8X^{11/2}}\left(\frac{11z_2^2}{X}-3\right)
\left(\frac{\delta_1}{z_2}\right)^4\right]\right\}.
\nonumber
\ees
\noindent
If the distortions on the surfaces are described by the stationary stochastic
functions with $\delta_i=$const, the result in the limit $L\to\infty$ is
\beq
\langle U_A(x_2,y_2,z_2)\rangle_{\! 1}=
-CN\frac{\pi}{10z_2^4}\left[1+10\left(\frac{\delta_1}{z_2}\right)^2
+105\left(\frac{\delta_1}{z_2}\right)^4\right].
\label{sV3.63}
\eeq
\noindent
In order to obtain the normalized potential of the Casimir force between the
plates let us integrate expression (\ref{sV3.63}) over the volume of the
second plate with the boundary function $z_2^{(s)}$ given by
Eq.~(\ref{sV3.59}). Then we calculate the mean value over all realizations
of the stochastic function $\{f_2\}$ and divide the result by the
normalization factor $K$ given by Eq.~(\ref{sIV3.11})
\beq
\langle\langle U^R(a)\rangle_{\! 1}\rangle_{\! 2}=
\frac{N}{K}\langle
\int\limits_{-L}^{L}dx_2
\int\limits_{-L}^{L}dy_2
\int\limits_{a+\delta_2f_2(x_2,y_2)}^{a+D}dz_2
\langle U_A(x_2,y_2,z_2)\rangle_{\! 1}\rangle_{\! 2}.
\label{sV3.64}
\eeq
\noindent
Substituting (\ref{sV3.64}) into (\ref{sV3.60}) we obtain for the Casimir
force per unit area
\beq
F_{ss}^R(a)=\frac{1}{(2L)^2}\frac{N}{K}
\int\limits_{-L}^{L}dx_2
\int\limits_{-L}^{L}dy_2
\langle\langle U_A\left(x_2,y_2,a+\delta_2f_2(x_2,y_2)\right)
\rangle_{\! 1}\rangle_{\! 2}.
\label{sV3.65}
\eeq
\noindent
Let us now expand the quantity $\langle U_A\rangle_{\! 1}$ defined by
(\ref{sV3.63}) as a series with respect to small parameter
$\delta_2/a$. Then substitute this series into Eq.~(\ref{sV3.65}).
Using Eqs.~(\ref{sV3.61}) and the condition $\delta_2=$const for the
stationary stochastic functions one obtains the final result
\bes
&&
F_{ss}^R(a)=F_{ss}(a)\left\{ 1+
10\left[\left(\frac{\delta_1}{a}\right)^2+
\left(\frac{\delta_2}{a}\right)^2\right]\right.
\nonumber\\
&&\phantom{aaaaaaaaaaaaa}\left.
+
105\left[\left(\frac{\delta_1}{a}\right)^2+
\left(\frac{\delta_2}{a}\right)^2\right]^2\right\}.
\label{sV3.66}
\ees
\noindent
It is seen that the correction to the Casimir force depends on the sum
$\delta_1^2+\delta_2^2$ only and does not depend on the correlation
radii $\rho_{1,2}$ of the stochastic functions describing the
distortions. This result coincides with the result of Ref.~\cite{sV3-14} 
where the corresponding corrections to the van der Waals and Casimir
forces were calculated (up to the second order in the dispersions only).
For a typical value of $\delta_{1,2}\approx 0.1$ the correction given
by Eq.~(\ref{sV3.66}) is 24\% of $F_{ss}$ where 4\% results from the 
fourth order.

The same calculation procedure can be applied to the configuration of
a lens (sphere) above a plate (disk) covered by the stochastic
roughness. In the case of stationary stochastic functions the Casimir
force, which is analogous to Eq.~(\ref{sV3.66}), is
\bes
&&
F_{dl}^R(a)=F_{dl}(a)\left\{ 1+
6\left[\left(\frac{\delta_1}{a}\right)^2+
\left(\frac{\delta_2}{a}\right)^2\right]\right.
\nonumber\\
&&\phantom{aaaaaaaaaaaaa}
\left.+
45\left[\left(\frac{\delta_1}{a}\right)^2+
\left(\frac{\delta_2}{a}\right)^2\right]^2\right\}.
\label{sV3.67}
\ees

The case of non-stationary stochastic functions describing roughness is 
more complicated. Some results for the Casimir force, however, were
obtained if the mean values of these stochastic functions do not depend on
$x_i,\,y_i$ but the dispersions are coordinate dependent:
$\delta_i=\delta_i(x_i,y_i)$ (see Ref.~\cite{sV3-7} for the case of two
plates and \cite{sV3-8} for a lens above a plate).

\subsection{Combined effect of different corrections}
\label{sec5.4}

As discussed above, the corrections due to nonzero temperature,
finite conductivity of the boundary material and surface roughness 
make important contributions to the value of the Casimir force and should
be taken into account when comparing theory and experiment.
Up to this point an assumption that each correction factor 
independently influences
the Casimir force has been made i.e. no multiplicative effects. 
Based on this assumption the combined effect of the several 
corrections can be calculated
by the additive summation of the results obtained in Secs.5.1--5.3.
The additivity of different corrections to the Casimir force holds,
however, in the first approximation only when all of them are small.
In the region where, e.g., two corrections are significant
(as finite conductivity and surface roughness at small separations)
more sophisticated methods to account for their combined effect are
needed. These methods are discussed below.

\subsubsection{Roughness and conductivity}
\label{sec5.4.1}

Finite conductivity corrections to the Casimir force were computed
in Sec.5.2 in the whole distance range $a\leq 1\,\mu$m. For larger
distances they should be considered together with the corrections
due to nonzero temperature (see Sec.5.4.2). For the separations of
$a\leq 1\,\mu$m the surface roughness makes important contributions 
to the value of the Casimir force. Surface roughness corrections were 
calculated in Sec.5.3 based on the retarded interatomic potential
of Eq.~(\ref{sIV3.8}) with a power index equal to seven. This
relationship is, however, specific for the pure Casimir regime only
(distances of order of $1\,\mu$m) and, broadly speaking, not
applicable to a wide transition region from the Casimir to the van der
Waals force. Because of this the results of Sec.5.3 may be not
be immediately applicable to separations $a< 1\,\mu$m.

At the same time the roughness corrections of Sec.5.3 demonstrate that 
the geometry of boundary surfaces covered by
roughness is more important than the specific type of the
interatomic potential. Let us consider two large plates of dimension
$L\times L$ whose boundary planes (perpendicular to the axis $z$)
are covered by small roughness. Let
\beq
a(x,y)=a_0+f(x,y)
\label{sV4.1}
\eeq
\noindent
be a distance between the points of boundary surfaces of both plates
with the coordinates ($x,y$). Here $a_0$ is the mean distance between
the plates which means that
\beq
\int\limits_{-L}^{L}dx
\int\limits_{-L}^{L}dy\,f(x,y)=0.
\label{sV4.2}
\eeq
\noindent
We remind the reader that for a wide range of surface distortions the 
Casimir force
is given by Eq.~(\ref{sV3.22}) (for configuration of two semispaces) 
and Eq.~(5.125) (for a lens above a disk). It is significant that
the same expression can be obtained by the integration of the Casimir
force, which takes into account the small distortions on the plates
in the surface separation, over the boundary surface
\beq
F_{ss}^{R}(a_0)=\frac{1}{L^2}
\int\limits_{-L}^{L}dx
\int\limits_{-L}^{L}dy\,F_{ss}\left(a(x,y)\right).
\label{sV4.3}
\eeq
\noindent
Here $F_{ss}(a)$ is given by Eq.~(\ref{sIV29}). 
In so doing the separation
function $f(x,y)$ introduced in Eq.~(\ref{sV4.1}) is connected with
the distortion functions of Sec.5.3 according to
\beq
f(x,y)=A_2f_2(x,y)-A_1f_1(x,y).
\label{sV4.4}
\eeq
\noindent
The analogous result can be obtained also for the configuration of
a spherical lens above a plate (this can be done, e.g., by 
application of the Proximity Force Theorem to Eq.~(\ref{sV4.3})).
The detailed example illustrating the above mentioned equivalence between 
the two methods in the case of the roughness correction is given below 
in Sec.6.4.1.
There the experimental configuration of a sphere (lens) above a disk
covered by roughness is considered and the roughness corrections are
calculated by both methods leading to the same result.

Although Eq.~(\ref{sV4.3}) is also applicable in the pure retarded,
Casimir, regime only, it can be simply generalized for the range
of smaller separations where the law of an interatomic interaction
changes and the finite conductivity corrections become significant.
To do so, it is enough to replace the ideal Casimir force 
between dielectrics
$F_{ss}$ in  Eq.~(\ref{sV4.3}) by the one from Sec.5.2 taking the
finite conductivity corrections into account. The resultant
expression
\beq
F_{ss}^{R,C}(a_0)=\frac{1}{L^2}
\int\limits_{-L}^{L}dx
\int\limits_{-L}^{L}dy\,F_{ss}^{C}\left(a(x,y)\right)
\label{sV4.5}
\eeq
\noindent
gives the possibility to evaluate the Casimir force with account of
both finite conductivity and surface roughness of boundary material.
Both the perturbation expression of Eq.~(\ref{sV2.8})
or the one of Sec.5.2.3 based on the optical tabulated data can be
substituted into Eq.~(\ref{sV4.5}) as a Casimir force $F_{ss}^{C}$
in the case of real metal.

In Sec.6.4.1 Eq.~(\ref{sV4.5}) is applied to find the combined effect of
surface roughness and finite conductivity corrections in the measurement
of the Casimir force by means of the Atomic Force Microscope
and the excellent agreement between theory and
experiment is demonstrated.

\subsubsection{Conductivity and temperature: two semispaces}
\label{sec5.4.2}

Both the finite conductivity of the boundary material and its
nonzero temperature are taken into account in the general Lifshitz formula
(\ref{foo}). Thus their combined effect can be calculated in a fundamental
way on the basis of this formula. The contribution of finite conductivity
into the Casimir force decreases with the increase of separation
distance whereas the contribution of nonzero temperature increases
with distance. Because of this, the two corrections are important
at the opposite ends of separation interval. Nevertheless at 
intermediate distances both corrections make important contributions 
and should be taken into account in precision experiments on the 
Casimir force. 
This subject is interesting also from a theoretical
point of view since it is connected with some delicate features of
Lifshitz formula which have been recognized only recently \cite{30}.

We start from the Lifshitz formula (\ref{foo}) and rewrite it more
conveniently in the form
\bes
&&
F_{ss}^{T}(a)=-\frac{k_BT}{2\pi}
\sum\limits_{l=-\infty}^{\infty}\int\limits_{0}^{\infty}
k_{\bot}\,dk_{\bot}\,q_l\left\{
\left[r_1^{-2}(\xi_l,k_{\bot})e^{2aq_l}-1\right]^{-1}\right.
\nonumber\\
&&\phantom{aaaaaaaaaaaaa}\left.
+\left[r_2^{-2}(\xi_l,k_{\bot})e^{2aq_l}-1\right]^{-1}\right\},
\label{sV42.1}
\ees
\noindent
where $r_{1,2}$ are the reflection coefficients with parallel
(perpendicular) polarization respectively given by
\beq
r_1^{-2}(\xi_l,k_{\bot})=\left[
\frac{\varepsilon(i\xi_l)q_l+k_l}{\varepsilon(i\xi_l)q_l-k_l}\right]^2,
\qquad
r_2^{-2}(\xi_l,k_{\bot})=\left(\frac{q_l+k_l}{q_l-k_l}\right)^2,
\label{sV42.2}
\eeq
\noindent
the other notations are introduced in Sec.5.1.1.

In terms of dimensionless variables
\beq
y=2aq_l=2a\sqrt{\frac{\xi_l^2}{c^2}+k_{\bot}^2},\qquad
{\tilde\xi}_l=2a\frac{\xi_l}{c}
\label{sV42.3}
\eeq
\noindent
Eqs.~(\ref{sV42.1}), (\ref{sV42.2}) can be rearranged to
\bes
&&
F_{ss}^{T}(a)=-\frac{k_BT}{16\pi a^3}
\sum\limits_{l=-\infty}^{\infty}\int\limits_{|{\tilde\xi}_l|}^{\infty}
y^2\,dy\left\{
\left[r_1^{-2}({\tilde\xi}_l,y)e^{y}-1\right]^{-1}\right.
\nonumber\\
&&\phantom{aaaaaaaaaaaaa}\left.
+\left[r_2^{-2}({\tilde\xi}_l,y)e^{y}-1\right]^{-1}\right\},
\label{sV42.4}
\ees
\noindent
where 
\bes
&&
r_1({\tilde\xi}_l,y)=
\frac{\varepsilon y-\sqrt{(\varepsilon-1){\tilde\xi}_l^2+
y^2}}{\varepsilon y+\sqrt{(\varepsilon-1){\tilde\xi}_l^2+y^2}},
\label{sV42.5}\\
&&
r_2({\tilde\xi}_l,y)=
\frac{y-\sqrt{(\varepsilon-1){\tilde\xi}_l^2+
y^2}}{y+\sqrt{(\varepsilon-1){\tilde\xi}_l^2+y^2}},
\qquad
\varepsilon\equiv\varepsilon(i{\xi}_l)=
\varepsilon\left(i\frac{c{\tilde\xi}_l}{2a}\right).
\nonumber 
\ees 

To calculate the combined effect of nonzero temperature and finite 
conductivity one should substitute some model dielectric function
$\varepsilon(i{\xi}_l)$ into Eqs.~(\ref{sV42.4}), (\ref{sV42.5}).
The simplest function of this kind is given by plasma model (see
Eq.~(\ref{sV2.2})). On the base of plasma model the combined effect
of nonzero temperature and finite conductivity was first examined
in \cite{29,30}.

It is common knowledge that the plasma model does not provide us with 
a correct behaviour of the dielectric function at small frequencies.
This is given by the more complete Drude model which takes relaxation
processes into account. The dielectric function of Drude model is
given by Eq.~(\ref{sV23.2}). There is, however, one major problem
when the dielectric function of  Eq.~(\ref{sV23.2}) is
substituted into Eqs.~(\ref{sV42.4}), (\ref{sV42.5}). Let us change the
discrete variable ${\tilde\xi}_l$ for continuous one $\tilde\xi$.
It is easily seen that the reflection coefficient $r_2({\tilde\xi}_l,y)$
is discontinuous as a function of the two variables at a point (0,0). 
Actually, substituting Eq.~(\ref{sV23.2}) into $r_2$ one obtains two limiting
values
\beq
\lim\limits_{\tilde\xi\to 0}r_2^2(\tilde\xi,0)=1,\qquad
\lim\limits_{y\to 0}r_2^2(0,y)=0,
\label{sV42.6}
\eeq
\noindent
which together demonstrate the presence of the discontinuity. 
As a result the zeroth
term of Eq.~(\ref{sV42.4}) becomes ambiguous.
What is more important, the perpendicular reflection coefficient
at zero frequency is discontinuous with respect to the relaxation
parameter $\gamma$ at a point $\gamma=0$ \cite{K-M}.
In \cite{sV23-9,sV42-1} the value $r_2^2(0,y)=0$ was used to calculate 
numerically the value of the Casimir force (\ref{sV42.4}). As a result
the temperature correction was obtained which changes its sign at different
separations. Also the value $r_2$ used in \cite{sV23-9,sV42-1} leads
to anomalously large by the modulus temperature corrections for 
the real metal at small separations and to 
wrong asymptotic values at large separations. These 
asymptotic 
values are two times smaller than that for the case of perfect metal, 
independent
of how high a conductivity is assumed for the real metal. What this means is
that the value of $r_2$ at zero frequency used in \cite{sV23-9,sV42-1}
is unacceptable. The use of the value $r_2^2(0,y)=1$ in \cite{SL,SL1}
as it is for
physical photons is unjustified also and leads to large temperature
corrections at small separations and to the absence of any conductivity
corrections at moderate separations of several nanometers with no
regard to the quality of a metal.
In fact the scattering problem which underlies the Lifshitz formula
(see Sec.~5.1.1) is not well defined at zero frequency in the presence
of relaxation. Because of this the zeroth term of Lifshitz
formula, when applied to real metals, should be corrected in appropriate
way like it was done in \cite{SRM} for the ideal metal.

To solve this problem one can use the representation of Lifshitz formula 
in terms of continuous variables instead of discrete summation.
Such a representation was suggested in
\cite{Mehra67,SRM} for other purposes. This approach is discussed below.

According to Poisson summation formula if $c(\alpha)$ is the Fourier
transform of a function $b(x)$
\beq
c(\alpha)=\frac{1}{2\pi}
\int\limits_{-\infty}^{\infty}b(x)e^{-i\alpha x}\,dx
\label{sV42.7}
\eeq
\noindent
then it follows [212]
\beq
\sum\limits_{l=-\infty}^{\infty}b(l)=
2\pi\sum\limits_{l=-\infty}^{\infty}c(2\pi l).
\label{sV42.8}
\eeq

Let us apply this formula to Eq.~(\ref{sV42.4}) using the 
identification
\beq
b_{ss}(l)\equiv -\frac{k_BT}{16\pi a^3}
\int\limits_{|l|\tau}^{\infty}y^2\,dy\,f_{ss}(l\tau,y),
\qquad
\tau\equiv\frac{4\pi ak_BT}{\hbar c},
\label{sV42.9}
\eeq
\noindent
where ${\tilde\xi}_l=\tau l$ and
\bes
&&
f_{ss}(l\tau,y)=f_{ss}^{(1)}(l\tau,y)+f_{ss}^{(2)}(l\tau,y)
\nonumber \\
&&\phantom{f_{ss}(l\tau,y)}
\equiv
\left(r_1^{-2}e^y-1\right)^{-1}+
\left(r_2^{-2}e^y-1\right)^{-1}
\label{sV42.10}
\ees
\noindent
is an even function of $l$.

Then the quantity $c_{ss}(\alpha)$ from Eq.~(\ref{sV42.7})
is given by
\beq
c_{ss}(\alpha)=-\frac{k_BT}{16\pi^2 a^3}
\int\limits_{0}^{\infty}dx\,\cos\alpha x
\int\limits_{x\tau}^{\infty}y^2\,dy\,f_{ss}(x\tau,y).
\label{sV42.11}
\eeq

Using Eqs.~(\ref{sV42.4}),  (\ref{sV42.8}), (\ref{sV42.11}) one
finally obtains the new representation of Lifshitz formula
\bes
&&
F_{ss}^{T}(a)=
\sum\limits_{l=-\infty}^{\infty}b_{ss}(l)
\label{sV42.12}\\
&&\phantom{aaa}
=-\frac{\hbar c}{16\pi^2 a^4}
\sum\limits_{l=0}^{\infty}{\vphantom{\sum}}^{\prime}
\int\limits_{0}^{\infty}d{\tilde\xi}\,
\cos\left(l{\tilde\xi}\frac{T_{eff}}{T}\right)
\int\limits_{\tilde\xi}^{\infty}y^2\,dy\,f_{ss}(\tilde\xi,y),
\nonumber
\ees
\noindent
where the continuous variable $\tilde\xi=\tau x$.
Note that in the representation (\ref{sV42.12}) the $l=0$ term gives the
force at zero temperature. 
As is shown in \cite{K-M}, for real metals Eq.~(\ref{sV42.12}) can be
transformed to the form
\bes
&&
F_{ss}^{T}(a)=-\frac{k_BT}{16\pi a^3}\left\{
\int\limits_{0}^{\infty}y^2\,dy\,
\left[f_{ss}^{(1)}(0,y)+f_{ss}^{(2)}(y,y)\right]
\right.
\nonumber \\
&&\phantom{aaaaa}
\left.
+2\sum\limits_{n=1}^{\infty}
\int\limits_{{\tilde\xi}_n}^{\infty}y^2\,dy\,
f_{ss}({\tilde\xi}_n,y)\right\}.
\label{5.157a}
\ees
\noindent
Here all terms with $n\geq 1$ coincide with the corresponding contributions
to (\ref{sV42.4}). The zeroth term of (\ref{sV42.4}) is modified
by the prescription generalizing the recipe used in \cite{SRM} for the
ideal metal (note that in the case of plasma dielectric function
Eq.~(\ref{5.157a}) is exactly equivalent to (\ref{sV42.4})).
The representation (\ref{5.157a}) of Lifshitz formula is not subject
of the above disadvantages and can be applied to calculate the temperature
Casimir force between real metals at all separations \cite{K-M}.

Here we present some analytical results which can be obtained in
frames of plasma model.
To obtain perturbation expansion of Eq.~(\ref{sV42.12}) in terms of
a small parameter of the plasma model $\delta_0/a$ (see Sec.5.2.1) it is
useful to change the order of integration and then rewrite it in terms
of the new variable $v\equiv\tilde\xi/y$ instead of $\tilde\xi$
\beq 
F_{ss}^{T}(a)
=-\frac{\hbar c}{16\pi^2 a^4}
\sum\limits_{l=0}^{\infty}{\vphantom{\sum}}^{\prime}
\int\limits_{0}^{\infty}y^3\,dy
\int\limits_{0}^{1}dv\,\cos\left(vyl\frac{T_{eff}}{T}\right)\,f_{ss}(v,y).
\label{sV42.13}
\eeq

Expanding the quantity $f_{ss}$ defined in (\ref{sV42.5}), (\ref{sV42.10})
up to the first order in powers of $\delta_0/a$ one obtains
\beq
f_{ss}(v,y)=\frac{2}{e^y-1}-
2\frac{ye^y}{(e^y-1)^2}(1+v^2)\frac{\delta_0}{a}.
\nonumber
\eeq

Substituting this into Eq.~(\ref{sV42.13}) we come to the
Casimir force including the effect of both the nonzero temperature and finite
conductivity (to underline this we have added index $C$)
\bes
&&
F_{ss}^{T,C}(a)=F_{ss}^{(0)}(a)\left\{
1+\frac{30}{\pi^4}\sum\limits_{n=1}^{\infty}
\left[\frac{1}{t_n^4}-\frac{\pi^3}{t_n}
\frac{\cosh(\pi t_n)}{\sinh^3(\pi t_n)}\right]\right.
\label{sV42.15}\\
&&\phantom{aaaaa}
-\frac{16}{3}\frac{\delta_0}{a}-
60\frac{\delta_0}{a}\sum\limits_{n=1}^{\infty}
\left[\frac{2\cosh^2(\pi t_n)+1}{\sinh^4(\pi t_n)}-
\frac{2\cosh(\pi t_n)}{\pi t_n \sinh^3(\pi t_n)}\right.
\nonumber\\
&&\phantom{aaaaaaaaaaa}\left.\left.
-\frac{1}{2\pi^2t_n^2\sinh^2(\pi t_n)}-
\frac{\coth(\pi t_n)}{2\pi^3t_n^3}\right]
\vphantom{\sum\limits_{n=1}^{\infty}}\right\},
\nonumber
\ees
\noindent
where $t_n\equiv nT_{eff}/T$. The first summation in (\ref{sV42.15}) is
exactly the temperature correction in the case of ideal metals 
(see Eq.~(\ref{sV1.24})). The second summation takes into account the effect 
of finite
conductivity.

In the limit of low temperatures $T\ll T_{eff}$ one has from 
(\ref{sV42.15}) up to exponentially small corrections 
\cite{30}
\bes
&&
F_{ss}^{T,C}(a)\approx F_{ss}^{(0)}(a)\left\{
1+\frac{1}{3}\left(\frac{T}{T_{eff}}\right)^4\right.
\label{sV42.16}\\
&&\phantom{aaaaaa}\left.
-\frac{16}{3}\frac{\delta_0}{a}\left[1-
\frac{45\zeta_R(3)}{8\pi^3}\left(\frac{T}{T_{eff}}\right)^3\right]
\right\}.
\nonumber
\ees
\noindent
For $\delta_0=0$ (perfect conductor) Eq.~(\ref{sV42.16}) turns into
Eq.~(\ref{sV1.25}), and for $T=0$ the result (\ref{sV2.8}) is
reproduced. Note that the first correction of mixing
finite conductivity and finite temperature is of order
$(T/T_{eff})^3$.  More significantly, 
note that  there are no temperature corrections of orders
$(T/T_{eff})^k$ with $k\leq 4$ 
in the higher order conductivity correction terms 
$(\delta_0/a)^i$
from the second up to the sixth order \cite{30}.

In the limit of high temperatures $T\gg T_{eff}$ Eq.~(\ref{sV42.15})
leads to
\beq
F_{ss}^{T,C}(a)\approx -\frac{k_BT}{4\pi a^3}\zeta_R(3)
\left(1-3\frac{\delta_0}{a}\right)
\label{sV42.17}
\eeq
\noindent
up to exponentially small terms.
For $\delta_0=0$ one obtains from (\ref{sV42.17}) the known result
(\ref{sV1.26}) for perfect conductors. Finite conductivity corrections
of higher orders are not essential at large separations.

In the next subsection the combined effect of finite conductivity and
nonzero temperature will be considered in the configuration of a spherical
lens (sphere) above a disk. There the results of numerical computation
are presented which give the possibility to observe the smooth
transition between the asymptotics of type (\ref{sV42.16})
and (\ref{sV42.17}).
Also the role of relaxation which is taken into account by the Drude model is
discussed there. 

\subsubsection{Conductivity and temperature: lens (sphere) above a disk}
\label{sec5.4.3}

Along the lines of the previous section the combined effect of nonzero 
temperature and finite conductivity can also be found in the configuration
of a sphere (or spherical lens) above a disk. Here one should start 
from the temperature Casimir force of Eqs.~(\ref{sV13.1}), (\ref{sV13.2})
acting in the abovementioned configurations. In terms of the reflection
coefficients introduced in Eq.~(\ref{sV42.2}) this force can be
represented as
\bes
&&
F_{dl}^{T}(a)=\frac{k_BTR}{2}
\sum\limits_{n=-\infty}^{\infty}\int\limits_{0}^{\infty}
k_{\bot}\,dk_{\bot}\,\left\{\ln\left[1-r_1^2(\xi_n,k_{\bot})e^{-2aq_n}
\right]\right.
\label{sV43.1}\\
&&\phantom{aaaaaaaaaaaaaaaaaaaaaa}\left.
+\ln\left[1-r_2^2(\xi_n,k_{\bot})e^{-2aq_n}
\right]\right\}.
\nonumber
\ees
\noindent
Introducing the dimensionless variables ${\tilde\xi}_n=2a\xi_n/c$ and
$y=2aq_n$ we rewrite  Eq.~(\ref{sV4.1}) in the form
\bes
&&
F_{dl}^{T}(a)=\frac{k_BTR}{8a^2}
\sum\limits_{n=-\infty}^{\infty}\int\limits_{|{\tilde\xi}_n|}^{\infty}
y\,dy\,\left\{\ln\left[1-r_1^2({\tilde\xi}_n,y)e^{-y}
\right]\right.
\label{sV43.2}\\
&&\phantom{aaaaaaaaaaaaaaaaaaaaaa}\left.
+\ln\left[1-r_2^2({\tilde\xi}_n,y)e^{-y}
\right]\right\}.
\nonumber
\ees
\noindent
The term of Eq.~(\ref{sV43.2}) with $n=0$ suffers exactly the same
ambiguity as the zeroth term of Eq.~(\ref{sV42.4}) for two plane
parallel plates if one uses the Drude model to describe the
dependence of dielectric permittivity on frequency. To eliminate
this ambiguity we rewrite Eq.~(\ref{sV43.2}) in the form 
analogical to (\ref{sV42.12}).
This is achieved by the Poisson summation formula of
Eqs.~(\ref{sV42.7}), (\ref{sV42.8}). Repeating the same transformations
as in Sec.5.4.2 the final result is obtained
\beq
F_{dl}^{T}(a)=\frac{\hbar c R}{8\pi a^3}
\sum\limits_{n=0}^{\infty}{\vphantom{\sum}}^{\prime}
\int\limits_{0}^{\infty}d{\tilde\xi}\,\cos\left(n{\tilde\xi}
\frac{T_{eff}}{T}\right)
\int\limits_{{\tilde\xi}}^{\infty}y\,dy\,f_{dl}({\tilde\xi},y),
\label{sV43.3}
\eeq
\noindent
where
\bes
&&
f_{dl}({\tilde\xi},y)=f_{dl}^{(1)}(\tilde\xi,y)+
f_{dl}^{(2)}(\tilde\xi,y)
\nonumber \\
&&\phantom{f_{dl}({\tilde\xi},y)}
\equiv
\ln\left(1-r_1^2e^{-y}\right)+
\ln\left(1-r_2^2e^{-y}\right).
\label{sV43.4}
\ees

Evidently, the zeroth term of Eq.~(\ref{sV43.3}) gives us the Casimir
force at zero temperature. The terms of (\ref{sV43.3}) with $n\geq 1$
represent the temperature correction. 
Following the prescription formulated in \cite{K-M} for real
metals Eq.~(\ref{sV43.3}) can be represented in the form
analogical to (\ref{5.157a})
\bes
&&
F_{dl}^{T}(a)=\frac{k_BTR}{8a^2}\left\{
\int\limits_{0}^{\infty}y\,dy\,
\left[f_{dl}^{(1)}(0,y)+f_{dl}^{(2)}(y,y)\right]
\right.
\nonumber \\
&&\phantom{aaaaa}
\left. +2
\sum\limits_{n=1}^{\infty}
\int\limits_{{\tilde\xi}_n}^{\infty}y\,dy\,
f_{dl}({\tilde\xi}_n,y)\right\}.
\label{5.165a}
\ees
\noindent
In the case of plasma dielectric function Eqs.~(\ref{sV43.3})
and (\ref{5.165a}) are equivalent. In the case of Drude model
Eq.~(\ref{5.165a}) avoids the difficulties connected with
the zeroth term of (\ref{sV43.2}) \cite{K-M}. 

Let us first calculate the temperature Casimir force (\ref{sV43.3}) in
the framework of the plasma model. Changing the order of integration and
introducing the new variable $v=\tilde\xi/y$ instead of $\tilde\xi$
one obtains
\beq
F_{dl}^{T}(a)=\frac{\hbar c R}{8\pi a^3}
\sum\limits_{n=0}^{\infty}{\vphantom{\sum}}^{\prime}
\int\limits_{0}^{\infty}y^2\,dy
\int\limits_{0}^{1}dv\,
\cos\left(n\frac{T_{eff}}{T}vy\right)
f_{dl}(v,y).
\label{sV43.5}
\eeq

The expansion of $f_{dl}$ up to first order in the small parameter
$\delta_0/a$ is
\beq
f_{dl}(v,y)=2\ln\left(1-e^{-y}\right)+2\frac{y}{e^y-1}(1+v^2)
\frac{\delta_0}{a}.
\nonumber
\eeq

Substitution of this into Eq.~(\ref{sV43.5}) leads to result
\bes
&&
F_{dl}^{T,C}(a)=F_{dl}^{(0)}(a)\left\{1+\frac{45}{\pi^3}
\sum\limits_{n=1}^{\infty}\left[
\frac{1}{t_n^3}\coth(\pi t_n)+\frac{\pi}{t_n^2\sinh^2(\pi t_n)}\right]
-\frac{1}{t_1^4}\right.
\nonumber\\
&&\phantom{aaa}
-4\frac{\delta_0}{a}+ \frac{180}{\pi^4}\frac{\delta_0}{a}
\sum\limits_{n=1}^{\infty}\left[
\frac{\pi\coth(\pi t_n)}{2t_n^3} -\frac{2}{t_n^4}+
\frac{\pi ^3\cosh(\pi t_n)}{t_n\sinh^3(\pi t_n)}
\right.
\nonumber\\
&&\phantom{aaaaaaaaaaaaaaaaaaaaaaaaaa}
\left.\left.
+\frac{\pi^2}{2t_n^2\sinh^2(\pi t_n)}
\right]\right\}.
\label{sV43.7}
\ees
\noindent
Remind that $t_n\equiv nT_{eff}/T$.

In the case of low temperatures $T\ll T_{eff}$ \cite{30}
\bes
&&
F_{dl}^{T,C}(a)\approx F_{dl}^{(0)}(a)\left\{1+\frac{45\zeta_R(3)}{\pi^3}
\left(\frac{T}{T_{eff}}\right)^3-\left(\frac{T}{T_{eff}}\right)^4\right.
\nonumber\\
&&\phantom{aaaaaa}\left.
-4\frac{\delta_0}{a}\left[1-\frac{45\zeta_R(3)}{2\pi^3}
\left(\frac{T}{T_{eff}}\right)^3+\left(\frac{T}{T_{eff}}\right)^4\right]
\right\}.
\label{sV43.8}
\ees
\noindent
For ideal metal $\delta_0=0$ and Eq.~(\ref{sV43.8}) coincides with
Eq.~(\ref{sV13.5}). At zero temperature the first perturbation order of
(\ref{sV2.13}) is reobtained. Once more the perturbation orders 
$(\delta_0/a)^i$ with $2\leq i\leq 6$ do not contain temperature
corrections of orders $(T/T_{eff})^3$ and $(T/T_{eff})^4$ 
or lower ones \cite{30}.

At high temperature $T\gg T_{eff}$
\beq
F_{dl}^{T,C}(a)\approx -\frac{\zeta_R(3)}{4a^2}Rk_BT
\left(1-2\frac{\delta_0}{a}\right).
\label{sV43.9}
\eeq
\noindent
For $\delta_0=0$ one reobtains Eq.~(\ref{sV13.6}).

In Ref.~\cite{30} the Casimir force was computed numerically by
Eq.~(\ref{sV43.5}) in the frames of plasma model with plasma
frequency $\omega_p=1.92\times 10^{16}\,$rad/s (as for aluminium),
$T=300\,$K, and $R=100\,\mu$m (see solid line in Fig.~\ref{figTemp}).
\begin{figure}[ht]
\setlength{\unitlength}{1cm}
\begin{picture}(13,19)
\epsfig{figure=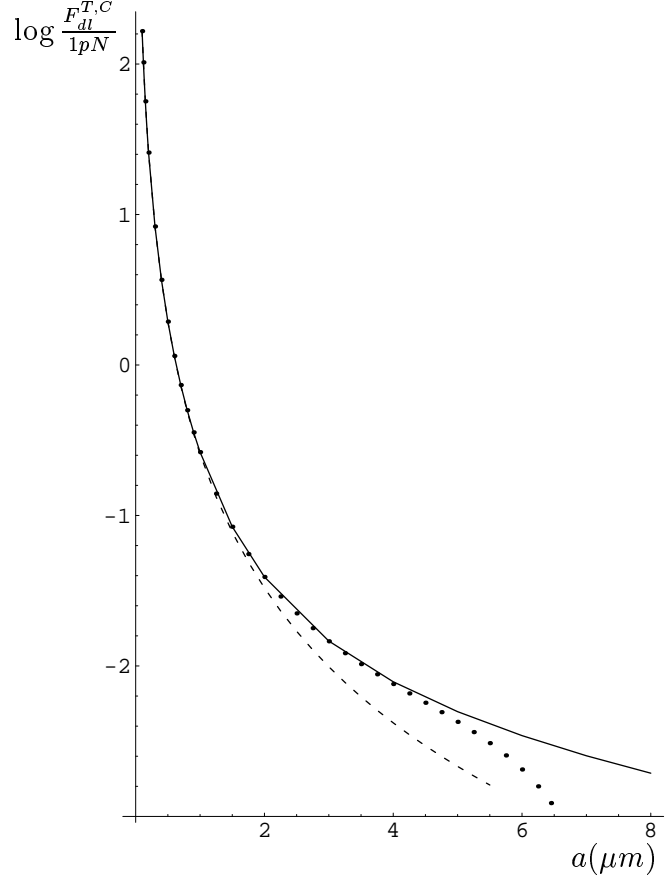}
\end{picture}
\vspace*{-7.5cm}
\caption{\label{figTemp}
The nonzero temperature Casimir force as a function of the surface
separation in the configuration of a sphere above a disk. The solid
line represents the result of numerical computations,
the dotted line is calculated by the perturbative
theory, the dashed line is the zero temperature 
result.
}
\end{figure}
The dotted line represents the perturbative results of low temperatures. 
The dashed line shows the Casimir force at zero temperature (but with
account of finite conductivity). It is seen from the figure that
perturbation theory works well within the range 
$0.1\,\mu\mbox{m}\leq a\leq 3.5\,\mu$m (note that six perturbative
orders in finite conductivity were used at small separations).
Starting from $a=6\,\mu$m the solid line represents the asymptotic
at high temperatures.

In the Drude model representation of the conductivity, 
the computations of the Casimir force
(\ref{5.165a}) were performed in Ref.\cite{K-M}. In addition to
above mentioned parameters the value of the relaxation frequency 
$\gamma=9.6\times 10^{13}\,$rad/s was used. At separation $a=8\,\mu$m
(which already corresponds to the high temperature asymptotic) the obtained
value is $F_{dl}^{T,C}\approx 1.9303\times 10^{-15}\,$N. This is
approximately 0.78\% lower compared to the asymptotic limiting
value for a perfect metal at large separations
$1.9454\times 10^{-15}\,$N. The same force
computed by the plasma model would be 
equal to $1.9378\times 10^{-15}\,$N which is
only 0.39\% lower compared to the
case of perfect metal. Thus there is a factor of 2 difference between  
the finite conductivity corrections to the high temperature Casimir force 
obtained by both models. It is apparent that at large separations
(where the low frequencies make important contributions) it is the
Drude model which gives the correct result for finite conductivity
correction. Note that this correction itself is rather small at large
separations.

\subsubsection{Combined effect of roughness, conductivity and
temperature}
\label{sec5.4.4}

As is seen from the previous subsection the combined effect of
conductivity and temperature is very nontrivial. It essentially 
depends on the behaviour of the reflection coefficients at small frequencies.
If this effect is  taken into account correctly, the temperature
effect for a real metal of finite conductivity is directly  analogous
to the case of the ideal metal. This means that the temperature corrections,
which at room temperature are rather small at the space separations
$a<1\,\mu$m,  increase monotonically with distance, and dominate the zero
temperature Casimir force starting at a separation distance of several 
micrometers. Note 
that in the
asymptotic regime of large separations $a>10\,\mu$m the finite
conductivity corrections are even smaller than the temperature corrections 
at small separations $a\sim 1\,\mu$m.

According to results of Sec.5.3 roughness corrections can make important 
contributions at $a\approx 1\,\mu$m, while for larger separations their
contribution decreases rapidly. The most important contribution of
surface roughness is given at the separation range $a<1\,\mu$m.
In this transition region to the van der Waals forces (as noted
in Sec.5.4.1)  the retarded interatomic potential is not applicable.
The combined effect of surface roughness and finite conductivity (which
are the two influential factors in this range) can be found, however,
by the geometrical averaging of Eq.~(\ref{sV4.5}) accounting for roughness
in the space separation entering 
the Casimir force law in the case of the real metallic
surfaces of finite conductivity.

Exactly the same, geometrical, approach can be applied to find the effect 
of roughness in addition to the combined effect of finite conductivity and
nonzero temperature. All one has to do is to replace the Casimir force
including only the effect of finite conductivity  in  Eq.~(\ref{sV4.5}) with
the Casimir force calculated at both nonzero temperature and finite
conductivity (see Secs.5.4.2, 5.4.3). The result is
\beq
F_{ss}^{R,T,C}(a_0)=\frac{1}{L^2}
\int\limits_{-L}^{L}dx
\int\limits_{-L}^{L}dy\,F_{ss}^{T,C}\left(a(x,y)\right).
\label{sV44.1}
\eeq
\noindent
Here the distance $a(x,y)$ between the interacting surfaces takes
roughness into account in accordance with Eqs.~(\ref{sV4.1}), (\ref{sV4.4}).

Eq.~(\ref{sV44.1}) can be used to calculate the combined effect of
surface roughness, nonzero temperature and finite conductivity of the
boundary metal in the experiments on the Casimir force  (see
Sec.6). It is necessary to stress, however, that up to the present,
none of the experiments has the precision necessary to measure the
temperature contribution to the Casimir force. The reason is that the
relative error of force measurements increases quickly with the increase 
in space separation and attains hundreds of percent at space separations 
where the temperature effects could be noticable (see the next section). 
Because of this up to the present only the finite conductivity and
roughness corrections to the Casimir force have been measured. 
As to the temperature corrections, its measurement is a problem to be
solved in the future.

\setcounter{equation}{0}
\section{Measurements of the Casimir force }
\label{sec6}  

In this section we review the experimental developments in the 
measurements of the Casimir force. Given that there have been only 
a few attempts at the measurement, a brief review of the older measurements
will also be provided.  It should be noted that the older measurements 
set the benchmark and the basis for improvement. Given two parallel 
plates of area $S$ and infinite conductivity, separated by a distance 
$a$ the Casimir force is given by Eq.~(\ref{FCas}).  
The force is a strong function of $a$ and is measurable only for 
$a \sim 1\,\mu$m. 
This force is on the order of $10^{-7}\,$N for flat surfaces of 
1\,cm${}^2$ area for a 
separation distance of $1\,\mu$m. Given the small value of the force for 
experimentally accessible surface areas, the force sensitivity of the 
experimental technique has been one of the most severe limitations on 
the accuracy of the various measurements.  Also given that the force has 
a very strong dependence on the separation distance, an accurate 
determination of the surface separation is necessary for good comparison 
to the theory. 

\subsection{General requirements for the Casimir force measurements}
\label{sec6.1}

The first experiments dealing with the measurements of the Casimir force 
were  done by M.J.~Sparnaay \cite{18,s6.3}.  The experimental 
technique based on a spring balance and parallel plates served 
to set the benchmarks.  They also clarified the problems associated with 
other Casimir force measurements. From the instrumental standpoint there 
are clear requirements, like an extremely high force 
sensitivity and the capability to reproducibly measure the surface 
separation between the two surfaces.   Other than these there are 
clear material requirements necessary for a good measurement of the 
Casimir force.  These fundamental requirements as spelled out 
by M.J.~Sparnaay \cite{18,s6.3} are:
\begin{enumerate}
\item
clean plate surfaces completely free of  chemical impurities and 
dust particles;
\item
precise and reproducible measurement of the separation between the 
two surfaces. In particular a measurement of the average distance 
on contact of the 
two surfaces which is nonzero due to the roughness of the metal 
surfaces and the presence of dust;
\item 
low electrostatic charges on the surface and low potential differences 
between the surfaces. Note that there can exist a large potential 
difference between clean and grounded metallic surfaces due to 
the work-function differences of the materials used and 
the cables used 
to ground the metal 
surfaces. Thus an independent measurement of the systematic error due to 
the residual electrostatic force is absolutely necessary. 
\end{enumerate}

Each of the above instrumental or material requirements is difficult to
obtain 
in practice and certainly very difficult to obtain together. They have 
bedeviled this field because, at least one or more of the above were 
neglected in the force measurements. Regarding the material requirements, 
as pointed out by M.J.~Sparnaay \cite{18}, requirement 1 was ignored in 
experiments with glass and quartz surfaces 
\cite{sIV3-3,sIV3-4,s6.6,s6.7,s6.8} 
where surface reactions with moisture and silicone oil from the vacuum 
apparatus lead to the formation of ``gel layer" \cite{18} on the surface.  
M.J.~Sparnaay expects this gel layer to completely 
modify the forces for surface 
separation distances less than 1.5$\,\mu$m.  The last two requirements are 
particularly difficult to meet in the case of non-conductive surfaces 
such as glass, quartz \cite{sIV3-3,sIV3-4,s6.6} or mica 
\cite{s6.9,s6.10,s6.11,s6.12}.  Yet all these early measurements 
possibly neglected 
the systematic correction due to the electrostatic force in their 
experiments.  Many other requirements such as the exact surface separation 
distance and the role of surface roughness were neglected in all but the 
most recent experiments.  Some experiments have tried to use an 
ionized environment \cite{s6.13} to neutralize the static charges  
reported additional electrostatic effects. 
Also all early measurements took the surface separation on contact to 
be zero. This can be a significant error 
for large flat surfaces or 
alternatively surfaces with large radius as the inevitable presence 
of obstacles prevent close contact of the two surfaces. 
As stated in \cite{18}, this is also true for some experiments with 
Pt metallic wires were the point of contact was assumed to be zero 
separation distance \cite{s6.6}. Thus independent checks of the surface 
separation are necessary for correct analysis of the data.  

Of the earlier experiments with metallic surfaces only two meet at least 
some of the stringent criteria set forth by Sparnaay
necessary for rigorous measurements of the 
Casimir force.  The first one is by M.J.~Sparnaay \cite{18}. The second is 
by P.~van Blokland and J.~Overbeek \cite{s6.8}. It should be mentioned 
that both experiments are a culmination of many years of improvements, 
references for which are provided in the respective publications.  
Some of the other significant older measurements on metal and dielectric 
surfaces such as those of B.~Derjaguin \cite{sIV3-3,sIV3-4,s6.6} 
(it should be noted that 
Derjaguin et al. were the first to use curved shaped bodies  which overcame 
the need to align the parallel surfaces) and with dielectric bodies 
by D.~Tabor, R.~Winterton and J.~Israelachvili 
\cite{s6.9,s6.10,s6.11,s6.12} and the dynamical measurements 
of Hunklinger et al. 
\cite {s6.12a,s6.12b} and those using 
micromechanical tunnelling transducers by Onofrio et al. 
\cite{s6.14,s6.15} will be briefly discussed. The experiments on van 
der Waals forces with liquid He films \cite{s6.16} are outside the 
confines of this review on retarded forces and accordingly will 
not be dealt in detail.  
Mention should also be made of the identification of van der Waals forces
with micron sized polystyrene spheres, using a tapping mode Atomic
Force Microscope [268]. This work is also outside the limits of this review.
Finally the more recent experiments using 
the torsion pendulum by S.K.~Lamoreaux \cite{32} and that using the Atomic 
Force Microscope by Mohideen et al. 
\cite{33,35,36} will be reviewed.  In each case the experimental  
technique and a discussion of the results will be provided. 

As discussed in Sec.4 the boundary dependence of the Casimir 
effect is one of its most intriguing properties. For example, 
the Casimir force is a strong function of geometry and that between 
two halves of thin metal spherical shells is repulsive. For two halves 
of a box it can be repulsive or attractive depending on the height 
to base ratio.  The sign and value of the Casimir force becomes even 
more interesting for complex topologies such as encountered with a torus.  
However given the difficulty of making unambiguous measurements  
there has been only one attempt at demonstrating the 
non-trivial boundary dependence of the Casimir force \cite{34}
for the case of corrugated plate. This experiment will also be reviewed.

\subsection{Primary achievements of the older measurements}
\label{sec6.2}

Here we briefly discuss the main experiments on the Casimir force
measurements which were performed until the year 1997 when the
modern stage in this field of research had started. In all cases, 
special mention is made of the necessary requirements 
for a good measurement that are
met by the experiment under consideration.
In many of them dielectric test bodies were used although
in several experiments the Casimir force between metallic surfaces
was also measured.

\subsubsection {Experiments with parallel plates by M.J.~Sparnaay}
\label {sec6.2.1}

M.J.~Sparnaay \cite{18} attempted to measure the Casimir 
force between two flat 
metal plates.  A force balance based on a spring balance was used 
in the final series of measurements.  The sensitivity of the spring 
balance was between (0.1--1)$\times 10^{-3}\,$dynes.  
The extension of the spring 
was measured through a measurement of the capacitance formed by the 
two flat plates. Calibration of this capacitance was done with the 
help of tungsten and Platinum wires (uncertainties in this 
calibration are not reported).  Care was taken with vibration isolation.
The author reports that 
the knife-edges and the springs used led to 
large hysteresis which made determination of the surface separation 
distance difficult.  This was reported to be the most severe drawback 
of the measurement technique.  The plates were mounted such that they 
were electrically insulated from the rest of the apparatus.  
Sparnaay realized that even a small potential difference of 17mV 
between the two parallel plates was sufficient to overwhelm the 
Casimir force.  

To take care of any potential differences between the surfaces the 
two plates were brought in contact together at the start of the 
experiment.  Three sets of metal plates, Al-Al, chromium-chromium  and 
chromium-steel were used in the measurements. Even with a variety of 
electrical and mechanical cleaning procedures, dust particles larger 
than 2--3$\,\mu$m were observed on the plates. 
The plates were aligned parallel 
by visual inspection with about a 10\% variation in the interplate 
distance from one of the plate to the other. Because of the presence 
of the dust particles it is estimated that even on contact, the plates 
are separated by 0.2$\,\mu$m (the procedure used to determine this 
was not provided).  
The chromium-steel and the chromium plates both led to attractive 
forces between them whereas the aluminium plates led to repulsive force. 
The peculiar repulsive force noticed in the case of the aluminium plates 
was thought to be due to presence of impurities on the aluminium surface.  
In the case of the attractive force for the chromium and chromium-steel 
plates, given the uncertainties in the measurement of the interplate 
distance only a general agreement with the Casimir force formula 
(here perfectly reflecting boundaries are assumed) could be achieved.
Barring 
the repulsive forces measured with aluminium plates necessary 
improvements other than the force sensitivity are:
\begin{enumerate}
\item
more accurate measurement of surface separation;
\item
more accurate measurement of the parallelism between the two 
surfaces (angles of less than 10${}^{-4}$ radians between plates 
of 1\,cm${}^2$ area are necessary);
\item
measurement of any residual electrostatic potential 
differences between the two surfaces given the presence 
of the dust particles. 
\end{enumerate}

In conclusion, these sets of measurements were the first indication 
of an attractive Casimir force between metallic surfaces, approximately 
in line with the 
expectations. (Note that the aluminium plates showed repulsive force 
and therefore the attractive force was not conclusive.) 
Most importantly from these
experiments, M.J.~Sparnaay clearly elucidated the 
problems that needed to be overcome for a rigorous and conclusive 
measurement of the Casimir force.  

\subsubsection {Experiments by Derjaguin et al.}
\label {sec6.2.2}

 One of the major improvements that was pioneered by the group of 
Derjaguin et al. \cite{sIV3-3} was the use of curved surfaces to avoid 
the need to maintain two flat plates perfectly parallel.  This was 
accomplished by replacing one or both of the plates by a curved surface 
such as a lens, sphere or cylinder. The first use of this technique 
was to measure the force between a silica lens and plate 
\cite{sIV3-3,sIV3-4,s6.4a2,s6.4a3,s6.4b}.  
Sparnaay [20] points out that this work 
did not take into account the presence of ``gel layer" which is 
usually present on such surfaces. Also the possible 
substantial electrostatic 
forces which will result in systematic errors are not reported in the 
experiment. These experiments will not be further discussed here. 

 There was also related work with metallic surfaces by Derjaguin et 
al \cite{sIV3-4,s6.6}.  In the case of metallic surfaces the forces between 
platinum fibers and gold beads were measured.  The force measurement 
was done by keeping one surface fixed and attaching the other surface 
to the coil of a galvanometer. The rotation of the galvanometer coil 
in response to the force led to the deflection of a light beam which 
was reflected off mirrors attached to the galvanometer coil.  
This deflected light beam was detected through a resistance bridge, 
two of whose elements could be photoactivated.  The measured forces 
indicated an unretarded van der Walls forces for distances below 
(50--80)\,nm 
and a retarded force region for larger distances. However more 
accurate modern theoretical results \cite{28}  predict an unretarded 
force below distances of 2nm in the case of gold. 
Derjaguin et al. report a discrepancy in the force measurements of 
around 60\% \cite{s6.6}.  Also any possible electrostatic 
forces due to the potential differences between the two surfaces appears 
to have been neglected.   While the distance on contact of the two 
surfaces appears to have been taken as the zero distance (ignoring the 
role of surface roughness), mention is made that surface roughness might 
have affected the experimental measurements and make the comparison to 
theory very difficult particularly for distances less than 30nm.

\subsubsection {Experiments of D.~Tabor, R.~Winterton and 
J.~Israelachvili using mica cylinders}
\label {sec6.2.3} 
  
In the inter-weaning years between the experiments of M.J.~Sparnaay and 
those of P.~van Blokland and J.~Overbeek  
using metallic surfaces, there were many force 
measurements on non-conductive surfaces.  Of these, the experiments 
on muscovite mica \cite{s6.9,s6.10,s6.11,s6.12} will be discussed here.  
The major improvement in these experiments was the use of atomically 
smooth surfaces from cleaved muscovite mica.  This provides the 
possibility of very close approach of the two surfaces.  
As a result it was possible
to measure the transition region between retarded and non-retarded 
van der Waals forces in those particular materials. Cylindrical 
surfaces of radii between 0.4\,cm--2\,cm obtained by wrapping the mica 
sheets on glass cylinders were used to measure the force.  
The procedure used in the making of the mica cylinder led to 50\% 
uncertainties in its radius. A spring type balance based on the jump 
method was used (for large separations modifications to this were done). 
Here the force of attachment of one of the cylinders to an extended 
spring is overcome by the attractive force from the opposite cylindrical 
surface. By using springs of different extensions and different spring 
constants a variety of distances could possibly be measured. 
Multiple beam interferometry was used for the measurement of the surface 
separation with a reported resolution of 0.3nm. In what appears to be the 
final work in this regard \cite{s6.10}, a sharp transition from the retarded 
to the non-retarded van der Waals force was found at 12nm 
(earlier work had measured a transition at larger surface 
separations \cite{s6.9}).  Later reanalysis of the data with more precise 
spectral properties for the mica revealed that the data could be 
reconciled with the theory only if errors of at least 30\% in the radius 
of curvature of the mica cylinders were introduced \cite{s6.12}.  
The possibility of 
changes in the spectral properties of the mica surface used to make 
the cylinders was also mentioned to explain the discrepancy.  
The separation on contact of 
the two surfaces was assumed to be zero, i.e., the surfaces were assumed 
to completely free of dust, impurities and any atomic steps on the 
cleaved surface. Additionally, as mica is a nonconductor 
which can easily accumulate static charges,  
the role of electrostatic forces between the cylinders 
is hard to estimate.

\subsubsection {Experiments of P.~van Blokland and J.~Overbeek}
\label {sec6.2.4}

The next major set of improved experiments with metallic surfaces were 
performed by P.~van Blokland and J.~Overbeek \cite{s6.8}.  
Here many of improvements 
achieved with dielectric surfaces were incorporated.  Also care was 
taken to address many of the concerns raised by M.J.~Sparnaay that were 
listed in the introductory notes. 
(Earlier measurements in the group \cite{s6.7} 
with dielectric surfaces did not report on the effects of 
chemical purity of the surface and the role of 
the electrostatic forces between the surfaces.) The final improved 
version of the experiment using metallic surfaces was done by P.~van 
Blokland and J.~Overbeek \cite{s6.8}. The experiment was done using 
a spring balance.  The force was measured between a lens and a flat 
plate coated with either (100$\pm$5)\,nm or (50$\pm$5)\,nm of chromium.  
The chromium 
surface was expected to be covered with (1--2)\,nm of surface oxide.  Water 
vapor was used to reduce the surface charges. This use of water vapor 
might have further affected the chemical purity of the metal surface. 
At the outset the authors recognize the outstanding problems in the 
Casimir force measurements as:
\begin{enumerate}
\item
the potential difference between the two surfaces leads to 
electrostatic forces which complicates the measurement;
\item
the exact determination of the separation distance between the 
two surfaces is required; 
\item
the exact determination of the nonzero surface separation on contact of 
the two surfaces should be performed. 
\end{enumerate}

The authors then try to address the above problems. 
The first was done by two methods:
by looking for a minimum in the 
Casimir force as a function of the applied voltage,
and by measuring the 
potential difference from the intersection point in the electrostatic 
force with application of positive and negative voltages.  The two 
methods yielded approximately consistent values for the potential 
difference of between (19--20)\,mV.   This large potential difference was 
equal to the Casimir force around 400\,nm surface separation.  Thus to 
measure the Casimir force the experiment had to be carried out with 
a compensating voltage present at all times.  

The separation distance between the two surfaces was measured 
through a measurement of the lens-plate capacitance using 
a Schering Bridge.  
This capacitance method is applicable for relative determination of distances 
as cables and stray capacitances are of the same order as that between the 
spherical surface and the plate.   Additional problems such as the tilt of 
the lens with respect to the plate were recognized by the authors.
The distance was calibrated 
with the help of the electrostatic force at a few points.  The force was 
measured for distances between 132\,nm to 670\,nm for the 100nm thick metal 
coating.  Only distances larger than 260\,nm could be probed for the 50\,nm 
metal coating. 

The theoretical treatment of chromium metal was 
noted to be problematic as it 
has two strong absorption bands around 600\,nm. Given this it was very hard 
to develop a complete theory based on the Lifshitz model and some empirical 
treatment was necessary. The imaginary part of the dielectric constant 
corresponding to this absorption was modeled as a Lorentz atom \cite{s6.23}. 
The two overlapping absorption 
bands were 
treated as a single absorption band.   The strength of this 
absorption band could only approximately 
be taken into account in the theoretical 
modeling. However this absorption band was found to make about 40\% of the 
total force.  The long wavelength response of chromium was modeled as 
Drude metal with a plasma frequency based on an electron number density 
of $1.15 \times 10^{22}\mbox{cm}^{-3}$.  With this theoretical 
treatment the measured force was shown to be consistent with the theory.  

The authors estimate the effect of surface roughness 
\cite{sIV3,sV3-1,sV3-2,sV3-8} 
which was neglected in the theoretical 
treatment, to make contributions of order 10\%.   The relative uncertainty 
in the measured force was reported to be around 25\% near 150\,nm 
separation but much 
larger around 500\,nm separation.  The authors report that the noise comes 
from the force measurement apparatus.  Given the above we can estimate 
the accuracy of the experiment to be of order 50\%. But it is worth noting 
that this was the first experiment to grapple with all the important 
systematics and other factors noted in Sec.6.1 which are 
necessary to make a clear measurement of the Casimir force.  
This experiment can therefore be considered as the first unambiguous
demonstration of the Casimir force between metallic surfaces.
Thus it is also the first measurement of surface forces in general were 
an independent estimate of the experimental precision can be attempted
(though 
none was provided by the authors).   

\subsubsection {Dynamical force measurement techniques}
\label {sec6.2.5}	
In theory, dynamical force measurements are more sensitive as 
the signal (and the noise) in a narrow bandwidth is monitored. 
A dynamical force measurement technique was first used by S.~Hunklinger,  
W.~Arnold and their co-authors
to measure the Casimir force between silica surfaces and 
silica surfaces coated with a thin layer of silicon \cite{s6.12a,s6.12b}. 
Here a glass lens of radius 2.5\,cm was attached 
to the metal coated top membrane of loudspeaker, while 
a glass flat plate surface was top surface of microphone. In one case the 
glass surfaces were coated with silicon \cite{s6.12b}. 
A sinusoidal voltage (at the microphone resonance frequency of 
3\,kHz) was applied to the loudspeaker such that 
the distance between the two surfaces also changed sinusoidally.  
This change resulted in the sinusoidal oscillation of the flat plate on the 
microphone due to the Casimir force. This oscillation on the 
microphone was detected.  The calibration was done by removing the 
plate and lens and applying an electrostatic voltage between the top of the 
loudspeaker and the microphone. A probable error of 20\% 
\cite{s6.12a} and 50\% \cite{s6.12b} 
in the force calibration is reported.  
A possible force sensitivity was of about $10 ^{-7}\,$dynes. 
The electrostatic force was minimized by use of water vapor and acetic 
acid vapor. No report of the residual electrostatic force was provided. 
Given the glass manufacturers roughness specifications of 50nm 
for the surfaces, the 
surface separation on contact was estimated to be around 80\,nm.
Deviations from the expected behavior where found for separation 
distances below 300nm and larger than 800\,nm. 
Such a deviation might be possible due to the presence of the
``gel layer'' pointed out by 
Sparnaay and the role of the electrostatic charges.

There is also a proposal to detect Casimir forces using a tunneling 
electromechanical transducer by  R. Onofrio and G. Carugno 
\cite{s6.14,s6.15}.  Here the force was to be measured between two 
parallel metal surfaces with one plate mounted on a cantilever and a 
disk mounted rigidly on a piezo electric stack. Small oscillations induced 
on the disk by the application of a sinusoidal voltage to the piezo will 
induce oscillations of the plate (and the cantilever) through the Casimir 
force.   This Casimir force induced cantilever motion 
should be detected by 
monitoring the tunneling current between a sharp needle and the cantilever.  
The sensitivity was increased by looking at the sidebands of the 
oscillation frequency as they were found to be less affected by seismic 
noise or $1/f$ electrical noise.  In this report only the noise limit was 
found by applying an electrostatic bias to the metal pieces. The 
noise limit at surface separations of 1$\,\mu$m 
was reported to be three orders of magnitude 
larger than that necessary for the detection of the Casimir force.  
Further improvements such as a resonator with larger mechanical 
quality factor, phase sensitive detection and vacuum operation 
were proposed.  

\subsection {Experiment by S.K.~Lamoreaux}
\label {sec6.3}

This experiment by S.K.~Lamoreaux \cite{32} was a landmark experiment, 
being the first in a modern phase of Casimir force 
measurements.  It particularly invigorated both the theoretical 
and experimental community by coinciding with 
the development of the modern unification theories on 
compactified dimensions (discussed below in Sec.7).  This 
coincidence heightened the awareness of the usefulness of 
the Casimir force measurements as being one of the 
most sensitive tests to new forces in the submillimeter distance range.  It 
was also the first time that most of the relevant parameters 
necessary for a careful calculation of the experimental 
precision was measured and reported.  Thus a quantification of 
the experimental precision could be consistently attempted 
without recourse to an arbitrary estimation of parameters.  

This experiment used a balance based on the 
torsion pendulum to measure the Casimir force between a gold coated 
spherical lens and f(lat plate. A lens with a radius of (11.3$\pm$0.1)\,cm 
(later corrected to (12.5$\pm$ 0.3)\,cm in \cite{sV23-1,s6.27}) was used.  
The two surfaces were first coated with 0.5$\,\mu$m of $Cu$ followed by 
a 0.5$\,\mu$m coating of $Au$. Both coatings were done by evaporation. 
The lens was mounted 
on a piezo stack and the plate on one arm of the torsion balance.  
The other arm of the torsion balance formed the center electrode of 
dual parallel plate capacitors $C_1$ and $C_2$. Thus the position of 
this arm and consequently  
the angle of the torsion pendulum could be controled by 
application of voltages to the plates of the dual capacitor.  
The Casimir force between the plate and lens surface would result in 
a torque, leading to a 
change in the angle of the torsion balance. This change in angle would 
result in changes of the capacitances $C_1$ and $C_2$ which were detected 
through a phase sensitive circuit.  Then compensating voltages were 
applied to the capacitors $C_1$ and $C_2$ through a feedback circuit to 
counteract the change in the angle of the torsion balance. These 
compensating voltages were a measure of the Casimir force. 
The calibration of this system was done electrostatically.  When the 
lens and plate surfaces were grounded, a ``shockingly large" [40]
potential difference  of 430\,mV 
was measured between the two surfaces.  This large electrostatic 
potential difference between the two surfaces was partially compensated 
with application of voltage to the lens (from the analysis there appears to
have been a 
residual electrostatic force even after this compensation).

The lens was moved towards the plate in sixteen steps by application of 
voltage to the piezo stack on which it was mounted.  At each step, 
the restoring force, given by 
the change in voltage required to keep the pendulum angle fixed 
was noted.  The maximum separation between the two surfaces was 
$12.3\,\mu$m.  The 
average displacement for a 5.75 volt step was about $0.75\,\mu$m.  
Considerable 
amount of hysteresis was noted between the up and down cycles, 
i.e. approach and retraction 
of the two surfaces. The displacement as a function of the sixteen 
applied voltages was reported as 
measured to 0.01$\,\mu$m accuracy with a laser interferometer.  
The total force was 
measured between separations of 10$\,\mu$m to contact of the two surfaces. 
The experiment was repeated and a 
total of 216 up/down sweeps were used in the final data set.  The total 
measured force data was binned into fifteen surface separation points.  
Two of the important experimental values needed (a) the residual 
electrostatic force and (b) the surface separation on contact of the 
two surfaces were obtained by curve fitting the expected Casimir force 
to total measured force in the following manner. The total measured 
force for separations greater than 2$\,\mu$m (12 of the 15 data points) was 
fit to the sum of the Casimir force (not including the conductivity 
corrections) and the electrostatic force.  Thus a fitting function 
was of the form 
\beq 
F^{m}(i)=F_{c}^{T}(a_i+a_0)+\frac{\beta}{a_i+a_0}+b,
\label{Lam7}
\eeq
where $F^{m}(i)$ is the total measured force at the  $i$-th step, 
$F_{c}^{T}(i)$ is the 
theoretical Casimir force including the temperature correction 
given by Eqs.~(\ref{sV13.5}), (\ref{sV13.6}).

The distance $a _ 0$ in Eq.~(\ref{Lam7}) is left as a
fit parameter which gives the absolute plate separation on contact of 
the two 
surfaces  and $b$ is a constant. Also $\beta$ which is  
a measure of the electrostatic force between the two surfaces was 
left to be 
determined by the fit.  The uncertainty in $a_{0}$ was noted to be 
less than 0.1$\,\mu$m and a typical drift of 0.1$\,\mu$m was noted 
between the up/down 
sweeps. About 10\% of the up/down sweeps were noted to be rejected because of 
poor convergence, non-physical value for $a_{0}$ or an inconsistent result for 
$\beta$.  Finally 216 up/down sweeps were retained.   The value of $a_{0}$
and 
the $\beta$
determined from this fit to the 12 data points were then used to subtract 
out the residual 
electrostatic force from the total measured force at all the 15 data 
points.  Thus,
\beq 
F_{c}^{m}(a_i)=F(a_i)-\frac{\beta}{a_i}-b,
\label{Lam8}
\eeq 
where $F_{c}^{m}(a_i)$ is the measured Casimir force.  Even
after 
application of the compensating voltages to the lens and plate 
surfaces, the Casimir force was 
noted to be only around 20\% of the residual electrostatic force at the 
point of closest approach.  

Next, a quantification of the experimental precision was attempted.  
Here the experimental measured force $F_{c}^{m}(a_{i})$ 
and the theoretical force $F^{T}_{c}$ where compared using the equation 
Eq.~(\ref{sIV3.7})
\beq 
F_{c}^{m}(a_i)=(1+\delta)F_{c}^{T}(a_i)+b^{\prime}.
\label{Lam9}
\eeq
In the above values of 
$b^{\prime}<5 \times 10^{-7}$\,dyn (95\% confidence level) 
were used in the fit. For the 216 sweeps considered the average of 
value of $\delta$  
determined from above was $0.01\pm 0.05$.  Based on this a 5\% degree 
of experimental precision was quoted for this experiment at all
surface separations from 0.6 to $6\,\mu$m.   
In the above 
determination of the precision, the conductivity correction of the 
first order from Eq.~(\ref{sV2.13})
was not included in the comparison.  No evidence of its influence on the
force value was gleaned from the measurement.  
In \cite {32} an 
upper limit of  3\% of any effect of  
the conductivity correction was placed. Surface roughness
corrections were not reported in \cite{32}.

 It was first realized in \cite{37} that 
finite conductivity corrections could amount to as much as 
20\% of the Casimir force at the separation of about $1\,\mu$m.  
Subsequently S.K.~Lamoreaux  
pointed out two errors \cite {s6.27} in his experimental 
measurement.  He calculated the finite conductivity correction for 
gold to be 22\% and 11\% for copper at 0.6$\,\mu$m.  Thus the 
theoretical value of the Casimir force would decrease by these 
percentage values. This calculation 
was based on the tabulated complex index of refraction for the two metals. 
A second error in the measurement of the radius of curvature was  noted.  
Here the lens surface was reported to be aspheric and the radius in the 
region where the Casimir force measurement was 
performed was reported to be (12.5$\pm$0.3)\,cm, 
in contrast with the (11.3$\pm$0.1)\,cm used earlier in \cite {32}.  
This change 
corresponds to a  10.6\% increase in the theoretical value of the Casimir 
force.  
It was noted in \cite {34} that only a pure copper film (with the 11\% 
conductivity correction) is consistent 
with a 5\% precision reported earlier in \cite {32}.  Thus an assumption 
of complete 
diffusion of the copper layer through the 0.5$\,\mu$m gold layer on 
both the lens and 
plate surfaces was thought necessary for the 
preservation of the experimental precision \cite{s6.27}.  Later work 
\cite{27,sV23-2} has shown that gold and copper surfaces lead to identical 
forces. 
  
In conclusion, this experiment introduced the modern phase sensitive 
detection of forces and thus brought possible increased sensitivity 
to the measurement of the Casimir forces.  
By using piezoelectric translation of the lens towards the plate,
reproducible measurements of the surface separation could be done.
The value of the electrostatic force and the surface separation on
contact could only be determined by curve fitting part of the experimental
data to the expected Casimir force. Such a procedure biases that part
of the experiment with the input value of the Casimir force. 
Also as noted by Lamoreaux in \cite{32}, the data is not 
of sufficient accuracy to demonstrate  the temperature corrections to 
the Casimir force.
It should be noted that the temperature corrections 
are 86\%, 129\% and 174\% of the zero temperature result
at surface separations of 4, 5 and 6$\,\mu$m respectively \cite{37}.
As it follows from the Fig.~4 of \cite{32}, the absolute error of
force measurements is around $\Delta F=1\times 10^{-11}\,$N resulting  
in the relative error of approximately 700\% at $a=6\,\mu$m. 
Thus the temperature corrections remain unconfirmed and the
(5--10)\% accuracy would probably apply 
at the smallest separations only.
Above all this work by applying modern positioning and force measurement
techniques to the Casimir effect promised rigorous tests of the theory
including the various corrections. Thus it stimulated a surge in
theoretical activity.
 
\subsection {Experiments with the Atomic Force Microscope by Mohideen et al.}
\label {sec6.4}

The increased sensitivity of the Atomic Force Microscope (AFM) was used 
by Mohideen et al. to perform the most definitive experiments on the 
measurement of the Casimir force \cite{33,35,36}.  
With the use of the AFM the authors report a statistical precision of 
1\% at smallest separations
in their measurement of the Casimir force. The three important 
requirements set forth by M.J.~Sparnaay, i.e. the use of non-reactive and 
clean metal surfaces,  the determination of the 
average surface separation on contact of the two surfaces only in [43] and
[44] and the minimization and independent measurement of 
the electrostatic potential differences were all done 
independent of the Casimir force measurement. In the case of the 
experiments with aluminium coating, a thin $Au/Pd$ coating was sputtered 
on top of the aluminium to reduce the effects of its oxidation.
In these experiments the $Au/Pd$ coating was treated 
phenomenologicaly as transparent. Complete theoretical treatment of 
thin metal coatings is complicated due to the wave vector dependence of 
the dielectric properties of metal layers \cite{28,sV23-5}.  
It should be 
noted that with regard to the initial work \cite{33} the complete 
conductivity and roughness corrections were described in detail in 
Ref.\cite{26}.  The last experiment \cite{36} using gold coating 
avoids all these ambiguities and the comparisons with theory are more 
solid.   Below we chronologically discuss the experiments and their 
improvements, culminating in the most accurate work with the gold surfaces.
In this section the configuration of a sphere above a flat disk is
considered. The case of a corrugated plate is a subject of Sec.6.5.

\subsubsection{First AFM experiment with aluminium surfaces}
\label {sec6.4.1}

A schematic diagram of the experiment \cite{33} is shown in Fig.~\ref{fig6.1}.
\begin{figure}[ht]
\epsfxsize=17cm\centerline{\epsffile{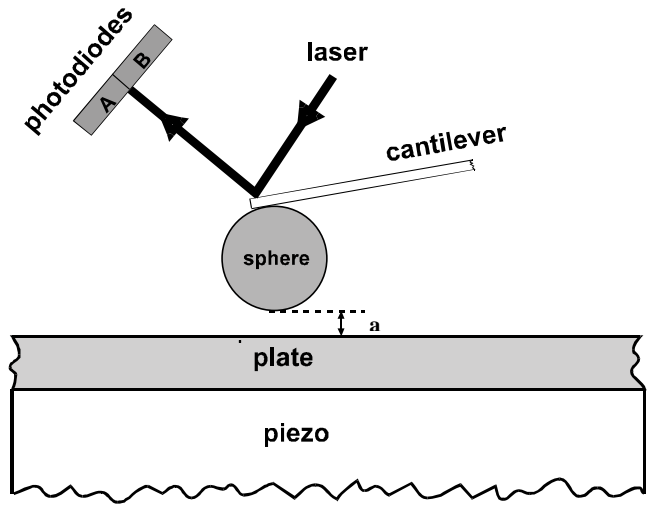} }
\vspace*{-9.5cm}
\caption{\label{fig6.1}
Schematic diagram of the experimental setup. Application of voltage
to the piezoelectric element results in the movement of the plate
towards the sphere.
}
\end{figure}
A force between the sphere and plate causes the cantilever to flex. 
This flexing of the cantilever is detected by the deflection of the 
laser beam leading to a difference signal between photodiodes A and B. 
This difference signal of the photodiodes was calibrated by means of 
an electrostatic force.  Polystyrene spheres 
of $(200\pm 4)\,\mu$m diameter were  mounted on the tip 
of the metal coated cantilevers with $Ag$ epoxy. A 1\,cm diameter optically 
polished sapphire disk is used as the plate.  The cantilever (with sphere) 
and plate were then coated by thermal evaporation with about 300\,nm of  
aluminium. To prevent the rapid oxidation of 
the aluminium coating 
and the development of space charges, the aluminium was sputter coated 
with a 60\%/40\% $Au/Pd$ coating of less than 20\,nm thickness.  
In the first experiments aluminium metal 
was used due to its high reflectivity at short wavelengths (corressponding 
to small surface separations). Aluminium coatings are also easy to apply 
due to the strong adhesion of the metal to a variety of surfaces and 
its low melting point. 

To measure the Casimir force between the sphere and 
the plate they are grounded together with the AFM.
The plate is then moved towards the sphere in 3.6\,nm steps 
and the corresponding photodiode difference
signal was measured. The signal
obtained for a typical scan is shown in Fig.~\ref{fig6.2}.
\begin{figure}[ht]
\epsfxsize=17cm\centerline{\epsffile{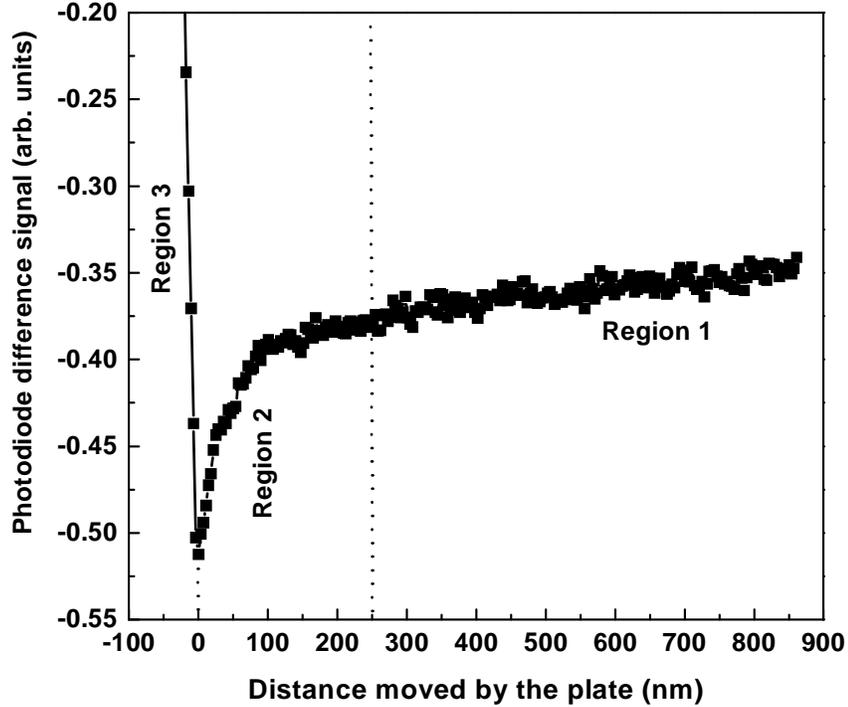} }
\vspace*{-9.5cm}
\caption{\label{fig6.2}
Typical force curve as a function of the distance moved by the plate.
}
\end{figure}
Here ``0" separation stands for contact of the sphere
and plate surfaces. It does not take into account the
absolute average separation between the $Au/Pd$
layers due to the surface roughness which is about
80\,nm . If one also takes into account  
the $Au/Pd$ cap layers which can be considered transparent at small
separations (see below) the absolute 
average separation at contact between $Al$ layers is
about 120\,nm. 
Note that in the experiment the separation distance on contact was found
by fitting the experimental data at large separations with the measured
electrostatic force and the theoretical Casimir force.
Region 1 shows that the force curve 
at large separations is dominated by a linear signal.
This is due to increased coupling of scattered light
into the diodes from the approaching flat surface.
Embedded in the signal is a long range attractive
electrostatic force from the contact potential
difference between the sphere and plate, and the
Casimir force (small at such large distances).
In region 2 (absolute separations vary from contact
to 350\,nm) the Casimir force is the dominant
characteristic far exceeding all the systematic
errors. Region 3 is 
the flexing of the cantilever resulting from the
continued extension of the piezo after contact of the
two surfaces. Given the distance moved by the flat
plate ($x$-axis), the difference signal of the
photodiodes can be calibrated to a cantilever
deflection in nanometers using the slope of the curve
in region 3.

Next, the force constant of the cantilever was
calibrated by an electrostatic measurement. The sphere
was grounded to the AFM and different voltages in the
range $\pm0.5\,$V to $\pm3\,$V were applied to the
plate. The force between a charged sphere and plate
is given as \cite{s6.33}
\beq
F=\frac{1}{2}(V_1-V_2)^2
\sum\limits_{n=1}^{\infty}
\mbox{csch}\,n\alpha(\coth\alpha-n\coth n\alpha).
\label{6.2}
\eeq
\noindent
Here $V_1$ is the applied voltage on the plate, and $V_2$
represents the residual potential on the grounded
sphere.
One more notation is 
$\alpha=\cosh^{-1}(1+a/R)$, where $R$ is the radius
of the sphere and $a$ is the separation between the
sphere and the plate. From the difference in force
for voltages $\pm V_1$ applied to the plate, we can
measure the residual potential on the grounded
sphere $V_2$ as 29\,mV. This residual potential is
a contact potential that arises from the different
materials used to ground the sphere. The electrostatic
force measurement was repeated at 5 different
separations and for 8 different voltages $V_1$.  While the force is 
electrostatically calibrated, one can derive an equivalent force
constant using Hooke's law and the force from Eq.~(\ref{6.2}). 
The average value thus derived was $k=$0.0182\,N/m.

The systematic error corrections to the force curve
of Fig.~\ref{fig6.2}, due to the residual potential on the
sphere and the true separations between the two
surfaces, are now calculated. Here the near linear
force curve in region 1, is fit to a function of the
form: 
$$F=F_c(\Delta a+a_0)+\frac{B}{\Delta a+a_0}+C\times(\Delta a+a_0)+E.$$
In this equation $a_0$ is the absolute separation at contact,
which is constrained to $120\pm5\,$nm, is the only
unknown to be completely obtained by the fit. The
second term represents the inverse linear dependence
of the electrostatic force between the sphere and
plate for $R\gg a$.
The constant $B=-2.8\,\mbox{nN}\cdot$nm corresponding
to $V_2=29\,$mV and $V_1=0$ is
used. The third term represents the linearly increasing
coupling of the scattered light into the photodiodes
and $E$ is the offset of the curve. Both $C$ and $E$
can be estimated from the force curve at large
separations. The best fit values of $C$, $E$ and
the absolute space separation $a_0$ are determined by
minimizing the $\chi^2$. The finite conductivity
correction and roughness correction (the largest
corrections) do not play a significant role in the
region 1 and thus the value of $a_0$
determined by the fitting is unbiased with respect
to these corrections. These values of $C$, $E$ and
$a_0$ are then used to subtract the systematic errors
from the force curve in region 1 and 2 to obtain the
measured Casimir force as $(F_c)_m=F_m-B/a-Ca-E$,
where $F_m$ is the measured total force.
This procedure was repeated for 26 scans in different
locations of the flat plate. The average measured
Casimir force $(F_c)_m$ as a function of sphere-plate
separations from all the scans is shown in 
Figs.~\ref{fig6.3} and \ref{fig6.4} as open squares
in the separation range $120\,\mbox{nm}\leq a\leq 950\,$nm
($a$ is the distance between $Al$ surfaces).
\begin{figure}[ht]
\epsfxsize=17cm\centerline{\epsffile{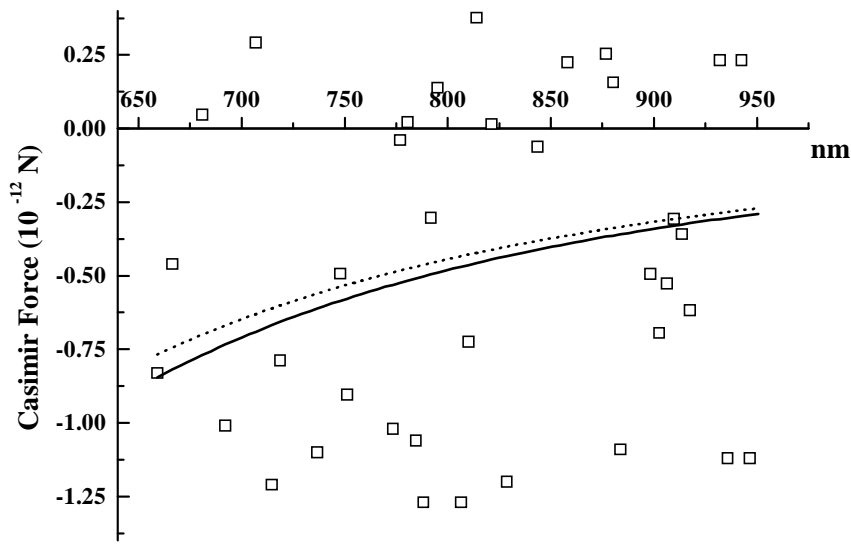} }
\vspace*{-9.5cm}
\caption{\label{fig6.3}
Measured average Casimir force for large distances
as a function of plate-sphere separation
is shown as open squares. The theoretical Casimir force with corrections
to surface roughness and finite conductivity is shown by the solid
line (when the space separation is defind as the distance between
$Al$ layers) and by the dashed line (with the distance between $Au/Pd$
layers).
}
\end{figure}
In Fig.~\ref{fig6.4} the dashed line represents the Casimir force
of Eq.~(\ref{sIV3.7}) which is in evident deviation from the
experimental data. Because of this, different corrections to the 
Casimir force in the above experimental configuration should be
estimated and taken into account.
\begin{figure}[ht]
\epsfxsize=17cm\centerline{\epsffile{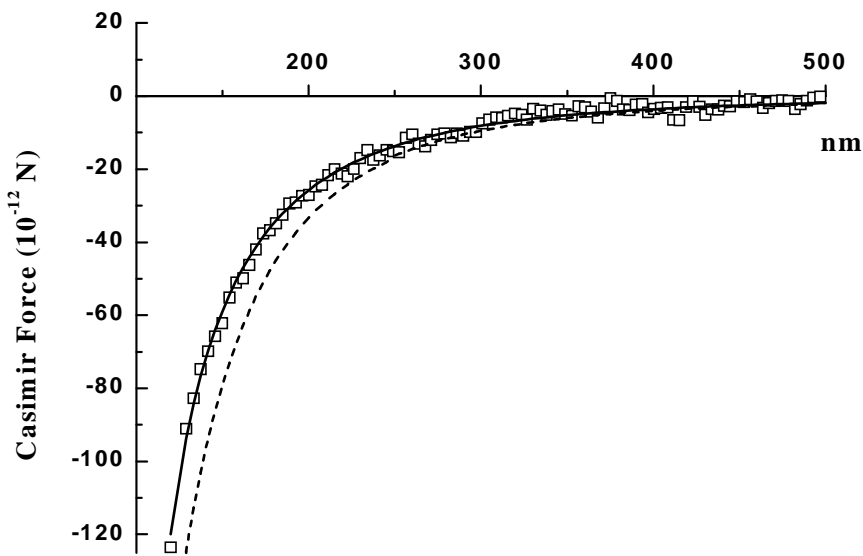} }
\vspace*{-9.5cm}
\caption{\label{fig6.4}
Measured average Casimir force for small distances
as a function of plate-sphere separation
is shown as open squares. The theoretical Casimir force with corrections
due to surface roughness and finite conductivity is shown by the solid
line, and without any correction
by the dashed line.
}
\end{figure}

Now let us compare the experimental results with a more precise theory 
taking into account surface roughness and finite conductivity
corrections to the Casimir force (temperature corrections are
negligibly small for the separations under consideration).
For  distances of $a\sim 1\mu$m between the
interacting bodies both the surface roughness and finite conductivity
of the boundary metal make important contributions
 to the value of the Casimir force.
Although the exact calculation is impossible, one can find the
corresponding corrections approximately with the
required accuracy.
In the first report \cite{33}  only the second order conductivity 
and roughness corrections were used for the comparison between theory 
and experiment. Using such a comparison, the authors found that the 
root mean square (rms) 
deviation of the experiment ($F_{exp}$) from the theory ($F_{th}$) is 
$\sigma=1.6\,$pN in complete measurement range.  
This is of order of 1\% of the forces measured at the closest 
separation and was used as the measure of precision. 
Second order perturbation theory is, however, insufficient to calculate
the force value with an accuracy of 1\%, because the third and fourth
perturbation orders contributions are generally larger than 1\%. Rather
good agreement between theory and experiment obtained in \cite{33} is
explained by the fact that both corrections are of different signs and 
partly compensate each other (see Secs.~5.2.2 and 5.3.5). As a result,
the value of $\sigma$ in \cite{33} was interval dependent, i.e. 
different if calculated separately at small separations, large
separations and in complete measurement range.  
Subsequently in 
Ref.~\cite{26} both the conductivity and roughness corrections were 
improved and a better approach to the theory was considered. 
This is discussed below starting with the  roughness correction.
 
We use the formalism presented in Sec.~5.3 to describe
a plane plate (disk) of dimension $2L$,
thickness $D$ and a sphere above it of
radius $R$ both covered by roughness.
The roughness on the plate is described by the
function (\ref{sV3.1}).
The roughness on the sphere is described by Eq.~(\ref{sV3.41}).
The values of the roughness amplitude are defined as specified
in Sec.~5.3. 
As is seen from the below experimental investigation the
characteristic lateral sizes of roughness on the plate ($T_d$) and
on the sphere ($T_l$) obey the inequality (\ref{sV3.51}).
In this case, following the development 
outlined in Sec.~5.3.5, the Casimir force with account of the 
roughness correction takes the form (see Eq.~(\ref{sV3.52}))
\bes
&&F_{dl}^R(a)=F_{dl}^{(0)}(a)\left\{1+
6\!\left[
\langle\!\langle f_1^2\rangle\!\rangle
\left(\frac{A_1}{a}\right)^2-
2\langle\!\langle f_1f_2\rangle\!\rangle
\frac{A_1}{a}
\frac{A_2}{a}+
\langle\!\langle f_2^2\rangle\!\rangle
\left(\frac{A_2}{a}\right)^2
\right]\right.
\nonumber\\
&&\phantom{a}+
10\left[
\langle\!\langle f_1^3\rangle\!\rangle
\left(\frac{A_1}{a}\right)^3-
3\langle\!\langle f_1^2f_2\rangle\!\rangle
\left(\frac{A_1}{a}\right)^2
\frac{A_2}{a}+
3\langle\!\langle f_1f_2^2\rangle\!\rangle
\frac{A_1}{a}
\left(\frac{A_2}{a}\right)^2
\right.
\nonumber\\
&&\phantom{a}-\left.
\langle\!\langle f_2^3\rangle\!\rangle
\left(\frac{A_2}{a}\right)^3
\right]+
15\left[
\langle\!\langle f_1^4\rangle\!\rangle
\left(\frac{A_1}{a}\right)^4-
4\langle\!\langle f_1^3f_2\rangle\!\rangle
\left(\frac{A_1}{a}\right)^3
\frac{A_2}{a}
\right.
\label{6.6}\\
&&\phantom{a}\left.\left.+
6\langle\!\langle f_1^2f_2^2\rangle\!\rangle
\left(\frac{A_1}{a}\right)^2
\left(\frac{A_2}{a}\right)^2-
4\langle\!\langle f_1f_2^3\rangle\!\rangle
\frac{A_1}{a}
\left(\frac{A_2}{a}\right)^3
+\langle\!\langle f_2^4\rangle\!\rangle
\left(\frac{A_2}{a}\right)^4
\right]\right\}.
\nonumber
\ees
\noindent
Here the double angle brackets denote two successive
averaging procedures with the first one performed over the surface area 
of interacting
bodies and the second one over all possible phase
shifts between the distortions situated on the
surfaces of interacting bodies against each other.
This second averaging is necessary because in the
experiment \cite{33} the measured Casimir force
was averaged over 26 scans done on different points on the plate surface.
$F_{dl}^{(0)}(a)$ is the Casimir force acting between perfect metals
in a perfectly shaped configuration.

The above result Eq.~(\ref{6.6}) was used to calculate
the roughness corrections to the Casimir
force in the experiment \cite{33}. 
The roughness of the metal 
surface was measured with the same AFM. After the
Casimir force measurement the cantilever with
sphere was replaced with a standard cantilever
having a sharp tip. Regions of the metal plate
differing in size from $1\mu\mbox{m}\times 1\mu$m  to
$0.5\mu\mbox{m}\times 0.5\mu$m  were scanned with the AFM.
A typical surface scan is shown in Fig.~\ref{fig6.5}.
\begin{figure}[ht]
\epsfxsize=15cm\centerline{\epsffile{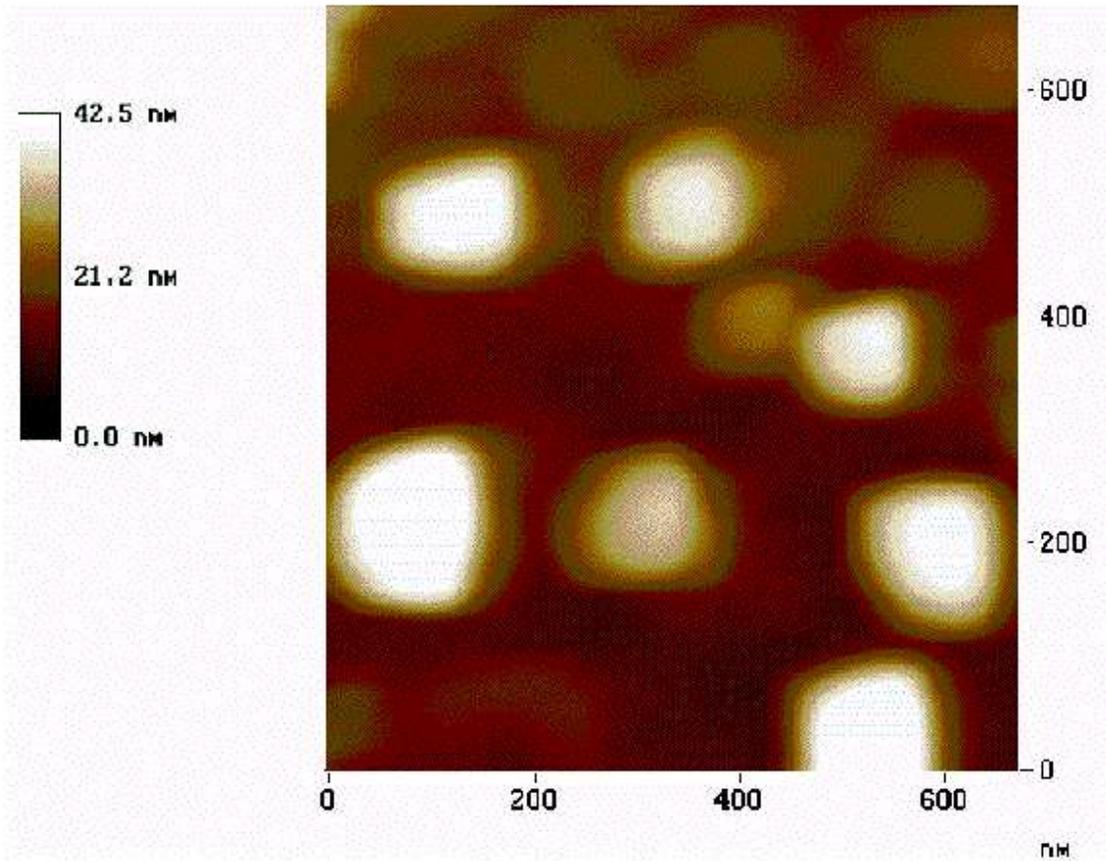} }
\vspace*{1.5cm}
\caption{\label{fig6.5}
Typical atomic force microscope scan of the metal surface. The lighter
tone corresponds to larger height as shown by the bar graph on the left.
}
\end{figure}
The roughness of the sphere was investigated with
a SEM and found to be
similar to the flat plate. In the surface scan of
Fig.~\ref{fig6.5}, the lighter tone corresponds to
larger height. 

As is seen from Fig.~\ref{fig6.5}
the major distortions are the large separate
crystals situated irregularly on the surfaces. They
can be modeled approximately by the parallelepipeds
of two heights. As the analysis of several AFM
images shows, the height of highest distortions is
about $h_1=40\,$nm  and of the intermediate ones --- about
$h_2=20\,$nm. Almost all surface between the
distortions is covered by the stochastic roughness
of height $h_0=10\,$nm. It consists of small
crystals which are not clearly visible in Fig.~\ref{fig6.5} due to the
vertical scale used.  All
together they form the  homogeneous background
of the averaged 
height $h_0/2$. The character of roughness 
on the plate and on the lens is quite similar. Note that in \cite{33}, 
only the highest distortions $h_1=40\,$nm were used to 
estimate the distortion amplitude.

Now it is possible to determine the height $H$
relative to which the mean value of the function,
describing the total roughness, is zero. It can be
found from the equation
\beq
(h_1-H)S_1+(h_2-H)S_2-
\left(H-\frac{h_0}{2}\right)S_0=0,
\label{6.7}
\eeq
\noindent
where $S_{1,2,0}$ are, correspondingly, the surface
areas occupied by distortions of the heights 
$h_1$, $h_2$ and stochastic roughness.
Dividing Eq.~(\ref{6.7}) into the area of interacting 
surface 
$S=S_1+S_2+S_0$ one gets
\beq
(h_1-H)v_1+(h_2-H)v_2-
\left(H-\frac{h_0}{2}\right)v_0=0,
\label{6.8}
\eeq
\noindent
where $v_{1,2,0}=S_{1,2,0}/S$ are the relative parts of
the surface occupied by the different kinds of
roughness. The analysis of the AFM pictures similar to
Fig.~\ref{fig6.5} gave us the values $v_1=0.11$, $v_2=0.25$,
$v_0=0.64$. Solving Eq.~(\ref{6.8}) we get the
height of the zero distortion level
$H=12.6\,$nm. The value of distortion amplitude
defined relatively to this level is
\beq
A=h_1-H=27.4\,\mbox{nm}.
\label{6.9}
\eeq

Below, two more parameters will also be used 
\beq
\beta_1=\frac{h_2-H}{A}\approx 0.231,
\qquad
\beta_2=\frac{H-h_0/2}{A}\approx 0.346.
\label{6.10}
\eeq
\noindent
With the help of them the distortion function from
Eq.~(\ref{sV3.1})  was represented as
\beq
f_1(x_1,y_1)=\left\{
\begin{array}{rcl}
1,&&(x_1,y_1)\in\Sigma_{S_1},\\
\beta_1,&&(x_1,y_1)\in\Sigma_{S_2},\\
-\beta_2,&&(x_1,y_1)\in\Sigma_{S_0},
\end{array}
\right.
\label{6.11}
\eeq
\noindent
where $\Sigma_{S_1,S_2,S_0}$ are the regions of the
first interacting body surface occupied by the
different kinds of roughness.

The same representation is valid for $f_2$ also
\beq
f_2(x_2,y_2)=\left\{
\begin{array}{rcl}
-1,&&(x_2,y_2)\in\tilde\Sigma_{S_1},\\
-\beta_1,&&(x_2,y_2)\in\tilde\Sigma_{S_2},\\
\beta_2,&&(x_2,y_2)\in\tilde\Sigma_{S_0},
\end{array}
\right.
\label{6.12}
\eeq
\noindent
$\tilde\Sigma_{S_1,S_2,S_0}$ are the regions of the
second interacting body surface occupied by the
distortions of
different kinds.

For the roughness under consideration
the characteristic lateral sizes of distortions are
$T_d,\,T_l\sim 200-300\,$nm as can be seen from Fig.~\ref{fig6.5}.
At the same time $\sqrt{aR}> 3000\,$nm. Thus the condition
(\ref{sV3.51}) is valid and Eq.~(\ref{6.6}) is really applicable to 
calculate the roughness corrections. 

Now it is not difficult to calculate the coefficients
of expansion (\ref{6.6}). One example is
\beq
\langle\!\langle f_1f_2\rangle\!\rangle=
-v_1^2-2\beta_1v_1v_2+2\beta_2v_1v_0
-\beta_1^2v_2^2+2\beta_1\beta_2 v_2v_0-
\beta_2^2 v_0^2=0,
\label{6.13}
\eeq
\noindent
which follows from Eqs.~(\ref{6.8})--(\ref{6.10}).
The results for the other coefficients are
\bes
&&
\langle\!\langle f_1^2\rangle\!\rangle =
\langle\!\langle f_2^2\rangle\!\rangle =
v_1+\beta_1^2 v_2+\beta_2^2 v_0,
\nonumber\\
&&
\langle\!\langle f_1^3\rangle\!\rangle =
-\langle\!\langle f_2^3\rangle\!\rangle =
v_1+\beta_1^3 v_2-\beta_2^3 v_0,
\quad
\langle\!\langle f_1f_2^2\rangle\!\rangle =
\langle\!\langle f_1^2 f_2\rangle\!\rangle =0,
\nonumber\\
&&
\langle\!\langle f_1^4\rangle\!\rangle =
\langle\!\langle f_2^4\rangle\!\rangle =
v_1+\beta_1^4 v_2+\beta_2^4 v_0,
\quad
\langle\!\langle f_1f_2^3\rangle\!\rangle =
\langle\!\langle f_1^3 f_2\rangle\!\rangle =0,
\nonumber\\
&&
\langle\!\langle f_1^2 f_2^2\rangle\!\rangle =
(v_1+\beta_1^2 v_2+\beta_2^2 v_0)^2.
\label{6.14}
\ees

Substituting Eq.~(\ref{6.14}) into Eq.~(\ref{6.6}) we get the
final expression for the Casimir force with 
surface distortions included up to the fourth order
in relative distortion amplitude
\bes
&&
F_{dl}^R(a)=F_{dl}^{(0)}(a)
\left\{
\vphantom{\left[\left(\beta_2^2v_0\right)^2\right]
\frac{A^4}{a^4}}
1+12\left(v_1+\beta_1^2v_2+\beta_2^2v_0\right)
\frac{A^2}{a^2}\right.
\nonumber\\
&&\phantom{aaaaa}
+20\left(v_1+\beta_1^3v_2-\beta_2^3v_0\right)
\frac{A^3}{a^3}
\label{6.15}\\
&&\phantom{aaaaa}\left.
+30\left[v_1+\beta_1^4v_2+\beta_2^4v_0+
3\left(v_1+\beta_1^2v_2+\beta_2^2v_0\right)^2\right]
\frac{A^4}{a^4}\right\}.
\nonumber
\ees

It should be noted that exactly the same result can be obtained in
a very simple way. To do this it is enough to calculate the value of
the Casimir force (4.108) 
for six different distances that are
possible between the distorted surfaces, multiply them by the
appropriate probabilities, and then summarize the results
\bes
&&F_{dl}^R(a)=
\sum\limits_{i=1}^{6}w_iF_{dl}^{(0)}(a_i)\equiv 
v_1^2F_{dl}^{(0)}(a-2A)
\nonumber\\
&&\phantom{aaa}
+2v_1v_2F_{dl}^{(0)}\left(a-A(1+\beta_1)\right)+
2v_2v_0F_{dl}^{(0)}\left(a-A(\beta_1-\beta_2)\right)
\label{6.15a}\\
&&\phantom{aaa}
+v_0^2F_{dl}^{(0)}(a+2A\beta_2)+v_2^2F_{dl}^{(0)}(a-2A\beta_1)
+2v_1v_0F_{dl}^{(0)}\left(a-A(1-\beta_2)\right).
\nonumber
\ees
\noindent
As was noted in Sec.~5.4 representations of this type immediately
follow from the Proximity Force Theorem and thereby, with an
appropriate interpretation of $F^{(0)}$, can be applied not only
inside the retarded (Casimir) regime, but also in the transition
region to the van der Waals force.

Now let us start with the corrections to the Casimir force due to
finite conductivity. 
The interacting bodies used in the experiment
\cite{33} were coated with 300\,nm of $Al$ in an
evaporator. The thickness of this metallic layer is
much larger than the penetration depth $\delta_0$
of electromagnetic oscillations into $Al$ for the
wavelengths (sphere-plate separations) of interest.
Taking $\lambda_p^{Al}= 100\,$nm as the 
approximative value
of the effective plasma wavelength of the electrons
in $Al$ \cite{sV23-3} one gets
$\delta_0=\lambda_p^{Al}/(2\pi)\approx 16\,$nm. What
this means is the interacting bodies can be 
considered as made of $Al$ as a whole.
Although $Al$ reflects more than 90\% of the
incident electromagnetic oscillations in the
complete measurement range 
$120\,\mbox{nm}< a <950\,$nm, 
some corrections to the Casimir force due to the
finiteness of its conductivity exist and should
be taken into account. In addition, as was already mentioned above,
to prevent the
oxidation processes, the surface of $Al$ in \cite{33}
was covered with less than $\Delta=20\,$nm layer of
$60\%Au/40\%Pd$. The reflectivity properties of this
alloy are much worse than of $Al$ (the effective
plasma wavelength of $Au$ is 
$\lambda_p^{Au}= 136\,$nm and the penetration
depth is $\tilde\delta_0\approx 21.6\,$nm).
Because of this, it is necessary
to take into account the finiteness of the metal
conductivity.

For large distances which are several times larger than
$\lambda_p^{Au}$ both $Al$ and
$Au/Pd$ are the good metals. In this case the
perturbation theory in the relative penetration depth
into both metals can be developed. The small parameter is the ratio
of an effective penetration depth $\delta_e$ 
(into both $Au/Pd$ and $Al$) and
a distance $a$ between the $Au/Pd$ layers. The
quantity $\delta_e$, in its turn, is understood as
a depth for which the electromagnetic oscillations
are attenuated by a factor of $e$. It takes into account
both the properties of $Al$  and of $Au/Pd$ layers.
The value of $\delta_e$ can be found from the
equation
\beq
\frac{\Delta}{\tilde\delta_0}+
\frac{\delta_e-\Delta}{\delta_0}=1,
\qquad
\delta_e=\left(1-\frac{\Delta}{\tilde\delta_0}\right)
\delta_0+\Delta\approx 21.2\,\mbox{nm}.
\label{6.16}
\eeq

The resultant finite conductivity correction is given by Eq.~(\ref{sV2.13}) 
where $\delta_e$ from Eq.~(\ref{6.16}) is substituted instead of
$\delta_0$.

Now we combine both corrections --- 
one due to the surface
roughness and the second
due to the finite conductivity of the
metals (see Sec.5.4.1). For this purpose we substitute the quantity
$F_{dl}^{\delta_e}(a_i)$ from Eq.~(\ref{sV2.13})
into Eq.~(\ref{6.15a}) instead of
$F_{dl}^{(0)}(a_i)$. The result is
\beq
F_{dl}^{R,C}(a)=\sum\limits_{i=1}^{6}
w_i F_{dl}^{\delta_e}(a_i),
\label{6.20}
\eeq
\noindent
where different possible distances between the
surfaces with roughness and their probabilities were
introduced in Eq.~(\ref{6.15a}). Eq.~(\ref{6.20})
describes the Casimir force between $Al$
bodies with $Au/Pd$ layers taking into account the
finite conductivity of the metals and surface
roughness for the distances several times larger than
$\lambda_p^{Au}$.

For the distances of order of $\lambda_p^{Au}$ or even
smaller a more simple, phenomenological, approach to calculation of the
Casimir force can be applied.
It uses the fact that the transmittance of 20\,nm
$Au/Pd$ films for the wavelength of around 300\,nm
is greater than 90\%. This transmission measurement
was made by taking the ratio of light transmitted
through a glass slide with and without the $Au/Pd$
coating in an optical spectrometer.

So  high transmittance gives the possibility to
neglect the $Au/Pd$ layers when calculating the
Casimir force and to enlarge the distance between the
bodies by $2\Delta=40\,$nm when comparing the
theoretical and experimental results. With this
approach for the distances 
$a<\lambda_p^{Au}$, instead of
Eq.~(\ref{6.20}), the following result is valid
\beq
F^{R,C}(a)=\sum\limits_{i=1}^{6}
w_i F_{dl}^{\delta_0}(a_i+2\Delta),
\label{6.21}
\eeq
\noindent
where the Casimir force with account of finite
conductivity is defined by the Eq.~(\ref{sV2.13}).

Now let us compare the theoretical Casimir force taking into account 
the fourth order roughness and conductivity
corrections with the experiment.
Let us first consider large surface separations 
(the distance
between the $Au/Pd$ layers changes in the interval
610\,nm$\leq a\leq 910\,$nm). 
We  compare the results given by Eq.~(\ref{6.20})
and Eq.~(\ref{6.21}) with experimental data.
In Fig.~\ref{fig6.3} the dashed
curve represents the results obtained by the 
Eq.~(\ref{6.20}), and solid curve --- by the
Eq.~(\ref{6.21}). 
The experimental points are shown as open squares.
For eighty experimental points,
which belong to the range of $a$ under consideration,
the root mean square average deviation between 
theory and experiment in both cases is 
$\sigma=1.5\,$pN. It is notable that for large $a$
the same result is also valid if we use the Casimir force from
Eq.~(\ref{sIV3.7}) (i.e. without any corrections) both for
$a$ and for $a+2\Delta$. By this is meant that for large $a$ the
problem of the proper definition of distance is not significant
due to the experimental uncertainty and the large scatter in 
experimental points. The same situation occurs with the corrections. 
At $a+2\Delta=950\,$nm the correction due to roughness (positive)
is about 0.2\% of $F_{dl}^{(0)}$, and the correction due to finite
conductivity (negative) is 6\%  of $F_{dl}^{(0)}$. Together they
give the negative contribution, which is also 6\%  of $F_{dl}^{(0)}$.
It is negligible if we take into account that the relative error of force 
measurement at the extreme distance of 950\,nm is of approximatelly
660\% (this is because the Casimir force is much less than the
experimental uncertainty at such distances).

Now we consider the range of smaller values of the
distance 80\,nm$\leq a\leq 460\,$nm (or, between
$Al$, 120\,nm$\leq a+2\Delta\leq 500\,$nm).
Here the Eq.~(\ref{6.21}) should be
used for the Casimir force. In Fig.~\ref{fig6.4} the Casimir
force $F_{dl}^{(0)}(a+2\Delta)$ from Eq.~(\ref{sIV3.7}) is shown by
the dashed curve. The solid curve represents the
dependence calculated according to Eq.~(\ref{6.21}).
The open squares are the experimental points.
Taking into account all one hundred experimental
points belonging to the range of smaller distances
we get for the solid curve the value of the root mean
square deviation between theory and experiment
$\sigma_{100}=1.5\,$pN. If we consider a more
narrow distance interval
80\,nm$\leq a\leq 200\,$nm which contains thirty
experimental points it turns out that 
$\sigma_{30}=1.6\,$pN
for the solid curve. In all the measurement range
80\,nm$\leq a\leq 910\,$nm the root mean square
deviation for the solid curves of Figs.~\ref{fig6.3},\,\ref{fig6.4} is
$\sigma_{223}=1.4\,$pN (223 experimental points). 
What this means is that 
the dependence
Eq.~(\ref{6.21}) gives equally good agreement with
experimental data in the region of small distances
(for the smallest ones 
the relative error of force measurement is about
1\%), in the region of large distances (where the
relative error is rather large) and in the whole
measurement range. If one uses less sophisticated expressions for
the corrections to the Casimir force due to surface roughness and
finite conductivity, the value of $\sigma$ calculated for small
$a$ would be larger than in the whole range.  

It is interesting to compare the obtained results with
those given by Eq.~(\ref{sIV3.7}), i.e. without taking
account of any corrections. In this case
for the interval
80\,nm$\leq a\leq 460\,$nm 
(one hundred experimental points) we have
$\sigma_{100}^0=8.7\,$pN. For the whole measurement
range
80\,nm$\leq a\leq 910\,$nm (223 points) there is
$\sigma_{223}^0=5.9\,$pN. 
It is evident that without 
appropriate treatment of the corrections to the Casimir
force the value of the root mean square deviation is
not only larger but also depends
significantly on the measurement range.

The comparative role of each correction is also quite
obvious. If we take into account only roughness
correction according to Eq.~(\ref{6.15}), then
one obtains
for the root mean square deviation in different
intervals: $\sigma_{30}^R=22.8\,$pN,
$\sigma_{100}^R=12.7\,$pN and 
$\sigma_{223}^R=8.5\,$pN. At $a+2\Delta=120nm$
the correction is 17\% of $F_{dl}^{(0)}$. For the single 
finite conductivity correction calculated by
Eq.~(\ref{sV2.13}) with $\delta_0$ instead of
$\delta_e$ it follows:
$\sigma_{30}^{\delta}=5.2\,$pN,
$\sigma_{100}^{\delta}=3.1\,$pN and 
$\sigma_{223}^{\delta}=2.3\,$pN.
At 120\,nm this correction contributes\ --34\%
of $F_{dl}^{(0)}$. (Note, that both corrections contribute\ 
--22\% of $F_{dl}^{(0)}$ at 120\,nm, so that their
nonadditivity is demonstrated most clearly.)

Several conlcusions can be reached on the first AFM experiment
on measuring the Casimir force.
This was the first experiment using the $10^{-12}\,$N force sensitivity 
of the AFM to make the Casimir force measurement.  In the 
initial report \cite{33} only the second order corrections to the 
conductivity and the roughness were considered. This was pointed out 
in Ref.\cite{s6.29}. Subsequently 
in Ref.\cite{26} both 
surface roughness and finite conductivity of the
metal were calculated up to the fourth order in
the respective small parameters. The obtained theoretical
results for the Casimir force with both corrections
were compared with the experimental data. 
The excellent agreement was demonstrated which is
characterized by almost the same value of the root
mean square deviation between theory and experiment
in the cases of small and large
space separations between the test bodies and in the complete
measurement range.  It was shown that the agreement between the theory 
and experiment is substantially worse if any one of 
the corrections is not taken into account. 
What this means is that the surface roughness
and finite conductivity corrections should be
taken into account in precision Casimir force measurements 
with space separations of
the order $1\mu$m and less.

Two of the three requirements set forth by M.J.~Sparnaay i.e. the use of a 
clean metal surface and the independent measurement of the electrostatic 
force between the two surfaces was met in this experiment. 
However, the value of the surface separation on contact of the two 
surfaces was done by fitting of the Casimir force to 
a part of the experimental curve at large separations, which as was the
case with [40] will bias the experimental curve. 
Also the roughness corrections were large, on the order of 17\% 
\cite{26,s6.29} at smallest separations. 
In the next set of experiments reported both these 
problems were eliminated.  Here an independent measurement of the surface 
separations was done and the roughness corrections were reduced to be 
order 1\% only.   These experiments are discussed below.

\subsubsection{Improved precision measurement with aluminium surfaces 
using the AFM}
\label {sec6.4.2}

The following year, Mohideen et al reported an improved version of 
the above experiment \cite{35}. The particular experimental improvements 
were (i) use of smoother metal coatings, which reduces the effect of 
surface roughness and allows for closer separations between the two 
surfaces (ii) vibration isolation which reduces the total noise (iii) 
independent electrostatic measurement of the surface separations and (iv) 
reductions in the systematic errors due to the residual electrostatic 
force, scattered light and instrumental drift. Also the complete dielectric 
properties of $Al$ is used in the theory along the lines of Sec.5.2.3. 
The average precision defined on 
the rms deviation between experiment and theory remained at the same 1\% 
of the forces measured at the closest separation. For a metal with a 
dielectric constant $\varepsilon(\omega)$ the force between a large 
sphere and  flat plate 
is given by the Lifshitz theory \cite{9}. 
Here the complete (extending from 0.04\,eV to 1000\,eV from 
Ref.\cite{sV23-3}
 along with the Drude model below 0.04\,eV is used to calculate 
$\varepsilon(i\omega)$. In the Drude model 
the dielectric constant of $Al$ $\varepsilon$  
along  the imaginary frequency axis is given by Eq.~(\ref{sV23.2}), 
where $\omega_p$ is the plasma frequency 
corresponding to a wavelength of 100\,nm and  
$\gamma$ is the relaxation frequency 
corresponding to 63\,meV \cite{sV23-3}.
Al metal was chosen because of its ease of fabrication 
and high reflectivity at short 
wavelengths (corresponding to the close surface separations). 

As in the previous experiment, the roughness of the metal surface is 
measured directly with the AFM.  The metal surface is composed of  
separate crystals on a smooth background.  The height of the highest 
distortions were 14\,nm and intermediate ones of 7nm both on a stochastic 
background of height 2\,nm with a fractional surface areas of 0.05, 0.11 
and 0.84 respectively.  The crystals are modeled as parallelepipeds.  
This leads to the complete Casimir force including roughness correction 
\cite{35}.   Here, $A=11.8\,nm$ is the effective height 
defined by requiring that the mean of the function describing the total 
roughness is zero and the numerical coefficients are the probabilities 
of different distance values between the interacting surfaces 
(see the previous subsection). As a result  
the roughness correction is  about 1.3\% of the measured force (a factor of 
nearly 20 improvement over the previous measurement).  There is also 
temperature correction which in this case was less than 1\% of the force 
at the closest separation.

Let us discuss now the measurement procedure of the improved
experiment.
The same technique for the attachment of the sphere to the AFM cantilever 
was done.  Then a 250\,nm aluminium metal coating was evaporated onto the 
sphere and a 1\,cm diameter sapphire plate. Next both surfaces were then 
sputter coated with $(7.9\pm 0.1)\,$nm layer of 60\%$Au$/40\%$Pd$. 
Thus here the $Au/Pd$ 
coating was made much thinner and also its thickness was precisely measured.  
The sphere diameter was measured using the Scanning Electron Microscope 
(SEM) to be $(201.7\pm 0.5)\,\mu$m.  The rms roughness amplitude of the 
$Al$ surfaces 
was measured using an AFM to be 3\,nm. The AFM was calibrated in the same 
manner as reported in the last section.  Next the residual potential of 
the grounded sphere was measured as $V_2=(7.9\pm 0.8)\,$mV by the AC
measurement 
technique again reported earlier  (factor of 3.5 improvement over the 
previous experiment). Minor corrections due to the piezo hysteresis and 
cantilever deflection were applied as reported.

To measure the Casimir force between the sphere and flat plate 
they are both grounded together with the AFM. The raw data from one scan 
is shown in Fig.~\ref{fig6.6}.
\begin{figure}[ht]
\epsfxsize=17cm\centerline{\epsffile{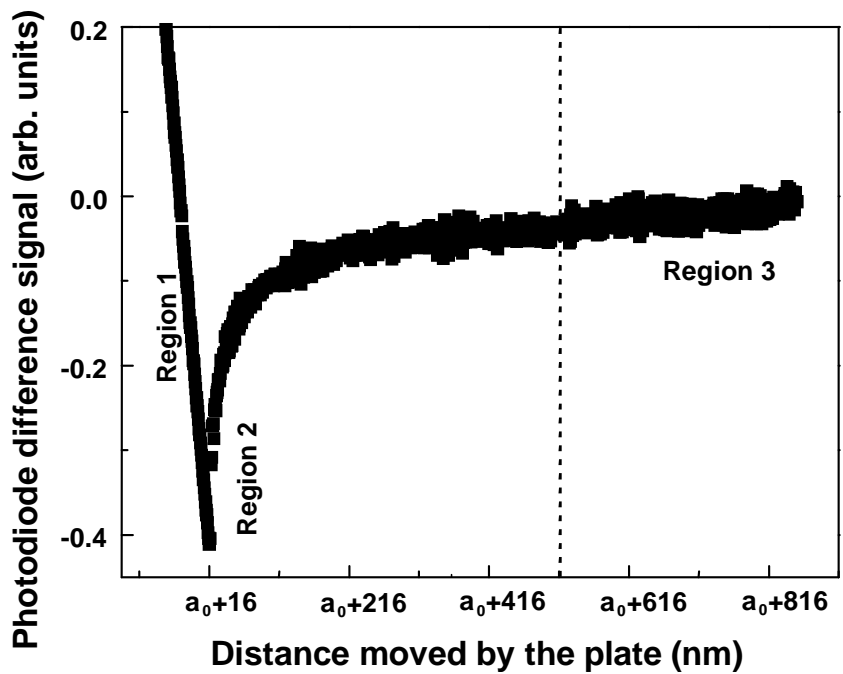} }
\vspace*{-9.5cm}
\caption{\label{fig6.6}
Typical force curve as a function of the distance moved by the plate.
}
\end{figure}
Region 1 is the flexing of the cantilever resulting 
from the continued extension of the piezo after contact of the two surfaces. 
In region 2 
($a_0+16\,\mbox{nm}<$surface separations$<a_0+516\,$nm) 
the Casimir force is the 
dominant characteristic far exceeding all systematic errors.  
The systematic effects are primarily from the residual electrostatic 
force ($<$1.5\% of the force at closest separation) and a linear contribution 
from scattered light.  This linear contribution due to scattered light 
(and some experimental drift) can be observed and measured in region 3.

In this experiment a key improvement is that, the electrostatic force between 
the sphere and flat plate was used to arrive at an independent and 
consistent measurement of $a_0$, the average surface separation on contact 
of the two surfaces. This was done immediately following the Casimir 
force measurement without breaking the vacuum and no lateral movement 
of the surfaces. The flat plate is connected to a DC voltage supply 
while the sphere remains grounded. The applied voltage $V_1$ in
Eq.~(\ref{6.2})
 is so chosen that the electrostatic force is $>10$ times the Casimir force.  
The open squares in Fig.~\ref{fig6.6} 
represent the measured total force for an 
applied voltage of 0.31\,V as a function of distance.  The force results 
from a sum of the electrostatic force and the Casimir force. The solid 
line which is a best $\chi^2$ fit for the data in 
Fig.~\ref{fig6.7} results in a 
$a_0=47.5\,$nm.   

This procedure was repeated for other voltages between 
(0.3--0.8)\,V leading to an average value of 
$a_0=(48.9\pm 0.6)\,$nm (the rms deviation 
is 3nm). Given the 7.9\,nm $Au/Pd$ coating on each surface this would 
correspond to an average surface separation 
$(48.9\pm 0.6+15.8)\,\mbox{nm}= (64.7\pm 0.6)\,$nm for the case of 
the Casimir force measurement. 
\begin{figure}[ht]
\epsfxsize=17cm\centerline{\epsffile{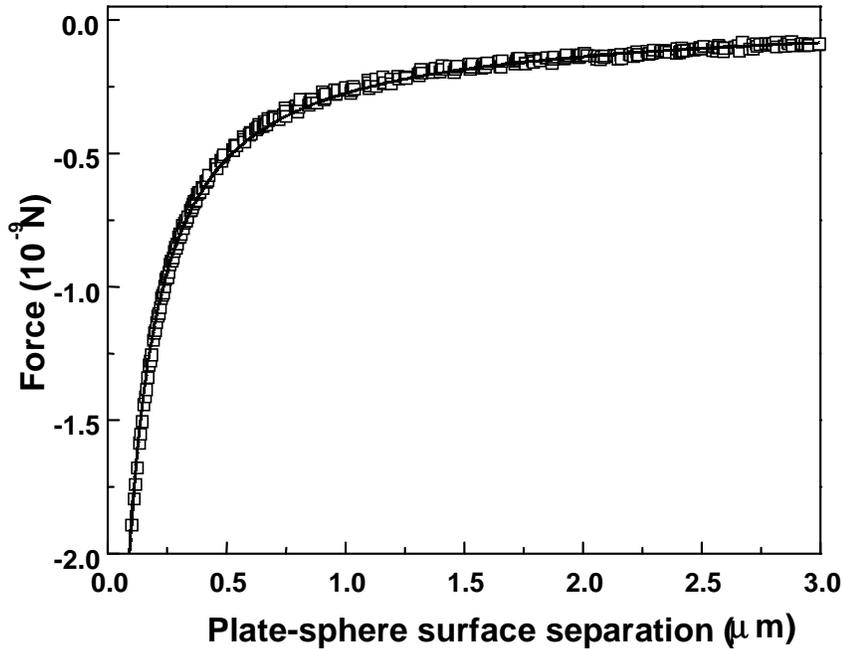} }
\vspace*{-9.5cm}
\caption{\label{fig6.7}
The measured electrostatic force for an applied voltage of 0.31\,V to
the plate. The best fit solid line shown leads to a $a_0=47.5\,$nm.
The average of many voltages leads to $a_0=48.9\pm0.6\,$nm.
}
\end{figure}
Now that we have presented the measurement procedure, we are coming
to the results.
The electrostatically determined value of $a_0$ was now used to apply 
the systematic error corrections to the force curve of Fig.~\ref{fig6.6}. 
Here the force curve in region 3, is fit to a function: 
$F= F_C(\Delta a+64.7\mbox{nm}) +F_e(\Delta a+48.9\,\mbox{nm}) + Ca$. 
The first term is the Casimir force contribution to the total force 
in region 3 and the second term represents the electrostatic force 
between the sphere and flat plate due to the residual potential 
difference of $V_2=7.9\,$mV.  The third term $C$ represents 
the linear coupling 
of scattered light from the moving plate into the diodes and experimental 
drift and corresponds to a force $<1\,$pN ($<$1\% of the forces at closest 
separation). The value of $C$ is determined by minimizing the $\chi^ {2}$. 
It is determined in region 3 and the electrostatic force 
corresponding to $V_2=7.9\,$mV and $V_1=0$ is used to subtract the systematic 
errors from the force curve in region 3 and 2 to obtain the measured 
Casimir force as:  $F_{C-m}= F_m-F_e- Ca$  where $F_m$ is the measured 
total force.  
Thus the measured Casimir force from region 2 has no adjustable parameters.

The experiment is repeated for 27 scans and the average Casimir force 
measured is shown as open squares in Fig.~\ref{fig6.8}.
\begin{figure}[ht]
\epsfxsize=17cm\centerline{\epsffile{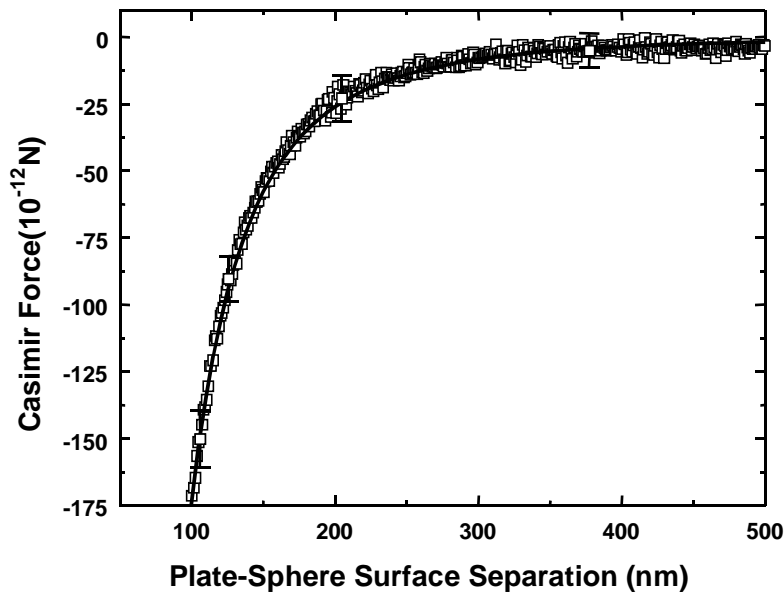} }
\vspace*{-9.5cm}
\caption{\label{fig6.8}
The measured average Casimir force as a function of plate-sphere separation
is shown as squares. The error bars represent the standard deviation
from 27 scans. The solid line is the theoretical Casimir force with
account of roughness and finite conductivity corrections.
}
\end{figure}
The error bars represent 
the standard deviation from the 27 scans at each data point.  The authors 
report that due to the surface roughness, the averaging procedure 
introduces $\pm 3\,$nm uncertainty in the surface separation on contact of 
the two surfaces. The theoretical curve is shown as a solid line.  
The authors used a variety of statistical measures to define the precision 
of the Casimir force measurement. They check the accuracy of the theoretical 
curve over the complete region between (100--500)\,nm with $N=441$ points 
(with an average of 27 measurements representing each point) with no 
adjustable parameters.  Given that the experimental standard deviation 
over this range is 7\,pN from thermal noise, the experimental uncertainty 
is $\leq7/\sqrt{27}=1.3\,$pN 
leading to a precision which is better than 1\% of the largest forces 
measured.  If one wished to consider the rms deviation of the experiment 
($F_{exp}$) from the theory ($F_{th}$), which is equal to 2.0\,pN as a 
measure of the 
precision, it is also on the order of 1\% of the forces measured at the 
closest separation. From the above definitions, the statistical measure
 of the experimental precision is of order 1\% of the forces at the closest 
separation.

These measurements of the Casimir force using an AFM and aluminium surface 
were conclusive to a statistical precision of 1\%. The second AFM experiment, 
met all three requirements by M.J.~Sparnaay noted in Sec.6.1
(in the first experiment the separation 
distance on contact was not independently determined).  
However, these aluminium surfaces required the use of a thin Au/Pd coating 
on top.  This coating could 
only be treated in a phenomenological manner.   A more complete theoretical 
treatment is complicated as non-local effects 
such as spatial dispersion need to be taken into account in the calculation 
of the Casimir force (see Sec.5.2.3). 
  
\subsubsection{Precision measurement with gold surfaces using the AFM}
\label {sec6.4.3}
 
This was the third in a series of precision measurements
using the AFM.  
The other two were discussed above. 
The primary differences here is the use of gold surfaces and the related 
experimental changes.  The use of a thin Au/Pd coating on top of the 
aluminium surface to reduce effects of oxidation 
in the above two experiments prevented a complete 
theoretical treatment of the properties of the metal coating. 
Thus it is important to use 
chemically inert materials such as gold for the measurement of the 
Casimir force. The complete dielectric properties of $Au$ is used 
in the theory. Here the complete (extending from 0.125\,eV to 9919\,eV 
from Ref.~\cite{sV23-3} along with the Drude model below 0.125\,eV is used 
to calculate $\varepsilon(i\xi)$. In the Drude representation 
$\omega_p=11.5\,$eV is the plasma 
frequency and  $\gamma$ is the relaxation frequency corresponding to
50\,meV.  
These values of $\omega_p$ and $\gamma$ 
are obtained in the manner detailed in \cite{27,sV23-2}. 
The temperature correction is $\ll 1$\% of the Casimir force for the surface 
separations reported here and can be neglected. 

The fabrication procedures had to be modified, given the different material 
properties of gold as compared to the aluminium coatings used previously 
in Refs. \cite{33,35}. The 320$\,\mu$m long AFM cantilevers were first 
coated with about 200\,nm of aluminium to improve their thermal conductivity. 
This metal coating on the cantilever decreases the thermally induced 
noise when the AFM is operated in vacuum.  Aluminium coatings are better, 
as applying thick gold coatings directly to these Silicon Nitride 
cantilevers led their curling due to the mismatch in the thermal expansion 
coefficients. Next polystyrene spheres were  mounted on the tip of the metal 
coated cantilevers with $Ag$ epoxy. A 1\,cm diameter optically polished 
sapphire 
disk is used as the plate.  The cantilever (with sphere) and plate were then 
coated with gold in an evaporator. The sphere diameter after the metal 
coating was measured using the Scanning Electron Microscope (SEM) to be 
$(191.3\pm 0.5)\,\mu$m.  The rms roughness amplitude $A$ of the gold surface 
on the 
plate was measured using an AFM to be $(1.0\pm 0.1)\,$nm. The thickness of the 
gold coating was measured using the AFM to be $(86.6\pm 0.6)\,$nm.  
Such a coating 
thickness is sufficient to reproduce the properties of an infinitely thick 
metal for the precision reported here.  To reduce the development of 
contact potential differences between the sphere and the plate, great 
care was taken to follow identical procedures in making the electrical 
contacts.  This is necessary given the large difference in the work 
function of aluminium and gold.  As before with the application of 
voltages $\pm\,V_1$ to the plate, 
the residual potential difference between the grounded sphere and the 
plate was measured to be $V_2=(3\pm 3)\,$mV. This residual potential leads to 
forces which are $\ll 1$\% of the Casimir forces at the closest separations 
reported here. 

To measure the Casimir force between the sphere and flat plate they are 
both grounded together with the AFM. The raw data from a scan is shown 
in Fig.~\ref{fig6.9}. 
\begin{figure}[ht]
\epsfxsize=17cm\centerline{\epsffile{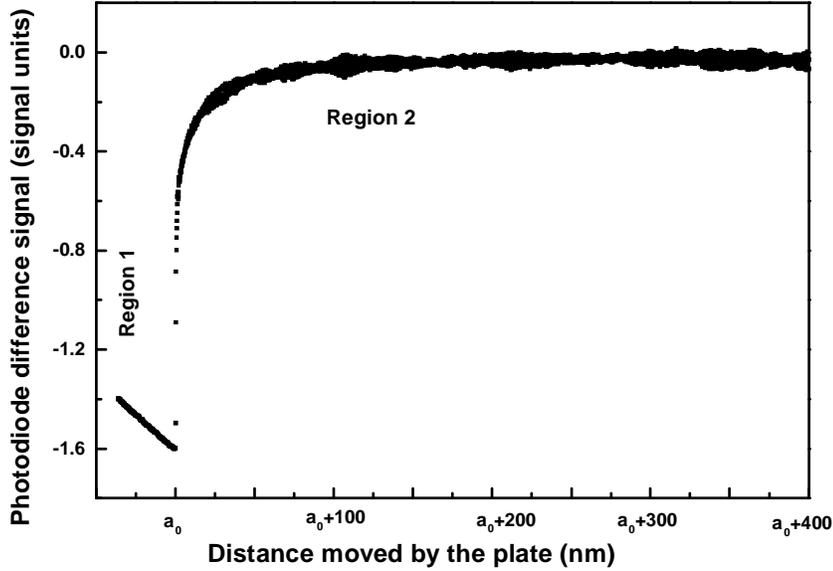} }
\vspace*{-9.5cm}
\caption{\label{fig6.9}
The raw data of the force measured as a photodiode difference signal
as a function of the distance moved by the plate.
}
\end{figure}
Region 1 is the flexing of the cantilever resulting from 
the continued extension of the piezo after contact of the two surfaces. 
Region 2 ($a_0<$surface separations$<a_0+400\,$nm) 
clearly shows the Casimir 
force as a function of separation distance. The Casimir force measurement 
is repeated for 30 scans.  The only systematic error associated with the 
Casimir force in these measurements is that due to the residual 
electrostatic force which is less than 0.1\% of the Casimir force 
at closest separation. For surface separations exceeding 400\,nm the 
experimental uncertainty in the force exceeds the value of the Casimir 
force.  The surface separation on contact, $a_0$, is a priori unknown due 
to the roughness of the metal surface and is determined independently as 
described below. A small additional correction to the separation distance 
results from the deflection of the cantilever in response to the attractive 
Casimir force.  As can be observed from the schematic in 
Fig.~\ref{fig6.1}, this leads 
to a decrease in the distance of separation of the two surfaces. This 
``deflection correction" modifies the separation distance between the two 
surfaces. This is given by: 
$a=a_0+a_{piezo}-F_{pd} m$, where $a$ is the correct 
separation between the two surfaces, $a_{piezo}$ is the distance moved by the 
plate due to the application of voltage applied to the piezo, i.e. 
along the 
horizontal axis of Fig.~\ref{fig6.9} and $F_{pd}$ is the photodiode 
difference signal 
shown along the vertical axis in Fig.~\ref{fig6.9}.  
Here $m$ is deflection coefficient 
corresponding to the rate of change of seperation distance per unit 
photodiode difference signal (from the cantilever deflection) and is 
determined independently as discussed below. The slope of the line in 
region 1 of the force curve shown in Fig.~\ref{fig6.9} cannot be used 
to determine $m$ 
as the free movement of the sphere is prevented on contact of the two 
surfaces (due to the larger forces encountered here).

Next the authors used the electrostatic force between the sphere and 
flat plate to arrive at an independent measurement of the constant $m$ in 
the deflection correction and $a_0$ the average surface separation on contact 
of the two surfaces. The flat plate is connected to a DC voltage supply 
while the sphere remains grounded. The applied voltage $V_1$ in 
Eq.~(\ref{6.2})
 is so chosen that the electrostatic force is much greater than the 
Casimir force. As can be observed from Fig.~\ref{fig6.1}, 
at the start of the force 
measurement, the plate and the sphere are separated by a fixed distance 
and the plate is moved towards the sphere in small steps with the help of 
the piezoelectric tube.  When different voltages $V_1$ are applied to the 
plate, the point of contact between the plate and sphere varies 
corresponding to the different cantilever deflections.  This is shown in 
Fig.~\ref{fig6.10} for three different applied voltages 
0.256, 0.202 and 0.154\,V.
\begin{figure}[ht]
\epsfxsize=17cm\centerline{\epsffile{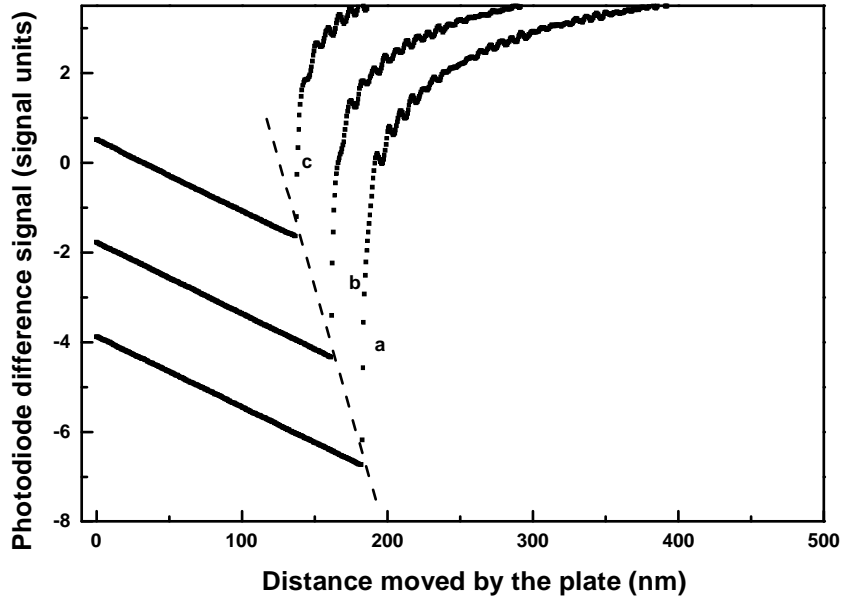} }
\vspace*{-9.5cm}
\caption{\label{fig6.10}
The measured electrostatic force curves for three different voltages 
0.256\, V (a), 0.202\,V (b), and 0.154\,V (c). The rate of change
of separation distance per unit photodiode difference signal
corresponding to the slope of the dashed line which connects the
vertices yeilds the deflection coefficient $m$.
}
\end{figure}
The vertex in each curve identifies the contact point between sphere 
and plate. The deflection coefficient $m$ can be determined from the slope 
of the dashed line connecting the vertices. The slope corresponds to an 
average value of $m=(8.9\pm 0.3)\,$nm per unit photodiode difference signal.  
The separation distance is then corrected for this cantilever deflection.  
Next the surface separation on contact $a_0$ was determined from the same 
electrostatic force curves. As explained previously in Sec.6.4.2, 
a best $\chi ^ {2}$ fit was done to the electrostatic force curves to
obtain an average value of 
$a_0=31.7\,$nm. 

The average Casimir force measured from the 30 scans is shown as open 
squares in Fig.~\ref{fig6.11}. 
\begin{figure}[ht]
\epsfxsize=17cm\centerline{\epsffile{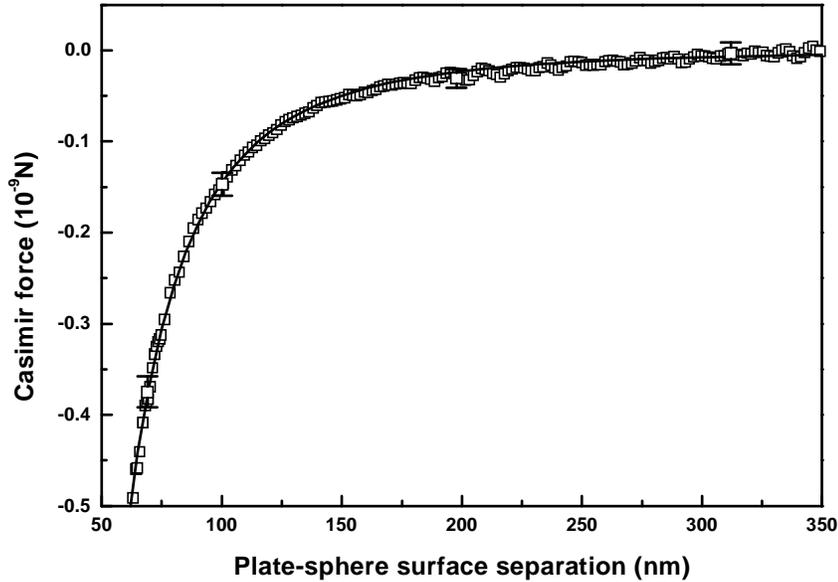} }
\vspace*{-9.5cm}
\caption{\label{fig6.11}
The measured average Casimir force as a function of plate-sphere separation
is shown as squares. For clarity only 10\% of the experimental points
are shown in the figure. 
The error bars represent the standard deviation
from 30 scans. The solid line is the theoretical Casimir force with
account of roughness and finite conductivity corrections.
}
\end{figure}
The theoretical curve including all 
the corrections 
is shown as a solid line.  For clarity only 10\% of the 2583 data points 
are shown in the figure. Thus the accuracy of the theoretical curve is 
checked over the complete region between (62--350)\,nm with $N=2583$ points 
(with an average of 30 measurements representing each point).  Given 
that the experimental standard deviation around 62\,nm is 19\,pN, the 
experimental uncertainty is $\leq19/\sqrt{30}=3.5\,$pN 
leading to a precision  which is better 
than 1\% of the largest forces measured.  If one wished to consider 
the rms deviation of the experiment ($F_{exp}$) from the theory 
($F_{th}$), $\sigma=3.8\,$pN as a measure of the precision, it is 
also on the order 
of 1\% of the forces measured at the closest separation. 
The authors note that the uncertainties of 3.8\,pN measured here are 
larger than the 2\,pN in previous AFM measurements due to the poor thermal 
conductivity of the cantilever resulting from the thinner metal coatings 
used.  Thus experiments at cryogenic temperatures should substantially 
reduce this noise.

In conclusion the above measurement of the Casimir force with the AFM met 
all three criteria set forth by M.J.~Sparnaay and thus is the most definite 
experiment to date. The complete conductivity properties of the metal 
coating were used in the comparison of the theory and experiment.  
Here the corrections to the surface roughness and 
that due to the electrostatic force were both reduced to $<<$ 1\% 
of the Casimir force at the closest separation.  The separation distance 
on contact of the two surfaces was determined independently of the Casimir 
force measurement.  In this experiment the 
temperature corrections were less than 1\% of the Casimir force at
the smallest separations.

\subsection {Demonstration of the nontrivial boundary properties of the 
Casimir force}
\label {sec6.5}

One of the most exciting and unique aspects of the Casimir force is its 
ability to change the sign and value of the force, given changes in the 
geometry and topology of the boundary.  However given the difficulties 
in the measurement of the Casimir force, it has only now become possible 
to measure Casimir forces for simple deviations from the flat boundary case. 
The first and only experiment to date in this regard is the measurement of 
Casimir forces between a sphere and plate with periodic uniaxial sinusoidal 
corrugations (PUSC) by A.~Roy and U.~Mohideen \cite{34}. 
Such PUSC surfaces have 
been theoretically shown to exhibit a rich variety of surface interactions 
\cite{40,sV3-3,sV3-8}.   
R.~Golestanian and M.~Kardar \cite{sV3-4} point out that such corrugated 
surfaces might lead to lateral forces and AC Josephson-like mechanical 
forces.  Also creation of photons (dynamic Casimir effect discussed in 
Secs.2.4 and 4.4) is thought to result from the lateral movement of two such 
corrugated surfaces.  In the present case of a measurement between a 
large sphere and the PUSC surface, the intent was to explore the effect 
of diffraction and any lateral forces introduced by the uniform corrugation.  
Thus the force 
resulting from a PUSC plate can be expected to be substantially different 
from that resulting from a flat one.  

The Casimir force was measured between the large sphere and the PUSC for 
surface separations between 0.1$\,\mu$m to 0.9$\,\mu$m using an AFM.  
The amplitude of the corrugation is much smaller than 
the separation. Yet the measured force shows significant deviations from 
a perturbative theory which only takes into account the small periodic 
corrugation of the plate in the surface separation. 
The authors also compare the 
measured Casimir force between the same sphere and identically coated 
flat plate and show that it agrees well to the same theory in the limit 
of zero amplitude of corrugation.  The authors point out that these 
results considered together, demonstrate the non-trivial boundary 
dependence of the Casimir force.

\subsubsection {Measurement of the Casimir force
due to the corrugated plate}
\label {sec6.5.1}

The experiment of \cite{34} deals with
 the configuration of polystyrene sphere above a 
$(7.5\times 7.5)\,\mbox{mm}^2$ plate with periodic uniaxial sinusoidal
corrugations. Both the sphere and the plate were coated with 250\,nm of
$Al$, and 8\,nm layer of $Au/Pd$. For the outer $Au/Pd$ layer 
transparencies greater than 90\% were measured at characteristic frequences 
contributing into the Casimir force. The diameter of the sphere was
$2R=(194.6\pm 0.5)\,\mu$m. The surface of the corrugated plate is
described by the function
\beq
z_s(x,y)=A\sin\frac{2\pi x}{L},
\label{6.22}
\eeq
\noindent
where the amplitude of the corrugation is $A=(59.4\pm 2.5)\,$nm and
its period is $L=1.1\,\mu$m. The mean amplitude of the stochastic
roughness on the corrugated plate was $A_p=4.7\,$nm, and on the
sphere bottom --- $A_s=5\,$nm.

As is seen from Eq.~(\ref{6.22}) the origin of the $z$-axis is taken such 
that the mean value of the corrugation is zero. 
The separation between the zero corrugation level and the sphere bottom
is given by $a$. 
The minimum value of $a$ 
is given by
$a_0\approx A+A_p+A_s+2h\approx 130\,$nm, where $h\approx 30\,$nm is 
the height of the highest occasional rare $Al$ crystals which prevent
the intimate contact between the sphere bottom and the maximum point
of the corrugation.  

As in the standard measurement of 
Casimir forces, boundary dependence of the Casimir force can be easily 
obscured by errors in the measurement of the surface separation.  
To eliminate this ambiguity, the authors used the electrostatric method 
described in Sec.6.4.2 to independently determine the exact surface 
separation and establish procedures for consistent comparison to theory. 
The electrostatic force between the sphere and the PUSC surface is given by:
\beq
F_e=-\frac{(V_1-V_2)^2}{4(a_0+\Delta a)}R
\sum\limits_{m=0}^{\infty}D_m\left(\frac{A}{a_0+\Delta a}\right)^m,
\label{6.23}
\eeq
\noindent
where $\Delta a$ is distance between the surfaces measured from contact 
and as before $a_0$ is the true average separation on contact of the two 
surfaces due to the periodic corrugation and stochastic roughness of the 
aluminium coating (note that $a=a_0+\Delta a$).  
The non-zero even power coefficients in 
Eq.~(\ref{6.23}) are: 
$D_0=1$, $D_2=1/2$, $D_4=3/8$, $D_6=5/16$,... 
$V_1$ and $V_2$ are voltages on the corrugated 
plate and sphere respectively.  The above expression was obtained 
using the Proximity Force Theorem,
by starting from the 
electrostatic energy between parallel flat plates. 

      First the residual potential of the grounded sphere was measured. 
The sphere is grounded and the electrostatic force between the sphere 
and the corrugated plate was measured for four different voltages and 
five different surface separations  $a\gg A$. 
With Eq.~(\ref{6.23}), from the 
difference in force for voltages  $+V_1$ and $-V_1$ applied to the corrugated 
plate, one can measure the residual potential on the grounded sphere 
$V_2$ as 14.9\,mV.  This residual potential is a contact potential that 
arises from the different materials used to fabricate the sphere and 
the corrugated plate.

As previously done, to measure the Casimir force between the sphere 
and the corrugated plate they are both grounded together with the AFM.  
The plate is then moved towards the sphere in 3.6\,nm steps and the 
corresponding photodiode difference signal was measured.  The signal 
obtained for a typical scan is shown in Fig.~\ref{fig6.12}.
\begin{figure}[ht]
\epsfxsize=17cm\centerline{\epsffile{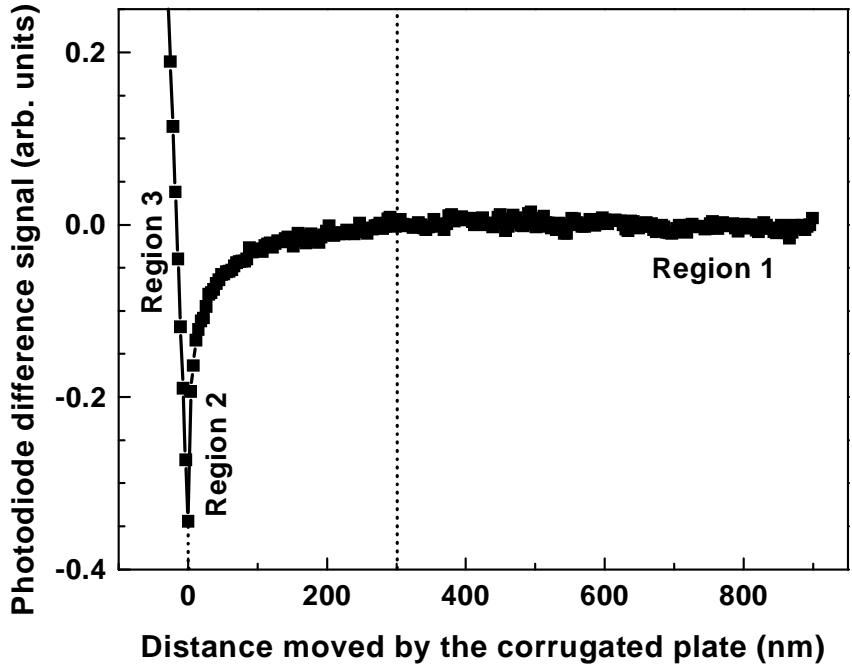} }
\vspace*{-9.5cm}
\caption{\label{fig6.12}
A typical force curve as a function of the distance moved by the plate.
The ``0'' distance stands for point of contact and does not take into 
account the amplitude of the corrugation and the roughness of the
metallic coating.
}
\end{figure}
Here ``0'' separation 
stands for contact of the sphere and corrugated plate surfaces, i.e., 
$\Delta a=0$.  
It does not take into account $a_0$. Region 1 can be used to subtract the 
minor ($<1$\% ) experimental systematic due to scattered laser light 
without biasing the results in region 2. In region 2 (absolute 
separations between contact and 450\,nm) the Casimir force is the dominant 
characteristic far exceeding all systematic errors (the electrostatic force 
is $<2$\% of the peak Casimir force).  
Region 3 is the flexing of the cantilever 
resulting from the continued extension of the piezo after contact of the 
two surfaces. 

Next by applying a DC voltage between the corrugated plate and sphere 
an independent and consistent measurement of $a_0$, the average surface 
separation on contact of the two surfaces, is arrived. 
Here the procedure outlined in 
Sec.6.4.2 is followed. The applied voltage $V_1$ in Eq.~(\ref{6.23}) is so 
chosen that the electrostatic force is $>$20 times the Casimir force. 
The experiment is repeated for other voltages between (0.4-0.7)\,V leading 
to an average value of $a_0=(132\pm 5)\,$nm. 
Given the 8\,nm $Au/Pd$ coating on each 
surface this would correspond to a average surface separation 
$(132\pm 5+8+8)\,\mbox{nm}= (148\pm 5)\,$nm 
for the case of the Casimir force measurement. 

\begin{figure}[h]
\vspace*{-3.5cm}
\epsfxsize=15cm\centerline{\epsffile{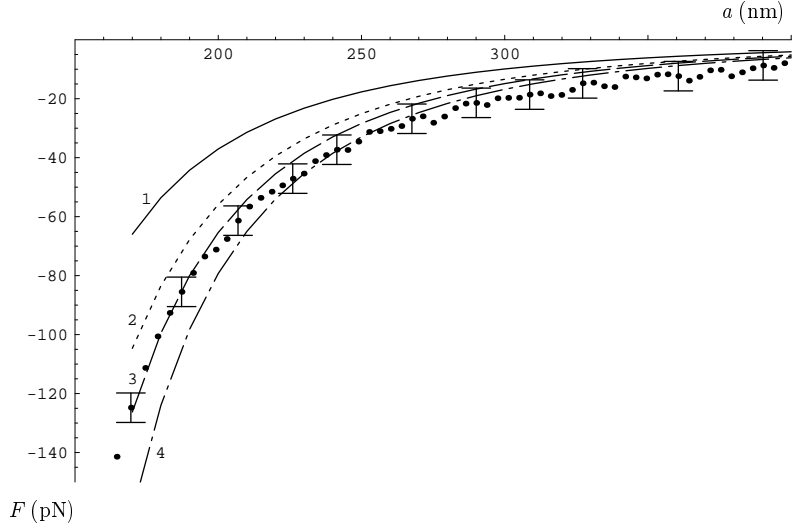} }
\vspace*{-6.5cm}
\caption{\label{fig6.14}
The measured Casimir force $F(a)$ 
as a function of the surface separation in
configuration of a sphere above a corrugated disk is shown by 
solid circles.
The error bars represent the standard deviation
from 15 scans.
Curve 1 represents 
the computational results obtained with the uniform probability
distribution, curve 2 --- with the distribution, when the sphere is located
with equal probability above the convex part of corrugations,
curve 3 --- with a distribution  which increases linearly when
the sphere approaches the points of a stable equilibrium, and curve 4 
--- for a sphere situated above the points of stable equilibrium. 
}
\end{figure}

The electrostatically determined value of $a_0$ can now be used to apply 
the systematic error corrections to the force curve considered, as above,
in three regions.  
The force curve in region 1, is fit to a function: 
$F= F_C(\Delta a+148) +F_e(\Delta a+132) + Ca$. The first term is the Casimir 
force contribution to the total force in region 1. The second term 
represents the electrostatic force between the sphere and corrugated 
plate as given by Eq.~(\ref{6.23}).  The third term $C$ represents the linear 
coupling of scattered light from the moving plate into the diodes 
and corresponds to a force $<1\,$pN ($<1$\% effect). Here again the
difference 
in $a_0$ in the electrostatic term and the Casimir force is due to the 
8\,nm $Au/Pd$ coating on each surface. The value of $C$ is determined by 
minimizing the $\chi^{2}$. The value of $C$ determined in region 1 
and the electrostatic force corresponding to $V_2=14.9\,$mV and $V_1=0$ in 
Eq.~(\ref{6.23}) is used to subtract the systematic errors from the force 
curve in regions 1 and 2 to obtain the measured Casimir force as:  
$F_{C-m}= F_m-F_e- Ca$  where $F_m$ is the measured total force.  
Thus the measured Casimir force from region 2 has no adjustable parameters.  

The experiment was repeated for 15 scans and the average Casimir 
force measured is presented as solid circles in Fig.~\ref{fig6.14}  
together with
the experimental uncertainties shown by the 
error bars.
To gain a better understanding of the accord between theory and experiment 
the distance range $a\leq 400\,$nm is considered here where the Casimir force
is measured with higher accuracy. As it is seen from the next subsection, 
there is significant deviation between the measured force and the perturbative
theory which takes into account the periodic corrugations of the plate in the
surface separation.

The experiment and analysis was repeated in \cite{34} for 
the same sphere and an 
identically coated flat plate (i.e. without corrugations).  
The average measured Casimir force from 15 scans  
shows good agreement with perturbation theory of Sec.~6.4.1. 
The reasons for this are discussed below.
 
\subsubsection{Possible explanation of the nontrivial boundary
dependence of the Casimir force}
\label{sec6.5.2}

In Ref.~\cite{s6.54} the perturbative calculations for both
vertical and lateral 
Casimir force acting in the configuration of a sphere situated above 
a corrugated plate is presented
(note that the lateral force arises due to the absence of
translational symmetry on a plate with corrugations). 
The study of \cite{s6.54} revealed
that a lateral force acts upon the sphere (in contrast to the 
usual vertical Casimir force) in such a way that it tends 
to change the bottom position of the sphere 
in the direction of a nearest maximum point
of the vertical Casimir force (which coincides with the maximum point of 
corrugations). In consequence of this, the assumption 
of a simple perturbation theory that 
the locations of the sphere above different points of a corrugated surface 
are equally probable can be violated. As indicated in \cite{s6.54}, 
the diverse 
assumptions on the probability distribution describing location of the
sphere above different points of the plate result in an essential change
in force-distance relation. Thus the perturbation theory
taking the lateral force into account may work for a case of corrugated
plate.

Let us take up first the vertical Casimir force acting between a corrugated
plate and a sphere. A sinusoidal corrugation of Eq.~(\ref{6.22}) leads to
the modification of the Casimir force between a flat plate and a sphere. 
The modified force can be calculated by 
averaging over the period 
\beq
F(a)=\int\limits_{0}^{L}dx\rho(x)\,F_{dl}^{C}\left(d(a,x)\right).
\label{6.24}
\eeq
\noindent
Here $d(a,x)$ is the separation between the sphere (lens) bottom and
the point $x$ on the surface of corrugated plate (disk)
\beq
d(a,x)=a-A_p-A_s-A\sin\frac{2\pi x}{L}.
\label{6.25}
\eeq
\noindent
$F_{dl}^{C}$ is the Casimir force acting between a flat plate and 
a sphere with
account of corrections due to finite conductivity of the boundary
metal given by Eq.~(\ref{sV2.13}). 

The quantity $\rho(x)$ from Eq.~(\ref{6.24}) describes the probability
distribution of the sphere positions above different points $x$
belonging to one corrugation period. 
If the plate corrugation is taken into account in the surface
separation only, then the uniform
distribution is assumed $\left(\rho(x)=1/L\right)$ which is to say
that the sphere is located above all points $x$ belonging to the
interval $0<x<L$ with equal probability. The right-hand side of
Eq.~(\ref{6.24}) can be expanded in powers of a small parameter
$A/(a-A_p-A_s)$. This expansion had shown significant deviations from
the measured data of \cite{34}
while Eq.~(\ref{sV2.13}) is in excellent agreement with
data for all $d\geq\lambda_p$  in the limit of zero
amplitude of corrugation.

Now we turn to the lateral projection of the Casimir force.
The lateral projection is nonzero only in the
case of nonzero corrugation amplitude. Let us find it in the
simplest case of the ideal metal.

This can be achieved by applying the additive summation method of the
retarded interatomic potentials over the volumes of a corrugated plate
and a sphere with subsequent normalization of the interaction constant.
Alternatively the same result is obtainable by the
Proximity Force Theorem.
Let an atom of a sphere be situated at a point with the coordinates
($x_A,y_A,z_A$) in the coordinate system described above. Integrating
the interatomic potential $U=-C/r_{12}^7$ over the volume of
corrugated plate ($r_{12}$ is a distance between this atom and the atoms
of a plate) and calculating the lateral force projection according to
$-\partial U/\partial x_A$ one obtains \cite{sV3-10,s6.55a}
\beq
F_x^{(A)}(x_A,y_A,z_A)=\frac{4\pi^2N_pC}{5z_A^5}\,\frac{A}{z_A}
\frac{z_A}{L}
\left[\cos\frac{2\pi x_A}{L}
+\frac{5}{2}\frac{A}{z_A}
\sin\frac{4\pi x_A}{L}\right],
\label{6.26}
\eeq
\noindent
where $N_p$ is the atomic density of a corrugated plate. Eq.~(\ref{6.26}) is
obtained by perturbation expansion of the integral (up to second order)
in small parameter $A/z_A$.

We can represent $x_A=x_0+x$, $y_A=y_0+y$, $z_A=z_0+z$ where 
($x_0,y_0,z_0$) are the coordinates of the sphere bottom in the above
coordinate system, and ($x,y,z$) are the coordinates of the sphere atom
in relation to the sphere bottom. The lateral Casimir force acting
upon a sphere is calculated  by the integration of (\ref{6.26}) over the
sphere volume and subsequent division by the normalization factor
$K=24CN_pN_s/(\pi\hbar c)$ obtained by comparison of additive and exact
results for the configuration of two plane parallel plates 
(see Sec.4.3)
\beq
F_x(x_0,y_0,z_0)=\frac{N_s}{K}
\int\limits_{V_s}d^3r\,F_x^{(A)}(x_0+x,y_0+y,z_0+z),
\label{6.27}
\eeq
\noindent
where $N_s$ is the atomic density of sphere metal.

Let us substitute Eq.~(\ref{6.26}) into Eq.~(\ref{6.27})  neglecting the
small contribution of the upper semisphere which is of order
$z_0/R<4\times 10^{-3}$ comparing to unity. In a cylindrical coordinate
system the lateral force acting upon a sphere rearranges to the form
\bes
&&
F_x(x_0,y_0,z_0)=\frac{\pi^3\hbar c}{30}\frac{A}{L}
\label{6.28}\\
&&\phantom{aa}\times
\left[\cos\frac{2\pi x_0}{L}
\int\limits_{0}^{R}\rho d\rho
\int\limits_{0}^{R-\sqrt{R^2-\rho^2}}
\frac{dz}{(z_0+z)^5}
\int\limits_{0}^{2\pi}
d\varphi\cos\left(\frac{2\pi\rho}{L}\cos\varphi\right)\right.
\nonumber \\
&&\phantom{aaa}
+\frac{5}{2}A\left.
\sin\frac{4\pi x_0}{L}
\int\limits_{0}^{R}\rho d\rho
\int\limits_{0}^{R-\sqrt{R^2-\rho^2}}
\frac{dz}{(z_0+z)^6}
\int\limits_{0}^{2\pi}
d\varphi\cos\left(\frac{4\pi\rho}{L}\cos\varphi\right)\right].
\nonumber
\ees

Using the standard formulas from \cite{sV2-6} the integrals with respect to
$\varphi$ and $z$ are taken explicitly. Preserving only the lowest
order terms in small parameter $x_0/R<10^{-2}$ we arrive at
\bes
&&
F_x(x_0,y_0,z_0)=-\frac{\pi^4\hbar c}{60z_0^4}\frac{A}{L}
\left[\cos\frac{2\pi x_0}{L}
\int\limits_{0}^{R}\rho d\rho
J_0\left(\frac{2\pi\rho}{L}\right)\right.
\nonumber\\
&&\phantom{aaaaaa}\left.
+2\frac{A}{z_0}
\sin\frac{4\pi x_0}{L}
\int\limits_{0}^{R}\rho d\rho
J_0\left(\frac{4\pi\rho}{L}\right)\right],
\label{6.29}
\ees
\noindent
where $J_n(z)$ is Bessel function.

Integrating in $\rho$ the final result is obtained \cite{s6.54}
\bes
&&
F_x(x_0,y_0,z_0)=3F_{dl}^{(0)}(z_0)\frac{A}{z_0}
\left[\cos\frac{2\pi x_0}{L}\,J_1\left(\frac{2\pi R}{L}\right)
\right.
\nonumber\\
&&\phantom{aaaaaa}\left.
+\frac{A}{z_0}
\sin\frac{4\pi x_0}{L}\,J_1\left(\frac{4\pi R}{L}\right)\right],
\label{6.30}
\ees
\noindent
where the vertical Casimir force $F_{dl}^{(0)}$ for ideal metal was
defined in Eq.~(\ref{sIV3.7}).

As is seen from Eq.~(\ref{6.30}) the lateral Casimir force takes zero value 
at the extremum points of the corrugation described by Eq.~(\ref{6.22}).
The lateral force achieves maximum at the points $x_0=0,\,L/2$ where the
corrugation function is zero. If the sphere is situated to the left of
a point $x_0=L/4$ (maximum of corrugation) it experiences a positive
lateral Casimir force. If it is situated to the right of $x_0=L/4$ the
lateral Casimir force is negative. In both cases the sphere tends to change 
its position in the direction of a corrugation maximum which is the
position of stable equilibrium. The situation here is the same as for
an atom near the wall covered by the large-scale roughness \cite{sV3-10}.
That is the reason why the different points of a corrugated plate are
not equivalent and the assumption that the locations of the sphere
above them are described by the uniform probability distribution
may be too simplistic.

On this basis, one may suppose that the probability distribution under
consideration is given by
\beq
\rho(x)=\left\{
\begin{array}{rl}
\frac{2}{L}, \quad&kL\leq x\leq \left(k+\frac{1}{2}\right)L,\\
&\\
0,\quad&\left(k+\frac{1}{2}\right)L\leq x\leq (k+1)L,
\end{array}
\right.
\label{6.31}
\eeq
\noindent
where $k=0,\,1,\,2,\ldots$ This would mean that in the course of the
measurements the sphere is located with equal probability above 
different points of the
convex part of corrugation but cannot be located above the concave one.

It is even more reasonable to suppose that the function 
$\rho$ increases linearly when the sphere approaches 
the points of a stable
equilibrium. In this case the functional dependence is given by
\beq
\rho(x)=\left\{
\begin{array}{l}
\frac{16}{L^2}x, \quad kL\leq x\leq \left(k+\frac{1}{4}\right)L,\\
\\
\frac{16}{L^2}\left(\frac{L}{2}-x\right),\quad
\left(k+\frac{1}{4}\right)L\leq x\leq \left(k+\frac{1}{2}\right)L,\\
\\
0,\quad \left(k+\frac{1}{2}\right)L\leq x\leq (k+1)L.
\end{array}
\right.
\label{6.32}
\eeq

By way of example in Fig.~\ref{fig6.14}  
the theoretical results 
computed by Eq.~(\ref{6.24}) and shown by the curves 1 
(uniform distribution), 2 (distribution of Eq.~(\ref{6.31})), 3
(distribution of Eq.~(\ref{6.32})), and 4 (the bottom of the sphere 
is directly over the maximum of corrugation
at all times) are compared with the measured Casimir force. 
It is seen that the curve 3 is
in agreement with experimental data in the limits of given
uncertainties $\Delta F=5\,$pN, $\Delta a=5\,$nm. The root
mean square average deviation between theory and experiment within
the range 169.5\,nm$\leq a\leq$400\,nm (62 experimental points) where the
perturbation theory is applicable is
$\sigma=20.28\,$pN for the curve 1,
$\sigma=8.92\,$pN (curve 2),
$\sigma=4.73\,$pN (curve 3),
and $\sigma=9.17\,$pN (curve 4).
By this means perturbation theory with account of the lateral Casimir force
can be made consistent with experimental data and might explain
the observed nontrivial
boundary dependence of the Casimir force \cite{s6.54}. 
The complete solution of the
problem may be achieved with an experiment where both the vertical
and the lateral Casimir forces are measured.

\subsection {The outlook  for measurements of the Casimir force}  
\label {sec6.6}

With the advent of sensitive force detection techniques such as phase 
sensitive detection and the AFM, the measurement of the Casimir force 
has become conclusive.  However the present measurements are only 
senisitive to forces at surface separations above 32nm and below 1000nm. 
The importance of these two separation limits and the possibility for 
their improvement will be discussed next.  

First, there is a clear 
need to measure Casimir (van der Waals) forces
at smaller and smaller surface separations 
given the possible presence of compactified dimensions as postulated by
modern 
unified theories. The measurement of the Casimir force at
separations above $1\,\mu$m is also of great interest as a test for some 
predictions of supersymmetry and string theory (see Sec.7). 
Such measurements would substantially improve the limits 
on the coupling constants of these hypothetical forces.  In order to 
make such measurements at smaller surface separations, one needs to use 
smoother 
metal coatings.  Atomic layer by layer growth of metal coatings by 
technologies like Molecular Beam Epitaxy might be the best suited. 
Even here, single atomic lattice steps of size $\pm$0.5\,nm are 
unavoidable. So the smallest separation distance that can possibly be 
achieved is on the order of (1-2)\,nm.  

In this regard, recently, a template stripped method  for the 
growth of smooth gold surfaces for the use in Casimir force measurements 
has been reported \cite{s6.57}.  
As a result an rms roughness of 0.4\,nm with 
pits of height (3-4)\,nm was achieved.  Here a separation distance of 
20\,nm is estimated for the contact of the two surfaces and a rms
deviation of 1\% of the Casimir force at the closest separation
distance was reported.  
However in this particular case, 
a hydrocarbon coating was necessary on top of the gold which might 
complicate the theoretical analysis and  
the independent measurement of the residual electrostatic force.   Also 
a substantial 
deformation of the gold coating appears to have prevented an independent 
and exact determination of surface separation \cite{s6.57}. Probably both 
the hydrocarbon layer and the deformation can be eliminated in future 
experiments leading to surface separations of (5-10)\,nm on contact of the 
two surfaces.  

Given the implications of the Casimir force measurements for the detection 
of new forces in the submillimeter distance scale, it is increasingly 
important that consistent values of the experimental precision be 
measured.  It should be noted that such a precision can only be 
attempted if an independent measurement of all the parameters 
particularly the surface separation on contact of the two surfaces 
is performed. All ambiguities such as thin protective 
coatings  should preferably be avoided.  Also an independent 
measurement of the systematics such as the electrostatic force 
between the two surfaces is also necessary.  A value for the 
effect of the roughness of the surface needs to be given in 
order to estimate the roughness correction. If all the above are 
indeed provided then one can devise methods to unambiguously measure 
the deviation from the theoretical values of the Casimir force.   The 
reasonable method is to report at each and every separation distance, the 
deviation of the average value of the force from the theoretical value
at appropriate confidence level and the absolute error of the
experimental measurement. 
One also needs to measure the errors introduced due 
to the uncertainties in the measurement of the surfaces separation.  
  
The second case of extending Casimir force measurements 
beyond 1000\,nm is also vitally important in order to measure the finite 
temperature corrections to the Casimir force.   The temperature 
corrections make a significant contribution to the Casimir 
force only for separation exceeding a micrometer.  Such measurements are 
needed given the controversies surrounding the theoretical 
treatments of the Casimir force 
between real metals (explained in Sec.5.4.2).   
It appears that techniques similar to the AFM would be best suited 
to increase the sensitivity. In the case of the AFM, the authors propose to
increase the 
sensitivity by (a) lithographic fabrication of cantilevers with large 
radius of curvature and (b) interferometric detection of cantilever 
deflection.  Lower temperatures can also be used to reduce the thermal 
noise. This would however also reduce the temperature corrections. 
In addition a dynamic measurement might lead to 
substantial increase in the sensitivity.  With regard to the boundary 
dependences, only one demonstration experiment has so far been done. 
Other boundaries, such as for example with cubical and spherical 
cavities might be feasible with the sensitivity available with the 
AFM technique. Also the influence of small material effects such as 
the polarization dependence of the material properties on the Casimir 
force might also be measurable in the future.

\setcounter{equation}{0}
\section{Constraints for non-Newtonian gravity from the Casimir effect}
\label{sec7}

According to the predictions of unified gauge theories, supersymmetry,
supergravity, and string theory, there would exist a number of
light and massless elementary particles (for example, the axion, scalar
axion, graviphoton, dilaton, arion, and others \cite{s7-1}).
The exchange of such particles between two atoms gives rise to an
interatomic force described by a Yukawa or power-law effective
potentials. The interaction range of this force is to be considered
from 1\,{\AA} to cosmic scales. Because of this, it is called a
``long-range force'' (in comparison with the nuclear size).

The long-range hypothetical force, or ``fifth force'' as it is often 
referred to \cite{s7-2}, may be considered as some specific correction to the
Newtonian gravitational interaction \cite{s7-3}. That is why the
experimental constraints for this force are known also as constraints 
for non-Newtonian gravity \cite{s7-4}. Constraints for the constants
of hypothetical long-range interactions, or on non-Newtonian gravity
are obtainable from Galileo-, E\"{o}tv{o}s-, and Cavendish-type
experiments. The gravitational experiments lead to rather strong
constraints over a distance range 
$10^{-2}\,\mbox{m}<\lambda <10^6\,$km \cite{s7-5}.
In submillimeter range, however, no constraints on hypothetical
long-range interactions follow from the gravitational experiments, and
the Newtonian law is not experimentally confirmed in this range.

The pioneering studies in applying the Casimir force measurements
to the problem of long-range interactions were made in 
\cite{s7-6,s7-7,s7-7a,s7-8,s7-8a}.
There it was shown that the Casimir effect leads to the strongest
constraints on the constants of Yukawa-type interactions with a range of
action of $10^{-8}\,\mbox{m}<\lambda <10^{-4}\,$m (see also
\cite{s7-9}). This means that the Casimir effect becomes a new
non-accelerator test for the search of hypothetical forces and
associated light and massless elementary particles. Tests of this type
take on great significance in the light of the exciting new ideas that
the gravitational and gauge interactions may become unified at the weak
scale (see, e.g., \cite{s7-10}). As a consequence, there should exist
extra spatial dimensions compactified at a relatively large scale of
$10^{-3}\,$m or less. Also, the Newtonian gravitational law acquires 
Yukawa-type corrections in the submillimeter range \cite{s7-10a,s7-10b}
like those predicted earlier from other considerations. These
corrections can be constrained and even discovered in experiments
on precision measurements of the Casimir force.

Below we demonstrate the application of the Casimir force measurements 
for obtaining stronger constraints on the parameters of hypothetical
long-range interactions and light elementary particles starting from 
historical experiments and finishing with the most modern ones.

\subsection{Constraints from the experiments with dielectric test bodies}
\label{sec7.1}

As was noted in Sec.6.2, the Casimir force measurements between the
dielectric disk and spherical lens were not very precise, and some
important factors were not taken into account. Although, 
the final accuracy of the force measurement was not calculated on a solid
basis in those experiments, it may be estimated liberally as
$\delta\sim 10\%$ in the separation distance range 
$0.1\,\mu\mbox{m}<a<1\,\mu$m (in this value of $\delta$ the factor of 
about 2--5 is implied which is not important for the discussion below).

The Casimir force acting between a disk (plate) and a lens of curvature
radius $R$ is given by Eq.~(\ref{sIV3.13a}), where $\varepsilon_{02}$
is the dielectric constant of a lens and a plate material. The effective
gravitational interaction between two atoms including an 
additional Yukawa-type term is given by
\beq
V(r_{12})=
-\frac{G M_1 M_2}{r_{12}}\left(1+\alpha_G e^{-r_{12}/\lambda}\right).
\label{7.1}
\eeq
\noindent
Here $M_{1,2}$ are the masses of the atoms, 
$r_{12}$ is the distance between them, $G$ is the Newtonian gravitational
constant, $\alpha_G$ is a dimensionless constant of hypothetical
interaction, $\lambda$ is the interaction range. In the case that
the Yukawa-type interaction is mediated by a light particle of mass
$m$ the interaction range is given by the Compton wave length of this
particle, so that
$\lambda=\hbar /(mc)$.

Let the thickness of the dielectric plate be $D$ and of the lens $H$. 
It is easily seen \cite{39}
that the gravitational force acting between a lens and a plate is several 
orders smaller than the Casimir force and can be neglected
(see also Sec.7.3).
The hypothetical force can be computed in the following way. We first find
the interaction potential between the whole plate and one atom of the lens 
located at a height $l$ above the plate center
\beq
v(l)=-2\pi GM_1M_2\alpha_G N\lambda^2e^{-l/\lambda}
\left(1- e^{-D/\lambda}\right),
\label{7.2}
\eeq
\noindent
where $N$ is the number of atoms per unit plate and lens volume, and
$a\leq l\leq a+H$. The density of atoms  in a thin horizontal
section of the lens at a height $l\geq a$ is given by
\beq
\sigma(l)=\pi N\left[2R(l-a)-(l-a)^2\right].
\label{7.3}
\eeq
\noindent
The interaction potential between a lens and a plate is found by integration 
of Eq.~(\ref{7.2}) weighted with (\ref{7.3}) in the limits $a$ and
$a+H$. Then the hypothetical force is obtained by differentiating
with respect to $a$. The result is
\bes
&&
F^{hyp}(a)=-4\pi^2G\rho^2\lambda^3R\alpha_G\left(1- e^{-D/\lambda}\right)
e^{-a/\lambda}
\label{7.4}\\
&&\phantom{aaaaa}
\times\left[1-\frac{\lambda}{R}+e^{-H/\lambda}
\left(\frac{H}{R}-1+\frac{\lambda}{R}+\frac{H^2}{2R\lambda}-
\frac{H}{\lambda}\right)\right],
\nonumber
\ees
\noindent
where $\rho$ is the density of plate and sphere material.
In the above calculations the inequalities $L,R,H,D\gg a$ were based
on the experimental configuration under consideration.
For $\lambda$ belonging to a submillimeter range it is also valid 
$D,H,R\gg\lambda$ so that Eq.~(\ref{7.4}) can be substantially 
simplified
\beq
F^{hyp}(a)=-4\pi^2G\rho^2\lambda^3R\alpha_Ge^{-a/\lambda}.
\label{7.5}
\eeq

The constraints on the constants of the hypothetical long-range
interaction $\alpha_G$ and $\lambda$ were obtained from the condition
\cite{s7-7,s7-8}
\beq
\left|F^{hyp}(a)\right|<\frac{\delta}{100\%}\left| F_{dl}(a)\right|,
\label{7.6}
\eeq
\noindent
where the Casimir force $F_{dl}(a)$ is given by Eq.~(\ref{sIV3.13a}).
This condition has the meaning that the hypothetical force was not
observed within the limits of experimental accuracy. The constraints
following from the inequality (\ref{7.6}) are shown by the curve 2
in Fig.~\ref{fig7.1} drown in a logarithmic scale.
\begin{figure}[ht]
\epsfxsize=17cm\centerline{\epsffile{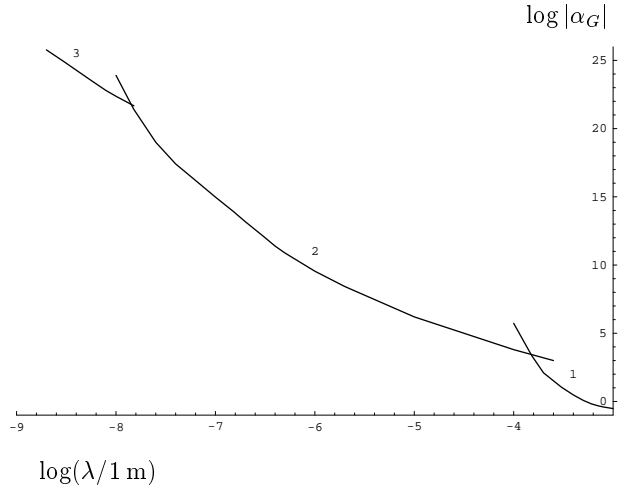} }
\vspace*{-13.5cm}
\caption{\label{fig7.1}
Constraints on the constants of Yukawa-type hypothetical
interaction following from the measurement of the Casimir (curve 2)
and van der Waals (curve 3) forces. 
Curve 1 represents the constraints following  from
one of the Cavendish-type experiments.
}
\end{figure}
The regions below the curves in this figure are permitted by the results of
force measurements, and those above the curves are prohibited. The
curve 2 was the strongest result restricting the value of hypothetical
force in the range
$10^{-8}\,\mbox{m}<\lambda <10^{-4}\,$m 
till the new measurements of the Casimir force between metallic surfaces
which were started in 1997. For slightly larger $\lambda$ the best 
constraints on
$\alpha_G,\lambda$ follow from the Cavendish-type experiment of
Ref.~\cite{s7-11} (see curve 1 in Fig.~\ref{fig7.1}). Note that in
Refs.~\cite{s6.15,s7-12,s7-13} the constraints were obtained based on the
search for displacements induced on a micromechanical resonator in
the presence of a dynamic gravitational field (see Sec.6.2.5).
For the moment
these constraints are less stringent than those from the Casimir force
measurements between dielectrics and from the Cavendish-type experiment
of \cite{s7-11}.

For $10^{-9}\,\mbox{m}<\lambda <10^{-8}\,$m the best constraints on
$\alpha_G,\lambda$ were found \cite{s7-14} from the measurements of
the van der Waals forces between the dielectric surfaces (lens above
a plate and two crossed cylinders). To do this the same procedure was
used as in the case of the Casimir force. The main difference is
that the van der Waals force of Eq.~(\ref{sIV3.6}) was substituted
into Eq.~(\ref{7.6}) as $F_{dl}$. The obtained constraints are shown
by the curve 3 in Fig.~\ref{fig7.1}).

For extremely short wave lengths 
$10^{-10}\,\mbox{m}<\lambda <10^{-9}\,$m the best constraints on
$\alpha_G,\lambda$ were obtained in \cite{s7-15,s7-15a} by measuring 
the van der Waals force acting between a plate and a tip of an Atomic
Force Microscope made of $Al_2O_3$.

As discussed above, the Yukawa-type corrections to Newtonian
gravitational law are caused by the exchange of light but massive
particles. In the case of massless particles (e.g., two-arion exchange) 
the corrections are of power-type, so that instead of Eq.~(\ref{7.1})
one has
\beq
V(r_{12})=
-\frac{G M_1 M_2}{r_{12}}\left[1+\lambda_n^G 
\left(\frac{r_0}{r_{12}}\right)^{n-1}\right].
\label{7.7}
\eeq
\noindent
Here $\lambda_n^G$  with $n=2,3,...$ are the dimensionless constants,
$r_0=1\,\mbox{F}=10^{-15}\,$m is introduced for the proper
dimensionality of potentials with different $n$ \cite{s7-16}.

Casimir force measurement between dielectrics has also led to rather
strong constraints on $\lambda_n^G$  with $n=2,\,3,\,4$ (see \cite{s7-7}).
At the present time, however, the best constraints for the constants
of power-type hypothetical interactions follow from the Cavendish-
and E\"{o}tv{o}s-type experiments \cite{s7-5,s7-17,s7-17a,s7-18}.

\subsection{Constraints from S.~K.~Lamoreaux experiment}
\label{sec7.2}

In \cite{32} the Casimir force between two metallized surfaces of a flat
disk and a spherical lens was measured with the use of torsion pendulum
(see Sec.6.3). The radius of the disk was $L=1.27\,$cm and its thickness
$D=0.5\,$cm. The final radius of curvature of the lens reported
was $R=12.5\,$cm and its height $H=0.18\,$cm. The separation between them
was varied from $a=0.6\,\mu$m up to $6\,\mu$m. Both bodies were made out
of quartz and covered by a continuous layer of copper with 
$\Delta=0.5\,\mu$m thickness. The surfaces facing each other were
additionally covered with a layer of gold of the same thickness.

The experimental data obtained in \cite{32} were found to be in agreement
with the ideal theoretical result of Eq.~(\ref{sIV3.7}) in the limits
of the absolute error of force measuremets $\Delta F=10^{-11}\,$N
for the distances $1\,\mu\mbox{m}\leq a\leq 6\,\mu$m (note 
that this $\Delta F$ is around 3\% of $F_{dl}^{(0)}$ at $a=1\,\mu$m).
No corrections due to surface roughness, finite conductivity of the
boundary metal or nonzero temperature were recorded. These corrections,
however, may not lie within the limits of the absolute error $\Delta F$ (see
Sec.6.3). The correction due to the finiteness of
the disk diameter $L$ (which is even smaller than the size of the lens)
was shown to be negligible \cite{37,37a,sV2-7}.

The constraints on the constants of hypothetical long-range interactions
following from this experiment \cite{32} can be obtained as follows. First, the
Casimir force acting in the configuration under consideration should be 
computed theoretically. It is the force $F^{R,T,C}(a)$ taking into
account surface roughness, nonzero temperature and finite conductivity in
accordance with Sec.5.4. In the first approximation the different
corrections are additive so that
\beq
F^{R,T,C}(a)=F_{dl}^{(0)}(a)+\Delta_{\Sigma}F_{dl}^{(0)}(a)
+\Delta_{L}F_{dl}^{(0)}(a),
\label{7.8}
\eeq
\noindent
where the total correction to the force due to nonzero temperature,
finite conductivity and short scale roughness 
\beq
\Delta_{\Sigma}F_{dl}^{(0)}(a)\approx \Delta_{T}F_{dl}^{(0)}(a)
+\Delta_{C}F_{dl}^{(0)}(a)+\Delta_{R}F_{dl}^{(0)}(a)
\label{7.9}
\eeq
\noindent
can be estimated theoretically. The quantity $\Delta_{L}F_{dl}^{(0)}(a)$
in (\ref{7.8}) is the correction due to the large scale roughness
(see, e.g., Eq.~(\ref{sV3.56}). It cannot be estimated theoretically
because the actual shape of interacting bodies was not investigated
experimentally in \cite{32}. To a first approximation, however,
for two different separations we can say that 
\beq
\Delta_{L}F_{dl}^{(0)}(a_2)=\frac{1}{k_{21}} 
\Delta_{L}F_{dl}^{(0)}(a_1), \qquad
k_{21}\equiv\left(\frac{a_2}{a_1}\right)^4.
\label{7.10}
\eeq

Next, the hypothetical force acting in the experimental configuration
should be computed. As in Sec.7.1, gravitational force is small
compared to the Casimir force. In view of the layer structure of interacting
bodies the Yukawa potential of their interaction takes the form
\bes
&&
U^{hyp}=-\alpha_G G\sum\limits_{i,j=1}^{3}
\rho_i^{\prime}\rho_j U_{ij}^{hyp},
\label{7.11}\\
&&
U_{ij}^{hyp}=\int\limits_{V_i^{\prime}}d^3r_1
\int\limits_{V_j}d^3r_2\frac{1}{r_{12}}e^{-r_{12}/\lambda}.
\nonumber
\ees
\noindent
Here $\rho_i^{\prime},\,V_i^{\prime}$ ($i=1,\,2\,3$) are the densities
and volumes of the lens and the covering metallic layers ($\rho_j,\,V_j$
are the same for the disk). In numerical calculations below the values
$\rho_1^{\prime}=2.23\,$g/cm${}^3$,  $\rho_1=2.4\,$g/cm${}^3$,
$\rho_2^{\prime}=\rho_2=8.96\,$g/cm${}^3$,
$\rho_3^{\prime}=\rho_3=19.32\,$g/cm${}^3$ are used. As usual the force
is obtained by $F^{hyp}=-\partial U^{hyp}/\partial a$.

In \cite{37} the hypothetical force was computed analytically in two
limiting cases: $\lambda <H$ and $\lambda\gg R$.
For example, if $\lambda$ is not only less than $H$ but also
$\lambda <a$ the particularly simple result is obtained \cite{37,37a}
\bes
&&
F^{hyp}(a)=-4\pi^2\lambda^3\alpha_GGRe^{-a/\lambda}
\left(\rho_1^{\prime}e^{-2\Delta/\lambda}+\rho_2e^{-\Delta/\lambda}
+\rho_3\right)
\nonumber\\
&&\phantom{aaaaaaaaa}
\times\left(\rho_1e^{-2\Delta/\lambda}+\rho_2e^{-\Delta/\lambda}
+\rho_3\right).
\label{7.12}
\ees
\noindent
In the intermediate range between $\lambda <H$ and $\lambda\gg R$
the integration in Eq.~(\ref{7.11}) was performed numerically \cite{37}.

By virtue of the fact that the values of different corrections to the
Casimir force surpass the absolute error of force measurements
$\Delta F$, their partial cancellation appears very likely. In this situation,
bearing in mind that the expression for $F_{dl}^{(0)}$ was confirmed with
an absolute error $\Delta F$, the constraints on the parameters
of hypothetical interaction $\alpha_G,\,\lambda$ can be calculated from
the inequality \cite{37}
\beq
\left|F^{R,T,C}(a)+F^{hyp}(a)-F_{dl}^{(0)}(a)\right|\leq\Delta F.
\label{7.13}
\eeq
\noindent
Here $F^{R,T,C}(a)$ is the theoretical Casimir force value with account
of all corrections given by Eq.~(\ref{7.8}). Substituting (\ref{7.8})
into (\ref{7.13}) one obtains
\beq
\left|F^{hyp}(a)+\Delta_{\Sigma}F_{dl}^{(0)}(a)
+\Delta_{L}F_{dl}^{(0)}(a)\right|\leq\Delta F.
\label{7.14}
\eeq

According to the above results, the value of the hypothetical force is
proportional to the interaction constant 
$F^{hyp}(a_i)=\alpha_G K_{\lambda}(a_i)$. Considering Eq.~(\ref{7.14})
for two different values of distance $a_1,\,a_2$ with account of
Eq.~(\ref{7.10}) and excluding the unknown quantity 
$\Delta_{L}F_{dl}^{(0)}(a_1)$ the desired constraints on $\alpha_G$
are obtained
\bes
&&
-(k_{21}+1)\Delta F-\Delta_{\Sigma}(a_1)+k_{21}\Delta_{\Sigma}(a_2)
\leq \alpha_G\left[K_{\lambda}(a_1)-k_{21}K_{\lambda}(a_2)\right],
\nonumber\\
&&
\alpha_G\left[K_{\lambda}(a_1)-k_{21}K_{\lambda}(a_2)\right]\leq
(k_{21}+1)\Delta F-\Delta_{\Sigma}(a_1)+k_{21}\Delta_{\Sigma}(a_2).
\label{7.15}
\ees

The specific values of $a_1,\,a_2$ in Eqs.~(\ref{7.15}) were chosen
in the interval $1\,\mu\mbox{m}\leq a\leq 6\,\mu$m in order to obtain
the strongest constraints on $\alpha_G,\,\lambda$. For the upper limit
of the distance interval ($a\approx 6\,\mu$m), the Casimir force
$F^{T}(a)$ from Eq.~(\ref{sV13.4}), i.e. together with the temperature
correction should be considered as the force under measurement. All
corrections to it are much smaller than $\Delta F=10^{-11}\,$N.
Thus for such values of $a$, the constraints on the hypothetical
interaction may be obtained from the
simplified inequality (instead of from Eq.~(\ref{7.14}))
\beq
\left|F^{hyp}(a)\right|= \left|\alpha_G K_{\lambda}(a)\right|
\leq\Delta F.
\label{7.16}
\eeq

The results of numerical computations with use of Eq.~(\ref{7.15}) are
shown in Fig.~\ref{fig7.2} by the curves 4,a ($\alpha_G>0$) and
4,b ($\alpha_G<0$). In the same figure the curves 1,\,2,\,3
show the previously known constraints following from the Cavendish-type
experiments and Casimir (van der Waals) force measurements between
dielectrics (Fig.~\ref{fig7.1}).
\begin{figure}[ht]
\epsfxsize=10cm\centerline{\epsffile{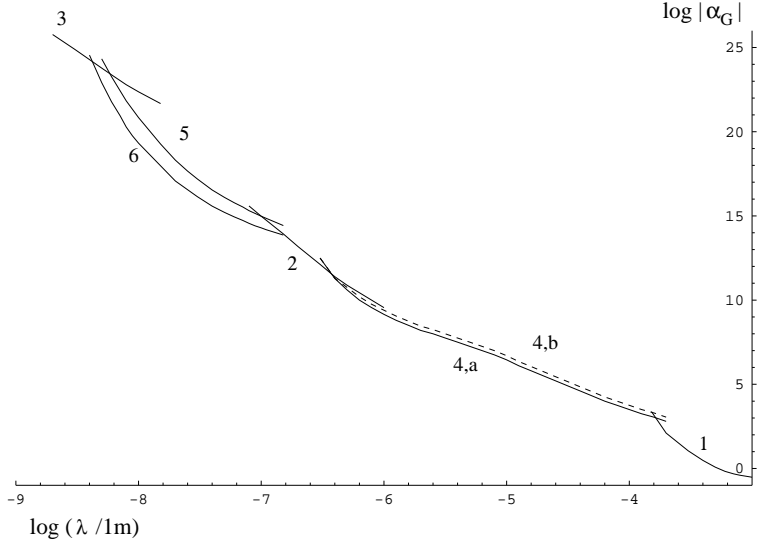} }
\caption{\label{fig7.2}
Constraints for the  Yukawa-type hypothetical interaction  
following from S.~K.~Lamoreaux experiment are shown by
curves 4,a ($\alpha_G>0$) and  4,b ($\alpha_G<0$). Constraints
following from the experiments by U.~Mohideen et al are shown by
curves 5,\,6. 
Curves 1--3 are the same as in the previous figure.
}
\end{figure}
For different $\lambda$ the values of $a_1=1\,\mu$m and $a_2=(1.5-3)\,\mu$m
were used. The complicated character of curves 4,a,b (nonmonotonic
behaviour of their first derivatives) is explained by the 
multilayered structure
of the test bodies. 
For $\lambda>10^{-5}\,$m the metallic layers do not contribute
much to the total value of the force which is determined mostly by quartz.
But for $\lambda<10^{-5}\,$m the contribution of metal layers is 
the predominant one.

It is seen that the new constarints following from \cite{32} are the best
ones within  a wide range
$2.2\times 10^{-7}\,\mbox{m}<\lambda <1.6\times 10^{-4}\,$m \cite{37}
(a slightly different result was obtained in Ref.~\cite{s7-19} where,
however, the corrections to the ideal Casimir force $F_{dl}^{(0)}$ were
not taken into account). They surpass the results obtained from the
Casimir force measurements between dielectrics up to a factor of 30.
For $\lambda <2.2\times 10^{-7}\,$m the latter lead to better constraints
than the new ones. This is caused by the small value of the Casimir force
between dielectrics compared to the case of metals and also by the fact that 
the force was measured for smaller values of $a$.

The constraints for power-type hypothetical interactions of Eq.~(\ref{7.7})
following from the experiment \cite{32} were calculated in \cite{s7-19a}.
They turned out to be weaker than the best ones obtained from the
Cavendish- and E\"{o}tvos-type experiments.

In Ref.~\cite{37} the possible strengthening by up to $10^4$ times 
of the obtained constraints
is proposed due to some
modifications of the experimental setup. The prospective constraints
would give the possibility to restrict the masses of the graviphoton
and dilaton. They also fall within the range $\alpha_G\sim 1$ predicted by
the theories with the weak unification scale. Because of this the obtaining 
of such strong constraints is gaining in importance.

\subsection{Constraints following from the 
atomic force microscope measurements of the Casimir force}
\label{sec7.3}

The results of the Casimir force measurements by means of an atomic force
microscope are presented in Sec.6.4. They were shown to be in good
agreement with the theory taking into account the finite conductivity
and roughness corrections. Temperature corrections are not important in
the interaction range
$0.1\,\mu\mbox{m}<a<0.9\,\mu$m as in \cite{33} or
$0.1\,\mu\mbox{m}<a<0.5\,\mu$m  \cite{35}.
In both experiments the test bodies (sapphire disk and polystyrene
sphere) were covered by the aluminium layer of 300\,nm thickness
\cite{33} (or 250\,nm thickness \cite{35}) and $Au/Pd$ layer of 20\,nm
or 7.9\,nm thickness respectively. This layer was demonstrated to be
transparent for electromagnetic oscillations of the characteristic frequency.
The absolute error of force measurements in \cite{33} was
$\Delta F=2\times 10^{-12}\,$N and almost two times lower in \cite{35}
due to use of vibration isolation, lower systematic errors, and independent
measurement of surface separation.

The confident experimental confirmation of the complete Casimir force
including corrections has made it possible to find the constraints on
the parameters $\alpha_G,\,\lambda$ of hypothetical interaction from the
inequality
\beq
\left|F_R^{hyp}(a)\right|\leq\Delta F.
\label{7.17}
\eeq
\noindent
Index $R$ indicates that the hypothetical force should be calculated 
with account of surface roughness. Note that here, as opposed to
Eq.~(\ref{7.15}), all the corrections are included in the Casimir force. 
By this reason the constraints on $|\alpha_G|$ rather
than on $\alpha_G$ are obtained.

The gravitational and hypothetical forces described by the potential
(\ref{7.1}) can be calculated as follows (we substitute the numerical
parameters of the improved experiment \cite{35}). The diameter
$2R=201.7\,\mu$m 
of the sphere is much smaller than the diameter of the disk 
$2L=1\,$cm. Because of this, each atom of the sphere can be considered as 
if it
would be placed above the center of the disk. 
Let an atom of the sphere with mass $M_1$
be at  height $l\ll L$ above the center of the disk.  
The vertical component of the Newtonian
gravitational force acting between this atom and the disk can be calculated 
as
\bea
&&
f_{N,z}(l)=\frac{\partial}{\partial l}
\left[GM_1 \rho 2\pi
\int\limits_{0}^{L}r\,dr
\int\limits_{l}^{l+D}
\frac{dz}{\sqrt{r^2+z^2}}\right]
\nonumber\\
&&\phantom{aaa}
\approx -2\pi GM_1\rho D\left[1-\frac{D+2l}{2L}\right],
\label{7.18}
\eea
\noindent
where $\rho=4.0\times 10^3\,$kg/m${}^3$ is the sapphire density,
$D=1\,$mm is the thickness of sapphire disk, and only the first-order terms
in $D/L$ and $l/L$ are retained. 

Integrating Eq.~(\ref{7.18}) over the volume of the sphere one obtains the
Newtonian gravitational force acting between a sphere and a disk
\beq
F_{N,z}\approx -\frac{8}{3}\pi^2 G\rho\rho^{\prime}DR^3
\left(1-\frac{D}{2L}-\frac{R}{L}\right),
\label{7.19}
\eeq
\noindent
where $\rho^{\prime}=1.06\times 10^3\,$kg/m${}^3$ is the polystyrene
density. Note that this force does not depend on distance $a$ between
the disk and the sphere because of 
$a=(0.1-0.5)\,\mu$m$\ll R$.

Substituting the values of parameters given above into (\ref{7.19}) we
arrive to the value $F_{N,z}\approx 6.7\times 10^{-18}\,$N which 
is much smaller than
$\Delta F$.
The value of Newtonian
gravitational force between the test bodies remains  nearly unchanged
when taking into account 
the contributions of $Al$ and $Au/Pd$ layers on the sphere and the disk. 
The corresponding result can be simply obtained by the
combination of several expressions of type (\ref{7.19}). The additions to
the force due to layers are suppressed by the small multiples
$\Delta_i/D$ and $\Delta_i/R$. That is why the Newtonian gravitational
force is negligible in the Casimir force measurement by means
of atomic force microscope
(note that for the configuration of two plane parallel plates gravitational
force can play more important role \cite{s7-20}).

Now we consider hypothetical force acting between a disk and a sphere
due to the Yukawa-type term from Eq.~(\ref{7.1}). It can be calculated
using the same procedure which was already applied in Sec.7.1
for the configuration of a lens above a disk and here in the case of
gravitational force. The only complication is that the contribution
of the two covering layers of thickness $\Delta_1$ ($Al$) and $\Delta_2$
($Au/Pd$) should be taken into account. Under the conditions
$a,\,\lambda\ll R$ the result is \cite{38,39}
\bea
&&F^{hyp}(a)=-4\pi^2G \alpha_G 
\lambda^3
e^{-\frac{a}{\lambda}}\,R
\left[\rho_2-(\rho_2-\rho_1)
e^{-\frac{\Delta_2}{\lambda}}
\right.
\nonumber\\
&&\phantom{aaaaa}\left.
-(\rho_1-\rho)
e^{-\frac{\Delta_2+\Delta_1}{\lambda}}\right]
\left[\rho_2 -
(\rho_2-\rho_1)
e^{-\frac{\Delta_2}{\lambda}}\right.
\nonumber \\
&&\phantom{aaaaa}
-\left.
(\rho_1-\rho^{\prime})
e^{-\frac{\Delta_2+\Delta_1}{\lambda}}
\right].
\label{7.20}
\eea
\noindent
Here $\rho_1=2.7\times 10^3\,$kg/m${}^3$ is the density of $Al$ and
$\rho_2=16.2\times 10^3\,$kg/m${}^3$ is the density of 60\%$Au$/40\%$Pd$.

As it was shown in \cite{38}, the surface distortions can significantly
influence the value of hypothetical force in the nanometer range. 
In \cite{35}  smoother metal coatings than in \cite{33} were used.
The roughness of the metal surface was measured with the atomic force
microscope. The major distortions both on the disk and on the sphere
can be modeled by parallelepipeds of two heights
$h_1=14\,$nm (covering the fraction $v_1=0.05$ of the surface) and
$h_2=7\,$nm (which cover the fraction 
$v_1=0.11$  of the surface). The surface 
between these distortions is covered by a stochastic roughness of
height $h_0=2\,$nm ($v_0=0.84$ fraction of the surface). It consists
of small crystals which form a homogeneous background of the average
height $h_0/2$.

The detailed calculation of roughness corrections to the Casimir force
for this kind of roughness was performed in Sec.6.4.1. Using similar 
methods, a result in perfect analogy with Eq.~(\ref{6.15a}) is
obtained
\bea
&&
F_R^{hyp}(a)=
\sum\limits_{i=1}^{6}w_iF^{hyp}(a_i)\equiv
v_1^2F^{hyp}(a-2A)
\nonumber\\
&&\phantom{a}
+2v_1v_2F^{hyp}\left(a-A(1+\beta_1)\right)
\label{7.21}\\
&&\phantom{a}
+
2v_2v_0F^{hyp}\left(a-A(\beta_1-\beta_2)\right)+
v_0^2F^{hyp}(a+2A\beta_2)
\nonumber\\
&&\phantom{a}
+v_2^2F^{hyp}(a-2A\beta_1)
+2v_1v_0F^{hyp}\left(a-A(1-\beta_2)\right).
\nonumber
\eea
\noindent
Here the value of amplitude is defined relative to the 
zero distortion level which is given by
$A=11.69\,$nm, $\beta_1=0.4012$, $\beta_2=0.1121$.

According to the results of \cite{35} no hypothetical force was observed.
The constraints on it are obtained by the substitution of Eq.~(\ref{7.21})
into Eq.~(\ref{7.17}).
The strongest constraints on $\alpha_G$ follow for
the smallest possible value of $a$. As was told above there is 
$a_{\min}=100\,$nm in the Casimir force measurement for the
experiment under consideration. 
This distance is between $Al$ layers
because the $Au/Pd$ layers of $\Delta_2=7.9\,$nm thickness were shown
to be transparent for the frequencies of order $c/a$.
Considering the Yukawa-type hypothetical interaction
this means that $a_{\min}^{Yu}=100\,\mbox{nm}-2\Delta_2=84.2\,$nm.
Substituting this value into Eqs.~(\ref{7.17}), (\ref{7.20}),
(\ref{7.21}) one obtains constraints on $\alpha_G$ 
for different $\lambda$ \cite{39}. The computational results
are represented by curve 5 of Fig.~\ref{fig7.2}. As follows from
Fig.~\ref{fig7.2}, the new constraints turned out to be up to 560 times
stronger than the constraints obtained from the Casimir and van der Waals
force measurements between dielectrics (curves 2,\,3 in
Figs.~\ref{fig7.1},\,\ref{fig7.2}). The strengthening takes place within the 
interaction range 
$5.9\times 10^{-9}\,\mbox{m}\leq\lambda\leq 1.15\times 10^{-7}\,$m.
The largest strengthening takes place for $\lambda=$(10--15)\,nm.
Note that the same calculations using the results of the experiment \cite{33}
lead to approximately four times less strong constraints.

Recently one more measurement of the Casimir force was performed using
the atomic force microscope \cite{36}. The test bodies (sphere and a disk)
were coated with gold instead of aluminium which removes some difficulties
connected with the additional thin $Au/Pd$ layers used in the
previous measurements to reduce the effect of oxidation processes 
on $Al$ surfaces. The polystyrene sphere used which was coated by gold layer
was of diameter $2R=191.3\,\mu$m and a sapphire disk had a diameter
$2L=1\,$cm, and a thickness $D=1\,$mm. The thickness of the gold
coating on both test bodies was $\Delta=86.6\,$nm. This can be considered 
as infinitely thick for the case of the Casimir force measurements.
The root mean square roughness amplitude of the gold surfaces was
decreased until 1\,nm which makes roughness corrections negligibly
small. The measurements were performed at smaller separations, i.e.
$62\,\mbox{nm}\leq a\leq 350\,$nm. The absolute error of force measurements
was, however, $\Delta F=3.8\times 10^{-12}\,$N, i.e., a bit larger
than in the previous experiments. The reason is the thinner gold
coating used in \cite{36} which led to poor thermal conductivity of
the cantilever. At smaller separations of about 65\,nm this error
is less than 1\% of the measured Casimir force.

The gravitational force given by Eq.~(\ref{7.19}) is once more negligible.
Even if the sphere and disk were made of the vacuo-distilled gold 
with $\rho=\rho^{\prime}=\rho_1=18.88\times10^3\,$kg/m${}^3$ one arrives from
(\ref{7.19}) at the negligibly small value of
$F_{N,z}\approx 6\times 10^{-16}\,\mbox{N}\ll\Delta F$ for the gravitational 
force.

The Yukawa-type addition to the Newtonian gravity which is due to the
second term of the potential (\ref{7.1}) should be calculated including the 
effects of the  
true materials of the test bodies.
It can be easily obtained using the same procedure which was applied 
above. The result is
\bes
&&F^{hyp}(a)=-4\pi^2G\alpha_G\lambda^3e^{-a/\lambda}R
\nonumber \\
&&\phantom{aaa}\times\left[\rho_1-(\rho_1-\rho)e^{-\Delta/\lambda}
\right]\, \left[\rho_1-(\rho_1-\rho^{\prime})e^{-\Delta/\lambda}
\right].
\label{7.22}
\ees

According to \cite{36} the theoretical value of the Casimir force was
confirmed within the limits of $\Delta F=3.8\times 10^{-12}\,$N and
no hypothetical force was observed. In such a situation, the
constraints on $\alpha_G$ can be obtained from the inequality
(\ref{7.16}). The strongest constraints follow for the smallest possible
values of $a\approx 65\,$nm. The computational results are presented
in Fig.~\ref{fig7.2} by curve 6 \cite{s7-21}.

As is seen from Fig.~\ref{fig7.2}, the Casimir force measurement 
between the gold
surfaces by means of an atomic force microscope gives the possibility
to strengthen the previously known constraints (curve 5) up to 19 times
within a range 
$4.3\times 10^{-9}\,\mbox{m}\leq\lambda\leq 1.5\times 10^{-7}\,$m.
The largest strengthening takes place for $\lambda=$(5-10)\,nm.
Comparing the constraints obtained from the Casimir and van der
Waals force measurements between dielectrics (curves 2 and 3) the
strengthening up to 4500 was achieved with the Casimir force measurement
between gold surfaces using the atomic force microscope. 
Note that there still persists a gap between the new constraints
of the curves 4 and 6 where the old results 
of curve 2 obtained from dielectric
surfaces are valid.

The Casimir force measurement between two crossed cylinders with gold
surfaces (\cite{s6.57}, see Sec.~6.6) also gives the possibility
to strengthen constraints on the Yukawa-type interaction. The
maximal strengthening in 300 times is achieved from this
experiment at $\lambda=4.26\,$nm \cite{306a}.

As indicated above, there is abundant evidence that the gravitational
interaction at small distances undergoes deviations from the Newtonian law.
These deviations can be described by the Yukawa-type potential.
They were predicted in theories with the quantum
gravity scale both of order $10^{18}\,$GeV and $10^3\,$GeV. 
In the latter case the problem of experimental search for such
deviations takes on even greater significance.
The existence of large extra dimensions can radically alter many
concepts of space-time, elementary particle physics, astrophysics and
cosmology. According to Sec.7.2 the improvement of the experiment
of Ref.~\cite{32} by a factor of $10^4$ in the range around 
$\lambda=10^{-4}\,$m gives the possibility to attain the values of
$\alpha_G\sim 1$. As to the experiments using the atomic force
microscope to measure the Casimir force, there remains almost
fourteen orders more needed to achieve these values. Thus, in the
experiments with an atomic force microscope it is desirable not only to
increase the strength of constraints but also to make a shift of the
interaction range to larger $\lambda$. For this purpose the sphere
radius and the space separation to the disk should be increased.
The other experimental schemes are of interest also, for example,
dymanical ones [294,295,308] (note that the gravitational experiments
on the search of hypothetical interactions in the submillimeter range
also suggest some promise [309]).
In any case the Casimir effect offers important advantages as a new test 
for fundamental physical theories. Further strengthening of constraints
on non-Newtonian gravity from the Casimir effect is expected in the
future.

\section{Conclusions and discussion}
\label{sec8}

The foregoing proves that the Casimir effect has become the subject of
diverse studies of general physical interest in a variety of fields.
It is equally interesting and important for Quantum Field Theory,
Condensed Matter Physics, Gravitation, Astrophysics and Cosmology,
Atomic Physics, and Mathematical Physics. Currently the Casimir
effect has been advanced as a new powerful test for hypothetical
long-range interactions, including corrections to Newtonian gravitational
law at small distances, predicted by the unified gauge theories, supersymmetry,
supergravity and string theory. It is also gaining in technological 
importance in vital
applications such as in nanoelectromechanical devices
\cite{nano,tech}.

From the field-theoretical standpoint all the most complicated problems 
arising from the theory of the Casimir effect are already solved 
(see Sections 3 and 4).  There are no more problems (at least fundamentally) 
with singularities.
Their general structure was determined by the combination of two powerful
tools --- heat kernel expansion and zeta-functional regularization.
Calculation procedures of the finite Casimir energies and forces are
settled as long as there is separation of variables. 
Otherwise, 
approximate methods should be applied or numerical computation by
brute force done, which is sensible only for the purposes of some specific
application.

The numerous illustrations of the Casimir effect in various configurations
are given in Sec.4. Here both the flat and curved boundaries are
considered and the important progress with the case of a sphere is illustrated.
Much importance is given to the additive methods and proximity forces
which provide us with simple alternative possibilities to calculate the
value of the Casimir force with high accuracy. In Sec.4 the new
developments in the dynamical Casimir effect are briefly discussed and
the results of calculations of the radiative corrections to the Casimir
force are presented. Special attention is paid to the Casimir effect
in spaces with non-Euclidean topology where cosmological models,
topological defects and Kaluza-Klein theories are considered. The subject
which is beyond the scope of our report is the atomic Casimir effect.
As was shown in Secs.3.5,\,4.5 the presence of cavity walls leads to the
modifications of the propagators. If the atom is situated between two
mirrors there arise specific modifications in the spontaneous emission
rates \cite{Morawitz,MiKn,Barton,AFT}. 
The role of Casimir-type retardation effects in
the atomic spectra is the subject of extensive study 
(see, e.g., \cite{KeSp,BaSp,FeSuAu,TREAS,MiSp} and references therein).

The important new developments are in the study of the Casimir effect for real
media, i.e. with account of nonzero temperature, finite conductivity of
the boundary metal and surface roughness. Not only to each of these
influential factors by itself contribute 
but also their combined effect should be
taken into account in order to make possible the reliable comparison of 
theory and experiment. The theoretical results obtained here during last
few years are presented in Sec.5. It turns out that the value of the Casimir
force depends crucially on the electrical and mechanical properties
of the boundary material. It was also discussed that the combination of
such factors as nonzero temperature and finite conductivity presents
a difficult theoretical problem, so that one should use extreme caution in
calculating their combined effect on the basis of the general Lifshitz
theory.

As was already noted in Introduction the most striking development
of the last years on this subject is the precision measurement of the
Casimir force between metallic surfaces. In Sec.6 the review of both older
and recent experiments on measuring the Casimir force is given and along with 
a comparison to modern theoretical results taking into account of
all corrections arising in real media. Excellent agreement between the
experiment by means of atomic force microscope
and theory at smallest separations is demonstrated.

The measure of the agreement between theoretical and experimental Casimir
force gives the possibility to obtain stronger constraints for the
corrections to Newtonian gravitational law and other hypothetical
long-range interactions predicted by the modern theories of fundamental
interactions. These results are reviewed in Sec.7 and represent
the Casimir effect as a new test for fundamental interactions.

In spite of quick progress in both theory and experiment during last
few years, the Casimir effect is on the verge of potentially 
exciting developments. Although the fundamental
theoretical foundations are already laid, much should be done in
the development of the approximate methods with controlled accuracy.
In future the investigation of the real media will involve spatial
dispersion which is needed for applications of the obtained results
to thin films. Some actual interest is connected with non-smooth
background (for example, when the derivative of a metric has a jump).
Also, as was noticed in Sec.4.2.3, the problem of the Casimir effect
for the dielectric sphere is still open.

Most future progress in the field is expected in connection with the new
Casimir force measurements. Here the accuracy will be undoubtedly
increased by several orders of magnitude, and the separation range will be
expanded. As a result the first measurement of the nonzero temperature
Casimir force will be performed very shortly. The increased accuracy
and fabrication of an array of many open boxes should give the
possibility to observe the repulsive Casimir force which will
have profound impact in nanotechnology (see Appendix A). 
The other experimental
advance to be expected in the near future is the observation
of the dynamical Casimir effect.

It is anticipated that the strength of the constraints on the constants
of hypothetical long-range interactions obtained by means of the
Casimir effect will be increased at least by four orders of magnitude
in the next two to three years. This would mean that the Casimir effect
gives the possibility to check Newtonian gravitational law in
the submillimeter range which was not possible by other methods in the last
300 years. As a result the exceptional predictions concerning the
structure of space-time at short scales should be confirmed or rejected
by the measurement of the Casimir force.

To conclude we would like to emphasize that although more than fifty
years have passed after its discovery, the Casimir effect is
gaining greater and greater importance on the development of modern
physics. In our opinion, at present the Casimir effect is on the
threshold of becoming a tool of exceptional importance both in
fundamental physics and in technological applications.

\section*{Acknowledgements}\label{ackn}

The authors are greatly indebted to Prof.~G.L.~Klimchitskaya for
numerous helpful discussions and collaboration. V.M.M is grateful to
the Institute of Theoretical Physics (Leipzig University) and
Brazilian Center of Physical Research (Rio de Janeiro) for kind
hospitality. His work was partly supported by DFG  (Germany) 
under the reference 436 RUS 17/19/00 and
by FAPERJ under the number E-26/150.867/2000 and CNPq (process
300106/98-0), Brazil.

\renewcommand{\thesubsection}{A.\arabic{subsection}}
\section*{Appendix A. 
Applications of the Casimir force in nanotechnology}
\label{appA}

      The first paper anticipating the dominant role of Casimir forces in
nanoscale devices appeared over 15 years ago [321], 
but was largely ignored,
as then the silicon chip fabrication dimensions were on the order of many
microns.  More recently, given the shrinking device dimensions to
nanometers,  the important role of Casimir forces present in nanoscale
devices is now well recognized [310,311].  
The important role of the Casimir
forces in both the device performance and device fabrication have been
acknowledged.  Very recently, even an actuator based on the Casimir force
has been fabricated using silicon nanofabrication technology[322]. 

\subsection{Casimir force and nanomechanical devices}
\label{A1} 
Most present day
nanomechanical devices are based on thin cantilever beams above a silicon
substrate fabricated by photolithography followed by dry and wet chemical
etching [310,311,322--324].  
Such cantilevers are usually suspended about 100\,nm above
the silicon substrate [323,324].  
The cantilevers move in response to the
Casimir force, applied voltages on the substrate, or in response to
incoming radio-frequency signals [323,324]. 
In the case of radio-frequency transmitters and receivers, a
high Quality factor ($Q$) is necessary for the narrow bandwidth 
operation of
these devices. However, due to the coupling to the substrate and
neighbouring cantilevers through the Casimir force, the vibration 
energy of
the cantilever can be dissipated.   This dissipation of mechanical 
energy
leads to a decrease in the $Q$ and crosstalk with neighboring 
receivers, both
leading to degradation in the signal. The problem will be exacerbated in
dense arrays of high $Q$ transmitters and 
receivers needed for future mobile
communication.  Thus effective incorporation of the Casimir force is
necessary in these devices to optimize their performance.

Recently the first actuator based on the Casimir force was developed
by the researchers at Bell labs. It is a
silicon based device that provides mechanical motion as a
result of controlling vacuum fluctuations [322].  
The device is fabricated by standard nanofabrication techniques such
as photolithography and chemical etching on a silicon substrate.
This device consists of 3.5$\,\mu$m thick, 500$\,\mu$m square heavily
doped polysilicon plate freely suspended on two of its opposite sides
by thin torsional rods as shown in Fig.~29A. The other ends of the
torsional rods are anchored to the substrate via support posts as
shown in Fig.~29B. There is a 2$\,\mu$m gap between the top plate and
the underlying substrate, which is created by etching a $SiO_2$
sacrificial layer.
\begin{figure}[ht]
\centerline{\epsffile{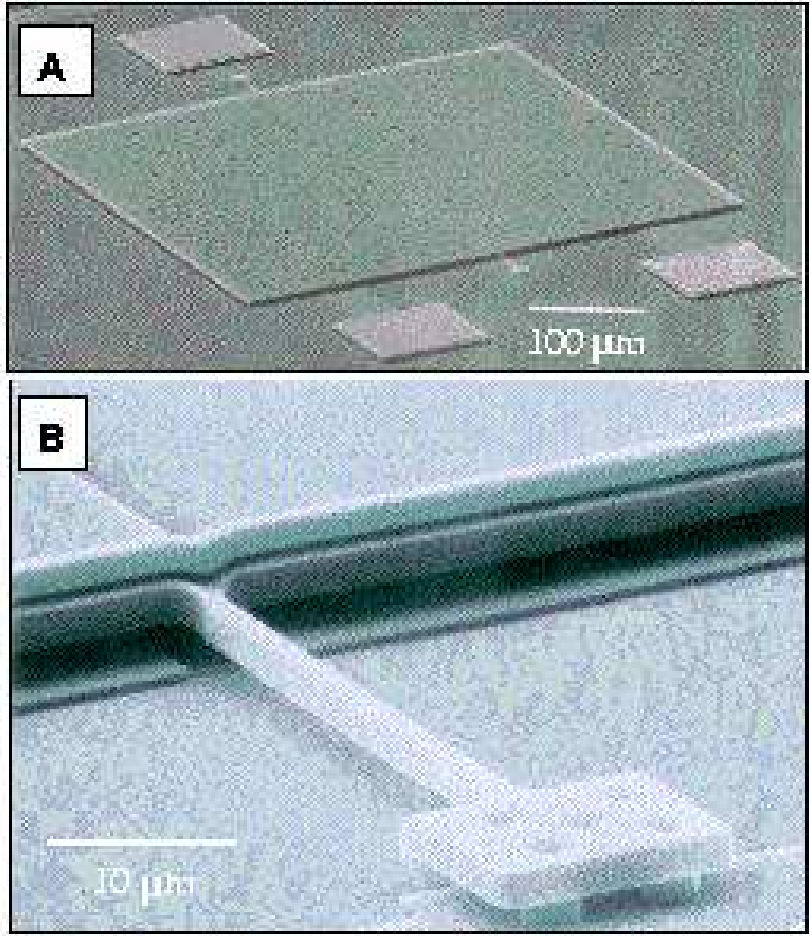} }
\caption{Scanning electron micrographs of (A) the nanofabricated
torsional device and (B) a close-up of one of the torsional rods
anchored to the substrate. Courtesy of Federico Capasso, Bell Labs,
Lucent Technologies.
}
\end{figure}
 To apply the Casimir force, the researchers suspended a gold coated ball
just above one side of the plate.  As the plate was moved closer to the
ball,  the Casimir force acting on the plate, tilted the plate about its
central axis towards the sphere.  Thus the Casimir force led to the
mechanical motion of the nanofabricated silicon plate.  This is the first
case of microelectromechanical 
device which shows actuation by the Casimir force.
         
The measured tilt angle of the plate was calibrated and found to
correspond accurately to the Casimir force.  The experiments were done at
room temperature and at a pressure less than 1\,mTorr.   The roughness
amplitude of the gold films was 30\,nm.  Due to the large roughness, the
closest approach of the two surfaces was 75.7\,nm.  Even though the rms
deviation of the experimental results from the theory was on  the order of
0.5\% of the forces at closest separation, the authors point out that the
large roughness correction and limited understanding of the metal coating,
prevents a better than 1\% accurate comparison to the theory.  However, an
unambiguous mechanical movement of the silicon plate in response to the
Casimir force was demonstrated. 
  
\subsection{Casimir force in nanoscale device fabrication}
\label{A2}

The Casimir force
dominates over other forces at distances of a few nanometers.  Thus movable
components in nanoscale devices fabricated at distances less than 100\,nm
between each other often stick together due to the strong Casimir force.
This process referred to as ``stiction" leads to the collapse of movable
elements to the substrate or the collapse of neighboring components during
nanoscale device operation.  This sometimes leads to permanent adhesion of
the device components [310,311]. 
Thus this phenomena severely restricts the
yield and operation of the devices.  This stiction process is complicated
by capillary forces that are present during fabrication.  These together
lead to poor yield in the microelectromechanical systems
fabrication process. 

              From the above it is clear that the Casimir forces
fundamentally influence the performance and yield of nanodevices.  The
Casimir forces might well set fundamental limits on the performance and
the possible density of devices that can be optimized on a single chip. 
On the grounds of
the above discussion even actuators based entirely on the Casimir
force, or using a combination of the Casimir force and electrostatic forces
will be possible in the near future. Thus a complete understanding of the
material and shape dependences of the Casimir effect will be necessary to
improve the yield and performance of the nanodevices.

\end{document}